\documentclass[12pt]{article}
\usepackage{jheppub}

\pdfoutput=1

\usepackage[table]{xcolor}
\usepackage{amsmath,bbm,array,amsfonts,graphicx,wrapfig,lscape,float,mathtools,multirow,longtable}
\usepackage{stackrel}
\usepackage[all]{xy}
\usepackage{subcaption}

\newcommand{\be}{\begin{equation}}
\newcommand{\ee}{\end{equation}}
\newcommand{\beq}{\begin{equation}}
\newcommand{\beql}[1]{\begin{equation}\label{#1}}
\newcommand{\eeq}{\end{equation}}
\newcommand{\ba}{\begin{array}}
\newcommand{\ea}{\end{array}}
\newcommand{\bea}{\begin{eqnarray}}
\newcommand{\beal}[1]{\begin{eqnarray}\label{#1}}
\newcommand{\eea}{\end{eqnarray}}
\newcommand{\ben}{\begin{enumerate}}
\newcommand{\een}{\end{enumerate}}
\newcommand{\bean}{\begin{eqnarray*}}
\newcommand{\eean}{\end{eqnarray*}}

\newcommand{\sref}[1]{\S\ref{#1}}

\newcommand{\fref}[1]{Figure \ref{#1}}
\newcommand{\btab}[1]{\begin{tabular}{#1}}
\newcommand{\etab}{\end{tabular}}

\newcommand{\master}{{}^{\text{Irr}}\mathcal{F}^\flat}

\newcommand{\comment}[1]{}

\newcommand{\qed}{\nobreak \ifvmode \relax \else
      \ifdim\lastskip<1.5em \hskip-\lastskip
      \hskip1.5em plus0em minus0.5em \fi \nobreak
      \vrule height0.75em width0.5em depth0.25em\fi}

\newcommand{\V}{\mathcal{V}}
\newcommand{\wV}{\widetilde{\mathcal{V}}}
\newcommand{\wmaster}{\widetilde{\master}}
\newcommand{\Tr}{\text{Tr}}
\newcommand{\ra}{\rightarrow}
\newcommand{\cp}{\mathbb{C}\mathbb{P}}
\definecolor{darkspringgreen}{rgb}{0.09, 0.45, 0.27}
\definecolor{forestgreen}{rgb}{0.13, 0.55, 0.13}

\usepackage{array}

\newcolumntype{C}[1]{>{\centering\let\newline\\\arraybackslash\hspace{0pt}}m{#1}}

\newcommand{\drawsquare}[2]{\hbox{%
\rule{#2pt}{#1pt}\hskip-#2pt%  left vertical
\rule{#1pt}{#2pt}\hskip-#1pt%  lower horizontal
\rule[#1pt]{#1pt}{#2pt}}\rule[#1pt]{#2pt}{#2pt}\hskip-#2pt%  upper horizontal
\rule{#2pt}{#1pt}}% right vertical

\newcommand{\fund}{~\raisebox{-.5pt}{\drawsquare{6.5}{0.4}}~}
\newcommand{\antifund}{~\overline{\raisebox{-.5pt}{\drawsquare{6.5}{0.4}}}~}

\newcommand{\symm}{~\raisebox{-.5pt}{\drawsquare{6.5}{0.4}}\hskip-0.4pt%
        \raisebox{-.5pt}{\drawsquare{6.5}{0.4}}~}%  symmetric second rank

\newcommand{\asymm}{~\raisebox{-3.5pt}{\drawsquare{6.5}{0.4}}\hskip-6.9pt%
        \raisebox{3pt}{\drawsquare{6.5}{0.4}}~}%  antisymmetric second rank

\newcommand{\antiasymm}{~\overline{\raisebox{-3.5pt}{\drawsquare{6.5}{0.4}}\hskip-6.9pt%
        \raisebox{3pt}{\drawsquare{6.5}{0.4}}}~}%  antisymmetric second rank

\newcommand{\antisymm}{~\overline{\raisebox{-.5pt}{\drawsquare{6.5}{0.4}}\hskip-0.4pt%
        \raisebox{-.5pt}{\drawsquare{6.5}{0.4}}}~}%  symmetric second rank
\title{Charting the Triality Webs for All Smooth Fano 3-Folds} 
\author[a,b]{Mario Carcamo}
\author[a,b,c]{and Sebasti\'an Franco}

\affiliation[a]{Physics Department, The City College of the CUNY\\
	160 Convent Avenue, New York, NY 10031, USA}
	
\affiliation[b]{Physics Program and \textsuperscript{$c$}Initiative for the Theoretical Sciences\\
	The Graduate School and University Center, The City University of New York\\
	365 Fifth Avenue, New York NY 10016, USA}

\emailAdd{mcarcamo@ccny.cuny.edu,sfranco@ccny.cuny.edu}

\begin{document}

\abstract{We determine all toric phases for the $2d$ $(0,2)$ theories on D1-branes probing the complex cones over the 18 smooth Fano 3-folds, whose toric diagrams correspond to the regular reflexive polytopes in 3 dimensions. These results significantly expand the list of explicitly known gauge theories on D1-branes over toric CY 4-folds. We go beyond the classification of toric phases and map the corresponding triality webs, establishing how the toric phases are connected by triality. The size and complexity of the webs constructed in this work far surpass any previously known examples, both in the contexts of Calabi–Yau 3-folds and 4-folds—with several of these CY 4-folds exhibiting hundreds of toric phases. We propose various new approaches for characterizing triality webs. Our work lays the foundation for a comprehensive exploration of the structure of triality webs.}

\maketitle

%=================================================================
\section{Introduction}
%=================================================================

Engineering quantum field theories (QFTs) via D-branes in string theory is a powerful approach to studying their dynamics. In this context, a fruitful class of setups consists of D-branes probing Calabi-Yau (CY)  singularities \cite{Douglas:1996sw,Klebanov:1998hh,Acharya:1998db,Franco:2005rj}.

In recent years, the engineering of $2d$ $\mathcal{N}=(0,2)$ gauge theories via D1-branes probing toric CY 4-folds has been thoroughly studied  \cite{Franco:2015tna,Franco:2015tya,Franco:2016tcm,Franco:2016qxh,Franco:2017cjj,Franco:2021ixh,Franco:2022iap,Franco:2023tyf,Franco:2024lxs}. In this case, {\it brane brick models}, a class of Type IIA configurations connected to the D1-branes at the toric CY$_4$ via T-duality, both encode the corresponding gauge theories and simplify the connection to geometry \cite{Franco:2015tna,Franco:2015tya}.

Several methods for obtaining the brane brick models (hence, determining the corresponding $2d$ (0,2) gauge theories) for toric CY 4-folds have been developed,\footnote{The gauge theories can have enhanced SUSY in non-generic cases.} including partial resolution \cite{Franco:2015tna,Franco:2015tya}, mirror symmetry \cite{Franco:2016qxh}, the topological B-model \cite{Closset:2017yte,Closset:2018axq}, orbifold reduction \cite{Franco:2016fxm}, $3d$ printing \cite{Franco:2018qsc}, and CY products \cite{Franco:2020avj}. These methods have been used to construct the brane brick models for a wide range of toric CY 4-folds, including infinite classes of them (see e.g. \cite{Franco:2022isw}).

Interestingly, $2d$ $(0,2)$ gauge theories exhibit {\it triality} \cite{Gadde:2013lxa}, an IR equivalence analogous to Seiberg for $4d$ $\mathcal{N}=1$ gauge theories \cite{Seiberg:1994pq}. The term ``triality” stems from the fact that, in its simplest form, it relates three SQCD-like theories in the IR. Alternatively, applying three consecutive triality transformations to the same gauge group returns the theory to its original form. Brane brick models \cite{Franco:2016nwv} and mirror symmetry \cite{Franco:2016tcm, Franco:2016qxh} provide an elegant geometric understanding of triality.\footnote{More generally, the $m$-graded quivers with potentials exhibit order-$(m+1)$ dualities. For $m\leq 3$, this corresponds to supersymmetric gauge theories in $6-2m$ dimensions. Specifically, $m=0$, 1, 2 and 3 correspond to $6d$ $\mathcal{N}=(0,1)$, $4d$ $\mathcal{N}=1$, $2d$ $\mathcal{N}=(0,2)$ and $0d$ $\mathcal{N}=1$ field theories, respectively \cite{Franco:2017lpa,Closset:2018axq,Franco:2019bmx}.}

The relations among theories connected by triality can be beautifully encoded into {\it triality webs}, similar to the {\it Seiberg duality webs or trees} that have been studied for $4d$ $\mathcal{N}=1$ theories \cite{Cachazo:2001sg,Franco:2003ja}. In these webs, every node corresponds to a different UV gauge theory and a link between two of them indicates they are related by triality (see \cite{Franco:2016nwv,Franco:2018qsc} for some early studies of triality webs). Contrary to the Seiberg duality case, links in a triality web are oriented, since two consecutive triality transformations do not amount to the identity. Preliminary studies suggest the existence of {\it triality cascades}, a type of RG-flow in which, as we flow to low energies, we switch to a triality dual description every time a gauge group becomes strongly coupled \cite{Closset:2018axq}. In this context, the triality webs become a chart of possible RG-flow trajectories. We expect the geometry and topology of the webs to capture interesting information. For instance, non-trivial closed cycles in the web correspond to periodic duality cascades.\footnote{By non-trivial cycles, we mean those that do not correspond to three consecutive triality transformations on the same gauge group.} Interestingly, when geometrically engineered using branes, the theories in a triality web can be globally characterized using Diophantine equations \cite{Franco:2020ijt}. 

In this paper, we conduct a comprehensive analysis of the triality webs associated with $2d$ $(0,2)$ gauge theories on D1-branes probing toric CY 4-folds. In particular, we focus on constructing the parts of the triality webs comprising {\it toric phases}.\footnote{A toric phase is a gauge theory whose structure is captured by a brane brick model or, equivalently, a periodic quiver on $\mathbb{T}^3$ \cite{Franco:2015tya}. In Section \sref{section constructing webs}, we discuss the conditions for trialities to remain within this class of theories.} We develop computational tools that streamline the execution of triality transformations, taking into account both quivers and their $J$- and $E$-terms. Additionally, we present various methodologies for characterizing the resulting triality webs.

The CY 4-folds we will focus on are the complex cones over the 18 smooth Fano 3-folds, whose toric diagrams are the 18 regular reflexive polytopes in 3 dimensions. Using a variety of techniques, a toric phase for each of these geometries has been explicitly constructed in \cite{Franco:2022gvl}. The CY$_3$ analogs of this family of geometries are the complex cones over $dP_n$ ($n=0,\ldots,3$) and $F_0$, which have served as a fertile testing ground for exploring various connections between $4d$ theories and geometry (see \cite{Beasley:1999uz,Feng:2001xr,Beasley:2001zp,Feng:2001bn,Feng:2002zw} for a representative, though not exhaustive, list of references). For each of these CY 4-folds, we will construct the complete toric component of the triality web.\footnote{In Section \sref{section constructing webs}, we will provide a more precise characterization of this part of the web.} Previous investigations into (portions of) triality webs can be found in \cite{Franco:2016nwv,Franco:2018qsc}. However, the scope of the results presented in this paper significantly exceeds these earlier efforts. To illustrate the scale of this improvement, the largest triality web containing all toric phases for a geometry previously known is that of $Q^{1,1,1}/\mathbb{Z}_2$, constructed in \cite{Franco:2018qsc}, which comprises 14 toric phases. By contrast, the largest web constructed in this paper consists of 831 phases. Furthermore, the level of detail we provide on the structure and properties of these webs is considerably deeper than in previous studies.

For each toric phase in the triality webs we will present in this paper, we have determined the corresponding quiver as well as the $J$- and $E$-terms. While, for space reasons, we only present part of this data in this paper, such as numbers of chiral and Fermi fields, these results represent the largest collection of $2d$ $(0,2)$ quiver theories for D1-branes on toric CY 4-folds explicitly constructed in the literature to date, aside from the infinite families of theories in \cite{Franco:2022isw}.

The paper is organized as follows. Section \sref{section reflexive polytopes and 2d gauge theories} provides a brief overview of the 18 regular reflexive polytopes in three dimensions and the prior classification of associated $2d$ $(0,2)$ gauge theories. Section \sref{section constructing webs} outlines our algorithm for constructing triality webs. Section \sref{section triality webs} classifies all toric phases for the 18 regular reflexive polytopes and their connections under triality. In Section \sref{section checks forward algorithm} we double check some of the new theories we constructed using the forward algorithm. We present our conclusions in \sref{section conclusions}. Appendix \sref{appendix further details web construction} provides additional details on the methodology used to determine toric phases. Finally, Appendix \sref{appendix additional toric phases} presents the quivers and $J$- and $E$-terms for new toric phases for all the polytopes with multiple toric phases, which were used in the consistency checks of Section \sref{section checks forward algorithm}.

%=================================================================
\section{Regular Reflexive Polytopes and $2d$ $(0,2)$ Gauge Theories}
%=================================================================

\label{section reflexive polytopes and 2d gauge theories}

In this paper, we focus on the class of CY 4-folds given by complex cones over Gorenstein Fano varieties whose toric diagrams are reflexive polytopes. Thanks to the Kreuzer–Skarke classification \cite{Kreuzer:1998vb,Kreuzer:2000xy,Kreuzer:2000qv}, we know that there are 4,319 reflexive polytopes in three dimensions, up to $GL(3, \mathbb{Z})$ equivalence. Moreover, we restrict our attention to the subset of 18 polytopes that are both reflexive and regular. The Gorenstein Fano varieties associated to regular reflexive polytopes are smooth.

%=================================================================
\begin{figure}[ht]
\begin{center}
\includegraphics[width=\textwidth]{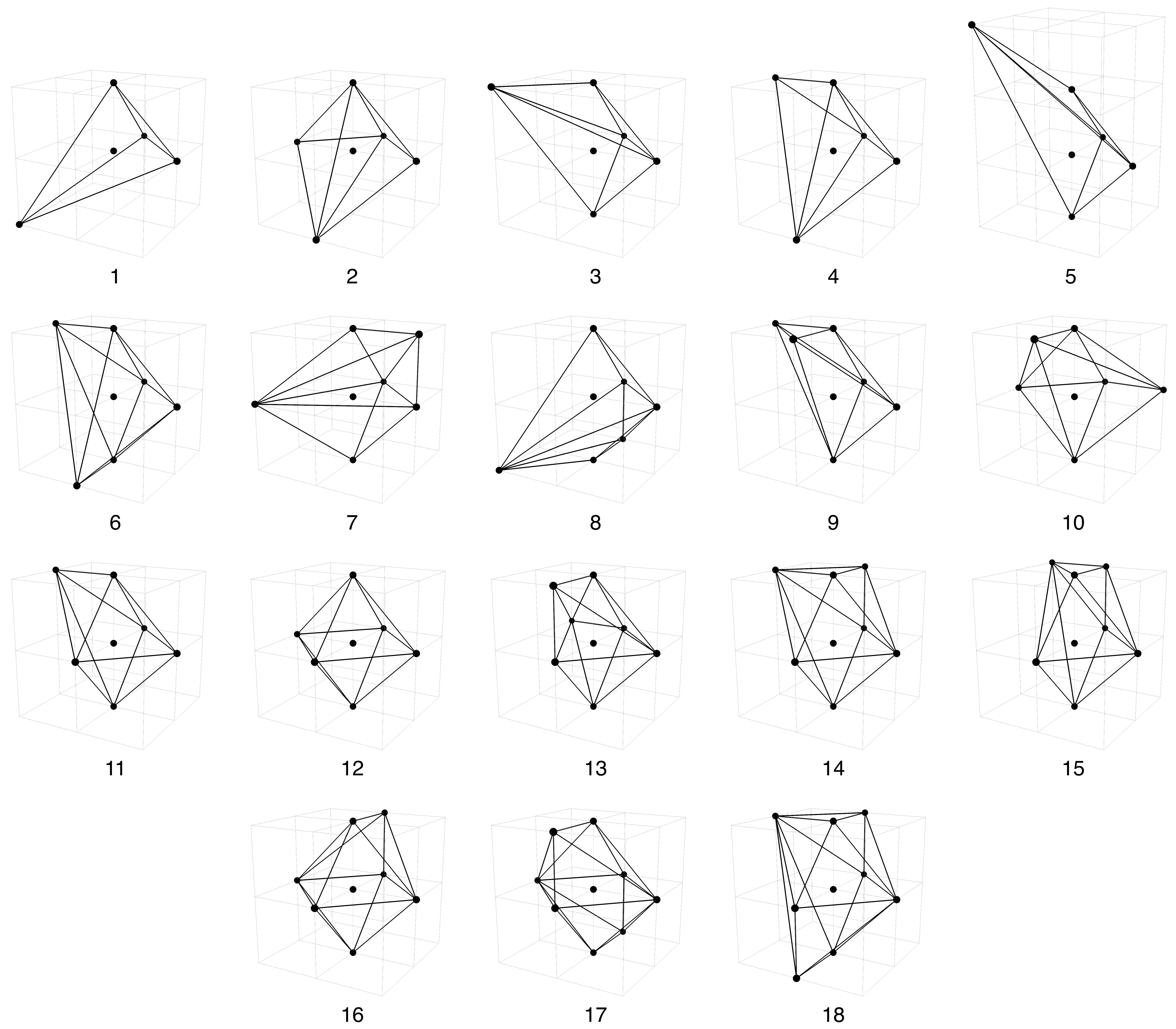} 
\caption{The 18 regular reflexive polytopes in dimension 3 corresponding to the toric diagrams of the non-compact Calabi-Yau 4-folds considered in this paper.}
\label{Toric_diagrams_reflexive}
 \end{center}
 \end{figure}
 %=================================================================

The systematic study of the $2d$ $(0,2)$ gauge theories for these CY 4-folds was first undertaken in \cite{Franco:2022gvl}, where a toric phase for each geometry was determined using various approaches, including partial resolution, orbifold reduction, and $3d$ printing. In the coming sections, we will use these theories as seeds for constructing the corresponding triality webs.\footnote{The figures we use for their quivers are taken from \cite{Franco:2022gvl}.}

%=================================================================
\section{Constructing the Triality Webs}
%=================================================================

\label{section constructing webs}

One of the primary goals of this work is to chart the space of toric phases for the complex cones over all smooth Fano 3-folds.\footnote{These theories also have non-toric phases, which are interesting but not the focus of this paper.} We systematically map these spaces of triality dual theories as follows. For each of CY 4-fold, we start from the toric phase constructed in \cite{Franco:2022gvl}. Next, we perform triality or inverse triality on all {\it toric nodes}. A node is said to be toric under triality or inverse triality if applying either operation results in a toric phase. A node is toric under triality if it has exactly two incoming chiral fields, and toric under inverse triality if it has exactly two outgoing chiral fields. Triality is different from Seiberg duality for $4d$ theories in that a node maybe non-toric, toric only triality or inverse triality, or toric under both. Moreover, even if a node is toric under both triality and inverse triality, it is possible that they result in different toric phases. Generically, these features lead to a much richer space of toric phases. After applying all (inverse) triality transformations on toric nodes, we identify the new toric phases encountered (up to node relabeling and conjugation of all chiral fields\footnote{This transformation, often referred to as {\it chiral conjugation}, is a symmetry of the theory \cite{Franco:2016nwv}. It is a combination of the exchange of fundamental and antifundamental representations for every node in the quiver with the symmetry of $2d$ (0,2) theories that exchanges Fermi fields and their conjugates while swapping the corresponding $J$- and $E$-terms.}) and add them to the list of phases. This process is iterated until no new phases are found.

We have automated these searches, developing a computer code that efficiently keeps track of not only the quivers, but also their $J$- and $E$-terms, integrating out massive fields when necessary. This is the largest characterization of spaces of triality dual theories available in the literature and, to our knowledge, the largest automatic implementation of a search of this kind.\footnote{This is similar in spirit to recent automated studies of the space of dual BPS quivers \cite{Carta:2025asr}, albeit more complex and at a larger scale.} Additional details on how our code works are summarized in Appendix \sref{appendix further details web construction}.

%=================================================================
\subsection*{Toric Islands}
%=================================================================

\begin{figure}[ht]
\begin{center}
\includegraphics[width=9cm]{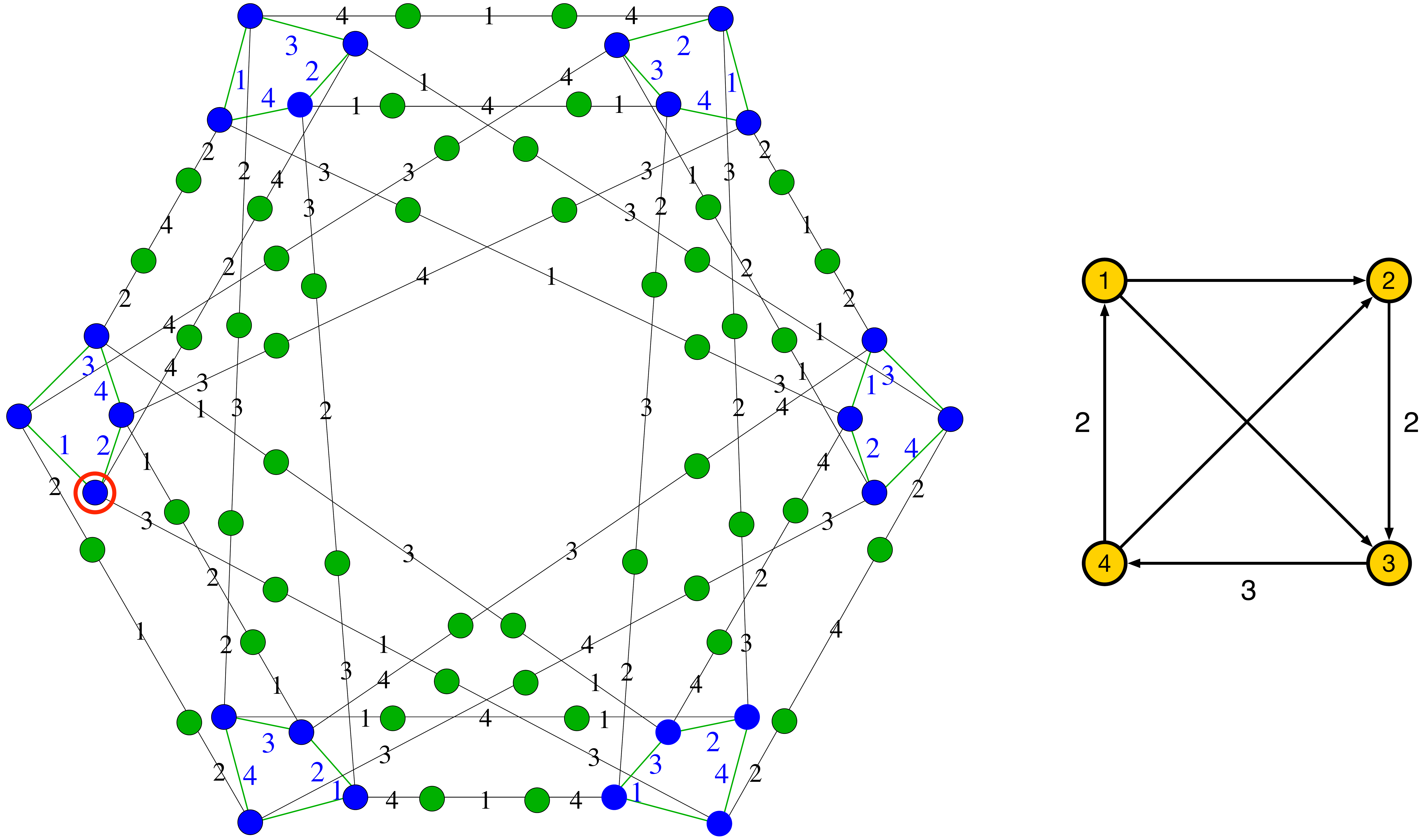} 
\caption{A portion of the web of Seiberg dual theories for $dP_1$, which has a single toric phase. The red circle indicates the toric phase in the form of the quiver on the right. Other blue nodes on the web correspond to permutations of this toric phase. The green dots correspond to permutations of a non-toric phase. The number on the lines denote the dualized node.}
\label{toric_islands_dP1}
 \end{center}
 \end{figure}

It is worth discussing more precisely what the algorithm outlined above constructs. The algorithm generates a component of the triality web consisting of toric phases connected by (inverse) triality transformations. In principle, it is possible for the triality web to contain multiple such components, where the transition between them involves trialities that pass through non-toric phases. Whenever this is the case, we refer to each of these individual toric components as a {\it toric island}. Toric islands were first discussed in the context of Seiberg duality webs for toric CY 3-folds in \cite{Franco:2003ja}. \fref{toric_islands_dP1}, partly taken from that paper, shows the 6 toric islands for $dP_1$.

%=================================================================
 %=================================================================
 
It is possible that, once we consider all toric islands in a web, not all possible node permutations of them are present.\footnote{This statement applies to duality webs, triality webs and, more generally, webs of order-$m$ dualities for toric CY $m+2$-folds.} This is clear in the $dP_1$ example shown in \fref{toric_islands_dP1}, where we see that the 6 toric islands contain 12 permutations of each of the two toric phases, while the number of permutations for 4 node quivers is 4!=24. If we start from a permutation of any of these toric phases that is not contained in the web, we should generate a duality web that is isomorphic to the one in the figure by a permutation of nodes. The theories in both webs, however, are not connected by dualities. In summary, all node permutations of a given phase (toric or not) may not be present on the same web. If this is the case, the different webs are isomorphic.

To conclude, we should keep in mind that two further scenarios are, in principle, logically possible. First, there could be additional toric phases that are not connected by trialities that go through other toric phases to the theory that we used as a seed in our analysis. Such theories would not arise in our analysis and would belong to additional toric phases with a different structure. Second, it could be possible that entirely new toric phases exist, which are not connected by any sequence of trialities—whether involving toric phases or not. This possibility would be particularly intriguing, as it would suggest a novel IR equivalence for $2d$ $(0,2)$ theories beyond the triality. 

Either of these scenarios would imply that a single toric island does not encompass all possible toric phases of a given theory. However, current evidence from known explicit examples, both for CY 3-folds and 4-folds, does not support these possibilities.

%=================================================================
\section{Triality Webs}
%=================================================================

\label{section triality webs}

Following the approach outlined in Section \sref{section constructing webs}, we constructed all toric phases for the 18 CY 4-folds associated with the smooth Fano 3-folds and established how they are connected by triality. We fully determined each toric phase, namely we determined the quiver along with the corresponding $J$- and $E$-terms. Due to space constraints, a full presentation of these results is impractical. Instead, we provide tables that summarize some of our key findings. The resulting tables are quite extensive. Readers primarily interested in the general features of these toric phases may skip directly to Section \sref{section detailed structure webs}, which provides a summary of key results from this section along with additional analyses of the data.

%=================================================================
\paragraph{The importance of $J$- and $E$-terms.}
%=================================================================
We emphasize the critical importance of explicitly tracking the $J$- and $E$-terms in our computer classification. At first glance, one might consider a simplified approach to classifying toric phases—or at least approximating such a classification—by focusing solely on the quiver and its behavior under triality and eliminating chiral-Fermi pairs of fields that connect the same pair of nodes whenever they arise. However, this approach is fundamentally flawed. Such chiral-Fermi pairs should only be integrated out only if they are massive, a determination that requires knowledge of the $J$- and $E$-terms. In fact, some of the toric phases we identify contain massless chiral-Fermi pairs, where the would-be mass terms are absent from the $J$- and $E$-terms. From an effective field theory perspective, these mass terms are inconsistent with the global symmetries of the gauge theory, which are ultimately dictated by the underlying geometry. Indeed, even the toric phases for Models 5 and 8 originally constructed in \cite{Franco:2022gvl}, which are going to serve as starting points of the studies in this paper, exhibit such massless chiral-Fermi pairs.

\medskip 

Below we present the results for the 18 models. We will refer to each of these geometries and the corresponding equivalence class of gauge theories as a {\it model}. For each model, we will systematically investigate all the corresponding toric phases. Additionally, we will provide alternative names commonly used for the CY 4-folds. Some of these names are standard, while others—such as $P_{+-}^1(dP_0)$ for Model 4—highlight connections between the toric diagram and that of a CY 3-fold via $3d$ printing \cite{Franco:2018qsc}. While this naming convention is not central to our work, interested readers can refer to \cite{Franco:2022gvl} for a more detailed explanation.

%=================================================================
\subsection{Introducing Conventions. Model 12: $Q^{1,1,1}/\mathbb{Z}_2$}%=================================================================

Instead of starting with Model 1, which has a single toric phase, we will first consider Model 12, $Q^{1,1,1}/\mathbb{Z}_2$, since it is better for introducing our conventions. Moreover, this is the only example for which all toric phases were previously classified \cite{Franco:2018qsc}.

\begin{figure}[H]
\begin{center}
\resizebox{0.3\hsize}{!}{
\includegraphics[height=6cm]{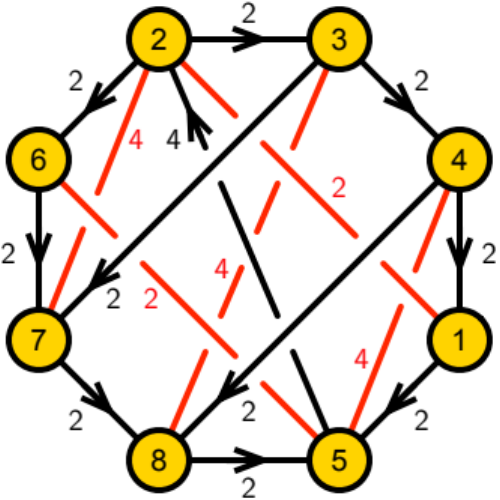} 
}
\caption{
Quiver for Phase 1 of Model 12.
\label{f_quiver_12}}
 \end{center}
 \end{figure}

Let us start with the toric phase for this model presented in \cite{Franco:2022gvl}. This theory was first constructed in \cite{Franco:2016nwv}, where it was denoted phase $C$. It also appeared in \cite{Franco:2018qsc}, where it was referred to as phase $F$. We refer to this theory as phase 1 and it will serve as the seed of our construction of the other toric phases. \fref{f_quiver_12} shows the quiver for this theory.

%=================================================================
 %=================================================================
 
 Its $J$- and $E$-terms are
\beq
{\footnotesize
\begin{array}{rrclrcl}
 &  & J &  & & E &  \\
\Lambda_{12}^{1} : &  Y_{23} Y_{34} X_{41}  &-& X_{23} Y_{34} Y_{41} & P_{15} X_{52}  &-&  Q_{15} R_{52} \\ 
\Lambda_{12}^{2} : &  X_{23} X_{34} Y_{41}  &-&  Y_{23} X_{34} X_{41} & P_{15} Y_{52}  &-&  Q_{15} S_{52} \\ 
\Lambda_{27}^{1} : &  X_{78} Y_{85} Y_{52}  &-&  Y_{78} Y_{85} X_{52} & P_{26} X_{67}  &-&  X_{23} P_{37} \\ 
\Lambda_{27}^{2} : &  Y_{78} X_{85} X_{52}  &-& X_{78} X_{85} Y_{52} & P_{26} Y_{67}  &-&  Y_{23} P_{37} \\ 
\Lambda_{27}^{3} : &  X_{78} Y_{85} S_{52}  &-&  Y_{78} Y_{85} R_{52} & X_{23} Q_{37}  &-& Q_{26} X_{67} \\ 
\Lambda_{27}^{4} : &  Y_{78} X_{85} R_{52}  &-& X_{78} X_{85} S_{52}  & Y_{23} Q_{37}  &-& Q_{26} Y_{67} \\ 
\Lambda_{38}^{1} : &  Y_{85} Y_{52} X_{23}  &-& X_{85} Y_{52} Y_{23} & P_{37} X_{78}  &-&  X_{34} P_{48} \\ 
\Lambda_{38}^{2} : &  X_{85} X_{52} Y_{23}  &-&  Y_{85} X_{52} X_{23} & P_{37} Y_{78}  &-&  Y_{34} P_{48} \\ 
\Lambda_{38}^{3} : &  Y_{85} S_{52} X_{23}  &-& X_{85} S_{52} Y_{23} & X_{34} Q_{48}  &-& Q_{37} X_{78}  \\ 
\Lambda_{38}^{4} : &  X_{85} R_{52} Y_{23}  &-&  Y_{85} R_{52} X_{23} &  Y_{34} Q_{48} &-& Q_{37} Y_{78} \\ 
\Lambda_{45}^{1} : &  X_{52} Y_{23} Y_{34}  &-&  Y_{52} Y_{23} X_{34} & P_{48} X_{85}  &-&  X_{41} P_{15} \\ 
\Lambda_{45}^{2} : &  Y_{52} X_{23} X_{34}  &-& X_{52} X_{23} Y_{34} & P_{48} Y_{85}  &-&  Y_{41} P_{15} \\ 
\Lambda_{45}^{3} : &  R_{52} Y_{23} Y_{34}  &-&  S_{52} Y_{23} X_{34} & X_{41} Q_{15}  &-& Q_{48} X_{85}  \\ 
\Lambda_{45}^{4} : &  S_{52} X_{23} X_{34}  &-& R_{52} X_{23} Y_{34} & Y_{41} Q_{15}  &-& Q_{48} Y_{85} \\ 
\Lambda_{56}^{1} : &  Y_{67} Y_{78} X_{85}  &-& X_{67} Y_{78} Y_{85} & R_{52} Q_{26}  &-&  X_{52} P_{26} \\ 
\Lambda_{56}^{2} : &  X_{67} X_{78} Y_{85}  &-&  Y_{67} X_{78} X_{85} & S_{52} Q_{26}  &-&  Y_{52} P_{26}
 \end{array} 
 }
\label{E_J_C_+-}
\eeq

Proceeding as explained, we find 14 toric phases for this geometry. We distinguish phases modulo relabeling of nodes. Table \ref{Table 1 - Model 12} summarizes some basic information characterizing these theories to facilitate their comparison. Compared to \cite{Franco:2018qsc}, we have used numbers instead of letters to label them. The phases are primarily ordered based on the sequence in which they were detected by the automated algorithm, rather than by their field content. For each phase, we provide a sequence of (inverse) triality transformations connecting it to Phase 1 in the form shown in \fref{f_quiver_12}. The numbers in the sequence specify the node to be mutated, with a preceding minus sign indicating inverse triality. Such sequences are, generically, not unique. We also present the number of Fermi fields $N_F$. For toric phases, the total number of fields is
\beq
N_{fields}=N_\chi+N_F+N_G \, ,
\label{N_fields}
\eeq
with $N_\chi$, $N_F$ and $N_G$ are the numbers of chiral fields, Fermi fields and nodes in the quivers (vector multiplets), respectively, can be expressed as 
\beq
N_{fields}=2(N_F+N_G) \, .
\eeq
Since $N_G$ is the same for all toric phases associated to a given CY$_4$, $N_F$ gives a measure of the number of UV degrees of freedom of the theory. In the “Fermi Multiplicities” column, we give the multiplicity of Fermi fields for the 8 nodes in the quiver. For example, $4 \times {\bf 2} + 4  \times {\bf 2}$ indicates that the corresponding theory has 4 nodes with 2 Fermis and 4 nodes with 4 Fermis. This basic structural information about the quiver is often sufficient for distinguishing toric phases.

%=================================================================
\begin{table}[h]
\begin{center}
\begin{tabular}{ |c|c|c|c|} 
\hline
Phase & Path & $N_F$ & Fermi Multiplicities\\
\hline
1&&16&2$\times\textbf{2}$+4$\times\textbf{4}$+2$\times\textbf{6}$\\ 
2&1&16&2$\times\textbf{2}$+5$\times\textbf{4}$+1$\times\textbf{8}$\\ 
3&-1&12&4$\times\textbf{2}$+4$\times\textbf{4}$\\ 
4&3&20&3$\times\textbf{2}$+1$\times\textbf{4}$+2$\times\textbf{6}$+1$\times\textbf{8}$+1$\times\textbf{10}$\\ 
5&4&20&2$\times\textbf{2}$+3$\times\textbf{4}$+1$\times\textbf{6}$+1$\times\textbf{8}$+1$\times\textbf{10}$\\ 
6&1,3&16&4$\times\textbf{2}$+2$\times\textbf{4}$+2$\times\textbf{8}$\\ 
7&1,-4&20&2$\times\textbf{2}$+4$\times\textbf{4}$+1$\times\textbf{8}$+1$\times\textbf{12}$\\ 
8&1,5&24&6$\times\textbf{4}$+2$\times\textbf{12}$\\ 
9&1,-8&24&2$\times\textbf{2}$+1$\times\textbf{4}$+2$\times\textbf{6}$+2$\times\textbf{8}$+1$\times\textbf{12}$\\ 
10&-1,3&16&4$\times\textbf{2}$+2$\times\textbf{4}$+2$\times\textbf{8}$\\ 
11&-1,-4&12&4$\times\textbf{2}$+4$\times\textbf{4}$\\ 
12&3,7&24&2$\times\textbf{2}$+4$\times\textbf{6}$+2$\times\textbf{10}$\\ 
13&4,-7&28&4$\times\textbf{4}$+4$\times\textbf{10}$\\ 
14&1,-4,-7&28&6$\times\textbf{4}$+2$\times\textbf{16}$\\ 
\hline
\end{tabular}
\end{center}
\caption{Basic information regarding the 14 toric phases of Model 12.}
\label{Table 1 - Model 12}
\end{table}
%=================================================================

As previously mentioned, we have explicitly constructed the quiver along with the $J$- and $E$-terms for each toric phase. Given the extensive volume of data, presenting it in full within this paper is impractical. However, interested readers can generate these phases themselves using the triality sequences provided in Table \ref{Table 1 - Model 12} and similar tables for other models.

Table \ref{Table 2 - Model 12} summarizes how the different phases are interconnected by triality. In this table, for each of the phases we consider the labeling of nodes obtained by acting on Phase 1 as shown in \fref{f_quiver_12} with the sequences of trialities in Table \ref{Table 1 - Model 12}.\footnote{Other sequences of trialities might lead to the same phases, but where the labels of nodes are permuted.} In each column, we indicate the phases obtained by acting with triality or inverse triality on the corresponding node. The underline indicates phases obtained by inverse triality while the blanks correspond to the nodes for which triality does not give a toric phase. Some entries contain only a single theory, as either triality or inverse triality— but not both—lead to a toric phase in those cases. The table also shows explicit examples in which acting with triality or inverse triality on a given node can lead to different toric phases.

%=================================================================
\begin{table}[H]
\begin{center}
\begin{tabular}{ |c|c|c|c|c|c|c|c|c|} 
\hline
Phase &1&2&3&4&5&6&7&8\\
\hline
1&2 , \underline{3}&&4&5&&3 , \underline{2}&\underline{5}&\underline{4}\\
2&3 , \underline{1}&&6&\underline{7}&8&3 , \underline{1}&\underline{7}&\underline{9}\\
3&1 , \underline{2}&5&10&3 , \underline{11}&2 , \underline{1}&11 , \underline{3}&\underline{10}&\underline{5}\\
4&6 , \underline{10}&&\underline{1}&&&5 , \underline{9}&12 , \underline{4}&\\
5&\underline{3}&&&\underline{1}&&10 , \underline{7}&\underline{13}&9 , \underline{4}\\
6&10 , \underline{4}&&\underline{2}&4 , \underline{10}&2&10 , \underline{4}&4 , \underline{10}&\\
7&11&&10 , \underline{5}&2&2&10 , \underline{5}&\underline{14}&\\
8&2&&2&\underline{2}&\underline{2}&2&\underline{2}&\\
9&5 , \underline{4}&&&&&5 , \underline{4}&&2\\
10&4 , \underline{6}&&\underline{3}&\underline{3}&7 , \underline{5}&7 , \underline{5}&4 , \underline{6}&\\
11&\underline{7}&7&3 , \underline{3}&3 , \underline{3}&3 , \underline{3}&3 , \underline{3}&\underline{7}&7\\
12&4 , \underline{4}&&&&&&4 , \underline{4}&\\
13&\underline{5}&&&\underline{5}&&5&5&\\
14&7&&7&7&7&7&7&\\\hline
\end{tabular}
\end{center}
\caption{Triality connections between the 14 toric phases of Model 12.}
\label{Table 2 - Model 12}
\end{table}
%=================================================================

In the following subsections, we present the remaining 17 models using the same format.

%=================================================================
\subsection{Model 1: $\mathbb{C}^4/\mathbb{Z}_4 ~(1,1,1,1)$}
%=================================================================
 
\begin{figure}[H]
\begin{center}
\resizebox{0.25\hsize}{!}{
\includegraphics[height=6cm]{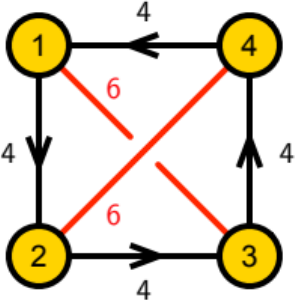} 
}
\caption{
Quiver for Model 1.
\label{f_quiver_01}}
 \end{center}
 \end{figure}

 Model 1 corresponds to the $\mathbb{C}^4/\mathbb{Z}_4$ orbifold with action $(1,1,1,1)$. Its quiver is shown in \fref{f_quiver_01}.
 
%=================================================================
%=================================================================
  
 The $J$- and $E$-terms are
  \beq
  {\footnotesize
\begin{array}{rrclrcl}
 &  & J &  & & E &  \\
\Lambda_{13}^{1}  : & Z_{34} Y_{41}  &-& Y_{34} Z_{41} &   P_{12} X_{23}  &-&  X_{12} P_{23} \\ 
\Lambda_{13}^{2}  : & X_{34} Z_{41}  &-&  Z_{34} X_{41}  &   P_{12} Y_{23}  &-&  Y_{12} P_{23} \\ 
\Lambda_{13}^{3}  : & Y_{34} X_{41}  &-& X_{34} Y_{41}  &   P_{12} Z_{23}  &-&  Z_{12} P_{23} \\ 
\Lambda_{24}^{1}  : & Z_{41} Y_{12}  &-& Y_{41} Z_{12}  &   P_{23} X_{34}  &-&  X_{23} P_{34} \\ 
\Lambda_{24}^{2}  : & X_{41} Z_{12}  &-&  Z_{41} X_{12}  &   P_{23} Y_{34}  &-&  Y_{23} P_{34} \\ 
\Lambda_{24}^{3}  : & Y_{41} X_{12}  &-& X_{41} Y_{12} &   P_{23} Z_{34}  &-&  Z_{23} P_{34} \\ 
\Lambda_{31}^{1}  : & Z_{12} Y_{23}  &-& Y_{12} Z_{23}  &   P_{34} X_{41}  &-&  X_{34} P_{41} \\ 
\Lambda_{31}^{2}  : & X_{12} Z_{23}  &-&  Z_{12} X_{23}  &   P_{34} Y_{41}  &-&  Y_{34} P_{41} \\ 
\Lambda_{31}^{3}  : & Y_{12} X_{23}   &-& X_{12} Y_{23} &   P_{34} Z_{41}  &-&  Z_{34} P_{41} \\ 
\Lambda_{42}^{1}  : & Z_{23} Y_{34}  &-& Y_{23} Z_{34}  &   P_{41} X_{12}  &-&  X_{41} P_{12} \\ 
\Lambda_{42}^{2}  : & X_{23} Z_{34}  &-&  Z_{23} X_{34}  &   P_{41} Y_{12}  &-&  Y_{41} P_{12} \\ 
\Lambda_{42}^{3}  : &  Y_{23} X_{34} &-& X_{23} Y_{34}  &   P_{41} Z_{12}  &-&  Z_{41} P_{12}
 \end{array} 
 }~.~
\label{E_J_C_+-}
 \eeq
 
 None of the nodes in this quiver are toric, so Phase 1 is the only toric phase for Model 1. Table \ref{Table 1 - Model 1} summarizes its Fermi content per node.
 
%=================================================================
\begin{table}[H]
\begin{center}
\begin{tabular}{ |c|c|c|} 
\hline
Phase & $N_F$ & Fermi Multiplicities\\
\hline
1 & 12 & 4$\times\textbf{6}$ \\ 
\hline
\end{tabular}
\end{center}
\caption{Basic information regarding the single toric phase of Model 1.}
\label{Table 1 - Model 1}
\end{table}
%=================================================================

%=================================================================
\subsection{Model 2:  $M^{3,2}$}
%=================================================================

\begin{figure}[H]
\begin{center}
\resizebox{0.25\hsize}{!}{
\includegraphics[height=6cm]{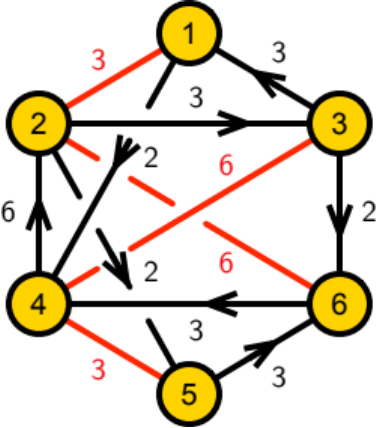} 
}
\caption{
Quiver for Phase 1 of Model 2.
\label{f_quiver_02}}
 \end{center}
 \end{figure}

\fref{f_quiver_02} shows the quiver for Phase 1 of Model 2.\footnote{Other toric phases for this model were previously found in \cite{Franco:2016qxh,Franco:2016fxm}.}

%=================================================================
%=================================================================

 The $J$- and $E$-terms are
  \beq
  {\footnotesize
\begin{array}{rrclrcl}
 &  & J &  & & E &  \\
\Lambda_{12}^{1}  : & X_{23} X_{31}  &-&  Z_{23} Y_{31}  &   P_{14} X_{42}  &-&  Q_{14} R_{42} \\
\Lambda_{12}^{2}  : & Y_{23} Y_{31}  &-& X_{23} Z_{31}  &   P_{14} Y_{42}  &-&  Q_{14} S_{42} \\
\Lambda_{12}^{3}  : & Z_{23} Z_{31}  &-& Y_{23} X_{31}  &   P_{14} Z_{42}  &-&  Q_{14} T_{42} \\
\Lambda_{26}^{1}  : & X_{64} X_{42}  &-&  Z_{64} Y_{42}  &   P_{25} X_{56}  &-&  X_{23} P_{36} \\
\Lambda_{26}^{2}  : & Y_{64} Y_{42}   &-& X_{64} Z_{42}  &   P_{25} Y_{56}  &-&  Y_{23} P_{36} \\
\Lambda_{26}^{3}  : & Z_{64} Z_{42}  &-& Y_{64} X_{42} &   P_{25} Z_{56}  &-&  Z_{23} P_{36} \\
\Lambda_{26}^{4}  : & X_{64} R_{42}  &-&  Z_{64} S_{42}  &  X_{23} Q_{36}  &-& Q_{25} X_{56} \\
\Lambda_{26}^{5}  : & Y_{64} S_{42}  &-& X_{64} T_{42}  &  Y_{23} Q_{36}   &-& Q_{25} Y_{56} \\
\Lambda_{26}^{6}  : & Z_{64} T_{42}  &-& Y_{64} R_{42} &   Z_{23} Q_{36}  &-& Q_{25} Z_{56}\\
\Lambda_{34}^{1}  : & X_{42} X_{23}  &-&  Z_{42} Y_{23}  &   P_{36} X_{64}  &-&  X_{31} P_{14} \\
\Lambda_{34}^{2}  : & Y_{42} Y_{23}  &-& X_{42} Z_{23}  &   P_{36} Y_{64}  &-&  Y_{31} P_{14} \\
\Lambda_{34}^{3}  : & Z_{42} Z_{23}   &-& Y_{42} X_{23} &   P_{36} Z_{64}  &-&  Z_{31} P_{14} \\
\Lambda_{34}^{4}  : & R_{42} X_{23}  &-&  T_{42} Y_{23}  &  X_{31} Q_{14}  &-& Q_{36} X_{64}\\
\Lambda_{34}^{5}  : & S_{42} Y_{23}  &-& R_{42} Z_{23}  &   Y_{31} Q_{14}  &-& Q_{36} Y_{64}\\
\Lambda_{34}^{6}  : & T_{42} Z_{23}  &-& S_{42} X_{23}  &   Z_{31} Q_{14}  &-& Q_{36} Z_{64} \\
\Lambda_{45}^{1}  : & X_{56} X_{64}  &-&  Z_{56} Y_{64}  &   R_{42} Q_{25}  &-&  X_{42} P_{25} \\
\Lambda_{45}^{2}  : & Y_{56} Y_{64}   &-& X_{56} Z_{64} &   S_{42} Q_{25}  &-&  Y_{42} P_{25} \\
\Lambda_{45}^{3}  : & Z_{56} Z_{64}  &-& Y_{56} X_{64}   &   T_{42} Q_{25}  &-&  Z_{42} P_{25}
 \end{array} 
 }~.~
\label{E_J_C_+-}
 \eeq

This model has 2 toric phases, which are summarized in Table \ref{Table 1 - Model 2}.

%=================================================================
\begin{table}[h]
\begin{center}
\begin{tabular}{ |c|c|c|c|} 
\hline
 Phase & Path & F & Fermi Multiplicities\\
\hline
1&&18&2$\times\textbf{3}$+2$\times\textbf{6}$+2$\times\textbf{9}$\\ 
2&-1&12&4$\times\textbf{3}$+2$\times\textbf{6}$\\ 
\hline
\end{tabular}
\end{center}
\caption{Basic information regarding the 2 toric phases of Model 2.}
\label{Table 1 - Model 2}
\end{table}
%=================================================================

Table \ref{Table 2 - Model 2} summarizes the connection between the toric phases under triality.

%=================================================================
\begin{table}[h]
\begin{center}
\begin{tabular}{ |c|c|c|c|c|c|c|} 
\hline
N&1&2&3&4&5&6\\
\hline
1&\underline{2}&&&&2&\\
2&1&&\underline{2}&\underline{1}&2&\\\hline
\end{tabular}
\end{center}
\caption{Triality connections between the 2 toric phases of Model 2.}
\label{Table 2 - Model 2}
\end{table}
%=================================================================

%=================================================================
\subsection{Model 3: $Y^{2,4}(\mathbb{CP}^2)$}
%=================================================================

\begin{figure}[H]
\begin{center}
\resizebox{0.25\hsize}{!}{
\includegraphics[height=6cm]{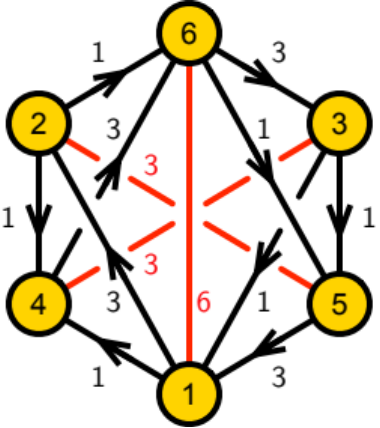} 
}
\caption{Quiver for Phase 1 of Model 3.
\label{f_quiver_03}}
 \end{center}
 \end{figure}

\fref{f_quiver_03} shows the quiver for Phase 1 of Model 3.

%=================================================================
%=================================================================

The $J$- and $E$-terms are
\beq
{\footnotesize
\begin{array}{rrclrcl}
 &  & J &  & & E &  \\
\Lambda_{16}^{1}  : & X_{63} P_{35} Z_{51}  &-&  Z_{63} P_{35} X_{51}   &    P_{12} X_{26}  &-&  Q_{14} X_{46} \\ 
\Lambda_{16}^{2}  : & P_{63} P_{35} X_{51}  &-&  X_{63} P_{35} Y_{51}   &    X_{12} X_{26}  &-& Q_{14} Y_{46} \\ 
\Lambda_{16}^{3}  : & P_{63} P_{35} Z_{51}  &-&  Z_{63} P_{35} Y_{51}   &    Q_{14} Z_{46}  &-&  Y_{12} X_{26} \\ 
\Lambda_{43}^{1}  : & P_{35} Z_{51} Q_{14}  &-&  Q_{31} X_{12} P_{24}   &    X_{46} X_{63}  &-&  Z_{46} P_{63} \\ 
\Lambda_{43}^{2}  : & P_{35} X_{51} Q_{14}  &-&  Q_{31} Y_{12} P_{24}   &    Y_{46} P_{63}  &-& X_{46} Z_{63} \\ 
\Lambda_{43}^{3}  : & P_{35} Y_{51} Q_{14}  &-&  Q_{31} P_{12} P_{24}   &    Z_{46} Z_{63}  &-& Y_{46} X_{63} \\ 
\Lambda_{52}^{1}  : & P_{24} X_{46} Q_{65}  &-&  X_{26} P_{63} P_{35}   &    X_{51} X_{12}  &-&  Z_{51} Y_{12} \\ 
\Lambda_{52}^{2}  : & P_{24} Y_{46} Q_{65}  &-&  X_{26} Z_{63} P_{35}   &    Y_{51} Y_{12}  &-& X_{51} P_{12} \\ 
\Lambda_{52}^{3}  : & P_{24} Z_{46} Q_{65}  &-&  X_{26} X_{63} P_{35}   &    Z_{51} P_{12}  &-& Y_{51} X_{12} \\ 
\Lambda_{61}^{1}  : & X_{12} P_{24} X_{46}  &-& P_{12} P_{24} Y_{46}   &    X_{63} Q_{31}  &-& Q_{65} X_{51}  \\ 
\Lambda_{61}^{2}  : & Y_{12} P_{24} Y_{46}  &-& X_{12} P_{24} Z_{46}   &    P_{63} Q_{31}  &-&  Q_{65} Y_{51} \\ 
\Lambda_{61}^{3}  : & P_{12} P_{24} Z_{46}  &-&  Y_{12} P_{24} X_{46}   &    Z_{63} Q_{31}  &-& Q_{65} Z_{51} 
 \end{array} 
 }~.~
\label{es0301}
 \eeq

This model has 2 toric phases, which are summarized in Table \ref{Table 1 - Model 3}.

%=================================================================
\begin{table}[h]
\begin{center}
\begin{tabular}{ |c|c|c|c|} 
\hline
 Phase & Path & F & Fermi Multiplicities\\
\hline
1&&12&4$\times\textbf{3}$+2$\times\textbf{6}$\\ 
2&-2&15&2$\times\textbf{3}$+4$\times\textbf{6}$\\ 
\hline
\end{tabular}
\end{center}
\vspace{-.4cm}\caption{Basic information regarding the 2 toric phases of Model 3.}
\label{Table 1 - Model 3}
\end{table}
%=================================================================

Table \ref{Table 2 - Model 3} summarizes the connection between the toric phases under triality.

%=================================================================
\begin{table}[H]
\begin{center}
\begin{tabular}{ |c|c|c|c|c|c|c|} 
\hline
N&1&2&3&4&5&6\\
\hline
1&&\underline{2}&\underline{2}&2&2&\\
2&&1&\underline{1}&&&\\\hline
\end{tabular}
\end{center}
\vspace{-.4cm}\caption{Triality connections between the 2 toric phases of Model 2.}
\label{Table 2 - Model 3}
\end{table}
%=================================================================

%=================================================================
\subsection{Model 4: $P_{+-}^{1}(\text{dP}_0)$}
%=================================================================
 
\begin{figure}[H]
\begin{center}
\resizebox{0.25\hsize}{!}{
\includegraphics[height=6cm]{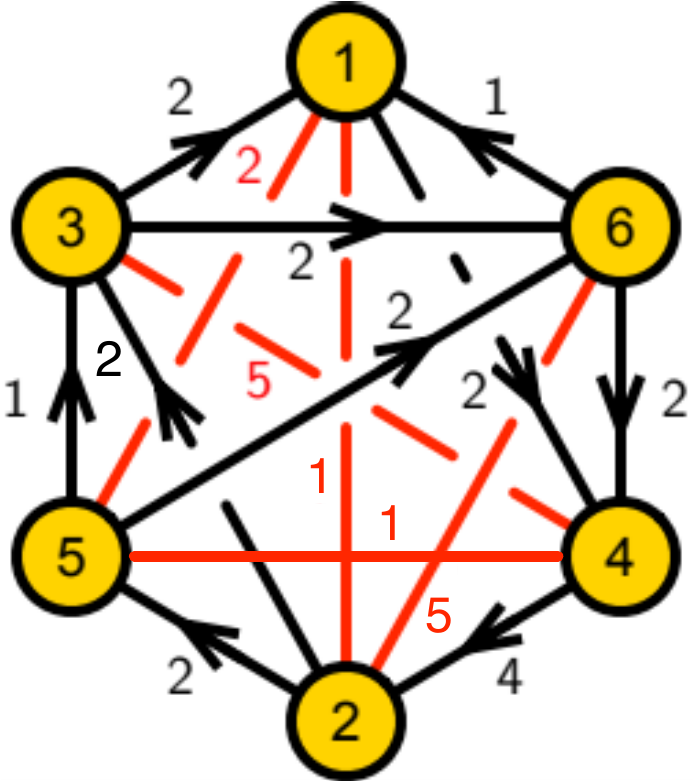} 
}
\caption{
Quiver for Phase 1 of Model 4.
\label{f_quiver_04}}
 \end{center}
 \end{figure}

 Model 4 has a single toric phase, whose quiver is shown in \fref{f_quiver_04}.
 
 %=================================================================
%=================================================================

\vspace{-.5cm}The $J$- and $E$-terms are
\beq
  {\footnotesize
\begin{array}{rrclrcl}
 &  & J &  & & E &  \\
\Lambda_{12}  : &  Z_{23} Z_{31}   &-& Y_{23} X_{31}  &   P_{14} Z_{42 }  &-&  Q_{14} T_{42} \\ 
\Lambda_{15}^{1}  : &  X_{53} X_{31}  &-&  Z_{56} Y_{61}  &   P_{14} X_{42} Q_{25 }  &-&  Q_{14} X_{42} P_{25} \\ 
\Lambda_{15}^{2}  : &  Y_{56} Y_{61}  &-& X_{53} Z_{31}  &   P_{14} Y_{42} Q_{25 }  &-&  Q_{14} Y_{42} P_{25} \\ 
\Lambda_{26}^{1}  : &  Y_{61} Q_{14} Y_{42}   &-& X_{64} Z_{42}  &   P_{25} Y_{56 }  &-&  Y_{23} P_{36} \\ 
\Lambda_{26}^{2}  : &  Z_{64} Z_{42 }  &-&  Y_{61} Q_{14} X_{42}  &   P_{25} Z_{56 }  &-&  Z_{23} P_{36} \\ 
\Lambda_{26}^{3}  : &   Y_{61} P_{14} Y_{42}  &-& X_{64} T_{42}   &   Y_{23} Q_{36}  &-& Q_{25} Y_{56} \\ 
\Lambda_{26}^{4}  : &  Z_{64} T_{42 }  &-&  Y_{61} P_{14} X_{42}  &    Z_{23} Q_{36}  &-& Q_{25} Z_{56}\\ 
\Lambda_{26}^{5}  : &  X_{64} X_{42 }  &-&  Z_{64} Y_{42}  &   P_{25} X_{53} Q_{36 }  &-&  Q_{25} X_{53} P_{36} \\ 
\Lambda_{34}^{1}  : &  X_{42} Q_{25} X_{53}   &-& Z_{42} Y_{23}  &   P_{36} X_{64 }  &-&  X_{31} P_{14} \\ 
\Lambda_{34}^{2}  : &  Z_{42} Z_{23 }  &-&  Y_{42} Q_{25} X_{53}  &   P_{36} Z_{64 }  &-&  Z_{31} P_{14} \\ 
\Lambda_{34}^{3}  : &  X_{42} P_{25} X_{53}  &-& T_{42} Y_{23}  &   X_{31} Q_{14}  &-& Q_{36} X_{64}  \\ 
\Lambda_{34}^{4}  : &  T_{42} Z_{23 }  &-&  Y_{42} P_{25} X_{53}  &   Z_{31} Q_{14}  &-& Q_{36} Z_{64} \\ 
\Lambda_{34}^{5}  : &  Y_{42} Y_{23}  &-& X_{42} Z_{23}  &   P_{36} Y_{61} Q_{14 }  &-&  Q_{36} Y_{61} P_{14} \\ 
\Lambda_{45}  : &  Z_{56} Z_{64}  &-& Y_{56} X_{64}  &   T_{42} Q_{25 }  &-&  Z_{42} P_{25}
 \end{array} 
 }~.~
\label{es0401}
 \eeq
 
 Table \ref{Table 1 - Model 4} summarizes its Fermi content per node.
 %=================================================================
\begin{table}[H]
\begin{center}
\begin{tabular}{ |c|c|c|} 
\hline
Phase & $N_F$ & Fermi Multiplicities\\
\hline
1 & 13 & 1 $\times\textbf{2}$ + 1 $\times\textbf{2}$ + 3 $\times\textbf{5}$ + 1 $\times\textbf{6}$ \\ 
\hline
\end{tabular}
\end{center}
\caption{Basic information regarding the single toric phase of Model 4.}
\label{Table 1 - Model 4}
\end{table}
%=================================================================

%=================================================================
\subsection{Model 5: $Y^{2,5}(\mathbb{CP}^2)$}
%=================================================================
 
\begin{figure}[H]
\begin{center}
\resizebox{0.25\hsize}{!}{
\includegraphics[height=6cm]{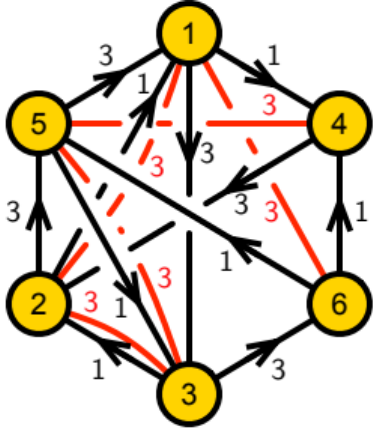} 
}
\caption{
Quiver for Phase 1 of Model 5.
\label{f_quiver_05}}
 \end{center}
 \end{figure}

 Model 5 has a single toric phase, whose quiver is shown in \fref{f_quiver_05}.
 
%=================================================================
%=================================================================

The $J$- and $E$-terms are

  \beq
  {\footnotesize
\begin{array}{rrclrcl}
 &  & J &  & & E &  \\
\Lambda_{12}^{1}  : &  X_{25} X_{51}  &-& Q_{25} Z_{51} &   P_{13} X_{32}  &-&  X_{14} P_{42} \\ 
\Lambda_{12}^{2}  : &  X_{25} R_{51}   &-& P_{25} Z_{51} &   X_{14} Q_{42}  &-& Q_{13} X_{32} \\ 
\Lambda_{16}^{1}  : &  X_{64} Q_{42} Y_{21}   &-& X_{65} X_{51}  &   P_{13} X_{36}  &-&  X_{13} P_{36} \\ 
\Lambda_{16}^{2}  : &  X_{65} Z_{51}  &-&  X_{64} X_{42} Y_{21}  &   P_{13} Q_{36}  &-&  Q_{13} P_{36} \\ 
\Lambda_{16}^{3}  : &  X_{64} P_{42} Y_{21}  &-& X_{65} R_{51}  &   X_{13} Q_{36}  &-& Q_{13} X_{36}  \\ 
\Lambda_{21}^{1}  : &  X_{14} X_{42}   &-& X_{13} X_{32} &   P_{25} X_{51}  &-&  Q_{25} R_{51} \\ 
\Lambda_{23}^{1}  : &  X_{36} X_{64} Q_{42}  &-& Q_{36} X_{64} X_{42}  &   P_{25} Y_{53}  &-&  Y_{21} P_{13} \\ 
\Lambda_{23}^{2}  : &  X_{36} X_{64} P_{42}  &-& P_{36} X_{64} X_{42}  &   Y_{21} Q_{13}  &-& Q_{25} Y_{53} \\ 
\Lambda_{32}^{1}  : &  Y_{21} X_{13}  &-& X_{25} Y_{53}  &   P_{36} X_{64} Q_{42}  &-&  Q_{36} X_{64} P_{42} \\ 
\Lambda_{35}^{1}  : &  Z_{51} Q_{13}   &-& X_{51} X_{13} &   P_{36} X_{65}  &-&  X_{32} P_{25} \\ 
\Lambda_{35}^{2}  : &  Z_{51} P_{13}   &-& R_{51} X_{13} &   X_{32} Q_{25}  &-& Q_{36} X_{65} \\ 
\Lambda_{45}^{1}  : &  X_{51} X_{14}  &-&  Y_{53} Q_{36} X_{64}  &   P_{42} X_{25}  &-&  X_{42} P_{25} \\ 
\Lambda_{45}^{2}  : &  Y_{53} X_{36} X_{64}  &-& Z_{51} X_{14}  &   P_{42} Q_{25}  &-&  Q_{42} P_{25} \\ 
\Lambda_{45}^{3}  : &  R_{51} X_{14}  &-&  Y_{53} P_{36} X_{64}  &   X_{42} Q_{25}  &-& Q_{42} X_{25} \\ 
\Lambda_{53}^{1}  : &  X_{32} X_{25}  &-&  X_{36} X_{65}  &   R_{51} Q_{13}  &-&  X_{51} P_{13}
 \end{array} 
 }~.~
\label{es0501}
 \eeq
 
Table \ref{Table 1 - Model 5} summarizes its Fermi content per node.
%=================================================================
\begin{table}[H]
\begin{center}
\begin{tabular}{ |c|c|c|} 
\hline
Phase & $N_F$ & Fermi Multiplicities\\
\hline
1 & 15 & 2 $\times\textbf{3}$ + 4 $\times\textbf{6}$ \\ 
\hline
\end{tabular}
\end{center}
\caption{Basic information regarding the single toric phase of Model 5.}
\label{Table 1 - Model 5}
\end{table}
%=================================================================

%=================================================================
\subsection{Model 6: $P^{1}_{+-}(\text{dP}_1)$}
%=================================================================

\begin{figure}[H]
\begin{center}
\resizebox{0.3\hsize}{!}{
\includegraphics[height=6cm]{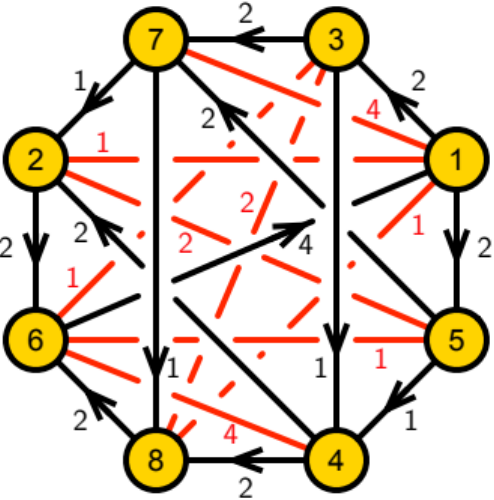} 
}
\caption{Quiver for Phase 1 of Model 6.
\label{f_quiver_06}}
 \end{center}
 \end{figure}

\fref{f_quiver_06} shows the quiver for Phase 1 of Model 6.

%=================================================================
%=================================================================

The $J$- and $E$-terms are
\beq
{\footnotesize
\begin{array}{rrclrcl}
 &  & J &  & & E &  \\
\Lambda_{17}^{1} : &  X_{78} Y_{86} Y_{61}  &-& X_{72} Q_{26} X_{61} &   P_{15} X_{57}  &-&  X_{13} P_{37} \\ 
\Lambda_{17}^{2} : &  X_{72} Q_{26} Z_{61}  &-&  X_{78} X_{86} Y_{61}  &   P_{15} Y_{57}  &-&  Y_{13} P_{37} \\ 
\Lambda_{17}^{3} : &  X_{78} Y_{86} R_{61}  &-& X_{72} P_{26} X_{61}  &  X_{13} Q_{37}   &-& Q_{15} X_{57} \\ 
\Lambda_{17}^{4} : &  X_{72} P_{26} Z_{61}  &-&  X_{78} X_{86} R_{61}  &  Y_{13} Q_{37}  &-& Q_{15} Y_{57} \\ 
\Lambda_{18}^{1} : &  X_{86} X_{61}  &-&  Y_{86} Z_{61}  &   P_{15} X_{54} Q_{48}  &-&  Q_{15} X_{54} P_{48} \\ 
\Lambda_{21}^{1} : &  X_{13} X_{34} Y_{42}  &-&  Y_{13} X_{34} X_{42}  &   P_{26} Y_{61}  &-&  Q_{26} R_{61} \\ 
\Lambda_{25}^{1} : &  X_{54} X_{42}  &-&  X_{57} X_{72}  &   P_{26} X_{61} Q_{15}  &-&  Q_{26} X_{61} P_{15} \\ 
\Lambda_{25}^{2} : &  Y_{57} X_{72}  &-& X_{54} Y_{42}  &   P_{26} Z_{61} Q_{15}  &-&  Q_{26} Z_{61} P_{15} \\ 
\Lambda_{36}^{1} : &  Z_{61} Y_{13}  &-& X_{61} X_{13} &   P_{37} X_{72} Q_{26}  &-&  Q_{37} X_{72} P_{26} \\ 
\Lambda_{38}^{1} : &  Y_{86} Y_{61} X_{13}  &-& X_{86} Y_{61} Y_{13}  &   P_{37} X_{78}  &-&  X_{34} P_{48} \\ 
\Lambda_{38}^{2} : &  Y_{86} R_{61} X_{13}  &-& X_{86} R_{61} Y_{13}  &   X_{34} Q_{48}  &-& Q_{37} X_{78} \\ 
\Lambda_{46}^{1} : &  X_{61} Q_{15} X_{54}  &-&  Y_{61} Y_{13} X_{34}  &   P_{48} X_{86}  &-&  X_{42} P_{26} \\ 
\Lambda_{46}^{2} : &  Y_{61} X_{13} X_{34}  &-&  Z_{61} Q_{15} X_{54}  &   P_{48} Y_{86}  &-&  Y_{42} P_{26} \\ 
\Lambda_{46}^{3} : &  X_{61} P_{15} X_{54}  &-& R_{61} Y_{13} X_{34}  &  X_{42} Q_{26}   &-& Q_{48} X_{86}  \\ 
\Lambda_{46}^{4} : &  R_{61} X_{13} X_{34}  &-&  Z_{61} P_{15} X_{54}  &   Y_{42} Q_{26}  &-& Q_{48} Y_{86} \\ 
\Lambda_{65}^{1} : &  X_{57} X_{78} Y_{86}  &-&  Y_{57} X_{78} X_{86}  &   R_{61} Q_{15}  &-&  Y_{61} P_{15}
 \end{array} 
 }~.~
\label{es0601}
 \eeq

This model has 6 toric phases, which are summarized in Table \ref{Table 1 - Model 6}.

%=================================================================
\begin{table}[H]
\begin{center}
\begin{tabular}{ |c|c|c|c|} 
\hline
 Phase & Path & F & Fermi Multiplicities\\
\hline
1&&16&4$\times\textbf{3}$+2$\times\textbf{4}$+2$\times\textbf{6}$\\ 
2&-2&16&1$\times\textbf{2}$+3$\times\textbf{3}$+3$\times\textbf{5}$+1$\times\textbf{6}$\\ 
3&4&20&2$\times\textbf{3}$+2$\times\textbf{4}$+2$\times\textbf{5}$+1$\times\textbf{6}$+1$\times\textbf{10}$\\ 
4&-2,3&16&4$\times\textbf{3}$+4$\times\textbf{5}$\\ 
5&-2,-4&18&2$\times\textbf{2}$+4$\times\textbf{5}$+2$\times\textbf{6}$\\ 
6&4,-7&28&2$\times\textbf{4}$+4$\times\textbf{7}$+2$\times\textbf{10}$\\ 
\hline
\end{tabular}
\end{center}
\caption{Basic information regarding the 6 toric phases of Model 6.}
\label{Table 1 - Model 6}
\end{table}
%=================================================================

Table \ref{Table 2 - Model 6} summarizes the connection between the toric phases under triality.

%=================================================================
\begin{table}[h]
\begin{center}
\begin{tabular}{ |c|c|c|c|c|c|c|c|c|} 
\hline
N&1&2&3&4&5&6&7&8\\
\hline
1&&\underline{2}&2&3&2&&\underline{3}&\underline{2}\\
2&&1&4&2 , \underline{5}&&&&\underline{3}\\
3&&\underline{2}&&\underline{1}&&&\underline{6}&\underline{2}\\
4&&2&\underline{2}&\underline{2}&&&2&\\
5&&&2 , \underline{2}&2 , \underline{2}&&&&\\
6&&&&\underline{3}&&&3&\\\hline
\end{tabular}
\end{center}
\caption{Triality connections between the 6 toric phases of Model 6.}
\label{Table 2 - Model 6}
\end{table}
%=================================================================

%=================================================================
\subsection{Model 7: $P_{++-}(\text{dP}_0)$}
%=================================================================
 
\begin{figure}[H]
\begin{center}
\resizebox{0.3\hsize}{!}{
\includegraphics[height=6cm]{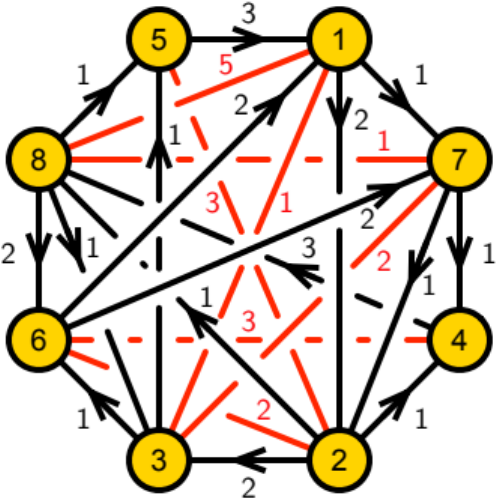} 
}
\caption{Quiver for Phase 1 of Model 7.
\label{f_quiver_07}}
 \end{center}
 \end{figure}

 \fref{f_quiver_07} shows the quiver for Phase 1 of Model 7.
 
%=================================================================
%=================================================================

The $J$- and $E$-terms are
\beq
{\footnotesize
\begin{array}{rrclrcl}
 &  & J &  & & E &  \\
\Lambda_{25}^{1} : &  X_{51} X_{12}  &-&  Z_{51} Y_{12} &  P_{24} X_{48} Q_{85}  &-&  X_{28} P_{83} P_{35} \\ 
\Lambda_{25}^{2} : &  Y_{51} Y_{12}  &-&  X_{51} Z_{17} P_{72} &  P_{24} Y_{48} Q_{85}  &-& Y_{23} P_{35} \\ 
\Lambda_{25}^{3} : &  Z_{51} Z_{17} P_{72}  &-& Y_{51} X_{12}  &  P_{24} Z_{48} Q_{85}  &-& Z_{23} P_{35} \\ 
\Lambda_{31}^{1} : &  Y_{12} Y_{23}  &-& X_{12} Z_{23} &  P_{35} Y_{51}  &-&  Y_{36} P_{61} \\ 
\Lambda_{62}^{1} : &  X_{28} X_{86}  &-&  Z_{23} Y_{36} &  P_{61} X_{12}  &-&  X_{67} P_{72} \\ 
\Lambda_{62}^{2} : &  Y_{23} Y_{36}  &-& X_{28} Z_{86}  &  P_{61} Y_{12}  &-&  Y_{67} P_{72} \\ 
\Lambda_{64}^{1} : &  X_{48} X_{86}  &-&  Z_{48} P_{83} Y_{36} &  X_{67} Q_{74}  &-&  Q_{61} X_{12} P_{24} \\ 
\Lambda_{64}^{2} : &  Y_{48} P_{83} Y_{36}  &-& X_{48} Z_{86} &  Y_{67} Q_{74}  &-&  Q_{61} Y_{12} P_{24} \\ 
\Lambda_{64}^{3} : &  Z_{48} Z_{86}  &-& Y_{48} X_{86}  &  P_{61} Z_{17} Q_{74}  &-&  Q_{61} Z_{17} P_{72} P_{24} \\ 
\Lambda_{73}^{1} : &  Y_{36} Y_{67}  &-&  P_{35} X_{51} Z_{17} &  P_{72} Y_{23}  &-&  Q_{74} Y_{48} P_{83} \\ 
\Lambda_{73}^{2} : &  P_{35} Z_{51} Z_{17}  &-& Y_{36} X_{67}  &  P_{72} Z_{23}  &-&  Q_{74} Z_{48} P_{83} \\ 
\Lambda_{78}^{1} : &  X_{86} X_{67}  &-&  Z_{86} Y_{67} &  P_{72} X_{28}  &-&  Q_{74} X_{48} \\ 
\Lambda_{81}^{1} : &  X_{12} P_{24} X_{48}  &-&  Z_{17} P_{72} P_{24} Y_{48} & X_{86} Q_{61}   &-& Q_{85} X_{51}\\ 
\Lambda_{81}^{2} : &  Y_{12} P_{24} Y_{48}  &-& X_{12} P_{24} Z_{48} &  P_{83} Y_{36} Q_{61}  &-& Q_{85} Y_{51} \\ 
\Lambda_{81}^{3} : &  Z_{17} P_{72} P_{24} Z_{48}  &-& Y_{12} P_{24} X_{48}  &  Z_{86} Q_{61}  &-& Q_{85} Z_{51} \\ 
\Lambda_{81}^{4} : &  X_{12} X_{28}  &-&  Z_{17} Q_{74} Y_{48} &  P_{83} P_{35} X_{51}  &-& X_{86} P_{61} \\ 
\Lambda_{81}^{5} : &  Z_{17} Q_{74} Z_{48}  &-& Y_{12} X_{28} &  P_{83} P_{35} Z_{51}  &-& Z_{86} P_{61} 
 \end{array} 
 }~.~
\label{E_J_C_+-}
\eeq

This model has 6 toric phases, which are summarized in Table \ref{Table 1 - Model 7}.

%=================================================================
\begin{table}[h]
\begin{center}
\begin{tabular}{ |c|c|c|c|} 
\hline
 Phase & Path & F & Fermi Multiplicities\\
\hline
1&&17&4$\times\textbf{3}$+2$\times\textbf{5}$+2$\times\textbf{6}$\\ 
2&-3&15&6$\times\textbf{3}$+1$\times\textbf{5}$+1$\times\textbf{7}$\\ 
3&4&20&4$\times\textbf{3}$+2$\times\textbf{6}$+2$\times\textbf{8}$\\ 
4&5&17&4$\times\textbf{3}$+2$\times\textbf{5}$+2$\times\textbf{6}$\\ 
5&-3,-2&17&4$\times\textbf{3}$+2$\times\textbf{5}$+2$\times\textbf{6}$\\ 
6&-3,4&20&3$\times\textbf{3}$+2$\times\textbf{5}$+2$\times\textbf{6}$+1$\times\textbf{9}$\\ 
\hline
\end{tabular}
\end{center}
\caption{Basic information regarding the 6 toric phases of Model 7.}
\label{Table 1 - Model 7}
\end{table}
%=================================================================

Table \ref{Table 2 - Model 7} summarizes the connection between the toric phases under triality.

%=================================================================
\begin{table}[h]
\begin{center}
\begin{tabular}{ |c|c|c|c|c|c|c|c|c|} 
\hline
N&1&2&3&4&5&6&7&8\\
\hline
1&&&\underline{2}&3&4&&\underline{2}&\\
2&&\underline{5}&1&6&2&\underline{5}&\underline{1}&\\
3&&&\underline{6}&\underline{1}&1&&6&\\
4&&&5&1&\underline{1}&&\underline{5}&\\
5&&2&&&4&\underline{2}&\underline{6}&\\
6&&&3&\underline{2}&5&&&\\\hline
\end{tabular}
\end{center}
\caption{Triality connections between the 6 toric phases of Model 7.}
\label{Table 2 - Model 7}
\end{table}
%=================================================================

\newpage
 
%=================================================================
\subsection{Model 8: $P_{++-}H_{+}(\text{dP}_0)$}
%=================================================================
 
\begin{figure}[H]
\begin{center}
\resizebox{0.3\hsize}{!}{
\includegraphics[height=6cm]{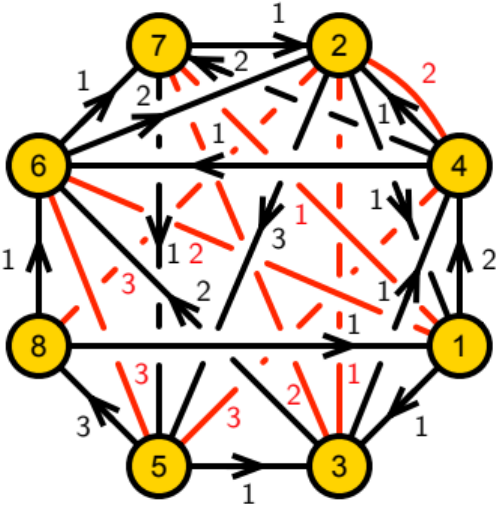} 
}
\caption{Quiver for Phase 1 of Model 8.
\label{f_quiver_08}}
 \end{center}
 \end{figure}

\fref{f_quiver_08} shows the quiver for Phase 1 of Model 8. 
 
%=================================================================
%=================================================================

The $J$- and $E$-terms are
\beq
{\footnotesize
\begin{array}{rrclrcl}
 &  & J &  & & E &  \\
\Lambda_{16}^{1}  : &  X_{67} Y_{75} Q_{58} X_{81}  &-& X_{62} X_{21} &   P_{14} X_{46}  &-&   X_{13} P_{36} \\ 
\Lambda_{16}^{2}  : &  X_{67} Y_{75} P_{58} X_{81}  &-& R_{62} X_{21} &   X_{13} Q_{36}  &-& Q_{14} X_{46}  \\ 
\Lambda_{17}^{1}  : &  X_{72} X_{21}  &-&   Y_{75} Y_{58} X_{81}  &   P_{14} Q_{47}  &-&   Q_{14} P_{47} \\ 
\Lambda_{24}^{1}  : &  Q_{47} X_{72}  &-&   X_{46} X_{62}  &   P_{25} X_{53} X_{34}  &-& X_{21} P_{14}  \\ 
\Lambda_{24}^{2}  : &  P_{47} X_{72}  &-&   X_{46} R_{62}  &   X_{21} Q_{14}  &-&   Q_{25} X_{53} X_{34} \\ 
\Lambda_{28}^{1}  : &  X_{86} X_{62}  &-&   X_{81} Q_{14} Y_{42}  &   P_{25} Y_{58}  &-&   Y_{25} P_{58} \\ 
\Lambda_{28}^{2}  : &  X_{81} X_{13} X_{34} Y_{42}   &-& X_{86} X_{67} X_{72} &   P_{25} Q_{58}  &-&   Q_{25} P_{58} 
 \end{array} 
 }~.~
\nonumber
\label{E_J_C_+-}
 \eeq

\beq
{\footnotesize
\begin{array}{rrclrcl}
 &  & J &  & & E &  \\
 \Lambda_{28}^{3}  : &  X_{86} R_{62}  &-&   X_{81} P_{14} Y_{42}  &    Y_{25} Q_{58} &-& Q_{25} Y_{58} \\ 
\Lambda_{32}^{1}  : &  Y_{25} X_{53}    &-& X_{21} X_{13}  &   P_{36} X_{62}  &-&   Q_{36} R_{62} \\ 
\Lambda_{37}^{1}  : &  Y_{75} Q_{58} X_{81} X_{13}   &-& X_{72} Q_{25} X_{53}  &   P_{36} X_{67}  &-&   X_{34} P_{47} \\ 
\Lambda_{37}^{2}  : &  Y_{75} P_{58} X_{81} X_{13}   &-& X_{72} P_{25} X_{53}  &    X_{34} Q_{47}  &-& Q_{36} X_{67} \\
 \Lambda_{45}^{1}  : &   Q_{58} X_{81} X_{13} X_{34}   &-& Y_{58} X_{81} Q_{14}  &   P_{47} Y_{75}  &-&   Y_{42} P_{25} \\ 
\Lambda_{45}^{2}  : &   P_{58} X_{81} X_{13} X_{34}  &-& Y_{58} X_{81} P_{14}  &   Y_{42} Q_{25}  &-& Q_{47} Y_{75} \\
\Lambda_{54}^{1}  : &   X_{46} X_{67} Y_{75}  &-& Y_{42} Y_{25}  &   P_{58} X_{81} Q_{14}  &-&   Q_{58} X_{81} P_{14} \\ 
\Lambda_{56}^{1}  : &  X_{62} Y_{25}  &-&   X_{67} X_{72} Q_{25}  &   P_{58} X_{86}  &-&   X_{53} P_{36} \\ 
\Lambda_{56}^{2}  : &  R_{62} Y_{25}  &-&   X_{67} X_{72} P_{25}  &   X_{53} Q_{36}  &-& Q_{58} X_{86}  \\ 
\Lambda_{65}^{1}  : &  Y_{58} X_{86}  &-&   X_{53} X_{34} X_{46}  &   R_{62} Q_{25}  &-&   X_{62} P_{25}
 \end{array} 
 }~.~
\label{E_J_C_+-}
 \eeq
 
\bigskip

This model has 4 toric phases, which are summarized in Table \ref{Table 1 - Model 8}.

%=================================================================
\begin{table}[h]
\begin{center}
\begin{tabular}{ |c|c|c|c|} 
\hline
 Phase & Path & F & Fermi Multiplicities\\
\hline
1&&17&4$\times\textbf{3}$+2$\times\textbf{5}$+2$\times\textbf{6}$\\ 
2&3&17&4$\times\textbf{3}$+2$\times\textbf{5}$+2$\times\textbf{6}$\\ 
3&-7&18&4$\times\textbf{3}$+1$\times\textbf{5}$+2$\times\textbf{6}$+1$\times\textbf{7}$\\ 
4&3,1&20&3$\times\textbf{3}$+1$\times\textbf{5}$+3$\times\textbf{6}$+1$\times\textbf{8}$\\ 
\hline
\end{tabular}
\end{center}
\caption{Basic information regarding the 4 toric phases of Model 8.}
\label{Table 1 - Model 8}
\end{table}
%=================================================================

Table \ref{Table 2 - Model 8} summarizes the connection between the toric phases under triality.

%=================================================================
\begin{table}[h]
\begin{center}
\begin{tabular}{ |c|c|c|c|c|c|c|c|c|} 
\hline
N&1&2&3&4&5&6&7&8\\
\hline
1&1&&2&&&&\underline{3}&\underline{3}\\
2&4&&\underline{1}&&&1&&\underline{4}\\
3&4&&&\underline{4}&&&1&\underline{1}\\
4&\underline{2}&&3&&&3&&\\\hline
\end{tabular}
\end{center}
\caption{Triality connections between the 4 toric phases of Model 8.}
\label{Table 2 - Model 8}
\end{table}
%=================================================================

\newpage

%=================================================================
\subsection{Model 9: $Y^{1,2}(\mathbb{CP}^1\times\mathbb{CP}^1)$}
%=================================================================
 
\begin{figure}[H]
\begin{center}
\resizebox{0.3\hsize}{!}{
\includegraphics[height=6cm]{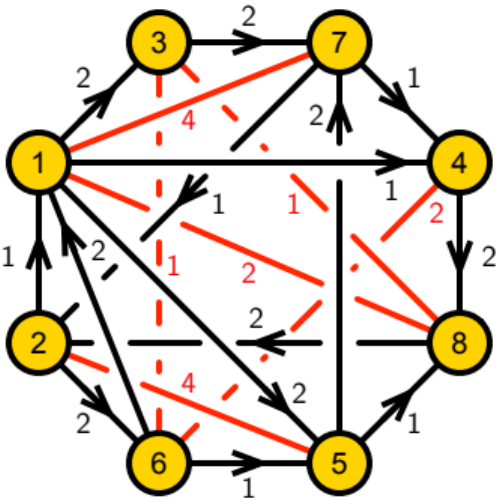} 
}
\caption{Quiver for Phase 1 of Model 9.
\label{f_quiver_09}}
 \end{center}
 \end{figure}

\fref{f_quiver_09} shows the quiver for Phase 1 of Model 9.
 
%=================================================================
%=================================================================

The $J$- and $E$-terms are
\beq
{\footnotesize
\begin{array}{rrclrcl}
 &  & J &  & & E &  \\
\Lambda_{17}^{1}   : & X_{74} Q_{48} Y_{82} Y_{21}   &-& X_{72} Q_{26} R_{61} &  P_{15} X_{57}  &-&  X_{13} P_{37} \\ 
\Lambda_{17}^{2}   : & X_{72} Q_{26} T_{61}  &-&  X_{74} Q_{48} X_{82} Y_{21}  &  P_{15} Y_{57}  &-&  Y_{13} P_{37} \\ 
\Lambda_{17}^{3}   : & X_{74} P_{48} Y_{82} Y_{21}   &-& X_{72} P_{26} R_{61} &  X_{13} Q_{37}  &-& Q_{15} X_{57}  \\ 
\Lambda_{17}^{4}   : & X_{72} P_{26} T_{61}  &-&  X_{74} P_{48} X_{82} Y_{21}  &  Y_{13} Q_{37}  &-& Q_{15} Y_{57} \\ 
\Lambda_{18}^{1}   : & X_{82} Q_{26} R_{61}  &-&  Y_{82} Q_{26} T_{61}  &  P_{15} X_{58}  &-&  X_{14} P_{48} \\ 
\Lambda_{18}^{2}   : & X_{82} P_{26} R_{61}  &-&  Y_{82} P_{26} T_{61}  &  X_{14} Q_{48}  &-& Q_{15} X_{58} \\ 
\Lambda_{25}^{1}   : & X_{58} X_{82}  &-& X_{57} X_{72}  &  P_{26} R_{61} Q_{15}  &-&  Q_{26} R_{61} P_{15} \\ 
\Lambda_{25}^{2}   : & X_{57} X_{74} Q_{48} Y_{82}  &-&  Y_{57} X_{74} Q_{48} X_{82}  &  P_{26} Y_{65}  &-&  Y_{21} P_{15} \\ 
\Lambda_{25}^{3}   : & Y_{57} X_{72}   &-& X_{58} Y_{82}  &  P_{26} T_{61} Q_{15}  &-&  Q_{26} T_{61} P_{15} \\ 
\Lambda_{25}^{4}   : & X_{57} X_{74} P_{48} Y_{82}  &-&  Y_{57} X_{74} P_{48} X_{82}  &  Y_{21} Q_{15}  &-& Q_{26} Y_{65}  \\ 
\Lambda_{36}^{1}   : & T_{61} Y_{13}  &-& R_{61} X_{13} &  P_{37} X_{72} Q_{26}  &-&  Q_{37} X_{72} P_{26} \\ 
\Lambda_{38}^{1}   : & Y_{82} Y_{21} X_{13}   &-& X_{82} Y_{21} Y_{13}&  P_{37} X_{74} Q_{48}  &-&  Q_{37} X_{74} P_{48} \\ 
\Lambda_{46}^{1}   : & R_{61} X_{14}  &-&  Y_{65} Y_{57} X_{74}  &  P_{48} X_{82} Q_{26}  &-&  Q_{48} X_{82} P_{26} \\ 
\Lambda_{46}^{2}   : & Y_{65} X_{57} X_{74}   &-& T_{61} X_{14} &  P_{48} Y_{82} Q_{26}  &-&  Q_{48} Y_{82} P_{26}
 \end{array} 
 }~.~
\label{es0901}
\eeq
 
 \newpage

This model has 8 toric phases, which are summarized in Table \ref{Table 1 - Model 9}.

%=================================================================
\begin{table}[H]
\begin{center}
\begin{tabular}{ |c|c|c|c|} 
\hline
 Phase & Path & F & Fermi Multiplicities\\
\hline
1&&14&2$\times\textbf{2}$+2$\times\textbf{3}$+3$\times\textbf{4}$+1$\times\textbf{6}$\\ 
2&3&12&4$\times\textbf{2}$+4$\times\textbf{4}$\\ 
3&-3&14&2$\times\textbf{2}$+6$\times\textbf{4}$\\ 
4&-4&16&2$\times\textbf{2}$+2$\times\textbf{4}$+4$\times\textbf{5}$\\ 
5&6&18&4$\times\textbf{3}$+2$\times\textbf{4}$+2$\times\textbf{8}$\\ 
6&-7&20&2$\times\textbf{3}$+3$\times\textbf{4}$+2$\times\textbf{6}$+1$\times\textbf{10}$\\ 
7&-8&18&4$\times\textbf{3}$+2$\times\textbf{4}$+2$\times\textbf{8}$\\ 
8&-7,-8&26&2$\times\textbf{3}$+1$\times\textbf{4}$+2$\times\textbf{5}$+1$\times\textbf{8}$+2$\times\textbf{12}$\\ 
\hline
\end{tabular}
\end{center}
\caption{Basic information regarding the 8 toric phases of Model 9.}
\label{Table 1 - Model 9}
\end{table}
%=================================================================

Table \ref{Table 2 - Model 9} summarizes the connection between the toric phases under triality.

%=================================================================
\begin{table}[h]
\begin{center}
\begin{tabular}{ |c|c|c|c|c|c|c|c|c|} 
\hline
N&1&2&3&4&5&6&7&8\\
\hline
1&&&2 , \underline{3}&1 , \underline{4}&&5&\underline{6}&\underline{7}\\
2&&&3 , \underline{1}&3 , \underline{1}&&1 , \underline{3}&1 , \underline{3}&\\
3&&&1 , \underline{2}&2 , \underline{1}&&&6&\underline{6}\\
4&&&1 , \underline{1}&1 , \underline{1}&&&&\\
5&&&1&7&1&\underline{1}&\underline{7}&\underline{1}\\
6&&&3&&&7&1&\underline{8}\\
7&&&\underline{6}&5&\underline{6}&1&\underline{8}&1\\
8&&&&&&6&7&6\\\hline
\end{tabular}
\end{center}
\caption{Triality connections between the 8 toric phases of Model 9.}
\label{Table 2 - Model 9}
\end{table}
%=================================================================

\newpage

%=================================================================
\subsection{Model 10: $P^{3}_{+-}(\text{dP}_1)$}
%=================================================================
 
\fref{f_quiver_02} shows the quiver for Phase 1 of Model 10. 
 
%=================================================================
\begin{figure}[H]
\begin{center}
\resizebox{0.3\hsize}{!}{
\includegraphics[height=6cm]{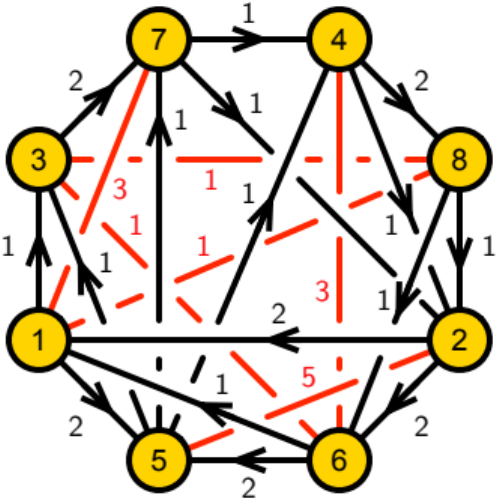} 
}
\caption{Quiver for Phase 1 of Model 10.
\label{f_quiver_10}}
 \end{center}
 \end{figure}
%=================================================================

The $J$- and $E$-terms are
\beq
{\footnotesize
\begin{array}{rrclrcl}
 &  & J &  & & E &  \\
\Lambda_{17}^{1}   : & X_{74} Q_{48} Y_{82} Y_{21}  &-& X_{72} Q_{26} X_{61}  &   P_{15} X_{57}  &-&  X_{13} P_{37} \\ 
\Lambda_{17}^{2}   : & X_{72} Z_{21}  &-&  X_{74} X_{42} Y_{21} &   P_{15} Y_{53} Q_{37}  &-&  Q_{15} Y_{53} P_{37} \\ 
\Lambda_{17}^{3}   : & X_{74} P_{48} Y_{82} Y_{21}  &-& X_{72} P_{26} X_{61} & X_{13} Q_{37}  &-& Q_{15} X_{57}  \\ 
\Lambda_{18}^{1}   : & X_{86} X_{61}  &-&  Y_{82} Z_{21} &   P_{15} X_{54} Q_{48}  &-&  Q_{15} X_{54} P_{48} \\ 
\Lambda_{25}^{1}   : & X_{54} X_{42}  &-&  X_{57} X_{72} &   P_{26} X_{61} Q_{15}  &-&  Q_{26} X_{61} P_{15} \\ 
\Lambda_{25}^{2}   : & X_{57} X_{74} Q_{48} Y_{82}  &-&  Y_{53} Q_{37} X_{74} X_{42} &   P_{26} Y_{65}  &-&  Y_{21} P_{15} \\ 
\Lambda_{25}^{3}   : & Y_{53} Q_{37} X_{72}  &-& X_{54} Q_{48} Y_{82} &   P_{26} Z_{65}  &-&  Z_{21} P_{15} \\ 
\Lambda_{25}^{4}   : & X_{57} X_{74} P_{48} Y_{82}  &-&  Y_{53} P_{37} X_{74} X_{42} &  Y_{21} Q_{15} &-& Q_{26} Y_{65} \\ 
\Lambda_{25}^{5}   : & Y_{53} P_{37} X_{72}  &-& X_{54} P_{48} Y_{82}  & Z_{21} Q_{15}   &-& Q_{26} Z_{65}  \\ 
\Lambda_{36}^{1}   : & Z_{65} Y_{53}  &-& X_{61} X_{13}  &   P_{37} X_{72} Q_{26}  &-&  Q_{37} X_{72} P_{26} \\ 
\Lambda_{38}^{1}   : & Y_{82} Y_{21} X_{13}  &-& X_{86} Y_{65} Y_{53} &   P_{37} X_{74} Q_{48}  &-&  Q_{37} X_{74} P_{48} \\ 
\Lambda_{46}^{1}   : & X_{61} Q_{15} X_{54}  &-&  Y_{65} Y_{53} Q_{37} X_{74} &   P_{48} X_{86}  &-&  X_{42} P_{26} \\ 
\Lambda_{46}^{2}   : & Y_{65} X_{57} X_{74}  &-& Z_{65} X_{54}  &   P_{48} Y_{82} Q_{26}  &-&  Q_{48} Y_{82} P_{26} \\ 
\Lambda_{46}^{3}   : & X_{61} P_{15} X_{54}  &-&  Y_{65} Y_{53} P_{37} X_{74} & X_{42} Q_{26}  &-& Q_{48} X_{86} 
 \end{array} 
 }~.~
\label{E_J_C_+-}
 \eeq
 
\newpage 
 
This model has 8 toric phases, which are summarized in Table \ref{Table 1 - Model 10}.

%=================================================================
\begin{table}[H]
\begin{center}
\begin{tabular}{ |c|c|c|c|} 
\hline
 Phase & Path & F & Fermi Multiplicities\\
\hline
1&&14&2$\times\textbf{2}$+2$\times\textbf{3}$+2$\times\textbf{4}$+2$\times\textbf{5}$\\ 
2&3&14&1$\times\textbf{2}$+5$\times\textbf{3}$+1$\times\textbf{5}$+1$\times\textbf{6}$\\ 
3&-3&16&1$\times\textbf{2}$+3$\times\textbf{3}$+3$\times\textbf{5}$+1$\times\textbf{6}$\\ 
4&4&18&1$\times\textbf{2}$+1$\times\textbf{3}$+2$\times\textbf{4}$+1$\times\textbf{5}$+3$\times\textbf{6}$\\ 
5&3,4&19&1$\times\textbf{2}$+2$\times\textbf{3}$+2$\times\textbf{5}$+2$\times\textbf{6}$+1$\times\textbf{8}$\\ 
6&3,-6&17&4$\times\textbf{3}$+2$\times\textbf{5}$+2$\times\textbf{6}$\\ 
7&3,-7&16&4$\times\textbf{3}$+2$\times\textbf{4}$+2$\times\textbf{6}$\\ 
8&3,8&18&1$\times\textbf{2}$+3$\times\textbf{3}$+1$\times\textbf{5}$+1$\times\textbf{6}$+2$\times\textbf{7}$\\ 
\hline
\end{tabular}
\end{center}
\caption{Basic information regarding the 8 toric phases of Model 10.}
\label{Table 1 - Model 10}
\end{table}
%=================================================================

Table \ref{Table 2 - Model 10} summarizes the connection between the toric phases under triality.

%=================================================================
\begin{table}[h]
\begin{center}
\begin{tabular}{ |c|c|c|c|c|c|c|c|c|} 
\hline
N&1&2&3&4&5&6&7&8\\
\hline
1&&&2 , \underline{3}&4&&&\underline{4}&3 , \underline{2}\\
2&\underline{2}&&3 , \underline{1}&5&&\underline{6}&\underline{7}&8\\
3&\underline{7}&&1 , \underline{2}&8&&&&\underline{8}\\
4&&&5 , \underline{8}&\underline{1}&&&&\underline{7}\\
5&\underline{6}&&8 , \underline{4}&\underline{2}&&&&\\
6&\underline{5}&&&&&2&\underline{2}&5\\
7&\underline{3}&4&\underline{4}&&&\underline{2}&2&3\\
8&&&\underline{3}&&&4 , \underline{5}&\underline{3}&\underline{2}\\\hline
\end{tabular}
\end{center}
\caption{Triality connections between the 8 toric phases of Model 10.}
\label{Table 2 - Model 10}
\end{table}
%=================================================================

\newpage

%=================================================================
\subsection{Model 11: $P^{0}_{+-}(\text{dP}_1)$}
%=================================================================

\begin{figure}[H]
\begin{center}
\resizebox{0.3\hsize}{!}{
\includegraphics[height=6cm]{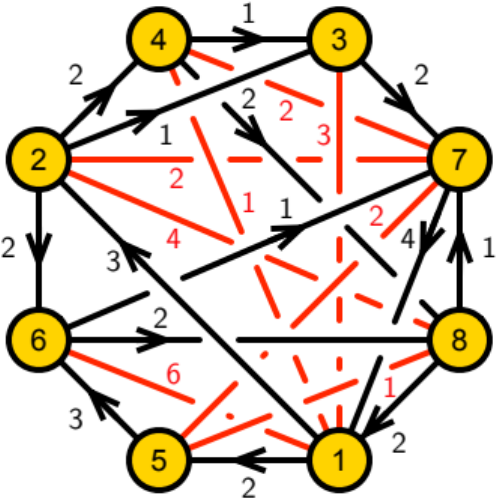} 
}
\caption{Quiver for Phase 1 of Model 11.
\label{f_quiver_11}}
 \end{center}
 \end{figure}

\fref{f_quiver_11} shows the quiver for Phase 1 of Model 11.

%=================================================================
%================================================================= 

The $J$- and $E$-terms are
\beq
{\footnotesize
\begin{array}{rrclrcl}
 &  & J &  & & E &  \\
\Lambda_{16}^{1}  : & Y_{68} X_{81}  &-& X_{67} X_{71}  &  P_{15} X_{56}  &-& X_{12} P_{26} \\ 
\Lambda_{16}^{2}  : & X_{67} Y_{71}  &-& X_{68} X_{81} &  P_{15} Y_{56}  &-& Y_{12} P_{26} \\ 
\Lambda_{16}^{3}  : & X_{68} X_{87} X_{71}  &-& Y_{68} X_{87} Y_{71} &  P_{15} Z_{56}  &-& Z_{12} P_{26} \\ 
\Lambda_{16}^{4}  : & Y_{68} S_{81}  &-& X_{67} S_{71} &  X_{12} Q_{26}  &-& Q_{15} X_{56} \\ 
\Lambda_{16}^{5}  : & X_{67} T_{71}  &-& X_{68} S_{81} &  Y_{12} Q_{26}  &-& Q_{15} Y_{56} \\ 
\Lambda_{16}^{6}  : & X_{68} X_{87} S_{71}  &-& Y_{68} X_{87} T_{71} &  Z_{12} Q_{26}  &-& Q_{15} Z_{56} \\ 
\Lambda_{27}^{1}  : & Y_{71} Y_{12}  &-& X_{71} X_{12}  &  P_{26} X_{67}  &-& X_{23} P_{37} \\ 
\Lambda_{27}^{2}  : & T_{71} Y_{12}  &-& S_{71} X_{12}  &  X_{23} Q_{37}  &-& Q_{26} X_{67} \\ 
\Lambda_{28}^{1}  : & X_{87} X_{71} Z_{12}  &-& X_{81} Y_{12} &  P_{26} X_{68}  &-& X_{24} P_{48} \\ 
\Lambda_{28}^{2}  : & X_{81} X_{12}  &-& X_{87} Y_{71} Z_{12} &  P_{26} Y_{68}  &-& Y_{24} P_{48} \\ 
\Lambda_{28}^{3}  : & X_{87} S_{71} Z_{12}  &-& S_{81} Y_{12}  &  X_{24} Q_{48}  &-& Q_{26} X_{68} \\ 
\Lambda_{28}^{4}  : & S_{81} X_{12}  &-& X_{87} T_{71} Z_{12} &  Y_{24} Q_{48}  &-& Q_{26} Y_{68}  \\ 
\Lambda_{31}^{1}  : & Z_{12} X_{24} X_{43}  &-& X_{12} X_{23}  &  P_{37} X_{71}  &-& Q_{37} S_{71} \\ 
\Lambda_{31}^{2}  : & Y_{12} X_{23}  &-& Z_{12} Y_{24} X_{43} &  P_{37} Y_{71}  &-& Q_{37} T_{71} \\ 
\Lambda_{41}  : & X_{12} Y_{24}  &-& Y_{12} X_{24} &  P_{48} X_{81}  &-& Q_{48} S_{81} \\ 
\Lambda_{47}^{1}  : & X_{71} Z_{12} X_{24}  &-& Y_{71} Z_{12} Y_{24} &  P_{48} X_{87}  &-& X_{43} P_{37} \\ 
\Lambda_{47}^{2}  : & S_{71} Z_{12} X_{24}  &-& T_{71} Z_{12} Y_{24} &  X_{43} Q_{37}  &-& Q_{48} X_{87} \\ 
\Lambda_{75}^{1}  : & Z_{56} X_{68} X_{87}  &-& X_{56} X_{67}  &  S_{71} Q_{15}  &-& X_{71} P_{15} \\ 
\Lambda_{75}^{2}  : & Y_{56} X_{67}  &-& Z_{56} Y_{68} X_{87} &  T_{71} Q_{15}  &-& Y_{71} P_{15} \\ 
\Lambda_{85}  : & X_{56} Y_{68}  &-& Y_{56} X_{68} &  S_{81} Q_{15}  &-& X_{81} P_{15}
 \end{array} 
 }
\label{E_J_C_+-}
 \eeq

This model has 17 toric phases, which are summarized in Table \ref{Table 1 - Model 11}.

%=================================================================
\begin{table}[h]
\begin{center}
\begin{tabular}{ |c|c|c|c|} 
\hline
 Phase & Path & F & Fermi Multiplicities\\
\hline
1&&20&1$\times\textbf{2}$+2$\times\textbf{3}$+1$\times\textbf{5}$+3$\times\textbf{6}$+1$\times\textbf{9}$\\ 
2&3&18&1$\times\textbf{2}$+2$\times\textbf{3}$+2$\times\textbf{4}$+1$\times\textbf{5}$+1$\times\textbf{6}$+1$\times\textbf{9}$\\ 
3&-3&16&3$\times\textbf{2}$+1$\times\textbf{3}$+2$\times\textbf{5}$+1$\times\textbf{6}$+1$\times\textbf{7}$\\ 
4&4&20&4$\times\textbf{3}$+2$\times\textbf{6}$+2$\times\textbf{8}$\\ 
5&5&14&1$\times\textbf{2}$+4$\times\textbf{3}$+2$\times\textbf{4}$+1$\times\textbf{6}$\\ 
6&3,2&22&2$\times\textbf{3}$+2$\times\textbf{4}$+2$\times\textbf{6}$+2$\times\textbf{9}$\\ 
7&3,-4&18&1$\times\textbf{2}$+3$\times\textbf{3}$+1$\times\textbf{4}$+1$\times\textbf{5}$+1$\times\textbf{6}$+1$\times\textbf{10}$\\ 
8&3,5&12&2$\times\textbf{2}$+4$\times\textbf{3}$+2$\times\textbf{4}$\\ 
9&3,7&20&2$\times\textbf{3}$+4$\times\textbf{4}$+2$\times\textbf{9}$\\ 
10&-3,4&14&2$\times\textbf{2}$+4$\times\textbf{3}$+2$\times\textbf{6}$\\ 
11&-3,-4&14&2$\times\textbf{2}$+4$\times\textbf{3}$+2$\times\textbf{6}$\\ 
12&-3,5&14&2$\times\textbf{2}$+2$\times\textbf{3}$+2$\times\textbf{4}$+2$\times\textbf{5}$\\ 
13&-3,7&18&2$\times\textbf{2}$+1$\times\textbf{3}$+1$\times\textbf{4}$+1$\times\textbf{5}$+1$\times\textbf{6}$+2$\times\textbf{7}$\\ 
14&-3,-7&20&2$\times\textbf{2}$+2$\times\textbf{5}$+2$\times\textbf{6}$+2$\times\textbf{7}$\\ 
15&5,-7&20&1$\times\textbf{2}$+1$\times\textbf{3}$+2$\times\textbf{4}$+2$\times\textbf{5}$+1$\times\textbf{8}$+1$\times\textbf{9}$\\ 
16&3,2,-4&18&1$\times\textbf{2}$+2$\times\textbf{3}$+1$\times\textbf{4}$+2$\times\textbf{5}$+1$\times\textbf{6}$+1$\times\textbf{8}$\\ 
17&-3,4,7&16&2$\times\textbf{2}$+2$\times\textbf{3}$+2$\times\textbf{5}$+2$\times\textbf{6}$\\ 
\hline
\end{tabular}
\end{center}
\caption{Basic information regarding the 17 toric phases of Model 11.}
\label{Table 1 - Model 11}
\end{table}
%=================================================================

Table \ref{Table 2 - Model 11} summarizes the connection between the toric phases under triality.

%=================================================================
\begin{table}[h]
\begin{center}
\begin{tabular}{ |c|c|c|c|c|c|c|c|c|} 
\hline
N&1&2&3&4&5&6&7&8\\
\hline
1&&&2 , \underline{3}&4&5&&&\\
2&&6&3 , \underline{1}&\underline{7}&8&&9&\\
3&&&1 , \underline{2}&10 , \underline{11}&12&&13 , \underline{14}&\\
4&&&\underline{10}&\underline{1}&10&&&1\\
5&9&&8 , \underline{12}&10&\underline{1}&11&\underline{15}&\underline{7}\\
6&&\underline{2}&&\underline{16}&16&&2&\\
7&&16 , \underline{15}&11&2&10&&5&\\
8&2&16&12 , \underline{5}&\underline{10}&\underline{2}&10&5 , \underline{12}&\underline{16}\\
9&&2&13&\underline{5}&5&\underline{13}&\underline{2}&\\
10&&\underline{7}&4&11 , \underline{3}&8&&17 , \underline{13}&5\\
11&&\underline{5}&\underline{7}&3 , \underline{10}&5&&10 , \underline{3}&7\\
12&&&5 , \underline{8}&8 , \underline{5}&\underline{3}&3&15&\underline{15}\\
13&&&\underline{9}&17 , \underline{10}&15&&14 , \underline{3}&\\
14&&&&13 , \underline{3}&&&3 , \underline{13}&\\
15&13&&12&&&7 , \underline{16}&5&\\
16&&15 , \underline{7}&&6&17&&8&\\
17&&\underline{16}&&10 , \underline{13}&16&&13 , \underline{10}&\\\hline
\end{tabular}
\end{center}
\caption{Triality connections between the 17 toric phases of Model 11.}
\label{Table 2 - Model 11}
\end{table}
%=================================================================

%=================================================================
\subsection{Model 13: $P^{1}_{+-}(\text{dP}_2)$}
%=================================================================
 
\begin{figure}[H]
\begin{center}
\resizebox{0.35\hsize}{!}{
\includegraphics[height=6cm]{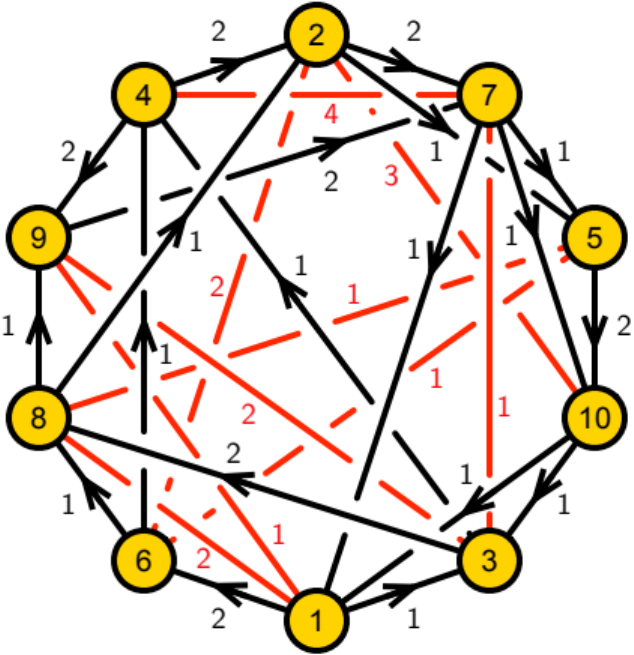} 
}
\caption{Quiver for Phase 1 of Model 13.
\label{f_quiver_13}}
 \end{center}
 \end{figure}

\fref{f_quiver_13} shows the quiver for Phase 1 of Model 13. 
 
%=================================================================
%=================================================================

The $J$- and $E$-terms are
\beq
 {\footnotesize
\begin{array}{rrclrcl}
 &  & J &  & & E &  \\
\Lambda_{18}^{1}  : & X_{89}  Y_{97}  X_{75}  Q_{5.10}  X_{10.1} &-&X_{82}  Q_{27}  X_{71}  &   P_{16}  X_{68} &-& X_{13}  P_{38} \\
\Lambda_{18}^{2}  : & X_{89}  Y_{97}  X_{75}  P_{5.10}  X_{10.1}  &-&X_{82}  P_{27}  X_{71}  &   X_{13}  Q_{38}  &-&Q_{16}  X_{68} \\
\Lambda_{19}^{1}  : & X_{97}  X_{71} &-& Y_{97}  Y_{7.10}  X_{10.1}  &   P_{16}  X_{64}  Q_{49} &-& Q_{16}  X_{64}  P_{49} \\
\Lambda_{26}^{1}  : & X_{64}  X_{42} &-& X_{68}  X_{82}  &   P_{27}  X_{71}  Q_{16} &-& Q_{27}  X_{71}  P_{16} \\
\Lambda_{2.10}^{1}  : & X_{10.1}  X_{13}  X_{34}  Y_{42}  &-&X_{10.3}  X_{34}  X_{42}  &   P_{27}  X_{75}  Q_{5.10} &-& Q_{27}  X_{75}  P_{5.10} \\
\Lambda_{2.10}^{2}  : & X_{10.3}  Q_{38}  X_{82} &-& X_{10.1} Q_{16}  X_{64}  Y_{42}  &   P_{27}  Y_{7.10} &-& Y_{25}  P_{5.10} \\
\Lambda_{2.10}^{3}  : & X_{10.3}  P_{38}  X_{82} &-& X_{10.1}  P_{16}  X_{64}  Y_{42}  &  Y_{25}  Q_{5.10}  &-&Q_{27}  Y_{7.10} \\
\Lambda_{37}^{1}  : & Y_{7.10}  X_{10.3} &-&X_{71}  X_{13}  &   P_{38}  X_{82}  Q_{27} &-& Q_{38}  X_{82}  P_{27} \\
\Lambda_{39}^{1}  : & Y_{97}  X_{75}  Q_{5.10}  X_{10.1}  X_{13} &-&X_{97}  X_{75}  Q_{5.10}  X_{10.3}  &   P_{38}  X_{89} &-& X_{34}  P_{49} \\
\Lambda_{39}^{2}  : & Y_{97}  X_{75}  P_{5.10}  X_{10.1}  X_{13}  &-&X_{97}  X_{75}  P_{5.10}  X_{10.3} &  X_{34}  Q_{49}  &-&Q_{38}  X_{89}  \\
\Lambda_{47}^{1}  : & X_{71}  Q_{16}  X_{64} &-& X_{75}  Q_{5.10}  X_{10.3}  X_{34}  &   P_{49}  X_{97} &-& X_{42}  P_{27} \\
\Lambda_{47}^{2}  : & X_{75}  Q_{5.10}  X_{10.1}  X_{13}  X_{34} &-&Y_{7.10}  X_{10.1}  Q_{16}  X_{64}  &   P_{49}  Y_{97} &-& Y_{42}  P_{27} \\
\Lambda_{47}^{3}  : & X_{71}  P_{16}  X_{64} &-& X_{75}  P_{5.10}  X_{10.3}  X_{34}  &   X_{42}  Q_{27} &-&Q_{49}  X_{97}  \\
\Lambda_{47}^{4}  : & X_{75}  P_{5.10}  X_{10.1}  X_{13}  X_{34} &-&Y_{7.10}  X_{10.1}  P_{16}  X_{64} & Y_{42}  Q_{27} &-&Q_{49}  Y_{97} \\
\Lambda_{56}^{1}  : & X_{68}  X_{89}  Y_{97}  X_{75} &-&X_{64}  Y_{42}  Y_{25}  &   P_{5.10}  X_{10.1}  Q_{16} &-& Q_{5.10}  X_{10.1}  P_{16} \\
\Lambda_{58}^{1}  : & X_{82}  Y_{25} &-& X_{89}  X_{97}  X_{75}  &   P_{5.10}  X_{10.3}  Q_{38} &-& Q_{5.10}  X_{10.3}  P_{38}
 \end{array} 
 }~.~
\label{E_J_C_+-}
 \eeq

This model has 90 toric phases, which are summarized in Table \ref{Table 1 - Model 13}.

%=================================================================
\begin{table}[h]
\begin{center}
\tiny
\resizebox{1 \textwidth}{!}{
\begin{tabular}{ |c|c|c|c|c|c|c|c| } 
\hline
 Phase & Path & F & Fermi Multiplicities & Phase & Path & F & Fermi Multiplicities \\
\hline
1&&16&2$\times\textbf{2}$+5$\times\textbf{3}$+2$\times\textbf{4}$+1$\times\textbf{5}$&46&1,3,-6&23&4$\times\textbf{3}$+3$\times\textbf{4}$+1$\times\textbf{5}$+1$\times\textbf{6}$+1$\times\textbf{11}$\\ 
2&1&20&1$\times\textbf{2}$+3$\times\textbf{3}$+3$\times\textbf{4}$+1$\times\textbf{5}$+2$\times\textbf{6}$&47&1,4,5&23&1$\times\textbf{2}$+3$\times\textbf{3}$+1$\times\textbf{4}$+3$\times\textbf{5}$+1$\times\textbf{6}$+1$\times\textbf{10}$\\ 
3&3&18&2$\times\textbf{2}$+3$\times\textbf{3}$+2$\times\textbf{4}$+3$\times\textbf{5}$&48&1,4,-5&23&1$\times\textbf{2}$+2$\times\textbf{3}$+3$\times\textbf{4}$+1$\times\textbf{5}$+2$\times\textbf{6}$+1$\times\textbf{9}$\\ 
4&4&21&1$\times\textbf{2}$+4$\times\textbf{3}$+1$\times\textbf{4}$+3$\times\textbf{5}$+1$\times\textbf{9}$&49&1,4,-9&21&1$\times\textbf{2}$+3$\times\textbf{3}$+1$\times\textbf{4}$+3$\times\textbf{5}$+2$\times\textbf{6}$\\ 
5&5&17&1$\times\textbf{2}$+7$\times\textbf{3}$+1$\times\textbf{4}$+1$\times\textbf{7}$&50&1,5,-9&23&1$\times\textbf{2}$+2$\times\textbf{3}$+1$\times\textbf{4}$+4$\times\textbf{5}$+1$\times\textbf{6}$+1$\times\textbf{8}$\\ 
6&-5&18&1$\times\textbf{2}$+5$\times\textbf{3}$+2$\times\textbf{4}$+1$\times\textbf{5}$+1$\times\textbf{6}$&51&1,-5,-9&22&1$\times\textbf{2}$+3$\times\textbf{3}$+1$\times\textbf{4}$+3$\times\textbf{5}$+2$\times\textbf{7}$\\ 
7&6&18&2$\times\textbf{2}$+4$\times\textbf{3}$+4$\times\textbf{5}$&52&1,-6,-7&21&1$\times\textbf{2}$+3$\times\textbf{3}$+2$\times\textbf{4}$+2$\times\textbf{5}$+1$\times\textbf{6}$+1$\times\textbf{7}$\\ 
8&-6&16&2$\times\textbf{2}$+6$\times\textbf{3}$+2$\times\textbf{5}$&53&1,-6,-8&19&1$\times\textbf{2}$+4$\times\textbf{3}$+2$\times\textbf{4}$+2$\times\textbf{5}$+1$\times\textbf{6}$\\ 
9&-9&17&3$\times\textbf{2}$+3$\times\textbf{3}$+2$\times\textbf{4}$+1$\times\textbf{5}$+1$\times\textbf{6}$&54&1,-6,-9&18&6$\times\textbf{3}$+2$\times\textbf{4}$+2$\times\textbf{5}$\\ 
10&-10&21&1$\times\textbf{2}$+3$\times\textbf{3}$+2$\times\textbf{4}$+2$\times\textbf{5}$+1$\times\textbf{6}$+1$\times\textbf{7}$&55&3,1,5&22&1$\times\textbf{2}$+4$\times\textbf{3}$+2$\times\textbf{5}$+2$\times\textbf{6}$+1$\times\textbf{8}$\\ 
11&1,3&23&1$\times\textbf{2}$+2$\times\textbf{3}$+2$\times\textbf{4}$+3$\times\textbf{5}$+1$\times\textbf{6}$+1$\times\textbf{9}$&56&3,1,-5&19&2$\times\textbf{2}$+4$\times\textbf{3}$+2$\times\textbf{5}$+2$\times\textbf{6}$\\ 
12&1,4&23&1$\times\textbf{2}$+2$\times\textbf{3}$+1$\times\textbf{4}$+4$\times\textbf{5}$+1$\times\textbf{6}$+1$\times\textbf{8}$&57&3,1,-6&25&3$\times\textbf{3}$+3$\times\textbf{4}$+1$\times\textbf{5}$+1$\times\textbf{6}$+1$\times\textbf{8}$+1$\times\textbf{10}$\\ 
13&1,5&21&1$\times\textbf{2}$+4$\times\textbf{3}$+1$\times\textbf{4}$+2$\times\textbf{5}$+1$\times\textbf{6}$+1$\times\textbf{8}$&58&3,1,-7&28&2$\times\textbf{3}$+2$\times\textbf{4}$+3$\times\textbf{6}$+1$\times\textbf{7}$+1$\times\textbf{8}$+1$\times\textbf{9}$\\ 
14&1,-5&20&1$\times\textbf{2}$+5$\times\textbf{3}$+1$\times\textbf{4}$+2$\times\textbf{6}$+1$\times\textbf{7}$&59&3,5,-2&19&6$\times\textbf{3}$+1$\times\textbf{4}$+2$\times\textbf{5}$+1$\times\textbf{6}$\\ 
15&1,-6&18&1$\times\textbf{2}$+5$\times\textbf{3}$+2$\times\textbf{4}$+1$\times\textbf{5}$+1$\times\textbf{6}$&60&3,5,6&25&3$\times\textbf{3}$+1$\times\textbf{4}$+3$\times\textbf{5}$+1$\times\textbf{6}$+1$\times\textbf{7}$+1$\times\textbf{9}$\\ 
16&1,-9&22&1$\times\textbf{2}$+1$\times\textbf{3}$+3$\times\textbf{4}$+3$\times\textbf{5}$+2$\times\textbf{6}$&61&3,6,-7&30&1$\times\textbf{2}$+2$\times\textbf{4}$+1$\times\textbf{5}$+5$\times\textbf{7}$+1$\times\textbf{10}$\\ 
17&3,1&21&1$\times\textbf{2}$+3$\times\textbf{3}$+3$\times\textbf{4}$+1$\times\textbf{5}$+1$\times\textbf{6}$+1$\times\textbf{8}$&62&3,-6,2&23&3$\times\textbf{3}$+2$\times\textbf{4}$+3$\times\textbf{5}$+1$\times\textbf{6}$+1$\times\textbf{8}$\\ 
18&3,5&19&1$\times\textbf{2}$+5$\times\textbf{3}$+3$\times\textbf{5}$+1$\times\textbf{6}$&63&3,-6,-7&28&1$\times\textbf{2}$+2$\times\textbf{4}$+3$\times\textbf{5}$+3$\times\textbf{7}$+1$\times\textbf{10}$\\ 
19&3,-5&18&2$\times\textbf{2}$+4$\times\textbf{3}$+4$\times\textbf{5}$&64&4,5,-8&25&3$\times\textbf{3}$+2$\times\textbf{4}$+2$\times\textbf{5}$+1$\times\textbf{6}$+1$\times\textbf{7}$+1$\times\textbf{10}$\\ 
20&3,6&22&1$\times\textbf{2}$+2$\times\textbf{3}$+2$\times\textbf{4}$+3$\times\textbf{5}$+1$\times\textbf{6}$+1$\times\textbf{7}$&65&4,-5,-9&20&2$\times\textbf{2}$+2$\times\textbf{3}$+2$\times\textbf{4}$+3$\times\textbf{5}$+1$\times\textbf{7}$\\ 
21&3,-6&20&1$\times\textbf{2}$+3$\times\textbf{3}$+2$\times\textbf{4}$+3$\times\textbf{5}$+1$\times\textbf{6}$&66&4,-9,1&20&2$\times\textbf{2}$+3$\times\textbf{3}$+3$\times\textbf{5}$+2$\times\textbf{6}$\\ 
22&3,-7&26&1$\times\textbf{2}$+2$\times\textbf{3}$+1$\times\textbf{4}$+1$\times\textbf{5}$+2$\times\textbf{6}$+2$\times\textbf{7}$+1$\times\textbf{9}$&67&5,-2,-4&19&3$\times\textbf{2}$+3$\times\textbf{3}$+2$\times\textbf{5}$+1$\times\textbf{6}$+1$\times\textbf{7}$\\ 
23&4,5&21&1$\times\textbf{2}$+5$\times\textbf{3}$+1$\times\textbf{4}$+2$\times\textbf{5}$+1$\times\textbf{11}$&68&5,-2,-8&19&2$\times\textbf{2}$+4$\times\textbf{3}$+2$\times\textbf{5}$+2$\times\textbf{6}$\\ 
24&4,-5&23&1$\times\textbf{2}$+2$\times\textbf{3}$+3$\times\textbf{4}$+2$\times\textbf{5}$+1$\times\textbf{6}$+1$\times\textbf{10}$&69&5,6,8&23&2$\times\textbf{2}$+1$\times\textbf{3}$+2$\times\textbf{4}$+3$\times\textbf{5}$+1$\times\textbf{7}$+1$\times\textbf{9}$\\ 
25&4,-9&18&2$\times\textbf{2}$+4$\times\textbf{3}$+1$\times\textbf{4}$+2$\times\textbf{5}$+1$\times\textbf{6}$&70&5,6,-8&21&2$\times\textbf{2}$+2$\times\textbf{3}$+2$\times\textbf{4}$+2$\times\textbf{5}$+2$\times\textbf{7}$\\ 
26&4,-10&25&3$\times\textbf{3}$+2$\times\textbf{4}$+1$\times\textbf{5}$+2$\times\textbf{6}$+1$\times\textbf{7}$+1$\times\textbf{9}$&71&5,6,-9&21&1$\times\textbf{2}$+5$\times\textbf{3}$+2$\times\textbf{5}$+1$\times\textbf{7}$+1$\times\textbf{8}$\\ 
27&5,-2&17&2$\times\textbf{2}$+6$\times\textbf{3}$+1$\times\textbf{5}$+1$\times\textbf{7}$&72&5,6,-10&22&1$\times\textbf{2}$+4$\times\textbf{3}$+2$\times\textbf{5}$+2$\times\textbf{6}$+1$\times\textbf{8}$\\ 
28&5,6&21&1$\times\textbf{2}$+5$\times\textbf{3}$+2$\times\textbf{5}$+1$\times\textbf{7}$+1$\times\textbf{8}$&73&5,-8,-9&19&6$\times\textbf{3}$+2$\times\textbf{4}$+2$\times\textbf{6}$\\ 
29&5,-8&19&5$\times\textbf{3}$+3$\times\textbf{4}$+1$\times\textbf{5}$+1$\times\textbf{6}$&74&5,-8,-10&22&4$\times\textbf{3}$+4$\times\textbf{4}$+2$\times\textbf{8}$\\ 
30&5,-9&17&2$\times\textbf{2}$+6$\times\textbf{3}$+1$\times\textbf{5}$+1$\times\textbf{7}$&75&5,-9,1&21&2$\times\textbf{2}$+2$\times\textbf{3}$+4$\times\textbf{5}$+2$\times\textbf{6}$\\ 
31&5,-10&22&4$\times\textbf{3}$+3$\times\textbf{4}$+1$\times\textbf{5}$+1$\times\textbf{6}$+1$\times\textbf{9}$&76&5,-9,-8&19&1$\times\textbf{2}$+5$\times\textbf{3}$+3$\times\textbf{5}$+1$\times\textbf{6}$\\ 
32&-5,-9&19&2$\times\textbf{2}$+4$\times\textbf{3}$+3$\times\textbf{5}$+1$\times\textbf{7}$&77&-5,-9,1&21&2$\times\textbf{2}$+3$\times\textbf{3}$+2$\times\textbf{5}$+2$\times\textbf{6}$+1$\times\textbf{7}$\\ 
33&6,8&20&2$\times\textbf{2}$+2$\times\textbf{3}$+1$\times\textbf{4}$+4$\times\textbf{5}$+1$\times\textbf{6}$&78&-5,-9,-4&22&2$\times\textbf{2}$+2$\times\textbf{3}$+3$\times\textbf{5}$+2$\times\textbf{6}$+1$\times\textbf{7}$\\ 
34&6,-9&19&1$\times\textbf{2}$+5$\times\textbf{3}$+3$\times\textbf{5}$+1$\times\textbf{6}$&79&-6,2,8&21&2$\times\textbf{2}$+2$\times\textbf{3}$+4$\times\textbf{5}$+2$\times\textbf{6}$\\ 
35&6,-10&21&2$\times\textbf{2}$+3$\times\textbf{3}$+3$\times\textbf{5}$+1$\times\textbf{6}$+1$\times\textbf{8}$&80&-6,2,-9&22&5$\times\textbf{3}$+3$\times\textbf{5}$+1$\times\textbf{6}$+1$\times\textbf{8}$\\ 
36&-6,2&19&1$\times\textbf{2}$+5$\times\textbf{3}$+3$\times\textbf{5}$+1$\times\textbf{6}$&81&-6,-8,-10&22&1$\times\textbf{2}$+3$\times\textbf{3}$+3$\times\textbf{5}$+3$\times\textbf{6}$\\ 
37&-6,8&18&2$\times\textbf{2}$+4$\times\textbf{3}$+4$\times\textbf{5}$&82&-6,-9,-10&21&1$\times\textbf{2}$+4$\times\textbf{3}$+3$\times\textbf{5}$+1$\times\textbf{6}$+1$\times\textbf{7}$\\ 
38&-6,-8&17&1$\times\textbf{2}$+6$\times\textbf{3}$+1$\times\textbf{4}$+2$\times\textbf{5}$&83&-9,1,-4&24&2$\times\textbf{2}$+1$\times\textbf{3}$+1$\times\textbf{4}$+2$\times\textbf{5}$+1$\times\textbf{6}$+3$\times\textbf{7}$\\ 
39&-6,-9&17&1$\times\textbf{2}$+7$\times\textbf{3}$+1$\times\textbf{5}$+1$\times\textbf{6}$&84&-9,-4,-6&20&3$\times\textbf{2}$+2$\times\textbf{3}$+1$\times\textbf{4}$+1$\times\textbf{5}$+2$\times\textbf{6}$+1$\times\textbf{7}$\\ 
40&-6,-10&21&2$\times\textbf{2}$+2$\times\textbf{3}$+4$\times\textbf{5}$+2$\times\textbf{6}$&85&1,3,-5,8&23&2$\times\textbf{2}$+1$\times\textbf{3}$+1$\times\textbf{4}$+3$\times\textbf{5}$+1$\times\textbf{6}$+2$\times\textbf{7}$\\ 
41&-9,1&21&2$\times\textbf{2}$+1$\times\textbf{3}$+2$\times\textbf{4}$+3$\times\textbf{5}$+2$\times\textbf{6}$&86&1,4,5,-9&21&1$\times\textbf{2}$+4$\times\textbf{3}$+1$\times\textbf{4}$+2$\times\textbf{5}$+1$\times\textbf{6}$+1$\times\textbf{8}$\\ 
42&-9,-4&20&3$\times\textbf{2}$+2$\times\textbf{3}$+1$\times\textbf{4}$+1$\times\textbf{5}$+2$\times\textbf{6}$+1$\times\textbf{7}$&87&3,1,5,-2&22&6$\times\textbf{3}$+2$\times\textbf{6}$+2$\times\textbf{7}$\\ 
43&-9,-10&21&2$\times\textbf{2}$+2$\times\textbf{3}$+2$\times\textbf{4}$+2$\times\textbf{5}$+2$\times\textbf{7}$&88&3,1,-6,-7&30&4$\times\textbf{4}$+2$\times\textbf{5}$+2$\times\textbf{7}$+2$\times\textbf{10}$\\ 
44&1,3,5&24&1$\times\textbf{2}$+3$\times\textbf{3}$+4$\times\textbf{5}$+1$\times\textbf{8}$+1$\times\textbf{9}$&89&3,-6,2,-7&31&2$\times\textbf{4}$+2$\times\textbf{5}$+4$\times\textbf{7}$+2$\times\textbf{8}$\\ 
45&1,3,-5&21&1$\times\textbf{2}$+4$\times\textbf{3}$+1$\times\textbf{4}$+2$\times\textbf{5}$+2$\times\textbf{7}$&90&-5,-9,1,-4&24&2$\times\textbf{2}$+2$\times\textbf{3}$+1$\times\textbf{5}$+2$\times\textbf{6}$+3$\times\textbf{7}$\\ 
\hline
\end{tabular}}
\end{center}
\caption{Basic information regarding the 90 toric phases of Model 13.}
\label{Table 1 - Model 13}
\end{table}
%=================================================================

Table \ref{Table 2 - Model 13} summarizes the connection between the toric phases under triality.

%=================================================================
\begin{center}
\tiny
\resizebox{1 \textwidth}{!}{
\begin{tabular}{ |c|c|c|c|c|c|c|c|c|c|c|} 
\hline
N&1&2&3&4&5&6&7&8&9&10\\
\hline
1&2&&3&4&5 , \underline{6}&7 , \underline{8}&&\underline{3}&\underline{9}&\underline{10}\\
2&\underline{1}&&11&12&13 , \underline{14}&\underline{15}&&&\underline{16}&\\
3&17&&\underline{1}&&18 , \underline{19}&20 , \underline{21}&\underline{22}&\underline{1}&&\\
4&12&&&\underline{1}&23 , \underline{24}&&&\underline{22}&\underline{25}&\underline{26}\\
5&13&\underline{27}&18&23&6 , \underline{1}&28&&\underline{29}&\underline{30}&\underline{31}\\
6&14&\underline{15}&19&24&1 , \underline{5}&\underline{14}&&\underline{20}&\underline{32}&\\
7&&&20&&28&8 , \underline{1}&&33 , \underline{19}&\underline{34}&\underline{35}\\
8&15&36&21&&\underline{14}&1 , \underline{7}&&37 , \underline{38}&\underline{39}&\underline{40}\\
9&41&&&25 , \underline{42}&30 , \underline{32}&34 , \underline{39}&&&1&\underline{43}\\
10&&&&26&31&35 , \underline{40}&&\underline{17}&\underline{43}&1\\
11&\underline{17}&&\underline{2}&&44 , \underline{45}&\underline{46}&&&&\\
12&\underline{4}&&&\underline{2}&47 , \underline{48}&&&&\underline{49}&\\
13&\underline{5}&\underline{36}&44&47&14 , \underline{2}&&&&\underline{50}&\\
14&\underline{6}&\underline{8}&45&48&2 , \underline{13}&\underline{6}&&&\underline{51}&\\
15&\underline{8}&27&46&&\underline{6}&2&\underline{52}&33 , \underline{53}&\underline{54}&\\
16&\underline{41}&&&49&50 , \underline{51}&\underline{54}&&&2&\\
17&\underline{3}&&11&&55 , \underline{56}&\underline{57}&\underline{58}&\underline{10}&&\\
18&55&\underline{59}&\underline{5}&&19 , \underline{3}&60&&\underline{38}&&\\
19&56&\underline{54}&\underline{6}&&3 , \underline{18}&\underline{51}&&33 , \underline{7}&&\\
20&&&\underline{7}&&60&21 , \underline{3}&\underline{61}&\underline{6}&&\\
21&57&62&\underline{8}&&\underline{51}&3 , \underline{20}&\underline{63}&\underline{29}&&\\
22&58&&&&&61 , \underline{63}&3&\underline{4}&&\\
23&47&\underline{30}&&\underline{5}&24 , \underline{4}&&&\underline{64}&\underline{27}&\underline{46}\\
24&48&\underline{53}&&\underline{6}&4 , \underline{23}&&&\underline{61}&\underline{65}&\\
25&66&&&42 , \underline{9}&27 , \underline{65}&\underline{66}&&&4&\underline{52}\\
26&&&&\underline{10}&46&&&\underline{58}&\underline{52}&4\\
27&36&5&59&30 , \underline{67}&15&&&\underline{68}&\underline{23}&25 , \underline{65}\\
28&&&60&&\underline{7}&\underline{5}&&69 , \underline{70}&\underline{71}&\underline{72}\\
29&&\underline{68}&38&64&\underline{21}&70&&5&\underline{73}&\underline{74}\\
30&75&\underline{23}&&27 , \underline{67}&32 , \underline{9}&71&&\underline{76}&5&\underline{53}\\
31&&\underline{65}&&46&\underline{10}&72&&\underline{74}&\underline{53}&5\\
32&77&\underline{46}&&65 , \underline{78}&9 , \underline{30}&\underline{45}&&&6&\\
33&&&&&69&37&&19 , \underline{7}&15 , \underline{53}&\underline{67}\\
34&&&&\underline{72}&71&39 , \underline{9}&&53&7&\underline{71}\\
35&&&&&72&40 , \underline{10}&&67 , \underline{56}&\underline{71}&7\\
36&27&\underline{8}&62&&\underline{13}&&&79 , \underline{76}&\underline{80}&\\
37&33&79&&&&\underline{33}&&38 , \underline{8}&8 , \underline{38}&\underline{79}\\
38&53&76&29&&\underline{48}&\underline{18}&&8 , \underline{37}&39&\underline{81}\\
39&59&80&&\underline{68}&\underline{45}&9 , \underline{34}&&38&8&\underline{82}\\
40&&&&&&10 , \underline{35}&&79 , \underline{81}&\underline{82}&8\\
41&\underline{9}&&&66 , \underline{83}&75 , \underline{77}&\underline{59}&&&16&\\
42&83&&&9 , \underline{25}&67 , \underline{78}&72 , \underline{84}&&&&\underline{69}\\
43&&&&52 , \underline{69}&53&71 , \underline{82}&&&10&9\\
44&\underline{55}&\underline{80}&\underline{13}&&45 , \underline{11}&&&&&\\
45&\underline{56}&\underline{39}&\underline{14}&&11 , \underline{44}&\underline{32}&&85&&\\
46&\underline{57}&23&\underline{15}&&\underline{32}&11&\underline{26}&\underline{31}&&\\
47&\underline{23}&\underline{76}&&\underline{13}&48 , \underline{12}&&&&\underline{86}&\\
48&\underline{24}&\underline{38}&&\underline{14}&12 , \underline{47}&60&&&\underline{70}&\\
49&\underline{66}&&&\underline{16}&86 , \underline{70}&\underline{52}&&&12&\\
50&\underline{75}&\underline{62}&&86&51 , \underline{16}&&&&13&\\\hline
\end{tabular}}
\end{center}                           
%=================================================================

%=================================================================
\begin{table}[H]
\begin{center}
\tiny
\resizebox{1 \textwidth}{!}{
\begin{tabular}{ |c|c|c|c|c|c|c|c|c|c|c|} 
\hline
N&1&2&3&4&5&6&7&8&9&10\\
\hline
51&\underline{77}&\underline{21}&&70&16 , \underline{50}&\underline{19}&&&14&\\
52&&25&26&&&&15&69 , \underline{43}&\underline{49}&\\
53&\underline{38}&30&31&&\underline{24}&&\underline{43}&15 , \underline{33}&34&\\
54&\underline{59}&59&&\underline{15}&\underline{19}&16&\underline{16}&19&15&\\
55&\underline{18}&\underline{87}&44&&56 , \underline{17}&&&\underline{81}&&\\
56&\underline{19}&\underline{59}&45&&17 , \underline{55}&\underline{77}&&67 , \underline{35}&&\\
57&\underline{21}&64&46&&\underline{77}&17&\underline{88}&\underline{74}&&\\
58&\underline{22}&&&&&\underline{88}&17&\underline{26}&&\\
59&87&18&\underline{27}&\underline{56}&54&&&\underline{39}&&41\\
60&&&\underline{28}&&\underline{20}&\underline{18}&&\underline{48}&&\\
61&&&&&&63 , \underline{22}&20&\underline{24}&&\\
62&64&\underline{21}&\underline{36}&&\underline{50}&&\underline{89}&\underline{73}&&\\
63&88&89&&&&22 , \underline{61}&21&\underline{64}&&\\
64&&\underline{75}&&\underline{29}&\underline{63}&&&23&\underline{62}&\underline{57}\\
65&84&\underline{31}&&78 , \underline{32}&25 , \underline{27}&\underline{83}&&&24&\\
66&\underline{25}&&&83 , \underline{41}&68 , \underline{84}&\underline{25}&&&49&\\
67&79&&56 , \underline{35}&27 , \underline{30}&33&&&\underline{85}&&42 , \underline{78}\\
68&&29&39&75 , \underline{85}&&&&27&\underline{86}&66 , \underline{84}\\
69&&&&&\underline{33}&&&70 , \underline{28}&52 , \underline{43}&\underline{42}\\
70&&&48&&\underline{51}&\underline{29}&&28 , \underline{69}&49 , \underline{86}&\underline{84}\\
71&&&&\underline{35}&\underline{34}&\underline{30}&&43 , \underline{82}&28&\underline{34}\\
72&&&&&\underline{35}&\underline{31}&&42 , \underline{84}&\underline{34}&28\\
73&&\underline{86}&76&62&\underline{62}&86&&\underline{76}&29&\underline{29}\\
74&&\underline{84}&\underline{31}&57&\underline{57}&84&&31&\underline{29}&29\\
75&\underline{30}&\underline{64}&&68 , \underline{85}&77 , \underline{41}&&&&50&\\
76&&\underline{47}&&36 , \underline{79}&&82&&30&73&\underline{38}\\
77&\underline{32}&\underline{57}&&84 , \underline{90}&41 , \underline{75}&\underline{56}&&&51&\\
78&90&&&32 , \underline{65}&42 , \underline{67}&\underline{90}&&&&\\
79&67&\underline{37}&&&&&&76 , \underline{36}&40 , \underline{81}&\\
80&87&\underline{39}&&&\underline{44}&&&81&36&\\
81&&&&&&\underline{55}&&40 , \underline{79}&80&38\\
82&&&&\underline{86}&&43 , \underline{71}&&76&40&39\\
83&\underline{42}&&&41 , \underline{66}&85 , \underline{90}&\underline{65}&&&&\\
84&65&&&68 , \underline{66}&77 , \underline{90}&42 , \underline{72}&&74&&\underline{70}\\
85&\underline{67}&75 , \underline{68}&&&&90 , \underline{83}&&\underline{45}&&\\
86&\underline{68}&\underline{73}&&\underline{50}&70 , \underline{49}&\underline{82}&&&47&\\
87&\underline{59}&55&80&\underline{55}&59&&&\underline{80}&&\\
88&\underline{63}&63&\underline{58}&&&58&57&\underline{57}&&\\
89&63&\underline{63}&&&&&62&\underline{62}&&\\
90&\underline{78}&&&77 , \underline{84}&83 , \underline{85}&\underline{78}&&&&\\\hline
\end{tabular}}
\end{center}
\caption{Triality connections between the 90 toric phases of Model 13.}
\label{Table 2 - Model 13}
\end{table}
%=================================================================

\newpage

%=================================================================
\subsection{Model 14: $P^{2}_{+-}(\text{dP}_2)$}
%=================================================================

\begin{figure}[H]
\begin{center}
\resizebox{0.35\hsize}{!}{
\includegraphics[height=6cm]{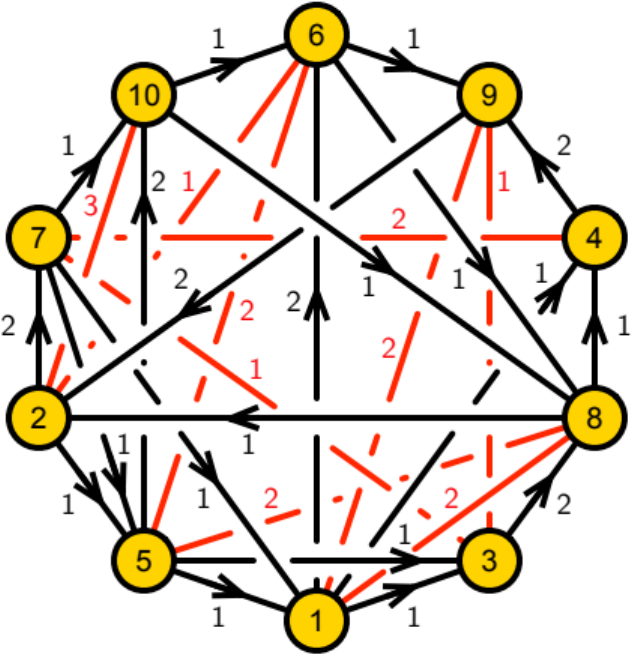} 
}
\caption{Quiver for Phase 1 of Model 14.
\label{f_quiver_14}}
 \end{center}
 \end{figure}

\fref{f_quiver_14} shows the quiver for Phase 1 of Model 14.

%=================================================================
%=================================================================

The $J$- and $E$-terms are
 \beq
 {\footnotesize
\begin{array}{rrclrcl}
 &  & J &  & & E &  \\
\Lambda_{18}^{1} : & X_{84} Q_{49} Y_{92} X_{25} X_{51} &-&X_{82} Q_{27} X_{71}  &  P_{16} X_{68} &-& X_{13} P_{38} \\ 
\Lambda_{18}^{2} : & X_{84} P_{49} Y_{92} X_{25} X_{51} &-&X_{82} P_{27} X_{71} &  X_{13} Q_{38} &-&Q_{16} X_{68} \\ 
\Lambda_{19}^{1} : & X_{92} Q_{27} X_{71} &-& Y_{92} Q_{27} Y_{75} X_{51} &  P_{16} X_{69} &-& X_{14} P_{49} \\ 
\Lambda_{19}^{2} : & X_{92} P_{27} X_{71} &-& Y_{92} P_{27} Y_{75} X_{51} &  X_{14} Q_{49} &-&Q_{16} X_{69}  \\ 
\Lambda_{26}^{1} : & X_{69} X_{92} &-&X_{68} X_{82} &  P_{27} X_{71} Q_{16} &-& Q_{27} X_{71} P_{16} \\ 
\Lambda_{2.10}^{1} : & X_{10.6} X_{68} X_{84} Q_{49} Y_{92} &-&X_{10.8} X_{84} Q_{49} X_{92} &  P_{27} X_{7.10} &-& X_{25} P_{5.10} \\ 
\Lambda_{2.10}^{2} : & X_{10.8} X_{82} &-& X_{10.6} X_{69} Y_{92} &  P_{27} Y_{75} Q_{5.10} &-& Q_{27} Y_{75} P_{5.10} \\ 
\Lambda_{2.10}^{3} : & X_{10.6} X_{68} X_{84} P_{49} Y_{92} &-&X_{10.8} X_{84} P_{49} X_{92} & X_{25} Q_{5.10} &-&Q_{27} X_{7.10} \\ 
\Lambda_{37}^{1} : & Y_{75} X_{53} &-&X_{71} X_{13}  &  P_{38} X_{82} Q_{27} &-& Q_{38} X_{82} P_{27} \\ 
\Lambda_{39}^{1} : & Y_{92} X_{25} X_{51} X_{13} &-&X_{92} X_{25} X_{53} &  P_{38} X_{84} Q_{49} &-& Q_{38} X_{84} P_{49} \\ 
\Lambda_{47}^{1} : & X_{71} X_{14} &-& X_{7.10} X_{10.8} X_{84} &  P_{49} X_{92} Q_{27} &-& Q_{49} X_{92} P_{27} \\ 
\Lambda_{47}^{2} : & X_{7.10} X_{10.6} X_{68} X_{84} &-&Y_{75} X_{51} X_{14}  &  P_{49} Y_{92} Q_{27} &-& Q_{49} Y_{92} P_{27} \\ 
\Lambda_{56}^{1} : & X_{68} X_{84} Q_{49} Y_{92} X_{25} &-&X_{69} Y_{92} Q_{27} Y_{75} &  P_{5.10} X_{10.6} &-& X_{51} P_{16} \\ 
\Lambda_{56}^{2} : & X_{68} X_{84} P_{49} Y_{92} X_{25} &-&X_{69} Y_{92} P_{27} Y_{75}  &  X_{51} Q_{16} &-&Q_{5.10} X_{10.6} \\ 
\Lambda_{58}^{1} : & X_{82} Q_{27} Y_{75} &-& X_{84} Q_{49} X_{92} X_{25} &  P_{5.10} X_{10.8} &-& X_{53} P_{38} \\ 
\Lambda_{58}^{2} : & X_{82} P_{27} Y_{75} &-& X_{84} P_{49} X_{92} X_{25} &  X_{53} Q_{38} &-&Q_{5.10} X_{10.8} 
 \end{array} 
 }~.~
\label{E_J_C_+-}
 \eeq

This model has 120 toric phases, which are summarized in Table \ref{Table 1 - Model 14}.

%=================================================================
\begin{table}[H]
\begin{center}
\tiny
\resizebox{1 \textwidth}{!}{
\begin{tabular}{ |c|c|c|c|c|c|c|c| } 
\hline
 Phase & Path & F & Fermi Multiplicities & Phase & Path & F & Fermi Multiplicities \\
\hline
1&&16&2$\times\textbf{2}$+4$\times\textbf{3}$+4$\times\textbf{4}$&61&1,-6,-8&23&1$\times\textbf{2}$+3$\times\textbf{3}$+2$\times\textbf{4}$+1$\times\textbf{5}$+2$\times\textbf{6}$+1$\times\textbf{10}$\\ 
2&1&20&2$\times\textbf{2}$+2$\times\textbf{3}$+2$\times\textbf{4}$+2$\times\textbf{5}$+2$\times\textbf{6}$&62&3,1,-4&17&1$\times\textbf{2}$+6$\times\textbf{3}$+2$\times\textbf{4}$+1$\times\textbf{6}$\\ 
3&3&14&4$\times\textbf{2}$+4$\times\textbf{3}$+2$\times\textbf{4}$&63&3,1,-6&15&3$\times\textbf{2}$+4$\times\textbf{3}$+3$\times\textbf{4}$\\ 
4&-3&16&2$\times\textbf{2}$+4$\times\textbf{3}$+4$\times\textbf{4}$&64&3,1,7&19&3$\times\textbf{2}$+2$\times\textbf{3}$+1$\times\textbf{4}$+2$\times\textbf{5}$+2$\times\textbf{6}$\\ 
5&4&17&2$\times\textbf{2}$+4$\times\textbf{3}$+3$\times\textbf{4}$+1$\times\textbf{6}$&65&3,1,-7&18&2$\times\textbf{2}$+3$\times\textbf{3}$+3$\times\textbf{4}$+1$\times\textbf{5}$+1$\times\textbf{6}$\\ 
6&-4&18&2$\times\textbf{2}$+3$\times\textbf{3}$+2$\times\textbf{4}$+3$\times\textbf{5}$&66&3,1,-10&22&4$\times\textbf{3}$+2$\times\textbf{4}$+4$\times\textbf{6}$\\ 
7&5&21&2$\times\textbf{2}$+2$\times\textbf{3}$+1$\times\textbf{4}$+2$\times\textbf{5}$+3$\times\textbf{6}$&67&3,4,-1&20&1$\times\textbf{2}$+4$\times\textbf{3}$+2$\times\textbf{4}$+1$\times\textbf{5}$+1$\times\textbf{6}$+1$\times\textbf{7}$\\ 
8&-6&16&3$\times\textbf{2}$+4$\times\textbf{3}$+2$\times\textbf{4}$+1$\times\textbf{6}$&68&3,4,5&18&1$\times\textbf{2}$+3$\times\textbf{3}$+5$\times\textbf{4}$+1$\times\textbf{5}$\\ 
9&7&20&1$\times\textbf{2}$+4$\times\textbf{3}$+2$\times\textbf{4}$+2$\times\textbf{5}$+1$\times\textbf{8}$&69&3,4,7&22&1$\times\textbf{2}$+3$\times\textbf{3}$+3$\times\textbf{4}$+1$\times\textbf{5}$+1$\times\textbf{6}$+1$\times\textbf{10}$\\ 
10&-8&22&4$\times\textbf{3}$+2$\times\textbf{4}$+4$\times\textbf{6}$&70&3,-4,-1&18&2$\times\textbf{2}$+4$\times\textbf{3}$+2$\times\textbf{4}$+1$\times\textbf{5}$+1$\times\textbf{7}$\\ 
11&-9&21&5$\times\textbf{3}$+3$\times\textbf{4}$+1$\times\textbf{6}$+1$\times\textbf{9}$&71&3,-4,5&18&1$\times\textbf{2}$+4$\times\textbf{3}$+3$\times\textbf{4}$+2$\times\textbf{5}$\\ 
12&-10&17&3$\times\textbf{2}$+2$\times\textbf{3}$+4$\times\textbf{4}$+1$\times\textbf{6}$&72&3,5,-9&21&2$\times\textbf{2}$+2$\times\textbf{3}$+2$\times\textbf{4}$+4$\times\textbf{6}$\\ 
13&1,3&19&2$\times\textbf{2}$+2$\times\textbf{3}$+3$\times\textbf{4}$+2$\times\textbf{5}$+1$\times\textbf{6}$&73&-3,4,-1&17&3$\times\textbf{2}$+3$\times\textbf{3}$+2$\times\textbf{4}$+1$\times\textbf{5}$+1$\times\textbf{6}$\\ 
14&1,-3&18&2$\times\textbf{2}$+3$\times\textbf{3}$+3$\times\textbf{4}$+1$\times\textbf{5}$+1$\times\textbf{6}$&74&-3,4,5&17&4$\times\textbf{2}$+2$\times\textbf{3}$+1$\times\textbf{4}$+2$\times\textbf{5}$+1$\times\textbf{6}$\\ 
15&1,4&20&2$\times\textbf{2}$+2$\times\textbf{3}$+3$\times\textbf{4}$+2$\times\textbf{5}$+1$\times\textbf{8}$&75&-3,4,9&18&2$\times\textbf{2}$+2$\times\textbf{3}$+4$\times\textbf{4}$+2$\times\textbf{5}$\\ 
16&1,-4&20&2$\times\textbf{2}$+3$\times\textbf{3}$+1$\times\textbf{4}$+2$\times\textbf{5}$+1$\times\textbf{6}$+1$\times\textbf{7}$&76&-3,4,-9&18&2$\times\textbf{2}$+4$\times\textbf{3}$+2$\times\textbf{4}$+1$\times\textbf{5}$+1$\times\textbf{7}$\\ 
17&1,-6&16&3$\times\textbf{2}$+4$\times\textbf{3}$+2$\times\textbf{4}$+1$\times\textbf{6}$&77&-3,5,-9&25&1$\times\textbf{2}$+2$\times\textbf{3}$+2$\times\textbf{4}$+1$\times\textbf{5}$+3$\times\textbf{6}$+1$\times\textbf{11}$\\ 
18&1,-10&23&2$\times\textbf{2}$+1$\times\textbf{3}$+1$\times\textbf{4}$+1$\times\textbf{5}$+5$\times\textbf{6}$&78&-3,-9,-6&25&3$\times\textbf{3}$+3$\times\textbf{4}$+2$\times\textbf{5}$+1$\times\textbf{9}$+1$\times\textbf{10}$\\ 
19&3,1&17&2$\times\textbf{2}$+4$\times\textbf{3}$+3$\times\textbf{4}$+1$\times\textbf{6}$&79&-3,-9,-10&25&4$\times\textbf{3}$+2$\times\textbf{4}$+2$\times\textbf{5}$+1$\times\textbf{8}$+1$\times\textbf{12}$\\ 
20&3,4&16&2$\times\textbf{2}$+4$\times\textbf{3}$+4$\times\textbf{4}$&80&4,1,5&22&4$\times\textbf{3}$+4$\times\textbf{4}$+1$\times\textbf{6}$+1$\times\textbf{10}$\\ 
21&3,-4&16&3$\times\textbf{2}$+4$\times\textbf{3}$+2$\times\textbf{4}$+1$\times\textbf{6}$&81&4,1,7&23&1$\times\textbf{2}$+3$\times\textbf{3}$+3$\times\textbf{4}$+1$\times\textbf{5}$+1$\times\textbf{8}$+1$\times\textbf{10}$\\ 
22&3,5&16&3$\times\textbf{2}$+3$\times\textbf{3}$+3$\times\textbf{4}$+1$\times\textbf{5}$&82&4,1,-9&24&5$\times\textbf{3}$+2$\times\textbf{4}$+1$\times\textbf{6}$+1$\times\textbf{9}$+1$\times\textbf{10}$\\ 
23&-3,4&15&4$\times\textbf{2}$+3$\times\textbf{3}$+2$\times\textbf{4}$+1$\times\textbf{5}$&83&4,1,-10&21&5$\times\textbf{3}$+2$\times\textbf{4}$+2$\times\textbf{6}$+1$\times\textbf{7}$\\ 
24&-3,5&19&2$\times\textbf{2}$+2$\times\textbf{3}$+3$\times\textbf{4}$+2$\times\textbf{5}$+1$\times\textbf{6}$&84&4,5,1&24&4$\times\textbf{3}$+1$\times\textbf{4}$+2$\times\textbf{5}$+2$\times\textbf{6}$+1$\times\textbf{10}$\\ 
25&-3,-9&23&4$\times\textbf{3}$+3$\times\textbf{4}$+1$\times\textbf{5}$+1$\times\textbf{6}$+1$\times\textbf{11}$&85&4,5,3&19&2$\times\textbf{2}$+3$\times\textbf{3}$+2$\times\textbf{4}$+1$\times\textbf{5}$+2$\times\textbf{6}$\\ 
26&-3,-10&19&2$\times\textbf{2}$+3$\times\textbf{3}$+2$\times\textbf{4}$+1$\times\textbf{5}$+2$\times\textbf{6}$&86&4,5,-9&23&1$\times\textbf{2}$+3$\times\textbf{3}$+2$\times\textbf{4}$+1$\times\textbf{5}$+1$\times\textbf{6}$+2$\times\textbf{8}$\\ 
27&4,1&19&6$\times\textbf{3}$+3$\times\textbf{4}$+1$\times\textbf{8}$&87&4,-6,2&24&2$\times\textbf{2}$+2$\times\textbf{3}$+4$\times\textbf{6}$+2$\times\textbf{7}$\\ 
28&4,5&21&2$\times\textbf{2}$+3$\times\textbf{3}$+1$\times\textbf{4}$+1$\times\textbf{5}$+2$\times\textbf{6}$+1$\times\textbf{8}$&88&4,-6,7&25&3$\times\textbf{3}$+3$\times\textbf{4}$+1$\times\textbf{5}$+1$\times\textbf{7}$+1$\times\textbf{8}$+1$\times\textbf{9}$\\ 
29&4,-6&19&2$\times\textbf{2}$+3$\times\textbf{3}$+2$\times\textbf{4}$+1$\times\textbf{5}$+2$\times\textbf{6}$&89&4,-6,-9&23&4$\times\textbf{3}$+2$\times\textbf{4}$+1$\times\textbf{5}$+1$\times\textbf{6}$+1$\times\textbf{7}$+1$\times\textbf{8}$\\ 
30&4,7&23&1$\times\textbf{2}$+3$\times\textbf{3}$+3$\times\textbf{4}$+1$\times\textbf{5}$+1$\times\textbf{8}$+1$\times\textbf{10}$&90&4,7,10&21&2$\times\textbf{2}$+3$\times\textbf{3}$+2$\times\textbf{4}$+1$\times\textbf{5}$+1$\times\textbf{7}$+1$\times\textbf{9}$\\ 
31&4,-9&22&5$\times\textbf{3}$+3$\times\textbf{4}$+1$\times\textbf{8}$+1$\times\textbf{9}$&91&-4,5,3&21&2$\times\textbf{2}$+1$\times\textbf{3}$+3$\times\textbf{4}$+3$\times\textbf{5}$+1$\times\textbf{8}$\\ 
32&4,-10&17&3$\times\textbf{2}$+3$\times\textbf{3}$+2$\times\textbf{4}$+1$\times\textbf{5}$+1$\times\textbf{6}$&92&-4,5,-10&20&3$\times\textbf{2}$+1$\times\textbf{3}$+2$\times\textbf{4}$+2$\times\textbf{5}$+1$\times\textbf{6}$+1$\times\textbf{7}$\\ 
33&-4,5&23&2$\times\textbf{2}$+1$\times\textbf{3}$+2$\times\textbf{4}$+2$\times\textbf{5}$+1$\times\textbf{6}$+1$\times\textbf{7}$+1$\times\textbf{8}$&93&-4,-10,-5&22&2$\times\textbf{2}$+1$\times\textbf{3}$+1$\times\textbf{4}$+4$\times\textbf{5}$+1$\times\textbf{6}$+1$\times\textbf{7}$\\ 
34&-4,-10&19&3$\times\textbf{2}$+2$\times\textbf{3}$+1$\times\textbf{4}$+2$\times\textbf{5}$+2$\times\textbf{6}$&94&5,3,-9&23&2$\times\textbf{3}$+4$\times\textbf{4}$+4$\times\textbf{6}$\\ 
35&5,3&19&2$\times\textbf{2}$+2$\times\textbf{3}$+3$\times\textbf{4}$+2$\times\textbf{5}$+1$\times\textbf{6}$&95&5,3,-10&16&4$\times\textbf{2}$+2$\times\textbf{3}$+2$\times\textbf{4}$+2$\times\textbf{5}$\\ 
36&5,-9&25&1$\times\textbf{2}$+2$\times\textbf{3}$+2$\times\textbf{4}$+1$\times\textbf{5}$+2$\times\textbf{6}$+1$\times\textbf{7}$+1$\times\textbf{10}$&96&5,-9,3&25&1$\times\textbf{2}$+1$\times\textbf{3}$+2$\times\textbf{4}$+2$\times\textbf{5}$+3$\times\textbf{6}$+1$\times\textbf{9}$\\ 
37&5,-10&18&3$\times\textbf{2}$+2$\times\textbf{3}$+2$\times\textbf{4}$+2$\times\textbf{5}$+1$\times\textbf{6}$&97&5,-10,-8&24&1$\times\textbf{2}$+1$\times\textbf{3}$+3$\times\textbf{4}$+2$\times\textbf{5}$+1$\times\textbf{6}$+1$\times\textbf{7}$+1$\times\textbf{8}$\\ 
38&-6,-1&18&2$\times\textbf{2}$+4$\times\textbf{3}$+2$\times\textbf{4}$+1$\times\textbf{5}$+1$\times\textbf{7}$&98&-6,-8,-9&26&1$\times\textbf{2}$+2$\times\textbf{3}$+1$\times\textbf{4}$+3$\times\textbf{5}$+1$\times\textbf{7}$+1$\times\textbf{8}$+1$\times\textbf{10}$\\ 
39&-6,2&21&3$\times\textbf{2}$+1$\times\textbf{3}$+1$\times\textbf{4}$+2$\times\textbf{5}$+2$\times\textbf{6}$+1$\times\textbf{7}$&99&-6,-9,-10&20&1$\times\textbf{2}$+3$\times\textbf{3}$+4$\times\textbf{4}$+1$\times\textbf{6}$+1$\times\textbf{7}$\\ 
40&-6,7&20&6$\times\textbf{3}$+2$\times\textbf{4}$+2$\times\textbf{7}$&100&7,-8,-2&28&1$\times\textbf{2}$+2$\times\textbf{3}$+1$\times\textbf{4}$+3$\times\textbf{5}$+1$\times\textbf{8}$+1$\times\textbf{9}$+1$\times\textbf{12}$\\ 
41&-6,-8&21&1$\times\textbf{2}$+3$\times\textbf{3}$+3$\times\textbf{4}$+1$\times\textbf{5}$+1$\times\textbf{6}$+1$\times\textbf{8}$&101&7,-8,-6&21&5$\times\textbf{3}$+3$\times\textbf{4}$+1$\times\textbf{7}$+1$\times\textbf{8}$\\ 
42&-6,-9&18&1$\times\textbf{2}$+5$\times\textbf{3}$+2$\times\textbf{4}$+1$\times\textbf{5}$+1$\times\textbf{6}$&102&7,-8,-9&22&1$\times\textbf{2}$+2$\times\textbf{3}$+3$\times\textbf{4}$+1$\times\textbf{5}$+2$\times\textbf{6}$+1$\times\textbf{7}$\\ 
43&7,-8&20&1$\times\textbf{2}$+4$\times\textbf{3}$+2$\times\textbf{4}$+2$\times\textbf{5}$+1$\times\textbf{8}$&103&7,-9,-6&17&2$\times\textbf{2}$+4$\times\textbf{3}$+2$\times\textbf{4}$+2$\times\textbf{5}$\\ 
44&7,-9&19&2$\times\textbf{2}$+3$\times\textbf{3}$+2$\times\textbf{4}$+2$\times\textbf{5}$+1$\times\textbf{7}$&104&-8,-6,-1&25&1$\times\textbf{2}$+1$\times\textbf{3}$+4$\times\textbf{4}$+1$\times\textbf{5}$+1$\times\textbf{6}$+2$\times\textbf{9}$\\ 
45&7,-10&19&1$\times\textbf{2}$+5$\times\textbf{3}$+2$\times\textbf{4}$+1$\times\textbf{5}$+1$\times\textbf{8}$&105&-8,-9,-10&26&3$\times\textbf{3}$+2$\times\textbf{4}$+3$\times\textbf{5}$+1$\times\textbf{8}$+1$\times\textbf{12}$\\ 
46&-8,-6&23&2$\times\textbf{3}$+5$\times\textbf{4}$+2$\times\textbf{6}$+1$\times\textbf{8}$&106&-9,-6,-1&23&1$\times\textbf{2}$+3$\times\textbf{3}$+1$\times\textbf{4}$+2$\times\textbf{5}$+1$\times\textbf{6}$+1$\times\textbf{7}$+1$\times\textbf{8}$\\ 
47&-8,-9&28&3$\times\textbf{3}$+2$\times\textbf{4}$+1$\times\textbf{5}$+1$\times\textbf{6}$+2$\times\textbf{8}$+1$\times\textbf{12}$&107&-9,-10,-6&22&1$\times\textbf{2}$+3$\times\textbf{3}$+3$\times\textbf{4}$+1$\times\textbf{5}$+1$\times\textbf{6}$+1$\times\textbf{10}$\\ 
48&-8,-10&22&3$\times\textbf{3}$+4$\times\textbf{4}$+1$\times\textbf{5}$+1$\times\textbf{6}$+1$\times\textbf{8}$&108&-9,-10,-8&23&1$\times\textbf{2}$+2$\times\textbf{3}$+3$\times\textbf{4}$+2$\times\textbf{5}$+2$\times\textbf{8}$\\ 
49&-9,-6&21&5$\times\textbf{3}$+2$\times\textbf{4}$+1$\times\textbf{5}$+1$\times\textbf{6}$+1$\times\textbf{8}$&109&-10,-5,3&18&3$\times\textbf{2}$+2$\times\textbf{3}$+3$\times\textbf{4}$+1$\times\textbf{5}$+1$\times\textbf{7}$\\ 
50&-9,-10&21&1$\times\textbf{2}$+4$\times\textbf{3}$+3$\times\textbf{4}$+1$\times\textbf{6}$+1$\times\textbf{10}$&110&-10,-5,-8&22&1$\times\textbf{2}$+3$\times\textbf{3}$+2$\times\textbf{4}$+2$\times\textbf{5}$+1$\times\textbf{7}$+1$\times\textbf{8}$\\ 
51&-10,-5&20&2$\times\textbf{2}$+2$\times\textbf{3}$+3$\times\textbf{4}$+1$\times\textbf{5}$+1$\times\textbf{6}$+1$\times\textbf{7}$&111&1,3,4,-5&26&3$\times\textbf{3}$+2$\times\textbf{4}$+1$\times\textbf{5}$+1$\times\textbf{6}$+1$\times\textbf{7}$+1$\times\textbf{8}$+1$\times\textbf{9}$\\ 
52&-10,-8&21&1$\times\textbf{2}$+2$\times\textbf{3}$+3$\times\textbf{4}$+2$\times\textbf{5}$+2$\times\textbf{6}$&112&1,3,4,-10&25&2$\times\textbf{2}$+1$\times\textbf{3}$+1$\times\textbf{5}$+5$\times\textbf{6}$+1$\times\textbf{8}$\\ 
53&1,3,4&20&2$\times\textbf{2}$+2$\times\textbf{3}$+2$\times\textbf{4}$+2$\times\textbf{5}$+2$\times\textbf{6}$&113&1,4,5,-10&27&1$\times\textbf{3}$+3$\times\textbf{4}$+1$\times\textbf{5}$+4$\times\textbf{6}$+1$\times\textbf{10}$\\ 
54&1,3,-10&24&2$\times\textbf{2}$+1$\times\textbf{3}$+1$\times\textbf{4}$+5$\times\textbf{6}$+1$\times\textbf{7}$&114&3,1,-4,-10&22&1$\times\textbf{2}$+3$\times\textbf{3}$+3$\times\textbf{5}$+3$\times\textbf{6}$\\ 
55&1,-3,-10&23&2$\times\textbf{2}$+2$\times\textbf{3}$+6$\times\textbf{6}$&115&3,1,-6,-7&16&2$\times\textbf{2}$+4$\times\textbf{3}$+4$\times\textbf{4}$\\ 
56&1,4,5&26&2$\times\textbf{3}$+3$\times\textbf{4}$+2$\times\textbf{5}$+2$\times\textbf{6}$+1$\times\textbf{12}$&116&3,4,5,-9&23&2$\times\textbf{3}$+4$\times\textbf{4}$+4$\times\textbf{6}$\\ 
57&1,4,-10&23&2$\times\textbf{2}$+1$\times\textbf{3}$+2$\times\textbf{4}$+1$\times\textbf{5}$+3$\times\textbf{6}$+1$\times\textbf{8}$&117&3,4,7,10&24&2$\times\textbf{2}$+1$\times\textbf{3}$+2$\times\textbf{4}$+3$\times\textbf{5}$+1$\times\textbf{7}$+1$\times\textbf{11}$\\ 
58&1,-4,-10&23&2$\times\textbf{2}$+2$\times\textbf{3}$+2$\times\textbf{5}$+2$\times\textbf{6}$+2$\times\textbf{7}$&118&-3,-9,-6,-10&28&3$\times\textbf{3}$+2$\times\textbf{4}$+3$\times\textbf{5}$+2$\times\textbf{12}$\\ 
59&1,-6,2&19&3$\times\textbf{2}$+2$\times\textbf{3}$+2$\times\textbf{4}$+1$\times\textbf{5}$+1$\times\textbf{6}$+1$\times\textbf{7}$&119&-3,-9,-10,-6&28&4$\times\textbf{3}$+1$\times\textbf{4}$+2$\times\textbf{5}$+1$\times\textbf{7}$+1$\times\textbf{11}$+1$\times\textbf{12}$\\ 
60&1,-6,-7&19&2$\times\textbf{2}$+4$\times\textbf{3}$+1$\times\textbf{4}$+2$\times\textbf{5}$+1$\times\textbf{8}$&120&-4,-10,-5,3&20&2$\times\textbf{2}$+2$\times\textbf{3}$+2$\times\textbf{4}$+2$\times\textbf{5}$+2$\times\textbf{6}$\\ 
\hline
\end{tabular}}
\end{center}
\caption{Basic information regarding the 120 toric phases of Model 14.}
\label{Table 1 - Model 14}
\end{table}
%=================================================================

Table \ref{Table 2 - Model 14} summarizes the connection between the toric phases under triality.

%=================================================================
\begin{center}
\tiny
\resizebox{1 \textwidth}{!}{
\begin{tabular}{ |c|c|c|c|c|c|c|c|c|c|c|} 
\hline
N&1&2&3&4&5&6&7&8&9&10\\
\hline
1&2&&3 , \underline{4}&5 , \underline{6}&7&\underline{8}&9&\underline{10}&\underline{11}&\underline{12}\\
2&\underline{1}&&13 , \underline{14}&15 , \underline{16}&&\underline{17}&&&&\underline{18}\\
3&19&&4 , \underline{1}&20 , \underline{21}&22&\underline{22}&21 , \underline{20}&1 , \underline{4}&&\underline{19}\\
4&14&&1 , \underline{3}&23 , \underline{17}&24&\underline{14}&&10&\underline{25}&\underline{26}\\
5&27&&20 , \underline{23}&6 , \underline{1}&28&\underline{29}&30&&\underline{31}&\underline{32}\\
6&16&&21 , \underline{17}&1 , \underline{5}&33&\underline{16}&&&&\underline{34}\\
7&&&35 , \underline{24}&28 , \underline{33}&\underline{1}&&&&\underline{36}&\underline{37}\\
8&17 , \underline{38}&39&22 , \underline{14}&29 , \underline{16}&&1&40&\underline{41}&\underline{42}&\\
9&&&21&30&&\underline{40}&\underline{1}&\underline{43}&\underline{44}&38 , \underline{45}\\
10&&&4&&&\underline{46}&43&1&\underline{47}&\underline{48}\\
11&&&\underline{25}&31&36&\underline{49}&44&\underline{47}&1&\underline{50}\\
12&18&&19 , \underline{26}&32 , \underline{34}&37 , \underline{51}&&45&\underline{52}&\underline{50}&1\\
13&\underline{19}&&14 , \underline{2}&53 , \underline{29}&&\underline{23}&&52&&\underline{54}\\
14&\underline{4}&&2 , \underline{13}&22 , \underline{8}&&\underline{4}&&48&&\underline{55}\\
15&\underline{27}&&53 , \underline{22}&16 , \underline{2}&56&\underline{21}&&&&\underline{57}\\
16&\underline{6}&&29 , \underline{8}&2 , \underline{15}&&\underline{6}&&&&\underline{58}\\
17&38 , \underline{8}&59&23 , \underline{4}&21 , \underline{6}&&2&\underline{60}&\underline{61}&\underline{43}&\\
18&\underline{12}&&54 , \underline{55}&57 , \underline{58}&&&&&&2\\
19&\underline{3}&&13&\underline{62}&&\underline{63}&64 , \underline{65}&12 , \underline{26}&&\underline{66}\\
20&\underline{67}&&23 , \underline{5}&21 , \underline{3}&68&\underline{53}&69&&&\underline{65}\\
21&62 , \underline{70}&&17 , \underline{6}&3 , \underline{20}&71&\underline{15}&\underline{69}&9&&\underline{64}\\
22&&&35&68 , \underline{71}&\underline{3}&&15 , \underline{53}&8 , \underline{14}&\underline{72}&\underline{63}\\
23&63 , \underline{73}&&5 , \underline{20}&17 , \underline{4}&74&\underline{13}&&&75 , \underline{76}&\underline{74}\\
24&&&7 , \underline{35}&74 , \underline{60}&\underline{4}&&&46&\underline{77}&\underline{75}\\
25&&&11&76&77&\underline{78}&&&4&\underline{79}\\
26&55&&12 , \underline{19}&74 , \underline{59}&75&&&&\underline{79}&4\\
27&\underline{5}&&\underline{63}&15&80&\underline{62}&81&&\underline{82}&\underline{83}\\
28&84&&85 , \underline{74}&33 , \underline{7}&\underline{5}&&&&\underline{86}&\underline{73}\\
29&62&87&53 , \underline{13}&16 , \underline{8}&&5&88&&\underline{89}&\\
30&81&&69&\underline{9}&&\underline{88}&\underline{5}&&\underline{51}&90 , \underline{50}\\
31&82&&\underline{76}&\underline{11}&86&\underline{82}&51&&5&\underline{45}\\
32&83&&65 , \underline{74}&34 , \underline{12}&73 , \underline{44}&&50&&\underline{45}&5\\
33&&&91 , \underline{60}&7 , \underline{28}&\underline{6}&&&&&\underline{92}\\
34&58&&64 , \underline{59}&12 , \underline{32}&92 , \underline{93}&&&&&6\\
35&&&24 , \underline{7}&85 , \underline{91}&\underline{22}&&&41&\underline{94}&\underline{95}\\
36&&&96 , \underline{77}&86&\underline{11}&&&&7&\underline{90}\\
37&&&95 , \underline{75}&73 , \underline{92}&51 , \underline{12}&&&\underline{97}&\underline{90}&7\\
38&8 , \underline{17}&86&73&70&&&9 , \underline{45}&\underline{90}&\underline{70}&\\
39&59 , \underline{86}&\underline{8}&72 , \underline{55}&87 , \underline{58}&&&&\underline{96}&&\\
40&\underline{9}&&71&88&&9&\underline{8}&\underline{88}&\underline{71}&8\\
41&61 , \underline{90}&96&35&&&\underline{46}&88&8&\underline{98}&\\
42&43 , \underline{70}&&\underline{48}&89&&\underline{49}&71&\underline{98}&8&\underline{99}\\
43&&\underline{100}&17&&&\underline{101}&\underline{10}&9&\underline{102}&70 , \underline{42}\\
44&&&&51 , \underline{93}&&\underline{103}&\underline{11}&\underline{102}&9&73 , \underline{32}\\
45&&&62&50&\underline{31}&60&\underline{12}&\underline{89}&\underline{32}&9 , \underline{38}\\
46&\underline{104}&&24&&&10&101&41&&\\
47&&&&&&&102&11&10&\underline{105}\\
48&&&14&&\underline{78}&&42&52&\underline{105}&10\\
49&\underline{106}&&\underline{78}&82&&11&103&&42&\underline{80}\\
50&&&\underline{79}&45&90 , \underline{30}&\underline{107}&32&\underline{108}&12&11\\
51&&&109&44 , \underline{93}&12 , \underline{37}&&31&\underline{110}&\underline{30}&\\
52&&&13&&97 , \underline{110}&&89&12&\underline{108}&\underline{48}\\
53&&&22 , \underline{15}&29 , \underline{13}&\underline{111}&\underline{20}&&&&\underline{112}\\
54&\underline{66}&&55 , \underline{18}&112 , \underline{87}&&&&&&13\\
55&\underline{26}&&18 , \underline{54}&72 , \underline{39}&&&&&&14\\
56&\underline{80}&&\underline{68}&&\underline{15}&\underline{69}&&&&\underline{113}\\
57&\underline{83}&&112 , \underline{72}&58 , \underline{18}&113&&&&&15\\
58&\underline{34}&&87 , \underline{39}&18 , \underline{57}&&&&&&16\\
59&86 , \underline{39}&\underline{17}&74 , \underline{26}&64 , \underline{34}&&&\underline{76}&\underline{77}&&\\
60&45&76&74 , \underline{24}&91 , \underline{33}&&&17&\underline{79}&\underline{101}&\\
61&90 , \underline{41}&77&75&&&&\underline{79}&17&\underline{100}&\\
\hline
\end{tabular}}
\end{center}                           
%=================================================================
%=================================================================
\begin{table}[h]
\begin{center}
\tiny
\resizebox{1 \textwidth}{!}{
\begin{tabular}{ |c|c|c|c|c|c|c|c|c|c|c|} 
\hline
N&1&2&3&4&5&6&7&8&9&10\\
\hline
62&70 , \underline{21}&&29&19&\underline{101}&\underline{27}&\underline{107}&45&&\underline{114}\\
63&\underline{22}&85&23 , \underline{73}&\underline{27}&&19&103 , \underline{115}&109 , \underline{95}&&\\
64&\underline{21}&&&&&\underline{103}&65 , \underline{19}&34 , \underline{59}&&106 , \underline{114}\\
65&\underline{20}&&&\underline{107}&&\underline{115}&19 , \underline{64}&32 , \underline{74}&&84\\
66&\underline{19}&&54&\underline{114}&&&114&\underline{54}&&19\\
67&20&&73&70&99 , \underline{102}&&&&&\underline{84}\\
68&\underline{99}&&85&71 , \underline{22}&\underline{20}&&56&&\underline{116}&\underline{115}\\
69&&&\underline{30}&\underline{21}&56&\underline{111}&\underline{20}&&&117 , \underline{107}\\
70&21 , \underline{62}&&38&\underline{67}&42 , \underline{43}&&\underline{117}&38&&\underline{106}\\
71&\underline{42}&&91&22 , \underline{68}&\underline{21}&&\underline{111}&40&&\underline{103}\\
72&&&94&116&&&57 , \underline{112}&39 , \underline{55}&22&\underline{85}\\
73&23 , \underline{63}&&\underline{67}&38&32 , \underline{44}&&&&37 , \underline{92}&\underline{28}\\
74&65 , \underline{32}&&28 , \underline{85}&60 , \underline{24}&\underline{23}&&&&26 , \underline{59}&\underline{23}\\
75&95 , \underline{37}&&&61&26&\underline{97}&&&76 , \underline{23}&\underline{24}\\
76&109 , \underline{92}&&31&\underline{25}&59&\underline{110}&&&23 , \underline{75}&\underline{60}\\
77&&&36 , \underline{96}&59&\underline{25}&&&&24&\underline{61}\\
78&&&49&110&&25&&&48&\underline{118}\\
79&&&50&60&61&\underline{119}&&&26&25\\
80&84&&\underline{115}&56&\underline{27}&\underline{107}&118&&\underline{49}&\underline{99}\\
81&\underline{30}&&&&118&\underline{101}&\underline{27}&&\underline{110}&104 , \underline{108}\\
82&\underline{31}&&\underline{109}&&49&\underline{31}&110&&27&\underline{89}\\
83&\underline{32}&&\underline{85}&57&99&&108&&\underline{89}&27\\
84&\underline{28}&&\underline{65}&&\underline{80}&&&&\underline{106}&\underline{67}\\
85&\underline{83}&&74 , \underline{28}&91 , \underline{35}&\underline{68}&&&&\underline{72}&\underline{63}\\
86&106&&39 , \underline{59}&\underline{36}&\underline{31}&&&&28&\underline{38}\\
87&114&\underline{29}&112 , \underline{54}&58 , \underline{39}&&&&&&\\
88&101&&111&\underline{40}&&30&\underline{29}&&\underline{91}&41\\
89&45&&\underline{52}&\underline{42}&&\underline{82}&91&&29&\underline{83}\\
90&104&&117&\underline{38}&\underline{36}&61 , \underline{41}&&&\underline{37}&50 , \underline{30}\\
91&\underline{89}&&60 , \underline{33}&35 , \underline{85}&\underline{71}&&&88&&\underline{109}\\
92&&&109 , \underline{76}&37 , \underline{73}&93 , \underline{34}&&&&&33\\
93&&&120&51 , \underline{44}&34 , \underline{92}&&&&&\\
94&&&\underline{96}&72&\underline{72}&&&96&35&\underline{35}\\
95&&&75 , \underline{37}&63 , \underline{109}&109 , \underline{63}&&&37 , \underline{75}&\underline{35}&35\\
96&&&77 , \underline{36}&39&&&&&94&\underline{41}\\
97&&&75&&110 , \underline{52}&&&37&\underline{104}&\\
98&100 , \underline{117}&&&&&&111&42&41&\\
99&102 , \underline{67}&&\underline{105}&83&\underline{113}&\underline{80}&68&&&42\\
100&&43&61&&&\underline{119}&&&&117 , \underline{98} \\
101&\underline{81}&\underline{119}&60&&&43&\underline{46}&88&&62\\
102&&&&&&&\underline{47}&44&43&67 , \underline{99}\\
103&\underline{64}&&&109 , \underline{120}&&44&\underline{49}&&71&63 , \underline{115}\\
104&46&&97&&&&81 , \underline{108}&90&&\\
105&&&&&\underline{118}&&99&108&48&47\\
106&49&&&86&&&64 , \underline{114}&&70&\underline{84}\\
107&&&\underline{119}&62&117 , \underline{69}&50&65&&&\underline{80}\\
108&&&&&104 , \underline{81}&&83&50&52&\underline{105}\\
109&&&\underline{51}&103 , \underline{120}&63 , \underline{95}&&82&92 , \underline{76}&\underline{91}&\\
110&&&76&&52 , \underline{97}&&82&51&\underline{81}&\underline{78}\\
111&&\underline{98}&71&88&53&\underline{69}&&&&\\
112&&&72 , \underline{57}&87 , \underline{54}&&&&&&53\\
113&\underline{99}&&\underline{116}&&\underline{57}&&&&&56\\
114&106 , \underline{64}&&87&66&&&&&&62\\
115&\underline{68}&68&\underline{65}&\underline{80}&&65&63 , \underline{103}&103 , \underline{63}&&80\\
116&\underline{113}&&72&\underline{72}&&&113&&68&\underline{68}\\
117&&&\underline{90}&\underline{70}&&100 , \underline{98}&&&&107 , \underline{69}\\
118&&&80&81&&119&&&105&78\\
119&&&107&101&100&79&&&&\underline{118}\\
120&&&\underline{93}&109 , \underline{103}&103 , \underline{109}&&&93&&\\\hline
\end{tabular}}
\end{center}
\caption{Triality connections between the 120 toric phases of Model 14.}
\label{Table 2 - Model 14}
\end{table}
%=================================================================

\newpage

%=================================================================
\subsection{Model 15: $P^{3}_{+-}(\text{dP}_2)$}
%=================================================================
 
\begin{figure}[H]
\begin{center}
\resizebox{0.35\hsize}{!}{
\includegraphics[height=6cm]{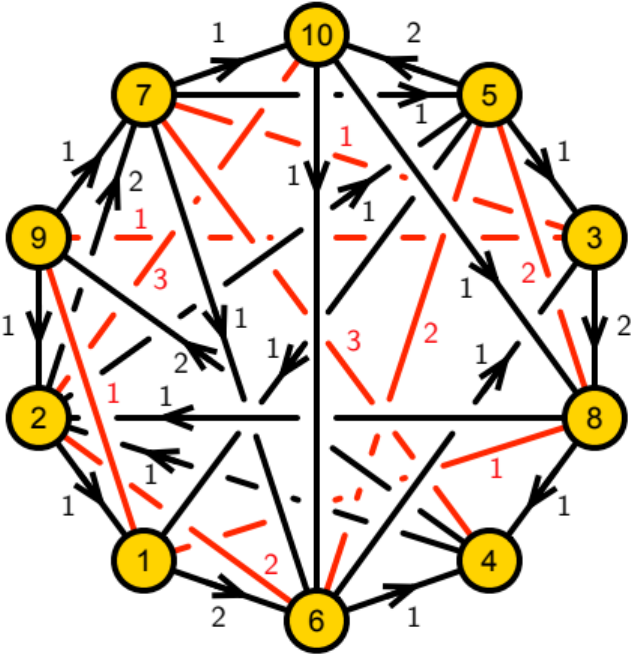} 
}
\caption{Quiver for Phase 1 of Model 15.
\label{f_quiver_15}}
 \end{center}
 \end{figure}

 \fref{f_quiver_15} shows the quiver for Phase 1 of Model 15.
 
%=================================================================
%=================================================================

The $J$- and $E$-terms are
\beq
{\footnotesize
\begin{array}{rrclrcl}
 &  & J &  & & E &  \\
\Lambda_{18}^{1} : & X_{84}  Y_{42}  X_{25}  X_{51} &-&X_{82}  X_{21}  &   P_{16}  X_{63}  Q_{38} &-& Q_{16}  X_{63}  P_{38}  \\    
\Lambda_{19}^{1} : & X_{92}  X_{21} &-& Y_{97}  Y_{75}  X_{51} &   P_{16}  X_{64}  Q_{49} &-& Q_{16}  X_{64}  P_{49}  \\    
\Lambda_{26}^{1} : & X_{64}  Q_{49}  X_{92} &-&X_{63}  Q_{38}  X_{82}  &   P_{27}  X_{76} &-& X_{21}  P_{16}  \\    
\Lambda_{26}^{2} : & X_{64}  P_{49}  X_{92} &-&X_{63}  P_{38}  X_{82} &   X_{21}  Q_{16} &-&Q_{27}  X_{76}  \\    
\Lambda_{2.10}^{1} : & X_{10.6}  X_{63}  Q_{38}  X_{84}  Y_{42} &-&X_{10.8}  X_{84}  Q_{49}  X_{92} &   P_{27}  X_{7.10} &-& X_{25}  P_{5.10}  \\    
\Lambda_{2.10}^{2} : & X_{10.8}  X_{82} &-& X_{10.6}  X_{64}  Y_{42} &   P_{27}  Y_{75}  Q_{5.10} &-& Q_{27}  Y_{75}  P_{5.10}  \\    
\Lambda_{2.10}^{3} : & X_{10.6}  X_{63}  P_{38}  X_{84}  Y_{42} &-&X_{10.8}  X_{84}  P_{49}  X_{92} & X_{25}  Q_{5.10} &-&Q_{27}  X_{7.10}  \\    
\Lambda_{37}^{1} : & Y_{75}  X_{53} &-&X_{76}  X_{63} &   P_{38}  X_{82}  Q_{27} &-& Q_{38}  X_{82}  P_{27}  \\    
\Lambda_{39}^{1} : & Y_{97}  X_{7.10}  X_{10.6}  X_{63} &-&X_{92}  X_{25}  X_{53}  &   P_{38}  X_{84}  Q_{49} &-& Q_{38}  X_{84} P_{49}  \\    
\Lambda_{47}^{1} : & X_{76}  X_{64} &-& X_{7.10}  X_{10.8}  X_{84} &   P_{49}  X_{92}  Q_{27} &-& Q_{49}  X_{92}  P_{27}  \\    
\Lambda_{47}^{2} : & X_{7.10}  X_{10.6}  X_{63}  Q_{38}  X_{84} &-&Y_{75}  Q_{5.10}  X_{10.6}  X_{64}  &   P_{49}  Y_{97} &-& Y_{42}  P_{27}  \\    
\Lambda_{47}^{3} : &  X_{7.10}  X_{10.6}  X_{63}  P_{38}  X_{84} &-&Y_{75}  X_{51}  P_{16}  X_{64} & Y_{42}  Q_{27} &-&Q_{49}  Y_{97}  \\    
\Lambda_{56}^{1} : & X_{63}  Q_{38}  X_{84}  Y_{42}  X_{25} &-&X_{64}  Y_{42}  Q_{27}  Y_{75} &   P_{5.10}  X_{10.6} &-& X_{51}  P_{16}  \\    
\Lambda_{56}^{2} : & X_{63}  P_{38}  X_{84}  Y_{42}  X_{25} &-&X_{64}  P_{49}  Y_{97}  Y_{75}  &   X_{51}  Q_{16}  &-&Q_{5.10}  X_{10.6}  \\    
\Lambda_{58}^{1} : & X_{82}  Q_{27}  Y_{75} &-& X_{84}  Q_{49}  X_{92}  X_{25} &   P_{5.10}  X_{10.8} &-& X_{53}  P_{38}  \\    
\Lambda_{58}^{2} : & X_{82}  P_{27}  Y_{75} &-& X_{84}  P_{49}  X_{92}  X_{25} &  X_{53}  Q_{38} &-& Q_{5.10}  X_{10.8} 
 \end{array} 
 }~.~
\label{E_J_C_+-}
 \eeq

This model has 75 toric phases, which are summarized in Table \ref{Table 1 - Model 15}.

%=================================================================
\begin{table}[H]
\begin{center}
\tiny
\resizebox{1 \textwidth}{!}{
\begin{tabular}{ |c|c|c|c|c|c|c|c| } 
\hline
 Phase & Path & F & Fermi Multiplicities & Phase & Path & F & Fermi Multiplicities \\
\hline
1&&16&3$\times\textbf{2}$+3$\times\textbf{3}$+3$\times\textbf{4}$+1$\times\textbf{5}$&39&-10,-5&20&2$\times\textbf{2}$+3$\times\textbf{3}$+1$\times\textbf{4}$+3$\times\textbf{5}$+1$\times\textbf{8}$\\ 
2&1&14&4$\times\textbf{2}$+4$\times\textbf{3}$+2$\times\textbf{4}$&40&-10,-8&19&2$\times\textbf{2}$+3$\times\textbf{3}$+2$\times\textbf{4}$+2$\times\textbf{5}$+1$\times\textbf{7}$\\ 
3&-1&16&2$\times\textbf{2}$+4$\times\textbf{3}$+4$\times\textbf{4}$&41&1,2,3&18&1$\times\textbf{2}$+5$\times\textbf{3}$+2$\times\textbf{4}$+1$\times\textbf{5}$+1$\times\textbf{6}$\\ 
4&3&16&2$\times\textbf{2}$+6$\times\textbf{3}$+2$\times\textbf{5}$&42&1,2,-8&21&2$\times\textbf{2}$+2$\times\textbf{3}$+2$\times\textbf{4}$+4$\times\textbf{6}$\\ 
5&-3&18&2$\times\textbf{2}$+4$\times\textbf{3}$+4$\times\textbf{5}$&43&1,2,-9&20&2$\times\textbf{2}$+3$\times\textbf{3}$+1$\times\textbf{4}$+1$\times\textbf{5}$+3$\times\textbf{6}$\\ 
6&4&21&2$\times\textbf{2}$+2$\times\textbf{3}$+2$\times\textbf{4}$+4$\times\textbf{6}$&44&-1,4,5&22&1$\times\textbf{2}$+5$\times\textbf{3}$+1$\times\textbf{4}$+1$\times\textbf{6}$+1$\times\textbf{8}$+1$\times\textbf{9}$\\ 
7&5&21&2$\times\textbf{2}$+3$\times\textbf{3}$+1$\times\textbf{4}$+1$\times\textbf{5}$+1$\times\textbf{6}$+2$\times\textbf{7}$&45&-1,4,6&23&3$\times\textbf{3}$+4$\times\textbf{4}$+2$\times\textbf{6}$+1$\times\textbf{9}$\\ 
8&-6&22&1$\times\textbf{2}$+2$\times\textbf{3}$+3$\times\textbf{4}$+1$\times\textbf{5}$+2$\times\textbf{6}$+1$\times\textbf{7}$&46&-1,4,-9&20&1$\times\textbf{2}$+5$\times\textbf{3}$+1$\times\textbf{4}$+1$\times\textbf{5}$+2$\times\textbf{7}$\\ 
9&-8&20&2$\times\textbf{2}$+2$\times\textbf{3}$+3$\times\textbf{4}$+3$\times\textbf{6}$&47&-1,4,-10&20&2$\times\textbf{2}$+4$\times\textbf{3}$+4$\times\textbf{6}$\\ 
10&9&18&1$\times\textbf{2}$+5$\times\textbf{3}$+2$\times\textbf{4}$+1$\times\textbf{5}$+1$\times\textbf{6}$&48&3,4,5&24&4$\times\textbf{3}$+2$\times\textbf{4}$+3$\times\textbf{6}$+1$\times\textbf{10}$\\ 
11&-9&17&2$\times\textbf{2}$+4$\times\textbf{3}$+3$\times\textbf{4}$+1$\times\textbf{6}$&49&3,4,-10&23&2$\times\textbf{2}$+3$\times\textbf{3}$+1$\times\textbf{5}$+1$\times\textbf{6}$+2$\times\textbf{7}$+1$\times\textbf{8}$\\ 
12&-10&17&4$\times\textbf{2}$+3$\times\textbf{3}$+2$\times\textbf{5}$+1$\times\textbf{7}$&50&3,5,-7&24&5$\times\textbf{3}$+2$\times\textbf{6}$+3$\times\textbf{7}$\\ 
13&1,2&18&3$\times\textbf{2}$+2$\times\textbf{3}$+2$\times\textbf{4}$+2$\times\textbf{5}$+1$\times\textbf{6}$&51&3,5,-9&20&1$\times\textbf{2}$+4$\times\textbf{3}$+2$\times\textbf{4}$+2$\times\textbf{5}$+1$\times\textbf{8}$\\ 
14&1,3&14&2$\times\textbf{2}$+8$\times\textbf{3}$&52&3,5,-10&17&2$\times\textbf{2}$+5$\times\textbf{3}$+1$\times\textbf{4}$+1$\times\textbf{5}$+1$\times\textbf{6}$\\ 
15&1,4&19&2$\times\textbf{2}$+2$\times\textbf{3}$+4$\times\textbf{4}$+2$\times\textbf{6}$&53&3,9,-10&23&5$\times\textbf{3}$+1$\times\textbf{4}$+1$\times\textbf{5}$+2$\times\textbf{7}$+1$\times\textbf{8}$\\ 
16&-1,4&19&2$\times\textbf{2}$+4$\times\textbf{3}$+1$\times\textbf{4}$+1$\times\textbf{5}$+1$\times\textbf{6}$+1$\times\textbf{7}$&54&-3,4,5&24&1$\times\textbf{2}$+5$\times\textbf{3}$+1$\times\textbf{6}$+1$\times\textbf{7}$+2$\times\textbf{9}$\\ 
17&-1,5&19&1$\times\textbf{2}$+5$\times\textbf{3}$+1$\times\textbf{4}$+1$\times\textbf{5}$+2$\times\textbf{6}$&55&-3,4,-9&20&1$\times\textbf{2}$+5$\times\textbf{3}$+1$\times\textbf{5}$+3$\times\textbf{6}$\\ 
18&3,4&22&2$\times\textbf{2}$+3$\times\textbf{3}$+1$\times\textbf{5}$+3$\times\textbf{6}$+1$\times\textbf{8}$&56&-3,4,-10&22&2$\times\textbf{2}$+4$\times\textbf{3}$+4$\times\textbf{7}$\\ 
19&3,5&18&1$\times\textbf{2}$+6$\times\textbf{3}$+1$\times\textbf{4}$+1$\times\textbf{5}$+1$\times\textbf{7}$&57&4,5,3&26&1$\times\textbf{2}$+3$\times\textbf{3}$+2$\times\textbf{4}$+1$\times\textbf{6}$+1$\times\textbf{8}$+1$\times\textbf{9}$+1$\times\textbf{10}$\\ 
20&3,-7&20&1$\times\textbf{2}$+3$\times\textbf{3}$+2$\times\textbf{4}$+3$\times\textbf{5}$+1$\times\textbf{6}$&58&4,-9,1&18&2$\times\textbf{2}$+6$\times\textbf{3}$+2$\times\textbf{7}$\\ 
21&3,9&20&1$\times\textbf{2}$+5$\times\textbf{3}$+1$\times\textbf{4}$+2$\times\textbf{6}$+1$\times\textbf{7}$&59&4,-9,-6&26&2$\times\textbf{3}$+3$\times\textbf{4}$+3$\times\textbf{6}$+1$\times\textbf{7}$+1$\times\textbf{9}$\\ 
22&3,-10&19&2$\times\textbf{2}$+4$\times\textbf{3}$+1$\times\textbf{4}$+1$\times\textbf{5}$+1$\times\textbf{6}$+1$\times\textbf{7}$&60&4,-9,7&22&2$\times\textbf{2}$+1$\times\textbf{3}$+2$\times\textbf{4}$+4$\times\textbf{5}$+1$\times\textbf{9}$\\ 
23&-3,4&21&2$\times\textbf{2}$+4$\times\textbf{3}$+2$\times\textbf{6}$+2$\times\textbf{7}$&61&5,3,9&23&1$\times\textbf{2}$+2$\times\textbf{3}$+2$\times\textbf{4}$+3$\times\textbf{5}$+1$\times\textbf{6}$+1$\times\textbf{9}$\\ 
24&-3,5&21&1$\times\textbf{2}$+5$\times\textbf{3}$+1$\times\textbf{5}$+1$\times\textbf{6}$+2$\times\textbf{7}$&62&5,3,-9&21&1$\times\textbf{2}$+4$\times\textbf{3}$+1$\times\textbf{4}$+2$\times\textbf{5}$+1$\times\textbf{6}$+1$\times\textbf{8}$\\ 
25&4,5&26&1$\times\textbf{2}$+3$\times\textbf{3}$+2$\times\textbf{4}$+3$\times\textbf{8}$+1$\times\textbf{9}$&63&5,3,-10&18&3$\times\textbf{2}$+4$\times\textbf{3}$+1$\times\textbf{5}$+1$\times\textbf{6}$+1$\times\textbf{7}$\\ 
26&4,-9&20&1$\times\textbf{2}$+3$\times\textbf{3}$+4$\times\textbf{4}$+1$\times\textbf{6}$+1$\times\textbf{7}$&64&5,-10,-8&22&2$\times\textbf{2}$+1$\times\textbf{3}$+2$\times\textbf{4}$+4$\times\textbf{5}$+1$\times\textbf{9}$\\ 
27&4,-10&20&3$\times\textbf{2}$+2$\times\textbf{3}$+1$\times\textbf{4}$+1$\times\textbf{5}$+2$\times\textbf{6}$+1$\times\textbf{7}$&65&-6,-7,-8&26&3$\times\textbf{3}$+2$\times\textbf{4}$+3$\times\textbf{5}$+1$\times\textbf{9}$+1$\times\textbf{11}$\\ 
28&5,3&21&2$\times\textbf{2}$+3$\times\textbf{3}$+1$\times\textbf{4}$+2$\times\textbf{5}$+1$\times\textbf{7}$+1$\times\textbf{8}$&66&-6,-8,-9&24&1$\times\textbf{2}$+2$\times\textbf{3}$+2$\times\textbf{4}$+1$\times\textbf{5}$+2$\times\textbf{6}$+1$\times\textbf{7}$+1$\times\textbf{8}$\\ 
29&5,9&23&1$\times\textbf{2}$+2$\times\textbf{3}$+2$\times\textbf{4}$+2$\times\textbf{5}$+2$\times\textbf{6}$+1$\times\textbf{8}$&67&-9,1,5&19&6$\times\textbf{3}$+2$\times\textbf{4}$+2$\times\textbf{6}$\\ 
30&5,-10&18&3$\times\textbf{2}$+4$\times\textbf{3}$+1$\times\textbf{5}$+1$\times\textbf{6}$+1$\times\textbf{7}$&68&-9,1,-10&20&1$\times\textbf{2}$+5$\times\textbf{3}$+2$\times\textbf{4}$+1$\times\textbf{7}$+1$\times\textbf{8}$\\ 
31&-6,-7&26&3$\times\textbf{3}$+1$\times\textbf{4}$+3$\times\textbf{5}$+1$\times\textbf{7}$+1$\times\textbf{8}$+1$\times\textbf{9}$&69&-9,-10,-8&18&2$\times\textbf{2}$+4$\times\textbf{3}$+1$\times\textbf{4}$+2$\times\textbf{5}$+1$\times\textbf{6}$\\ 
32&-6,-8&24&1$\times\textbf{2}$+2$\times\textbf{3}$+3$\times\textbf{4}$+1$\times\textbf{5}$+1$\times\textbf{6}$+1$\times\textbf{8}$+1$\times\textbf{9}$&70&-10,-5,3&20&2$\times\textbf{2}$+3$\times\textbf{3}$+1$\times\textbf{4}$+3$\times\textbf{5}$+1$\times\textbf{8}$\\ 
33&-6,-9&21&1$\times\textbf{2}$+2$\times\textbf{3}$+4$\times\textbf{4}$+3$\times\textbf{6}$&71&1,2,3,-7&20&4$\times\textbf{3}$+2$\times\textbf{4}$+4$\times\textbf{5}$\\ 
34&-8,-9&22&1$\times\textbf{2}$+3$\times\textbf{3}$+2$\times\textbf{4}$+1$\times\textbf{5}$+2$\times\textbf{6}$+1$\times\textbf{8}$&72&1,2,-8,-9&24&2$\times\textbf{2}$+2$\times\textbf{3}$+2$\times\textbf{5}$+2$\times\textbf{6}$+2$\times\textbf{8}$\\ 
35&-8,-10&20&2$\times\textbf{2}$+2$\times\textbf{3}$+3$\times\textbf{4}$+2$\times\textbf{5}$+1$\times\textbf{8}$&73&-1,4,5,6&24&3$\times\textbf{3}$+3$\times\textbf{4}$+2$\times\textbf{5}$+1$\times\textbf{6}$+1$\times\textbf{11}$\\ 
36&9,-10&19&2$\times\textbf{2}$+4$\times\textbf{3}$+1$\times\textbf{4}$+2$\times\textbf{5}$+1$\times\textbf{8}$&74&3,4,5,-9&26&2$\times\textbf{3}$+4$\times\textbf{4}$+2$\times\textbf{6}$+1$\times\textbf{7}$+1$\times\textbf{11}$\\ 
37&-9,1&17&1$\times\textbf{2}$+6$\times\textbf{3}$+2$\times\textbf{4}$+1$\times\textbf{6}$&75&-6,-7,-8,-2&34&3$\times\textbf{4}$+3$\times\textbf{5}$+2$\times\textbf{7}$+1$\times\textbf{13}$+1$\times\textbf{14}$\\ 
38&-9,-10&18&3$\times\textbf{2}$+4$\times\textbf{3}$+2$\times\textbf{5}$+1$\times\textbf{8}$& & & & \\ 
\hline
\end{tabular}}
\end{center}
\caption{Basic information regarding the 75 toric phases of Model 15.}
\label{Table 1 - Model 15}
\end{table}
%=================================================================

Table \ref{Table 2 - Model 15} summarizes the connection between the toric phases under triality.

%=================================================================
\begin{center}
\tiny
\resizebox{.92 \textwidth}{!}{
\begin{tabular}{ |c|c|c|c|c|c|c|c|c|c|c|} 
\hline
N&1&2&3&4&5&6&7&8&9&10\\
\hline
1&2 , \underline{3}&&4 , \underline{5}&6&7&\underline{8}&&\underline{9}&10 , \underline{11}&\underline{12}\\
2&3 , \underline{1}&13&14 , \underline{4}&15&&1 , \underline{3}&&\underline{15}&4 , \underline{14}&\underline{13}\\
3&1 , \underline{2}&&2 , \underline{1}&16&17&8&&\underline{8}&\underline{17}&\underline{16}\\
4&14 , \underline{2}&&5 , \underline{1}&18&19&&\underline{20}&\underline{10}&21&\underline{22}\\
5&4 , \underline{1}&&1 , \underline{4}&23&24&&&&\underline{24}&\underline{23}\\
6&15 , \underline{16}&&18 , \underline{23}&\underline{1}&25&&&&\underline{26}&\underline{27}\\
7&\underline{17}&&28 , \underline{24}&25&\underline{1}&&&&29 , \underline{27}&\underline{30}\\
8&3&&&&&1&\underline{31}&\underline{32}&9 , \underline{33}&\\
9&33 , \underline{8}&&10&&&\underline{32}&&1&21 , \underline{34}&\underline{35}\\
10&4&&21&&29&\underline{9}&10&\underline{21}&11 , \underline{1}&\underline{36}\\
11&37 , \underline{17}&&\underline{24}&26&27&\underline{33}&36&\underline{34}&1 , \underline{10}&\underline{38}\\
12&13 , \underline{16}&&22 , \underline{23}&27&30 , \underline{39}&&&\underline{40}&36 , \underline{38}&1\\
13&&\underline{2}&41 , \underline{20}&&&12 , \underline{16}&&\underline{42}&18 , \underline{43}&\\
14&2 , \underline{4}&41&4 , \underline{2}&43&\underline{19}&37&\underline{41}&\underline{37}&19&\underline{43}\\
15&16 , \underline{6}&&43 , \underline{22}&\underline{2}&&33&&&\underline{37}&\underline{42}\\
16&6 , \underline{15}&&13 , \underline{12}&\underline{3}&44&45&&&\underline{46}&\underline{47}\\
17&7&&37 , \underline{11}&44&\underline{3}&31&&&\underline{47}&\underline{46}\\
18&43 , \underline{13}&&23 , \underline{6}&\underline{4}&48&&&&&\underline{49}\\
19&\underline{14}&&28&48&\underline{4}&&\underline{50}&\underline{21}&35 , \underline{51}&\underline{52}\\
\hline
\end{tabular}}
\end{center}                           
%=================================================================

%=================================================================
\begin{table}[h]
\begin{center}
\tiny
\resizebox{.92 \textwidth}{!}{
\begin{tabular}{ |c|c|c|c|c|c|c|c|c|c|c|} 
\hline
N&1&2&3&4&5&6&7&8&9&10\\
\hline
20&41 , \underline{13}&&&&50&&4&\underline{36}&34&\\
21&19&&\underline{10}&&35&&9 , \underline{34}&\underline{10}&\underline{4}&\underline{53}\\
22&43 , \underline{15}&&23 , \underline{12}&49&52&&&\underline{53}&53&4\\
23&22 , \underline{12}&&6 , \underline{18}&\underline{5}&54&&&&\underline{55}&\underline{56}\\
24&\underline{11}&&7 , \underline{28}&54&\underline{5}&&&&\underline{56}&\underline{55}\\
25&\underline{44}&&57 , \underline{54}&\underline{7}&\underline{6}&&&&\underline{6}&\underline{7}\\
26&58 , \underline{46}&&\underline{55}&\underline{11}&6&\underline{59}&60&&6&\underline{11}\\
27&42 , \underline{47}&&49 , \underline{56}&\underline{12}&7 , \underline{29}&&&&\underline{11}&6\\
28&\underline{37}&&24 , \underline{7}&57&\underline{19}&&&&61 , \underline{62}&\underline{63}\\
29&&&61&&\underline{10}&&&&27 , \underline{7}&\underline{39}\\
30&\underline{46}&&63 , \underline{55}&7&39 , \underline{12}&&&\underline{64}&39 , \underline{12}&7\\
31&17&&&&&&8&\underline{65}&34&\\
32&45&&&&&9&\underline{65}&8&35 , \underline{66}&\\
33&15&&&\underline{45}&&11&53&\underline{66}&8 , \underline{9}&\\
34&\underline{31}&&&&&\underline{66}&20&11&9 , \underline{21}&\underline{51}\\
35&66 , \underline{32}&&21&&\underline{61}&&&40&19 , \underline{51}&9\\
36&20&&53&&39 , \underline{60}&&11&\underline{52}&38 , \underline{12}&10\\
37&17 , \underline{11}&&\underline{28}&58&67&\underline{15}&52&&14&\underline{68}\\
38&68 , \underline{44}&&\underline{54}&11&12 , \underline{36}&&38&\underline{69}&12 , \underline{36}&11\\
39&&&70&29&12 , \underline{30}&&&\underline{70}&60 , \underline{36}&\\
40&\underline{45}&&53&&64 , \underline{70}&&&12&52 , \underline{69}&\underline{35}\\
41&&\underline{14}&20 , \underline{13}&&\underline{51}&68&\underline{71}&\underline{67}&48&\\
42&&\underline{15}&67&&&27 , \underline{47}&&13&49 , \underline{72}&\\
43&&\underline{14}&\underline{50}&&&22 , \underline{15}&&\underline{72}&13 , \underline{18}&\\
44&25&&68 , \underline{38}&\underline{17}&\underline{16}&73&&&\underline{16}&\underline{17}\\
45&\underline{33}&&\underline{40}&32&73&\underline{16}&&&\underline{59}&\\
46&26 , \underline{58}&&\underline{30}&\underline{17}&16&59&&&16&\underline{17}\\
47&27 , \underline{42}&&42 , \underline{27}&\underline{16}&17&&&&\underline{17}&16\\
48&\underline{41}&&57&\underline{19}&\underline{18}&&&&\underline{74}&\underline{62}\\
49&72 , \underline{42}&&56 , \underline{27}&\underline{22}&62&&&&&18\\
50&\underline{43}&&&&\underline{20}&&19&\underline{53}&66&\\
51&\underline{41}&&62&74&&&65&\underline{34}&19 , \underline{35}&\underline{69}\\
52&\underline{37}&&63 , \underline{70}&62&\underline{22}&&&\underline{36}&40 , \underline{69}&19\\
53&50&&\underline{36}&&40&&33&\underline{22}&\underline{22}&21\\
54&\underline{38}&&25 , \underline{57}&\underline{24}&\underline{23}&&&&\underline{23}&\underline{24}\\
55&63 , \underline{30}&&26&\underline{24}&23&&&&23&\underline{24}\\
56&49 , \underline{27}&&27 , \underline{49}&\underline{23}&24&&&&\underline{24}&23\\
57&\underline{68}&&54 , \underline{25}&\underline{28}&\underline{48}&&&&\underline{48}&\underline{28}\\
58&46 , \underline{26}&&\underline{63}&\underline{37}&37&26 , \underline{46}&63&&37&\underline{37}\\
59&46&&&\underline{45}&&26&64&&&\underline{45}\\
60&63&&&39 , \underline{36}&&\underline{64}&\underline{26}&&&39 , \underline{36}\\
61&&&\underline{29}&&\underline{35}&&&&62 , \underline{28}&\underline{70}\\
62&\underline{67}&&\underline{49}&48&\underline{51}&&&&28 , \underline{61}&\underline{52}\\
63&\underline{58}&&55 , \underline{30}&28&70 , \underline{52}&&&\underline{60}&70 , \underline{52}&28\\
64&\underline{59}&&60&&70 , \underline{40}&&&30&70 , \underline{40}&\\
65&73&\underline{75}&&&&&32&31&51&\\
66&&&&&&34&50&33&32 , \underline{35}&\\
67&42&&\underline{62}&37&\underline{37}&\underline{42}&62&&41&\underline{41}\\
68&44 , \underline{38}&&\underline{57}&37&41&&69&&41&37\\
69&\underline{73}&&&\underline{51}&40 , \underline{52}&&68&38&40 , \underline{52}&\underline{51}\\
70&&&\underline{39}&61&52 , \underline{63}&&&\underline{39}&64 , \underline{40}&\\
71&&\underline{41}&&&\underline{74}&41&41&\underline{41}&74&\\
72&&\underline{43}&&&&49 , \underline{42}&&43&42 , \underline{49}&\\
73&&&\underline{69}&65&\underline{45}&\underline{44}&&&\underline{45}&65\\
74&\underline{71}&&48&\underline{51}&&&75&&48&\underline{51}\\
75&65&65&&&&&&&74&\\
\hline
\end{tabular}}
\end{center}      
\caption{Triality connections between the 75 toric phases of Model 15.}
\label{Table 2 - Model 15}
\end{table}             
%=================================================================

%=================================================================
\subsection{Model 16: $P^{0}_{+-}(\text{dP}_2)$}
%=================================================================

\begin{figure}[H]
\begin{center}
\resizebox{0.35\hsize}{!}{
\includegraphics[height=6cm]{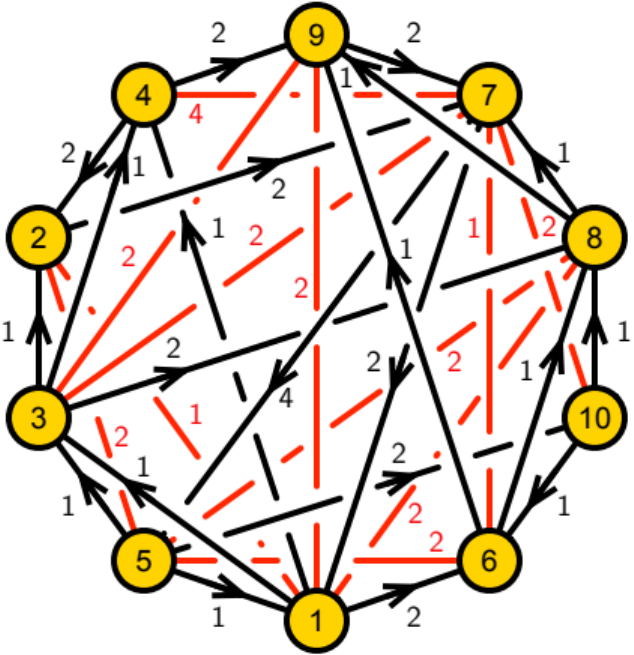} 
}
\caption{Quiver for Phase 1 of Model 16.
\label{f_quiver_16}}
 \end{center}
 \end{figure}

\fref{f_quiver_16} shows the quiver for Phase 1 of Model 16.

%=================================================================
%=================================================================

The $J$- and $E$-terms are
\beq
{\footnotesize
\begin{array}{rrclrcl}
 &  & J &  & & E &  \\
\Lambda_{18}^{1} : & X_{87} X_{71} &-& X_{89} Y_{97} X_{75} X_{51} & P_{16} X_{68} &-& X_{13} P_{38}  \\  
\Lambda_{18}^{2} : & X_{87} Y_{71} &-& X_{89} Y_{97} R_{75} X_{51} &  X_{13} Q_{38}  &-& Q_{16} X_{68}\\  
\Lambda_{19}^{1} : & Y_{97} Y_{75} X_{51} &-&X_{97} X_{71} & P_{16} X_{69} &-& X_{14} P_{49}  \\  
\Lambda_{19}^{2} : & Y_{97} S_{75} X_{51} &-&X_{97} Y_{71} &  X_{14} Q_{49} &-&Q_{16} X_{69} \\  
\Lambda_{21}  : & X_{13} X_{32} &-& X_{14} X_{42} & P_{27} X_{71} &-& Q_{27} Y_{71}  \\  
\Lambda_{25}^{1} : & X_{53} X_{34} X_{42} &-& X_{51} X_{13} X_{34} Y_{42} & P_{27} X_{75} &-& Q_{27} R_{75}  \\  
\Lambda_{25}^{2} : & X_{51} X_{14} Y_{42} &-&X_{53} X_{32} & P_{27} Y_{75} &-& Q_{27} S_{75}  \\  
\Lambda_{37}^{1} : & X_{71} X_{13} &-& Y_{75} X_{53} & P_{38} X_{87} &-& X_{32} P_{27}  \\  
\Lambda_{37}^{2} : & Y_{71} X_{13} &-&S_{75} X_{53} &  X_{32} Q_{27}  &-&Q_{38} X_{87} \\  
\Lambda_{39}^{1} : & X_{97} X_{75} X_{53} &-& Y_{97} X_{75} X_{51} X_{13} & P_{38} X_{89} &-& X_{34} P_{49}  \\  
\Lambda_{39}^{2} : & X_{97} R_{75} X_{53} &-& Y_{97} R_{75} X_{51} X_{13} &  X_{34} Q_{49} &-&Q_{38} X_{89}  \\  
\Lambda_{47}^{1} : & X_{75} X_{53} X_{34} &-&X_{71} X_{14} & P_{49} X_{97} &-& X_{42} P_{27}  \\  
\Lambda_{47}^{2} : & Y_{75} X_{51} X_{14} &-& X_{75} X_{51} X_{13} X_{34} & P_{49} Y_{97} &-& Y_{42} P_{27}  \\  
\Lambda_{47}^{3} : & R_{75} X_{53} X_{34} &-&Y_{71} X_{14} & X_{42} Q_{27} &-&Q_{49} X_{97}  \\  
\Lambda_{47}^{4} : & S_{75} X_{51} X_{14} &-& R_{75} X_{51} X_{13} X_{34} &  Y_{42} Q_{27} &-&Q_{49} Y_{97}  \\  
\Lambda_{56}^{1} : & X_{69} Y_{97} Y_{75} &-& X_{68} X_{89} Y_{97} X_{75} & P_{5.10} X_{10.6} &-& X_{51} P_{16}  \\  
\Lambda_{56}^{2} : & X_{69} Y_{97} S_{75} &-& X_{68} X_{89} Y_{97} R_{75} &  X_{51} Q_{16} &-&Q_{5.10} X_{10.6} \\  
\Lambda_{58}^{1} : & X_{89} X_{97} X_{75} &-&X_{87} Y_{75}  & P_{5.10} X_{10.8} &-& X_{53} P_{38}  \\  
\Lambda_{58}^{2} : & X_{89} X_{97} R_{75} &-&X_{87} S_{75} &  X_{53} Q_{38} &-&Q_{5.10} X_{10.8}  \\  
\Lambda_{76} : & X_{68} X_{87} &-& X_{69} X_{97} &  Y_{71} Q_{16} &-&X_{71} P_{16}  \\  
\Lambda_{7.10}^{1} : & X_{10.8} X_{89} X_{97} &-& X_{10.6} X_{68} X_{89} Y_{97} & R_{75} Q_{5.10} &-& X_{75} P_{5.10}  \\  
\Lambda_{7.10}^{2} : & X_{10.6} X_{69} Y_{97} &-&X_{10.8} X_{87}  &  S_{75} Q_{5.10} &-& Y_{75} P_{5.10}
 \end{array} 
 }~.~
\label{E_J_C_+-}
\eeq

This model has 143 toric phases, which are summarized in Table \ref{Table 1 - Model 16}.

%=================================================================
\begin{table}[h]
\begin{center}
\tiny
\resizebox{1 \textwidth}{!}{
\begin{tabular}{ |c|c|c|c|c|c|c|c| } 
\hline
 Phase & Path & F & Fermi Multiplicities & Phase & Path & F & Fermi Multiplicities \\
\hline
1&&22&1$\times\textbf{2}$+2$\times\textbf{3}$+4$\times\textbf{4}$+1$\times\textbf{5}$+1$\times\textbf{6}$+1$\times\textbf{9}$&73&3,2,-10&20&2$\times\textbf{2}$+2$\times\textbf{3}$+2$\times\textbf{4}$+2$\times\textbf{5}$+2$\times\textbf{6}$\\ 
2&-2&16&2$\times\textbf{2}$+4$\times\textbf{3}$+4$\times\textbf{4}$&74&3,-6,-8&21&3$\times\textbf{2}$+3$\times\textbf{4}$+1$\times\textbf{5}$+2$\times\textbf{6}$+1$\times\textbf{7}$\\ 
3&3&20&3$\times\textbf{2}$+1$\times\textbf{3}$+2$\times\textbf{4}$+2$\times\textbf{5}$+1$\times\textbf{6}$+1$\times\textbf{7}$&75&3,8,-6&23&3$\times\textbf{2}$+2$\times\textbf{4}$+1$\times\textbf{5}$+2$\times\textbf{6}$+1$\times\textbf{7}$+1$\times\textbf{8}$\\ 
4&4&23&2$\times\textbf{2}$+1$\times\textbf{3}$+4$\times\textbf{4}$+2$\times\textbf{6}$+1$\times\textbf{11}$&76&3,8,-10&20&2$\times\textbf{2}$+2$\times\textbf{3}$+2$\times\textbf{4}$+2$\times\textbf{5}$+2$\times\textbf{6}$\\ 
5&-6&22&4$\times\textbf{3}$+4$\times\textbf{4}$+2$\times\textbf{8}$&77&3,-8,-6&19&4$\times\textbf{2}$+2$\times\textbf{3}$+2$\times\textbf{5}$+2$\times\textbf{7}$\\ 
6&-8&20&2$\times\textbf{2}$+2$\times\textbf{3}$+2$\times\textbf{4}$+2$\times\textbf{5}$+2$\times\textbf{6}$&78&3,-10,6&25&1$\times\textbf{2}$+2$\times\textbf{3}$+2$\times\textbf{4}$+3$\times\textbf{5}$+1$\times\textbf{8}$+1$\times\textbf{11}$\\ 
7&-9&23&2$\times\textbf{2}$+2$\times\textbf{3}$+1$\times\textbf{4}$+2$\times\textbf{5}$+2$\times\textbf{6}$+1$\times\textbf{10}$&79&3,-10,-8&19&2$\times\textbf{2}$+2$\times\textbf{3}$+4$\times\textbf{4}$+1$\times\textbf{5}$+1$\times\textbf{7}$\\ 
8&10&18&3$\times\textbf{2}$+2$\times\textbf{3}$+3$\times\textbf{4}$+1$\times\textbf{5}$+1$\times\textbf{7}$&80&4,1,9&24&3$\times\textbf{2}$+1$\times\textbf{3}$+1$\times\textbf{4}$+2$\times\textbf{5}$+1$\times\textbf{7}$+1$\times\textbf{8}$+1$\times\textbf{10}$\\ 
9&-10&21&1$\times\textbf{2}$+3$\times\textbf{3}$+4$\times\textbf{4}$+1$\times\textbf{5}$+1$\times\textbf{10}$&81&4,1,-9&24&2$\times\textbf{2}$+2$\times\textbf{3}$+1$\times\textbf{4}$+2$\times\textbf{5}$+2$\times\textbf{7}$+1$\times\textbf{10}$\\ 
10&-2,1&20&2$\times\textbf{2}$+3$\times\textbf{3}$+1$\times\textbf{4}$+1$\times\textbf{5}$+3$\times\textbf{6}$&82&4,1,10&23&2$\times\textbf{2}$+1$\times\textbf{3}$+2$\times\textbf{4}$+3$\times\textbf{5}$+1$\times\textbf{7}$+1$\times\textbf{9}$\\ 
11&-2,3&16&3$\times\textbf{2}$+4$\times\textbf{3}$+2$\times\textbf{4}$+1$\times\textbf{6}$&83&4,1,-10&24&2$\times\textbf{2}$+1$\times\textbf{3}$+1$\times\textbf{4}$+4$\times\textbf{5}$+1$\times\textbf{7}$+1$\times\textbf{10}$\\ 
12&-2,4&15&3$\times\textbf{2}$+4$\times\textbf{3}$+3$\times\textbf{4}$&84&4,-6,-8&30&6$\times\textbf{4}$+2$\times\textbf{8}$+2$\times\textbf{10}$\\ 
13&-2,-4&16&3$\times\textbf{2}$+3$\times\textbf{3}$+3$\times\textbf{4}$+1$\times\textbf{5}$&85&4,-8,-6&28&2$\times\textbf{3}$+3$\times\textbf{4}$+2$\times\textbf{6}$+1$\times\textbf{8}$+2$\times\textbf{9}$\\ 
14&-2,5&23&1$\times\textbf{2}$+2$\times\textbf{3}$+3$\times\textbf{4}$+2$\times\textbf{5}$+1$\times\textbf{6}$+1$\times\textbf{10}$&86&4,-8,10&21&3$\times\textbf{2}$+1$\times\textbf{3}$+2$\times\textbf{4}$+1$\times\textbf{5}$+2$\times\textbf{6}$+1$\times\textbf{8}$\\ 
15&3,1&23&3$\times\textbf{2}$+1$\times\textbf{3}$+1$\times\textbf{4}$+1$\times\textbf{5}$+1$\times\textbf{6}$+2$\times\textbf{7}$+1$\times\textbf{8}$&87&4,-8,-10&23&1$\times\textbf{2}$+1$\times\textbf{3}$+5$\times\textbf{4}$+2$\times\textbf{6}$+1$\times\textbf{9}$\\ 
16&3,2&19&3$\times\textbf{2}$+1$\times\textbf{3}$+2$\times\textbf{4}$+3$\times\textbf{5}$+1$\times\textbf{6}$&88&4,9,-6&24&3$\times\textbf{2}$+2$\times\textbf{4}$+3$\times\textbf{6}$+2$\times\textbf{8}$\\ 
17&3,-6&22&1$\times\textbf{2}$+2$\times\textbf{3}$+3$\times\textbf{4}$+1$\times\textbf{5}$+2$\times\textbf{6}$+1$\times\textbf{7}$&89&4,-9,-6&20&4$\times\textbf{2}$+1$\times\textbf{3}$+1$\times\textbf{4}$+3$\times\textbf{6}$+1$\times\textbf{7}$\\ 
18&3,8&22&2$\times\textbf{2}$+1$\times\textbf{3}$+2$\times\textbf{4}$+2$\times\textbf{5}$+2$\times\textbf{6}$+1$\times\textbf{7}$&90&4,-9,10&18&5$\times\textbf{2}$+3$\times\textbf{4}$+1$\times\textbf{6}$+1$\times\textbf{8}$\\ 
19&3,-8&18&4$\times\textbf{2}$+1$\times\textbf{3}$+1$\times\textbf{4}$+3$\times\textbf{5}$+1$\times\textbf{6}$&91&4,10,6&21&3$\times\textbf{2}$+2$\times\textbf{3}$+2$\times\textbf{4}$+2$\times\textbf{6}$+1$\times\textbf{10}$\\ 
20&3,10&20&3$\times\textbf{2}$+1$\times\textbf{3}$+2$\times\textbf{4}$+2$\times\textbf{5}$+1$\times\textbf{6}$+1$\times\textbf{7}$&92&4,-10,6&26&1$\times\textbf{2}$+3$\times\textbf{3}$+3$\times\textbf{4}$+1$\times\textbf{6}$+1$\times\textbf{8}$+1$\times\textbf{15}$\\ 
21&3,-10&21&2$\times\textbf{2}$+2$\times\textbf{3}$+2$\times\textbf{4}$+2$\times\textbf{5}$+1$\times\textbf{6}$+1$\times\textbf{8}$&93&4,-10,-8&26&1$\times\textbf{3}$+6$\times\textbf{4}$+1$\times\textbf{6}$+1$\times\textbf{8}$+1$\times\textbf{11}$\\ 
22&4,1&23&2$\times\textbf{2}$+2$\times\textbf{3}$+1$\times\textbf{4}$+2$\times\textbf{5}$+1$\times\textbf{6}$+1$\times\textbf{7}$+1$\times\textbf{9}$&94&-8,-3,2&23&1$\times\textbf{2}$+4$\times\textbf{3}$+1$\times\textbf{5}$+2$\times\textbf{6}$+1$\times\textbf{7}$+1$\times\textbf{8}$\\ 
23&4,-6&25&2$\times\textbf{3}$+5$\times\textbf{4}$+1$\times\textbf{6}$+1$\times\textbf{8}$+1$\times\textbf{10}$&95&-8,-6,10&15&3$\times\textbf{2}$+4$\times\textbf{3}$+3$\times\textbf{4}$\\ 
24&4,-8&25&1$\times\textbf{2}$+1$\times\textbf{3}$+3$\times\textbf{4}$+4$\times\textbf{6}$+1$\times\textbf{9}$&96&-8,10,-7&23&3$\times\textbf{2}$+1$\times\textbf{4}$+3$\times\textbf{5}$+1$\times\textbf{6}$+1$\times\textbf{7}$+1$\times\textbf{8}$\\ 
25&4,9&26&2$\times\textbf{2}$+3$\times\textbf{4}$+4$\times\textbf{6}$+1$\times\textbf{12}$&97&-9,-4,-6&24&2$\times\textbf{2}$+2$\times\textbf{3}$+4$\times\textbf{6}$+2$\times\textbf{7}$\\ 
26&4,-9&22&3$\times\textbf{2}$+1$\times\textbf{3}$+2$\times\textbf{4}$+1$\times\textbf{5}$+2$\times\textbf{6}$+1$\times\textbf{10}$&98&-9,-4,10&22&4$\times\textbf{2}$+1$\times\textbf{4}$+1$\times\textbf{5}$+2$\times\textbf{6}$+1$\times\textbf{7}$+1$\times\textbf{8}$\\ 
27&4,10&19&4$\times\textbf{2}$+1$\times\textbf{3}$+3$\times\textbf{4}$+1$\times\textbf{6}$+1$\times\textbf{9}$&99&-9,-6,10&17&4$\times\textbf{2}$+2$\times\textbf{3}$+2$\times\textbf{4}$+2$\times\textbf{6}$\\ 
28&4,-10&22&2$\times\textbf{2}$+2$\times\textbf{3}$+4$\times\textbf{4}$+1$\times\textbf{6}$+1$\times\textbf{12}$&100&-9,-6,-10&18&2$\times\textbf{2}$+4$\times\textbf{3}$+2$\times\textbf{4}$+2$\times\textbf{6}$\\ 
29&-8,-3&22&2$\times\textbf{2}$+2$\times\textbf{3}$+2$\times\textbf{5}$+4$\times\textbf{6}$&101&-9,10,-5&22&3$\times\textbf{2}$+4$\times\textbf{4}$+2$\times\textbf{7}$+1$\times\textbf{8}$\\ 
30&-8,-6&21&1$\times\textbf{2}$+4$\times\textbf{3}$+2$\times\textbf{4}$+1$\times\textbf{5}$+1$\times\textbf{7}$+1$\times\textbf{8}$&102&-9,-10,6&20&3$\times\textbf{2}$+3$\times\textbf{3}$+1$\times\textbf{4}$+1$\times\textbf{5}$+1$\times\textbf{6}$+1$\times\textbf{10}$\\ 
31&-8,10&16&4$\times\textbf{2}$+2$\times\textbf{3}$+2$\times\textbf{4}$+2$\times\textbf{5}$&103&-9,-10,-8&22&2$\times\textbf{2}$+2$\times\textbf{3}$+2$\times\textbf{4}$+1$\times\textbf{5}$+1$\times\textbf{6}$+1$\times\textbf{7}$+1$\times\textbf{8}$\\ 
32&-8,-10&18&2$\times\textbf{2}$+2$\times\textbf{3}$+4$\times\textbf{4}$+2$\times\textbf{5}$&104&-10,-5,6&20&1$\times\textbf{2}$+4$\times\textbf{3}$+3$\times\textbf{4}$+1$\times\textbf{5}$+1$\times\textbf{9}$\\ 
33&-9,-4&26&2$\times\textbf{2}$+1$\times\textbf{3}$+2$\times\textbf{5}$+3$\times\textbf{6}$+1$\times\textbf{7}$+1$\times\textbf{10}$&105&-10,-5,-8&26&2$\times\textbf{3}$+6$\times\textbf{4}$+2$\times\textbf{11}$\\ 
34&-9,-6&21&2$\times\textbf{2}$+3$\times\textbf{3}$+1$\times\textbf{4}$+3$\times\textbf{6}$+1$\times\textbf{7}$&106&-10,6,-8&21&2$\times\textbf{2}$+1$\times\textbf{3}$+5$\times\textbf{4}$+1$\times\textbf{5}$+1$\times\textbf{10}$\\ 
35&-9,10&19&4$\times\textbf{2}$+1$\times\textbf{3}$+2$\times\textbf{4}$+1$\times\textbf{5}$+1$\times\textbf{6}$+1$\times\textbf{8}$&107&-10,-8,-9&24&2$\times\textbf{3}$+6$\times\textbf{4}$+2$\times\textbf{9}$\\ 
36&-9,-10&20&3$\times\textbf{2}$+2$\times\textbf{3}$+2$\times\textbf{4}$+1$\times\textbf{5}$+1$\times\textbf{6}$+1$\times\textbf{9}$&108&-2,1,3,-10&24&1$\times\textbf{2}$+2$\times\textbf{3}$+2$\times\textbf{4}$+3$\times\textbf{5}$+1$\times\textbf{8}$+1$\times\textbf{9}$\\ 
37&10,-5&21&2$\times\textbf{2}$+1$\times\textbf{3}$+4$\times\textbf{4}$+1$\times\textbf{5}$+2$\times\textbf{7}$&109&-2,1,4,5&25&2$\times\textbf{3}$+5$\times\textbf{4}$+1$\times\textbf{6}$+1$\times\textbf{8}$+1$\times\textbf{10}$\\ 
38&10,6&16&3$\times\textbf{2}$+4$\times\textbf{3}$+2$\times\textbf{4}$+1$\times\textbf{6}$&110&-2,1,4,-10&19&2$\times\textbf{2}$+1$\times\textbf{3}$+4$\times\textbf{4}$+3$\times\textbf{5}$\\ 
39&-10,-5&26&2$\times\textbf{3}$+5$\times\textbf{4}$+1$\times\textbf{5}$+1$\times\textbf{9}$+1$\times\textbf{12}$&111&-2,1,-4,9&23&2$\times\textbf{2}$+1$\times\textbf{3}$+4$\times\textbf{4}$+1$\times\textbf{5}$+1$\times\textbf{8}$+1$\times\textbf{10}$\\ 
40&-10,6&21&1$\times\textbf{2}$+4$\times\textbf{3}$+3$\times\textbf{4}$+1$\times\textbf{5}$+1$\times\textbf{11}$&112&-2,3,1,-10&22&2$\times\textbf{2}$+1$\times\textbf{3}$+4$\times\textbf{4}$+1$\times\textbf{5}$+2$\times\textbf{8}$\\ 
41&-10,-8&23&2$\times\textbf{3}$+6$\times\textbf{4}$+1$\times\textbf{7}$+1$\times\textbf{9}$&113&-2,3,5,-10&19&2$\times\textbf{2}$+2$\times\textbf{3}$+3$\times\textbf{4}$+2$\times\textbf{5}$+1$\times\textbf{6}$\\ 
42&-2,1,3&21&1$\times\textbf{2}$+4$\times\textbf{3}$+2$\times\textbf{4}$+1$\times\textbf{5}$+1$\times\textbf{7}$+1$\times\textbf{8}$&114&-2,4,1,5&20&1$\times\textbf{2}$+3$\times\textbf{3}$+4$\times\textbf{4}$+1$\times\textbf{5}$+1$\times\textbf{8}$\\ 
43&-2,1,4&18&2$\times\textbf{2}$+3$\times\textbf{3}$+3$\times\textbf{4}$+1$\times\textbf{5}$+1$\times\textbf{6}$&115&-2,4,1,9&18&3$\times\textbf{2}$+2$\times\textbf{3}$+4$\times\textbf{4}$+1$\times\textbf{8}$\\ 
44&-2,1,-4&18&3$\times\textbf{2}$+3$\times\textbf{3}$+1$\times\textbf{4}$+1$\times\textbf{5}$+2$\times\textbf{6}$&116&-2,4,1,-9&20&2$\times\textbf{2}$+3$\times\textbf{3}$+2$\times\textbf{4}$+2$\times\textbf{6}$+1$\times\textbf{7}$\\ 
45&-2,1,10&22&2$\times\textbf{2}$+1$\times\textbf{3}$+2$\times\textbf{4}$+3$\times\textbf{5}$+1$\times\textbf{6}$+1$\times\textbf{8}$&117&-2,4,1,-10&18&2$\times\textbf{2}$+2$\times\textbf{3}$+4$\times\textbf{4}$+2$\times\textbf{5}$\\ 
46&-2,1,-10&21&2$\times\textbf{2}$+1$\times\textbf{3}$+2$\times\textbf{4}$+3$\times\textbf{5}$+2$\times\textbf{6}$&118&-2,4,-3,9&18&3$\times\textbf{2}$+3$\times\textbf{3}$+2$\times\textbf{4}$+1$\times\textbf{6}$+1$\times\textbf{7}$\\ 
47&-2,3,1&19&3$\times\textbf{2}$+2$\times\textbf{3}$+3$\times\textbf{4}$+1$\times\textbf{6}$+1$\times\textbf{8}$&119&-2,4,-3,-9&20&3$\times\textbf{2}$+3$\times\textbf{3}$+1$\times\textbf{5}$+1$\times\textbf{6}$+2$\times\textbf{7}$\\ 
48&-2,3,-4&14&4$\times\textbf{2}$+4$\times\textbf{3}$+2$\times\textbf{4}$&120&-2,4,-3,-10&18&3$\times\textbf{2}$+4$\times\textbf{3}$+1$\times\textbf{4}$+2$\times\textbf{7}$\\ 
49&-2,3,5&20&2$\times\textbf{2}$+1$\times\textbf{3}$+3$\times\textbf{4}$+3$\times\textbf{5}$+1$\times\textbf{6}$&121&-2,4,5,1&22&2$\times\textbf{2}$+2$\times\textbf{3}$+2$\times\textbf{4}$+2$\times\textbf{5}$+2$\times\textbf{8}$\\ 
50&-2,3,-6&18&2$\times\textbf{2}$+4$\times\textbf{3}$+2$\times\textbf{4}$+2$\times\textbf{6}$&122&-2,4,5,3&20&3$\times\textbf{2}$+2$\times\textbf{3}$+1$\times\textbf{4}$+2$\times\textbf{5}$+1$\times\textbf{6}$+1$\times\textbf{8}$\\ 
51&-2,3,-10&17&2$\times\textbf{2}$+4$\times\textbf{3}$+3$\times\textbf{4}$+1$\times\textbf{6}$&123&-2,4,5,-9&23&2$\times\textbf{2}$+1$\times\textbf{3}$+3$\times\textbf{4}$+1$\times\textbf{5}$+1$\times\textbf{6}$+1$\times\textbf{7}$+1$\times\textbf{9}$\\ 
52&-2,4,1&17&2$\times\textbf{2}$+4$\times\textbf{3}$+3$\times\textbf{4}$+1$\times\textbf{6}$&124&-2,4,9,-6&22&2$\times\textbf{2}$+2$\times\textbf{3}$+2$\times\textbf{4}$+3$\times\textbf{6}$+1$\times\textbf{8}$\\ 
53&-2,4,-3&17&3$\times\textbf{2}$+5$\times\textbf{3}$+1$\times\textbf{6}$+1$\times\textbf{7}$&125&-2,4,-9,10&18&3$\times\textbf{2}$+2$\times\textbf{3}$+4$\times\textbf{4}$+1$\times\textbf{8}$\\ 
54&-2,4,5&20&3$\times\textbf{2}$+2$\times\textbf{3}$+2$\times\textbf{4}$+1$\times\textbf{5}$+1$\times\textbf{6}$+1$\times\textbf{9}$&126&-2,-4,5,9&30&1$\times\textbf{2}$+6$\times\textbf{4}$+1$\times\textbf{8}$+1$\times\textbf{10}$+1$\times\textbf{16}$\\ 
55&-2,4,9&18&2$\times\textbf{2}$+2$\times\textbf{3}$+5$\times\textbf{4}$+1$\times\textbf{6}$&127&3,1,2,-10&23&2$\times\textbf{2}$+2$\times\textbf{3}$+1$\times\textbf{4}$+2$\times\textbf{5}$+1$\times\textbf{6}$+2$\times\textbf{8}$\\ 
56&-2,4,-9&18&2$\times\textbf{2}$+3$\times\textbf{3}$+3$\times\textbf{4}$+1$\times\textbf{5}$+1$\times\textbf{6}$&128&3,1,8,-10&25&2$\times\textbf{2}$+1$\times\textbf{3}$+1$\times\textbf{4}$+1$\times\textbf{5}$+2$\times\textbf{6}$+2$\times\textbf{7}$+1$\times\textbf{8}$\\ 
57&-2,4,10&15&4$\times\textbf{2}$+3$\times\textbf{3}$+2$\times\textbf{4}$+1$\times\textbf{5}$&129&3,1,-10,-8&26&2$\times\textbf{2}$+1$\times\textbf{3}$+1$\times\textbf{4}$+2$\times\textbf{6}$+3$\times\textbf{7}$+1$\times\textbf{8}$\\ 
58&-2,4,-10&14&4$\times\textbf{2}$+4$\times\textbf{3}$+2$\times\textbf{4}$&130&3,2,-4,-10&22&2$\times\textbf{2}$+2$\times\textbf{3}$+2$\times\textbf{4}$+3$\times\textbf{6}$+1$\times\textbf{8}$\\ 
59&-2,-4,5&23&2$\times\textbf{2}$+1$\times\textbf{3}$+4$\times\textbf{4}$+1$\times\textbf{5}$+1$\times\textbf{6}$+1$\times\textbf{12}$&131&3,2,-6,7&28&6$\times\textbf{4}$+2$\times\textbf{7}$+2$\times\textbf{9}$\\ 
60&-2,-4,9&23&2$\times\textbf{2}$+1$\times\textbf{3}$+4$\times\textbf{4}$+1$\times\textbf{6}$+1$\times\textbf{8}$+1$\times\textbf{9}$&132&3,2,-6,-8&20&3$\times\textbf{2}$+4$\times\textbf{4}$+1$\times\textbf{5}$+1$\times\textbf{6}$+1$\times\textbf{7}$\\ 
61&-2,-4,10&16&4$\times\textbf{2}$+2$\times\textbf{3}$+2$\times\textbf{4}$+2$\times\textbf{5}$&133&3,2,-10,-8&18&2$\times\textbf{2}$+2$\times\textbf{3}$+4$\times\textbf{4}$+2$\times\textbf{5}$\\ 
62&-2,5,3&23&1$\times\textbf{2}$+2$\times\textbf{3}$+3$\times\textbf{4}$+1$\times\textbf{5}$+2$\times\textbf{6}$+1$\times\textbf{9}$&134&3,8,-6,-1&26&3$\times\textbf{2}$+1$\times\textbf{4}$+3$\times\textbf{6}$+3$\times\textbf{8}$\\ 
63&-2,5,-9&30&4$\times\textbf{4}$+2$\times\textbf{5}$+2$\times\textbf{7}$+2$\times\textbf{10}$&135&3,-8,-6,-1&22&4$\times\textbf{2}$+1$\times\textbf{3}$+2$\times\textbf{5}$+1$\times\textbf{7}$+2$\times\textbf{8}$\\ 
64&3,1,2&20&3$\times\textbf{2}$+2$\times\textbf{3}$+1$\times\textbf{4}$+2$\times\textbf{5}$+1$\times\textbf{6}$+1$\times\textbf{8}$&136&4,1,9,6&26&1$\times\textbf{2}$+3$\times\textbf{3}$+2$\times\textbf{5}$+1$\times\textbf{6}$+1$\times\textbf{7}$+2$\times\textbf{9}$\\ 
65&3,1,8&23&3$\times\textbf{2}$+1$\times\textbf{3}$+1$\times\textbf{4}$+1$\times\textbf{5}$+2$\times\textbf{6}$+1$\times\textbf{7}$+1$\times\textbf{9}$&137&4,1,-9,10&24&2$\times\textbf{2}$+1$\times\textbf{3}$+2$\times\textbf{4}$+2$\times\textbf{5}$+1$\times\textbf{6}$+1$\times\textbf{7}$+1$\times\textbf{10}$\\ 
66&3,1,-8&23&3$\times\textbf{2}$+1$\times\textbf{3}$+1$\times\textbf{4}$+2$\times\textbf{5}$+1$\times\textbf{6}$+1$\times\textbf{7}$+1$\times\textbf{10}$&138&4,-8,-10,-9&26&6$\times\textbf{4}$+2$\times\textbf{6}$+2$\times\textbf{8}$\\ 
67&3,1,10&25&2$\times\textbf{2}$+1$\times\textbf{3}$+2$\times\textbf{4}$+1$\times\textbf{5}$+1$\times\textbf{6}$+1$\times\textbf{7}$+1$\times\textbf{8}$+1$\times\textbf{9}$&139&4,-9,-6,10&16&6$\times\textbf{2}$+2$\times\textbf{4}$+2$\times\textbf{6}$\\ 
68&3,1,-10&26&2$\times\textbf{2}$+1$\times\textbf{3}$+1$\times\textbf{4}$+2$\times\textbf{5}$+1$\times\textbf{6}$+2$\times\textbf{8}$+1$\times\textbf{9}$&140&4,-10,6,-8&28&1$\times\textbf{2}$+6$\times\textbf{4}$+2$\times\textbf{8}$+1$\times\textbf{14}$\\ 
69&3,2,-4&21&3$\times\textbf{2}$+1$\times\textbf{3}$+2$\times\textbf{4}$+1$\times\textbf{5}$+2$\times\textbf{6}$+1$\times\textbf{8}$&141&-8,-3,2,-6&24&6$\times\textbf{3}$+2$\times\textbf{7}$+2$\times\textbf{8}$\\ 
70&3,2,-6&21&1$\times\textbf{2}$+2$\times\textbf{3}$+4$\times\textbf{4}$+1$\times\textbf{5}$+1$\times\textbf{6}$+1$\times\textbf{7}$&142&-9,-10,-8,-6&23&2$\times\textbf{2}$+2$\times\textbf{3}$+1$\times\textbf{4}$+1$\times\textbf{5}$+2$\times\textbf{6}$+1$\times\textbf{7}$+1$\times\textbf{8}$\\ 
71&3,2,-8&17&3$\times\textbf{2}$+3$\times\textbf{3}$+2$\times\textbf{4}$+1$\times\textbf{5}$+1$\times\textbf{6}$&143&-2,1,4,5,-10&26&6$\times\textbf{4}$+1$\times\textbf{6}$+2$\times\textbf{7}$+1$\times\textbf{8}$\\ 
72&3,2,10&19&3$\times\textbf{2}$+1$\times\textbf{3}$+2$\times\textbf{4}$+3$\times\textbf{5}$+1$\times\textbf{6}$& & & & \\ 
\hline
\end{tabular}}
\end{center}
\caption{Basic information regarding the 143 toric phases of Model 16.}
\label{Table 1 - Model 16}
\end{table}
%=================================================================

Table \ref{Table 2 - Model 16} summarizes the connection between the toric phases under triality.

%=================================================================
\begin{center}
\tiny
\resizebox{.98 \textwidth}{!}{
\begin{tabular}{ |c|c|c|c|c|c|c|c|c|c|c|} 
\hline
N&1&2&3&4&5&6&7&8&9&10\\
\hline
1&&\underline{2}&3&4&&\underline{5}&&\underline{6}&\underline{7}&8 , \underline{9}\\
2&10&1&11&12 , \underline{13}&14&\underline{11}&\underline{1}&\underline{10}&\underline{14}&13 , \underline{12}\\
3&15&16 , \underline{11}&\underline{1}&&&\underline{17}&&18 , \underline{19}&&20 , \underline{21}\\
4&22&\underline{12}&&\underline{1}&&\underline{23}&&\underline{24}&25 , \underline{26}&27 , \underline{28}\\
5&\underline{1}&\underline{11}&17&23&&1&&\underline{23}&\underline{17}&11\\
6&&\underline{10}&19 , \underline{29}&24&&\underline{30}&&1&&31 , \underline{32}\\
7&&\underline{14}&&26 , \underline{33}&&\underline{34}&&&1&35 , \underline{36}\\
8&&\underline{13}&20&27&18 , \underline{37}&38 , \underline{11}&&\underline{31}&\underline{35}&9 , \underline{1}\\
9&&\underline{12}&21&28&\underline{39}&40&&\underline{41}&\underline{36}&1 , \underline{8}\\
10&\underline{2}&&42&43 , \underline{44}&&\underline{38}&\underline{6}&&&45 , \underline{46}\\
11&47&3 , \underline{16}&\underline{2}&\underline{48}&49&\underline{50}&\underline{5}&8 , \underline{38}&&44 , \underline{51}\\
12&52&4&\underline{53}&13 , \underline{2}&54&\underline{51}&\underline{9}&\underline{46}&55 , \underline{56}&57 , \underline{58}\\
13&44&&48 , \underline{38}&2 , \underline{12}&59&\underline{44}&\underline{8}&\underline{45}&60&61 , \underline{57}\\
14&&&62&54 , \underline{59}&\underline{2}&&\underline{7}&&\underline{63}&\underline{56}\\
15&\underline{3}&64 , \underline{47}&&&&\underline{34}&&65 , \underline{66}&&67 , \underline{68}\\
16&64&11 , \underline{3}&&\underline{69}&&\underline{70}&&37 , \underline{71}&&72 , \underline{73}\\
17&34&70 , \underline{50}&\underline{5}&&&3&&\underline{74}&&49\\
18&65&37 , \underline{8}&&&&\underline{75}&&19 , \underline{3}&&\underline{76}\\
19&66&71 , \underline{38}&29 , \underline{6}&&&\underline{77}&&3 , \underline{18}&&61 , \underline{31}\\
20&67&72 , \underline{44}&\underline{8}&&&45 , \underline{49}&&\underline{61}&&21 , \underline{3}\\
21&68&73 , \underline{51}&\underline{9}&&&78&&\underline{79}&&3 , \underline{20}\\
22&\underline{4}&\underline{52}&&&&\underline{30}&&&80 , \underline{81}&82 , \underline{83}\\
23&30&\underline{51}&&\underline{5}&&4&&\underline{84}&\underline{74}&47\\
24&&\underline{46}&&\underline{6}&&\underline{85}&&4&&86 , \underline{87}\\
25&80&\underline{55}&&&&\underline{88}&&&26 , \underline{4}&26 , \underline{4}\\
26&81&\underline{56}&&33 , \underline{7}&&\underline{89}&&&4 , \underline{25}&90 , \underline{27}\\
27&82&\underline{57}&&\underline{8}&65 , \underline{36}&91 , \underline{47}&&\underline{86}&26 , \underline{90}&28 , \underline{4}\\
28&83&\underline{58}&&\underline{9}&\underline{9}&92&&\underline{93}&4 , \underline{27}&4 , \underline{27}\\
29&&94&6 , \underline{19}&&&\underline{94}&&&&19 , \underline{6}\\
30&\underline{22}&\underline{42}&77 , \underline{94}&85&&6&&23&&95\\
31&&\underline{44}&61 , \underline{19}&86&76 , \underline{79}&48 , \underline{95}&\underline{96}&8&&32 , \underline{6}\\
32&&\underline{43}&31 , \underline{6}&87&\underline{41}&43&&41&\underline{87}&6 , \underline{31}\\
33&&&&7 , \underline{26}&&\underline{97}&&&&98 , \underline{65}\\
34&\underline{15}&\underline{62}&&89 , \underline{97}&&7&&&17&99 , \underline{100}\\
35&&\underline{59}&&90 , \underline{98}&75 , \underline{101}&53 , \underline{99}&&&8&36 , \underline{7}\\
36&&\underline{54}&&27 , \underline{65}&\underline{37}&102 , \underline{64}&&\underline{103}&9&7 , \underline{35}\\
37&&\underline{60}&&36&8 , \underline{18}&71 , \underline{16}&&\underline{79}&\underline{101}&39\\
38&\underline{10}&\underline{38}&45&91&19 , \underline{71}&11 , \underline{8}&&13 , \underline{48}&\underline{53}&40\\
39&&\underline{55}&&9&9&104&&\underline{105}&\underline{37}&\underline{37}\\
40&&\underline{53}&78&92&\underline{104}&\underline{9}&&59 , \underline{106}&\underline{102}&\underline{38}\\
41&&\underline{52}&79&93&\underline{105}&106&&9&\underline{107}&\underline{32}\\
42&\underline{47}&&\underline{10}&\underline{51}&&\underline{53}&\underline{30}&82&&78 , \underline{108}\\
43&\underline{52}&&\underline{50}&44 , \underline{10}&109&\underline{48}&\underline{32}&&106&49 , \underline{110}\\
44&\underline{13}&&51 , \underline{11}&10 , \underline{43}&&\underline{13}&\underline{31}&&111&20 , \underline{72}\\
45&\underline{13}&&78&49 , \underline{20}&&\underline{38}&&&&46 , \underline{10}\\
46&\underline{12}&&108&110 , \underline{72}&&&\underline{24}&&&10 , \underline{45}\\
47&\underline{11}&15 , \underline{64}&42&\underline{95}&&\underline{99}&\underline{23}&27 , \underline{91}&&111 , \underline{112}\\
48&95 , \underline{31}&\underline{69}&38 , \underline{13}&11&43&\underline{43}&\underline{11}&13 , \underline{38}&69&31 , \underline{95}\\
49&&&62&110 , \underline{43}&\underline{11}&&\underline{17}&20 , \underline{45}&&\underline{113}\\
50&99&17 , \underline{70}&\underline{11}&\underline{43}&&11&70 , \underline{17}&\underline{99}&&43\\
51&112&21 , \underline{73}&\underline{12}&\underline{95}&113&42&\underline{23}&\underline{95}&&11 , \underline{44}\\
52&\underline{12}&22&\underline{99}&43&114&\underline{95}&\underline{41}&&115 , \underline{116}&113 , \underline{117}\\
53&99 , \underline{35}&&12&38&102&\underline{42}&\underline{40}&&118 , \underline{119}&77 , \underline{120}\\
54&121&&122 , \underline{102}&59 , \underline{14}&\underline{12}&&\underline{36}&&60 , \underline{123}&\underline{57}\\
55&115&25&\underline{118}&60&60&\underline{124}&\underline{39}&&56 , \underline{12}&56 , \underline{12}\\
56&116&26&\underline{119}&\underline{14}&123&\underline{116}&&&12 , \underline{55}&125 , \underline{57}\\
57&113&27&\underline{77}&61 , \underline{13}&\underline{54}&71 , \underline{95}&&\underline{72}&56 , \underline{125}&58 , \underline{12}\\
58&117&28&\underline{120}&57 , \underline{12}&57 , \underline{12}&120&\underline{28}&\underline{117}&12 , \underline{57}&12 , \underline{57}\\
59&&&106 , \underline{40}&14 , \underline{54}&\underline{13}&&\underline{35}&&126&\underline{125}\\
60&111&&69 , \underline{91}&\underline{55}&126&&\underline{37}&&\underline{13}&123 , \underline{54}\\
61&20&&31 , \underline{19}&13 , \underline{57}&\underline{123}&19 , \underline{31}&&\underline{20}&123&57 , \underline{13}\\
62&&&\underline{14}&114 , \underline{106}&\underline{49}&&\underline{34}&67&&\underline{116}\\
63&&&&123&\underline{14}&&&&14&\underline{123}\\
64&\underline{16}&47 , \underline{15}&&\underline{122}&&\underline{100}&&36 , \underline{102}&&121 , \underline{127}\\
65&\underline{18}&36 , \underline{27}&&&&33 , \underline{98}&&66 , \underline{15}&&\underline{128}\\
66&\underline{19}&102 , \underline{91}&&&&\underline{119}&&15 , \underline{65}&&123 , \underline{96}\\
67&\underline{20}&121 , \underline{111}&&&&\underline{62}&&\underline{123}&&68 , \underline{15}\\
68&\underline{21}&127 , \underline{112}&&&&&&\underline{129}&&15 , \underline{67}\\
69&122 , \underline{96}&48&&16&&\underline{109}&&60 , \underline{91}&&86 , \underline{130}\\
70&100&50 , \underline{17}&&\underline{109}&&16&131&\underline{132}&&110\\
71&102&38 , \underline{19}&94&\underline{91}&&\underline{120}&104&16 , \underline{37}&&57 , \underline{95}\\
72&121&44 , \underline{20}&&\underline{86}&&46 , \underline{110}&&\underline{57}&&73 , \underline{16}\\
73&127&51 , \underline{21}&&\underline{130}&&108&&\underline{133}&&16 , \underline{72}\\
74&89 , \underline{88}&132 , \underline{99}&\underline{23}&&&75 , \underline{77}&&17&&113\\
75&98 , \underline{134}&101 , \underline{35}&&&&18&&77 , \underline{74}&&\\
\hline
\end{tabular}}
\end{center}                           
%=================================================================
%=================================================================
\begin{table}[h]
\begin{center}
\tiny
\resizebox{.98 \textwidth}{!}{
\begin{tabular}{ |c|c|c|c|c|c|c|c|c|c|c|} 
\hline
N&1&2&3&4&5&6&7&8&9&10\\
\hline
76&128&79 , \underline{31}&&&\underline{18}&&&31 , \underline{79}&\underline{128}&18\\
77&119 , \underline{135}&120 , \underline{53}&94 , \underline{30}&&&19&&74 , \underline{75}&&57\\
78&&108 , \underline{42}&\underline{40}&&&\underline{21}&&\underline{111}&&\underline{45}\\
79&129&133 , \underline{95}&\underline{41}&&\underline{37}&111&&21&\underline{103}&76 , \underline{31}\\
80&\underline{25}&\underline{115}&&&&136 , \underline{135}&&&81 , \underline{22}&81 , \underline{22}\\
81&\underline{26}&\underline{116}&&&&\underline{119}&&&22 , \underline{80}&137 , \underline{82}\\
82&\underline{27}&\underline{113}&&&&\underline{42}&&&81 , \underline{137}&83 , \underline{22}\\
83&\underline{28}&\underline{117}&&&&&&&22 , \underline{82}&22 , \underline{82}\\
84&85&\underline{112}&&\underline{23}&&\underline{85}&&23&&112\\
85&&\underline{108}&&\underline{30}&&24&&84&&130\\
86&&\underline{72}&&\underline{31}&128 , \underline{103}&69 , \underline{130}&&27&&87 , \underline{24}\\
87&&\underline{110}&&\underline{32}&\underline{107}&109&&93&\underline{138}&24 , \underline{86}\\
88&135 , \underline{134}&\underline{124}&&&&25&&&89 , \underline{74}&89 , \underline{74}\\
89&119 , \underline{98}&\underline{116}&&97 , \underline{34}&&26&&&74 , \underline{88}&139 , \underline{99}\\
90&137&\underline{125}&&98 , \underline{35}&98 , \underline{35}&118 , \underline{139}&&&27 , \underline{26}&27 , \underline{26}\\
91&&\underline{71}&&\underline{38}&66 , \underline{102}&47 , \underline{27}&&60 , \underline{69}&\underline{118}&92\\
92&&\underline{120}&&\underline{40}&\underline{40}&\underline{28}&&126 , \underline{140}&\underline{91}&\underline{91}\\
93&&\underline{117}&&\underline{41}&\underline{41}&140&&28&\underline{87}&\underline{87}\\
94&&\underline{29}&\underline{71}&&&\underline{141}&136&&&77 , \underline{30}\\
95&\underline{52}&\underline{51}&57 , \underline{71}&130&79 , \underline{133}&31 , \underline{48}&\underline{122}&51&47&\underline{30}\\
96&&&123 , \underline{66}&&128 , \underline{129}&69 , \underline{122}&31&&&\\
97&\underline{33}&&&34 , \underline{89}&&33&&&&89 , \underline{34}\\
98&&&&35 , \underline{90}&134 , \underline{75}&119 , \underline{89}&&&&65 , \underline{33}\\
99&\underline{47}&\underline{106}&&139 , \underline{89}&74 , \underline{132}&35 , \underline{53}&&52&50&100 , \underline{34}\\
100&\underline{64}&\underline{114}&&99 , \underline{34}&\underline{70}&64&&114&70&34 , \underline{99}\\
101&&\underline{126}&&35 , \underline{75}&35 , \underline{75}&120 , \underline{132}&&&37&37\\
102&&\underline{102}&&91 , \underline{66}&\underline{71}&64 , \underline{36}&&54 , \underline{122}&40&\underline{53}\\
103&&\underline{121}&&86 , \underline{128}&\underline{79}&122 , \underline{142}&&36&107&\\
104&136&\underline{118}&&40&40&\underline{39}&&125 , \underline{115}&\underline{71}&\underline{71}\\
105&&\underline{115}&\underline{39}&41&41&115&&39&\underline{41}&\underline{41}\\
106&&\underline{99}&111&140&\underline{115}&\underline{41}&&40 , \underline{59}&62 , \underline{114}&\underline{43}\\
107&&\underline{114}&103&87&\underline{41}&114&&\underline{103}&41&\underline{87}\\
108&\underline{112}&&\underline{46}&\underline{73}&&&\underline{85}&&&42 , \underline{78}\\
109&\underline{114}&&\underline{70}&&\underline{43}&\underline{69}&\underline{87}&&140&\underline{143}\\
110&\underline{117}&&\underline{70}&72 , \underline{46}&143&&\underline{87}&&114&43 , \underline{49}\\
111&\underline{60}&&112 , \underline{47}&\underline{106}&&78&\underline{79}&&\underline{44}&67 , \underline{121}\\
112&\underline{51}&68 , \underline{127}&108&\underline{133}&&&\underline{84}&\underline{130}&&47 , \underline{111}\\
113&&&116 , \underline{124}&117 , \underline{52}&\underline{51}&82&\underline{74}&\underline{57}&&49\\
114&121&&\underline{100}&109&\underline{52}&\underline{122}&\underline{107}&&106 , \underline{62}&\underline{110}\\
115&\underline{55}&80&\underline{139}&106&106&104 , \underline{125}&\underline{105}&&116 , \underline{52}&116 , \underline{52}\\
116&\underline{56}&81&\underline{89}&&62&\underline{56}&&&52 , \underline{115}&124 , \underline{113}\\
117&\underline{58}&83&\underline{132}&110&110&&\underline{93}&&52 , \underline{113}&52 , \underline{113}\\
118&139 , \underline{90}&&55&91&91&\underline{137}&\underline{104}&&119 , \underline{53}&119 , \underline{53}\\
119&89 , \underline{98}&&56&&66&\underline{81}&&&53 , \underline{118}&135 , \underline{77}\\
120&132 , \underline{101}&&58&71&71&141&\underline{92}&&53 , \underline{77}&53 , \underline{77}\\
121&\underline{54}&&127 , \underline{64}&&\underline{114}&&\underline{103}&&111 , \underline{67}&\underline{72}\\
122&142 , \underline{103}&&102 , \underline{54}&114&&&\underline{64}&&69 , \underline{96}&\underline{95}\\
123&67&&96 , \underline{66}&\underline{63}&\underline{56}&&&&54 , \underline{60}&\underline{61}\\
124&125&88&\underline{137}&&&55&&&116 , \underline{113}&116 , \underline{113}\\
125&124&90&\underline{135}&\underline{59}&\underline{59}&104 , \underline{115}&&&57 , \underline{56}&57 , \underline{56}\\
126&&&140 , \underline{92}&\underline{60}&\underline{60}&&\underline{101}&&\underline{59}&\underline{59}\\
127&\underline{73}&112 , \underline{68}&&\underline{142}&&&&\underline{142}&&64 , \underline{121}\\
128&\underline{76}&103 , \underline{86}&&&&&&96 , \underline{129}&&65\\
129&\underline{79}&142 , \underline{130}&&&&&&68&&128 , \underline{96}\\
130&142 , \underline{129}&95&&73&&85&&\underline{112}&&69 , \underline{86}\\
131&70&70&&\underline{143}&&&\underline{70}&\underline{70}&&143\\
132&99 , \underline{74}&99 , \underline{74}&&\underline{140}&&101 , \underline{120}&70&70&&117\\
133&142&95 , \underline{79}&&\underline{112}&\underline{73}&112&&73&\underline{142}&79 , \underline{95}\\
134&75 , \underline{98}&75 , \underline{98}&&&&&&135 , \underline{88}&&\\
135&77 , \underline{119}&77 , \underline{119}&136 , \underline{80}&&&&&88 , \underline{134}&&125\\
136&&\underline{104}&&&&135 , \underline{80}&&&\underline{94}&\underline{94}\\
137&\underline{90}&\underline{124}&&&&\underline{118}&&&82 , \underline{81}&82 , \underline{81}\\
138&&\underline{143}&&\underline{87}&\underline{87}&143&&87&87&\\
139&118 , \underline{90}&\underline{115}&&89 , \underline{99}&89 , \underline{99}&90 , \underline{118}&&115&99 , \underline{89}&99 , \underline{89}\\
140&&\underline{132}&&\underline{106}&\underline{106}&\underline{93}&&92 , \underline{126}&\underline{109}&\underline{109}\\
141&\underline{94}&\underline{94}&\underline{120}&&&94&94&&&120\\
142&&\underline{127}&&130 , \underline{129}&\underline{133}&103 , \underline{122}&&127&&\\
143&\underline{110}&&\underline{131}&&\underline{110}&&\underline{138}&&109&109\\
\hline
\end{tabular}}
\end{center}
\caption{Triality connections between the 143 toric phases of Model 16.}
\label{Table 2 - Model 16}
\end{table}
%=================================================================

%=================================================================
\subsection{Model 17: $P^{0}_{+-}(\text{dP}_3)$}
%=================================================================

\fref{f_quiver_02} shows the quiver for Phase 1 of Model 17.

%=================================================================
\begin{figure}[H]
\begin{center}
\resizebox{0.40\hsize}{!}{
\includegraphics[height=6cm]{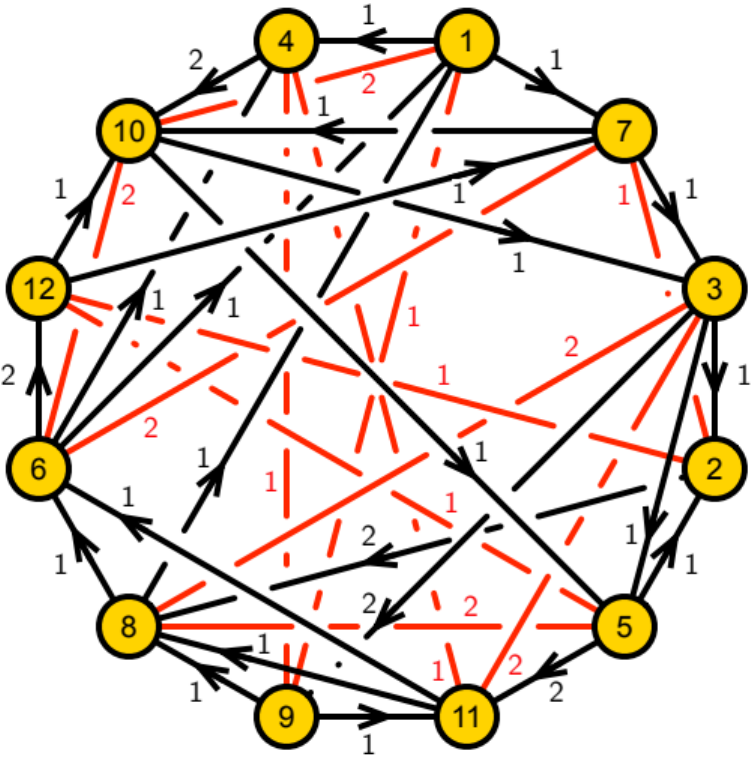} 
}
\caption{Quiver for Phase 1 of Model 17.
\label{f_quiver_17}}
 \end{center}
 \end{figure}
%=================================================================

The $J$- and $E$-terms are
\beq
{\footnotesize
\begin{array}{rrclrcl}
 &  & J &  & & E &  \\
\Lambda_{1.10}^{1} : & X_{10.5} X_{52} Q_{28} X_{81}  &-&  X_{10.3} X_{35} X_{52} Q_{28} X_{86} X_{61}  &  X_{14} P_{4.10}  &-& P_{17} X_{7.10}   \\  %%
\Lambda_{1.10}^{2} : & X_{10.5} P_{5.11} X_{11.8} X_{81}  &-&  X_{10.3} P_{39} X_{9.11} X_{11.8} X_{86} X_{61}  &  Q_{17} X_{7.10}  &-&  X_{14} Q_{4.10}  \\ %
\Lambda_{38}^{1} : & X_{81} Q_{17} X_{73}  &-&  X_{86} Q_{6.12} X_{12.10} X_{10.3}  &  P_{39} X_{98}  &-&  X_{32} P_{28}  \\ %
\Lambda_{38}^{2} : & X_{81} P_{17} X_{73}  &-&  X_{86} X_{64} P_{4.10} X_{10.3}  &   X_{32} Q_{28} &-& Q_{39} X_{98} \\  %%
\Lambda_{3.11}^{1} : & X_{11.6} Q_{6.12} X_{12.7} X_{73}  &-&  X_{11.8} X_{86} Q_{6.12} X_{12.7} X_{7.10} X_{10.3}  &  X_{35} P_{5.11}  &-&  P_{39} X_{9.11}  \\  %%
\Lambda_{3.11}^{2} : & X_{11.6} X_{61} P_{17} X_{73}  &-&  X_{11.8} X_{86} X_{61} X_{14} P_{4.10} X_{10.3}  &  Q_{39} X_{9.11}  &-&  X_{35} Q_{5.11}  \\ %
\Lambda_{58}^{1} : & X_{81} Q_{17} X_{7.10} X_{10.5}  &-&  X_{86} Q_{6.12} X_{12.7} X_{7.10} X_{10.3} X_{35}  &  X_{52} P_{28} &-& P_{5.11} X_{11.8}  \\
\end{array} 
}~.~
\label{E_J_C_+-}
\eeq

This model has 537 toric phases, which are summarized in Table \ref{Table 1 - Model 17}.

%=================================================================
\begin{center}
\tiny
\resizebox{1 \textwidth}{!}{
\begin{tabular}{ |c|c|c|c|c|c|c|c| } 
\hline
 Phase & Path & F & Fermi Multiplicities & Phase & Path & F & Fermi Multiplicities \\
\hline
1&&18&4$\times\textbf{2}$+4$\times\textbf{3}$+4$\times\textbf{4}$&76&2,-5,6&24&3$\times\textbf{2}$+2$\times\textbf{3}$+2$\times\textbf{4}$+2$\times\textbf{5}$+3$\times\textbf{6}$\\ 
2&1&21&3$\times\textbf{2}$+5$\times\textbf{3}$+1$\times\textbf{4}$+1$\times\textbf{5}$+2$\times\textbf{6}$&77&2,-5,-7&21&5$\times\textbf{2}$+1$\times\textbf{3}$+3$\times\textbf{4}$+1$\times\textbf{5}$+2$\times\textbf{6}$\\ 
3&2&17&4$\times\textbf{2}$+6$\times\textbf{3}$+2$\times\textbf{4}$&78&2,-5,-8&24&4$\times\textbf{2}$+1$\times\textbf{3}$+2$\times\textbf{4}$+3$\times\textbf{5}$+1$\times\textbf{6}$+1$\times\textbf{8}$\\ 
4&-2&18&4$\times\textbf{2}$+4$\times\textbf{3}$+4$\times\textbf{4}$&79&2,-5,-9&24&1$\times\textbf{2}$+5$\times\textbf{3}$+3$\times\textbf{4}$+2$\times\textbf{5}$+1$\times\textbf{9}$\\ 
5&3&25&3$\times\textbf{2}$+1$\times\textbf{3}$+4$\times\textbf{4}$+1$\times\textbf{5}$+2$\times\textbf{6}$+1$\times\textbf{8}$&80&2,6,-9&23&3$\times\textbf{2}$+3$\times\textbf{3}$+1$\times\textbf{4}$+3$\times\textbf{5}$+2$\times\textbf{6}$\\ 
6&1,2&20&2$\times\textbf{2}$+7$\times\textbf{3}$+1$\times\textbf{4}$+1$\times\textbf{5}$+1$\times\textbf{6}$&81&2,-7,5&21&4$\times\textbf{2}$+3$\times\textbf{3}$+2$\times\textbf{4}$+1$\times\textbf{5}$+2$\times\textbf{6}$\\ 
7&1,-2&19&4$\times\textbf{2}$+5$\times\textbf{3}$+1$\times\textbf{4}$+1$\times\textbf{5}$+1$\times\textbf{6}$&82&2,-7,-8&20&5$\times\textbf{2}$+2$\times\textbf{3}$+4$\times\textbf{4}$+1$\times\textbf{8}$\\
\hline
\end{tabular}}
\end{center}
%=================================================================
%=================================================================
\begin{center}
\tiny
\resizebox{1 \textwidth}{!}{
\begin{tabular}{ |c|c|c|c|c|c|c|c| } 
\hline
 Phase & Path & F & Fermi Multiplicities & Phase & Path & F & Fermi Multiplicities \\
\hline
151&1,3,-9,6&32&3$\times\textbf{2}$+5$\times\textbf{4}$+1$\times\textbf{6}$+1$\times\textbf{8}$+1$\times\textbf{10}$+1$\times\textbf{14}$&226&3,2,-5,-12&27&1$\times\textbf{2}$+1$\times\textbf{3}$+7$\times\textbf{4}$+2$\times\textbf{6}$+1$\times\textbf{9}$\\ 
152&1,3,-9,-7&26&4$\times\textbf{2}$+5$\times\textbf{4}$+2$\times\textbf{6}$+1$\times\textbf{12}$&227&3,2,8,1&26&2$\times\textbf{2}$+2$\times\textbf{3}$+5$\times\textbf{4}$+1$\times\textbf{6}$+2$\times\textbf{8}$\\ 
153&1,3,-9,11&26&4$\times\textbf{2}$+3$\times\textbf{4}$+2$\times\textbf{5}$+2$\times\textbf{6}$+1$\times\textbf{10}$&228&3,2,8,-9&26&2$\times\textbf{2}$+2$\times\textbf{3}$+5$\times\textbf{4}$+2$\times\textbf{6}$+1$\times\textbf{10}$\\ 
154&1,3,-9,-11&24&4$\times\textbf{2}$+5$\times\textbf{4}$+2$\times\textbf{5}$+1$\times\textbf{10}$&229&3,2,8,-12&26&2$\times\textbf{2}$+2$\times\textbf{3}$+3$\times\textbf{4}$+3$\times\textbf{5}$+1$\times\textbf{6}$+1$\times\textbf{9}$\\ 
155&1,3,-12,2&26&1$\times\textbf{2}$+4$\times\textbf{3}$+1$\times\textbf{4}$+4$\times\textbf{5}$+1$\times\textbf{6}$+1$\times\textbf{8}$&230&1,2,3,-5,-7&22&2$\times\textbf{2}$+4$\times\textbf{3}$+2$\times\textbf{4}$+4$\times\textbf{5}$\\ 
156&1,3,-12,5&28&4$\times\textbf{3}$+2$\times\textbf{4}$+4$\times\textbf{5}$+2$\times\textbf{8}$&231&1,2,3,-5,8&22&4$\times\textbf{2}$+6$\times\textbf{4}$+2$\times\textbf{6}$\\ 
157&1,-4,-6,-7&28&1$\times\textbf{2}$+3$\times\textbf{3}$+2$\times\textbf{4}$+3$\times\textbf{5}$+1$\times\textbf{6}$+1$\times\textbf{7}$+1$\times\textbf{9}$&232&1,2,3,-5,-9&32&8$\times\textbf{4}$+2$\times\textbf{6}$+2$\times\textbf{10}$\\
\hline
\end{tabular}}
\end{center}
%=================================================================
%=================================================================
\begin{center}
\tiny
\resizebox{1 \textwidth}{!}{
\begin{tabular}{ |c|c|c|c|c|c|c|c| } 
\hline
 Phase & Path & F & Fermi Multiplicities & Phase & Path & F & Fermi Multiplicities \\
\hline
301&1,-2,-9,-12,5&24&2$\times\textbf{2}$+8$\times\textbf{4}$+2$\times\textbf{6}$&376&2,-7,5,-8,-9&24&5$\times\textbf{2}$+2$\times\textbf{4}$+2$\times\textbf{5}$+2$\times\textbf{6}$+1$\times\textbf{8}$\\ 
302&1,3,2,-5,7&28&4$\times\textbf{3}$+3$\times\textbf{4}$+2$\times\textbf{5}$+1$\times\textbf{6}$+1$\times\textbf{7}$+1$\times\textbf{9}$&377&2,-7,5,-9,1&24&1$\times\textbf{2}$+2$\times\textbf{3}$+6$\times\textbf{4}$+2$\times\textbf{5}$+1$\times\textbf{6}$\\ 
303&1,3,2,-5,8&24&1$\times\textbf{2}$+2$\times\textbf{3}$+7$\times\textbf{4}$+2$\times\textbf{6}$&378&2,-7,5,-9,-10&26&2$\times\textbf{2}$+2$\times\textbf{3}$+3$\times\textbf{4}$+3$\times\textbf{5}$+1$\times\textbf{7}$+1$\times\textbf{8}$\\ 
304&1,3,2,-5,-9&30&2$\times\textbf{3}$+7$\times\textbf{4}$+1$\times\textbf{6}$+2$\times\textbf{10}$&379&2,-7,-8,1,-11&28&2$\times\textbf{2}$+2$\times\textbf{3}$+5$\times\textbf{4}$+2$\times\textbf{5}$+1$\times\textbf{16}$\\ 
305&1,3,2,-5,-12&30&2$\times\textbf{3}$+6$\times\textbf{4}$+1$\times\textbf{6}$+1$\times\textbf{7}$+1$\times\textbf{8}$+1$\times\textbf{9}$&380&2,-7,-8,-1,-11&28&5$\times\textbf{3}$+4$\times\textbf{4}$+2$\times\textbf{5}$+1$\times\textbf{15}$\\ 
306&1,3,2,8,12&32&2$\times\textbf{2}$+3$\times\textbf{4}$+2$\times\textbf{5}$+2$\times\textbf{6}$+2$\times\textbf{8}$+1$\times\textbf{10}$&381&2,-7,-8,-9,11&28&4$\times\textbf{3}$+6$\times\textbf{4}$+2$\times\textbf{10}$\\ 
307&1,3,2,8,-12&30&2$\times\textbf{2}$+4$\times\textbf{4}$+2$\times\textbf{5}$+1$\times\textbf{6}$+3$\times\textbf{8}$&382&2,-7,-9,1,6&26&2$\times\textbf{2}$+2$\times\textbf{3}$+4$\times\textbf{4}$+3$\times\textbf{6}$+1$\times\textbf{8}$\\
\hline
\end{tabular}}
\end{center}
%=================================================================
%=================================================================
\begin{table}[H]
\begin{center}
\tiny
\resizebox{1 \textwidth}{!}{
\begin{tabular}{ |c|c|c|c|c|c|c|c| } 
\hline
 Phase & Path & F & Fermi Multiplicities & Phase & Path & F & Fermi Multiplicities \\
\hline
451&1,3,2,-5,-9,-12&32&8$\times\textbf{4}$+2$\times\textbf{6}$+2$\times\textbf{10}$&495&3,2,-5,8,-9,6&30&3$\times\textbf{2}$+5$\times\textbf{4}$+1$\times\textbf{6}$+2$\times\textbf{7}$+1$\times\textbf{14}$\\ 
452&1,3,2,-9,6,11&34&3$\times\textbf{2}$+4$\times\textbf{4}$+2$\times\textbf{8}$+2$\times\textbf{9}$+1$\times\textbf{12}$&496&3,2,8,-9,6,-10&42&6$\times\textbf{4}$+2$\times\textbf{6}$+2$\times\textbf{9}$+1$\times\textbf{12}$+1$\times\textbf{18}$\\ 
453&1,3,2,-9,6,-11&30&3$\times\textbf{2}$+5$\times\textbf{4}$+1$\times\textbf{6}$+2$\times\textbf{8}$+1$\times\textbf{12}$&497&1,2,3,-5,-9,6,-12&52&6$\times\textbf{6}$+4$\times\textbf{10}$+2$\times\textbf{14}$\\ 
454&1,3,2,-9,11,-12&28&2$\times\textbf{2}$+2$\times\textbf{3}$+2$\times\textbf{4}$+2$\times\textbf{5}$+2$\times\textbf{6}$+2$\times\textbf{8}$&498&1,2,3,-9,6,11,-12&44&1$\times\textbf{2}$+2$\times\textbf{4}$+2$\times\textbf{6}$+2$\times\textbf{7}$+1$\times\textbf{8}$+1$\times\textbf{10}$+2$\times\textbf{11}$+1$\times\textbf{12}$\\ 
455&1,3,2,-9,-11,-12&26&2$\times\textbf{2}$+2$\times\textbf{3}$+3$\times\textbf{4}$+2$\times\textbf{5}$+2$\times\textbf{6}$+1$\times\textbf{8}$&499&1,2,3,-9,6,-11,-12&40&1$\times\textbf{2}$+3$\times\textbf{4}$+2$\times\textbf{6}$+2$\times\textbf{7}$+1$\times\textbf{8}$+2$\times\textbf{10}$+1$\times\textbf{12}$\\ 
456&1,3,7,5,-9,6&36&5$\times\textbf{3}$+3$\times\textbf{4}$+2$\times\textbf{6}$+1$\times\textbf{12}$+1$\times\textbf{21}$&500&1,2,-5,-3,8,7,-5&30&3$\times\textbf{2}$+2$\times\textbf{3}$+2$\times\textbf{4}$+2$\times\textbf{6}$+2$\times\textbf{8}$+1$\times\textbf{12}$\\
\hline
\end{tabular}}
\end{center}
\caption{Basic information regarding the 537 toric phases of Model 17.}
\label{Table 1 - Model 17}
\end{table}
%=================================================================

Table \ref{Table 2 - Model 17} summarizes the connection between the toric phases under triality.

%=================================================================
\begin{center}
\tiny
\resizebox{1 \textwidth}{!}{
\begin{tabular}{ |c|c|c|c|c|c|c|c|c|c|c|c|c|} 
\hline
N&1&2&3&4&5&6&7&8&9&10&11&12\\
\hline
1&2&3 , \underline{4}&5&3 , \underline{4}&2&5&\underline{2}&\underline{5}&4 , \underline{3}&\underline{5}&\underline{2}&4 , \underline{3}\\
2&\underline{1}&6 , \underline{7}&8&\underline{9}&10&&\underline{11}&&12 , \underline{13}&&\underline{14}&15 , \underline{16}\\
3&6&4 , \underline{1}&17&18 , \underline{3}&\underline{19}&20&\underline{21}&22 , \underline{23}&9 , \underline{24}&\underline{25}&\underline{16}&7\\
4&7&1 , \underline{3}&15&3 , \underline{1}&9 , \underline{26}&12&\underline{12}&27&26 , \underline{9}&\underline{27}&\underline{15}&\underline{7}\\
5&8&28 , \underline{15}&\underline{1}&20 , \underline{12}&&29&&&\underline{23}&&&27 , \underline{25}\\
6&\underline{3}&7 , \underline{2}&30&\underline{24}&\underline{31}&&\underline{32}&33&34 , \underline{35}&&\underline{36}&37\\
7&\underline{4}&2 , \underline{6}&37&\underline{3}&38 , \underline{19}&&39 , \underline{40}&41&42 , \underline{43}&&\underline{44}&\underline{45}\\
\hline
\end{tabular}}
\end{center}                           
%=================================================================

%=================================================================
\begin{center}
\tiny
\resizebox{1 \textwidth}{!}{
\begin{tabular}{ |c|c|c|c|c|c|c|c|c|c|c|c|c|} 
\hline
N&1&2&3&4&5&6&7&8&9&10&11&12\\
\hline
35&\underline{83}&43 , \underline{13}&98&\underline{83}&\underline{95}&116&\underline{108}&96&6 , \underline{34}&&117 , \underline{118}&46\\
36&\underline{16}&44 , \underline{14}&&119 , \underline{80}&&&\underline{64}&&115 , \underline{120}&&6&121\\
37&\underline{7}&\underline{15}&99&\underline{79}&&&\underline{110}&113&8 , \underline{46}&&\underline{121}&\underline{6}\\
38&\underline{9}&10&\underline{122}&\underline{18}&19 , \underline{7}&&19 , \underline{7}&123&124 , \underline{125}&&\underline{9}&\underline{126}\\
39&&\underline{103}&127&\underline{22}&19 , \underline{78}&&40 , \underline{7}&128&129 , \underline{130}&&\underline{61}&\\
40&\underline{12}&11 , \underline{59}&131&\underline{23}&7 , \underline{39}&&7 , \underline{39}&132&133 , \underline{134}&&\underline{12}&\\
41&89&\underline{33}&99&74 , \underline{17}&123 , \underline{77}&&128 , \underline{132}&\underline{7}&135 , \underline{136}&&&\\
\hline
\end{tabular}}
\end{center}                           
%=================================================================
%=================================================================
\begin{center}
\tiny
\resizebox{1 \textwidth}{!}{
\begin{tabular}{ |c|c|c|c|c|c|c|c|c|c|c|c|c|} 
\hline
N&1&2&3&4&5&6&7&8&9&10&11&12\\
\hline
125&\underline{165}&58&&\underline{67}&68 , \underline{43}&283&68 , \underline{43}&282&38 , \underline{124}&&\underline{165}&\underline{284}\\
126&\underline{53}&58&285 , \underline{280}&\underline{67}&\underline{142}&&\underline{142}&&224 , \underline{284}&&\underline{53}&38\\
127&&\underline{246}&\underline{39}&\underline{85}&202 , \underline{212}&&131 , \underline{113}&286&\underline{287}&&&\\
128&223&\underline{247}&286&196 , \underline{71}&77 , \underline{208}&&132 , \underline{41}&\underline{39}&288 , \underline{246}&&&\\
129&&\underline{249}&&&75 , \underline{206}&&133 , \underline{42}&288&130 , \underline{39}&&289 , \underline{137}&\\
130&&\underline{250}&287&\underline{71}&68 , \underline{187}&290&134 , \underline{43}&246&39 , \underline{129}&&291 , \underline{225}&\\
131&&60 , \underline{169}&\underline{40}&\underline{86}&113 , \underline{127}&&113 , \underline{127}&292&\underline{293}&&&\\
\hline
\end{tabular}}
\end{center}                           
%=================================================================
%=================================================================
\begin{center}
\tiny
\resizebox{1 \textwidth}{!}{
\begin{tabular}{ |c|c|c|c|c|c|c|c|c|c|c|c|c|} 
\hline
N&1&2&3&4&5&6&7&8&9&10&11&12\\
\hline
212&347&&349&202 , \underline{127}&87&&\underline{369}&202 , \underline{127}&78 , \underline{157}&&&\\
213&252&&81&345 , \underline{203}&205&&&357 , \underline{373}&&&\underline{343}&\\
214&374 , \underline{375}&\underline{154}&\underline{373}&81 , \underline{196}&208 , \underline{82}&&&81 , \underline{196}&325 , \underline{376}&\underline{154}&&295\\
215&377&\underline{118}&&\underline{100}&209 , \underline{83}&&&\underline{363}&81 , \underline{163}&\underline{378}&&116\\
216&257&\underline{176}&&178&210 , \underline{84}&&&229 , \underline{154}&326 , \underline{378}&81&\underline{342}&297\\
217&218 , \underline{82}&\underline{152}&86&105 , \underline{195}&374 , \underline{367}&&86&105 , \underline{195}&218 , \underline{82}&\underline{152}&\underline{379}&293\\
218&82 , \underline{217}&\underline{150}&160&106&375 , \underline{368}&&&106&324 , \underline{161}&\underline{150}&\underline{380}&171\\
\hline
\end{tabular}}
\end{center}                           
%=================================================================
%=================================================================
\begin{center}
\tiny
\resizebox{1 \textwidth}{!}{
\begin{tabular}{ |c|c|c|c|c|c|c|c|c|c|c|c|c|} 
\hline
N&1&2&3&4&5&6&7&8&9&10&11&12\\
\hline
299&\underline{280}&166&323&&&445&&&\underline{159}&&300 , \underline{142}&140\\
300&\underline{285}&178&326&\underline{180}&448&446&&&\underline{277}&&142 , \underline{299}&141\\
301&\underline{284}&179&448&\underline{179}&185 , \underline{142}&447&\underline{447}&&142 , \underline{185}&&\underline{448}&284\\
302&&202&\underline{186}&\underline{193}&144&&\underline{143}&209&\underline{188}&&&\\
303&270&209&\underline{231}&377 , \underline{220}&145&&209&\underline{143}&\underline{143}&&449&\underline{450}\\
304&\underline{381}&188&\underline{232}&\underline{200}&146&392&188&143&143&&&\underline{451}\\
305&\underline{226}&\underline{320}&\underline{233}&\underline{221}&148&&&450&\underline{451}&&&143\\
\hline
\end{tabular}}
\end{center}                           
%=================================================================

%=================================================================
\begin{center}
\tiny
\resizebox{1 \textwidth}{!}{
\begin{tabular}{ |c|c|c|c|c|c|c|c|c|c|c|c|c|} 
\hline
N&1&2&3&4&5&6&7&8&9&10&11&12\\
\hline
386&\underline{229}&&\underline{353}&455 , \underline{310}&\underline{436}&&&\underline{307}&\underline{229}&&&227 , \underline{383}\\
387&&&\underline{354}&428 , \underline{318}&\underline{495}&\underline{228}&&\underline{228}&&\underline{496}&&\underline{354}\\
388&&&\underline{355}&406&&496&&\underline{359}&&228&&460 , \underline{483}\\
389&412 , \underline{478}&449&402 , \underline{245}&423 , \underline{269}&&&402 , \underline{245}&&&&\underline{231}&\\
390&411 , \underline{479}&450&&422 , \underline{268}&236&&&\underline{233}&\underline{233}&&&231\\
391&\underline{451}&392&\underline{233}&\underline{240}&237&\underline{232}&\underline{233}&&&&&\underline{497}\\
392&304&391&\underline{274}&304&\underline{308}&234&\underline{274}&&&391&&\\
\hline
\end{tabular}}
\end{center}                           
%=================================================================
%=================================================================
\begin{table}[H]
\begin{center}
\tiny
\resizebox{1 \textwidth}{!}{
\begin{tabular}{ |c|c|c|c|c|c|c|c|c|c|c|c|c|} 
\hline
N&1&2&3&4&5&6&7&8&9&10&11&12\\
\hline
473&\underline{339}&\underline{520}&340&\underline{522}&\underline{337}&\underline{339}&\underline{341}&&342&&338&340\\
474&\underline{341}&\underline{521}&338&\underline{524}&&\underline{341}&338&&&526&\underline{341}&338\\
475&\underline{476}&&&&&526&&&343 , \underline{352}&\underline{341}&\underline{476}&\\
476&&475&\underline{345}&&&&&&&475&\underline{344}&\underline{527}\\
477&\underline{512}&344&&344&\underline{512}&&527&344&344&344&527&344\\
478&&&485 , \underline{372}&389 , \underline{412}&&&485 , \underline{372}&389 , \underline{412}&&&\underline{348}&\\
479&\underline{350}&&&390 , \underline{411}&358&&&390 , \underline{411}&\underline{350}&&&348\\
\hline
\end{tabular}}
\end{center}
\caption{Triality connections between the 537 toric phases of Model 17.}
\label{Table 2 - Model 17}
\end{table}                     
%=================================================================

\newpage

%=================================================================
\subsection{Model 18: $P^{1}_{+-}(\text{dP}_3)$}
%=================================================================
 
\begin{figure}[H]
\begin{center}
\resizebox{0.40\hsize}{!}{
\includegraphics[height=6cm]{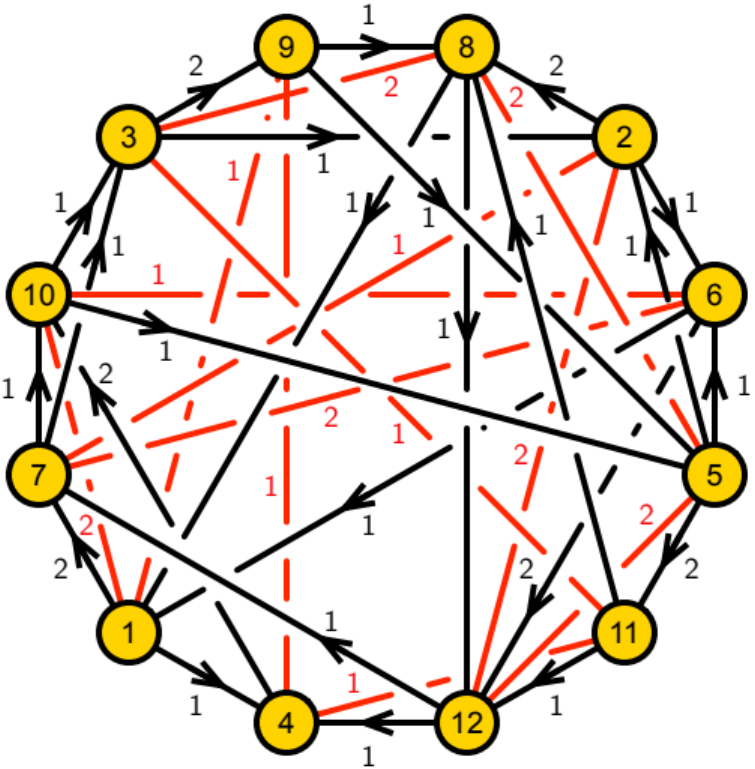} 
}
\caption{Quiver for Phase 1 of Model 18.
\label{f_quiver_18}}
 \end{center}
 \end{figure}

 \fref{f_quiver_18} shows the quiver for Phase 1 of Model 18.
  
%=================================================================
%=================================================================

The $J$- and $E$-terms are
\beq
{\footnotesize
\begin{array}{rrclrcl}
 &  & J &  & & E &  \\
 \Lambda_{19}^{1} : &  X_{95} X_{56} X_{61} &-&X_{98} X_{81}   & P_{17} X_{73} Q_{39} &-& Q_{17} X_{73} P_{39}  \\ 
 \Lambda_{1.10}^{1} : & X_{10.5} X_{52} Q_{28} X_{81} &-& X_{10.3} Q_{39} X_{95} X_{52} X_{26} X_{61}    &  P_{17} X_{7.10} &-& X_{14} P_{4.10}  \\ 
 \Lambda_{1.10}^{2} : &  X_{10.5} P_{5.11} X_{11.8} X_{81} &-&X_{10.3} P_{39} X_{95} X_{52} X_{26} X_{61}    &  X_{14} Q_{4.10}  &-&Q_{17} X_{7.10} \\ 
 \Lambda_{27}^{1} : & X_{7.10} X_{10.5} X_{52} &-&X_{73} X_{32}     &  P_{28} X_{81} Q_{17} &-& Q_{28} X_{81} P_{17}  \\ 
 \Lambda_{2.12}^{1} : & X_{12.4} Q_{4.10} X_{10.3} X_{32} &-&  X_{12.7} X_{7.10} X_{10.3} Q_{39} X_{95} X_{52}    &  P_{28} X_{8.12} &-& X_{26} P_{6.12}  \\ 
 \Lambda_{2.12}^{2} : & X_{12.4} P_{4.10} X_{10.3} X_{32} &-&  X_{12.7} X_{7.10} X_{10.3} P_{39} X_{95} X_{52}    & X_{26} Q_{6.12}  &-&Q_{28} X_{8.12} \\ 
 \Lambda_{38}^{1} : &  X_{8.12} X_{12.4} Q_{4.10} X_{10.3}&-&X_{81} Q_{17} X_{73}    &  P_{39} X_{98} &-& X_{32} P_{28}  \\
\end{array} 
}~.~
\label{E_J_C_+-}
\eeq

This model has 831 toric phases, which are summarized in Table \ref{Table 1 - Model 18}.

%=================================================================
\begin{center}
\tiny
\resizebox{1 \textwidth}{!}{
\begin{tabular}{ |c|c|c|c|c|c|c|c| } 
\hline
 Phase & Path & F & Fermi Multiplicities & Phase & Path & F & Fermi Multiplicities \\
\hline
1&&18&3$\times\textbf{2}$+6$\times\textbf{3}$+3$\times\textbf{4}$&76&-8,-9&23&2$\times\textbf{2}$+5$\times\textbf{3}$+2$\times\textbf{4}$+1$\times\textbf{5}$+1$\times\textbf{6}$+1$\times\textbf{8}$\\ 
2&1&21&2$\times\textbf{2}$+5$\times\textbf{3}$+3$\times\textbf{4}$+1$\times\textbf{5}$+1$\times\textbf{6}$&77&-8,-10&28&1$\times\textbf{2}$+4$\times\textbf{3}$+3$\times\textbf{4}$+1$\times\textbf{6}$+3$\times\textbf{8}$\\ 
3&2&19&4$\times\textbf{2}$+5$\times\textbf{3}$+1$\times\textbf{4}$+1$\times\textbf{5}$+1$\times\textbf{6}$&78&-9,1&21&1$\times\textbf{2}$+6$\times\textbf{3}$+3$\times\textbf{4}$+2$\times\textbf{5}$\\ 
4&3&23&1$\times\textbf{2}$+5$\times\textbf{3}$+3$\times\textbf{4}$+1$\times\textbf{5}$+2$\times\textbf{6}$&79&-9,-8&21&1$\times\textbf{2}$+8$\times\textbf{3}$+2$\times\textbf{4}$+1$\times\textbf{8}$\\ 
5&4&19&2$\times\textbf{2}$+7$\times\textbf{3}$+2$\times\textbf{4}$+1$\times\textbf{5}$&80&-9,-10&24&2$\times\textbf{2}$+4$\times\textbf{3}$+2$\times\textbf{4}$+2$\times\textbf{5}$+1$\times\textbf{6}$+1$\times\textbf{8}$\\ 
6&-4&20&2$\times\textbf{2}$+6$\times\textbf{3}$+2$\times\textbf{4}$+2$\times\textbf{5}$&81&-9,-11&18&2$\times\textbf{2}$+8$\times\textbf{3}$+2$\times\textbf{4}$\\ 
7&5&25&1$\times\textbf{2}$+4$\times\textbf{3}$+3$\times\textbf{4}$+4$\times\textbf{6}$&82&-9,-12&23&2$\times\textbf{2}$+5$\times\textbf{3}$+1$\times\textbf{4}$+3$\times\textbf{5}$+1$\times\textbf{8}$\\
\hline
\end{tabular}}
\end{center}
%=================================================================
%=================================================================
\begin{center}
\tiny
\resizebox{1 \textwidth}{!}{
\begin{tabular}{ |c|c|c|c|c|c|c|c| } 
\hline
 Phase & Path & F & Fermi Multiplicities & Phase & Path & F & Fermi Multiplicities \\
\hline
151&2,-9,1&24&1$\times\textbf{2}$+4$\times\textbf{3}$+3$\times\textbf{4}$+2$\times\textbf{5}$+2$\times\textbf{6}$&226&6,5,-12&26&1$\times\textbf{2}$+4$\times\textbf{3}$+2$\times\textbf{4}$+2$\times\textbf{5}$+1$\times\textbf{6}$+2$\times\textbf{7}$\\ 
152&2,-9,-8&20&2$\times\textbf{2}$+7$\times\textbf{3}$+2$\times\textbf{4}$+1$\times\textbf{7}$&227&6,-8,-9&26&2$\times\textbf{2}$+3$\times\textbf{3}$+2$\times\textbf{4}$+2$\times\textbf{5}$+1$\times\textbf{6}$+1$\times\textbf{7}$+1$\times\textbf{8}$\\ 
153&2,-9,-10&25&2$\times\textbf{2}$+5$\times\textbf{3}$+2$\times\textbf{5}$+1$\times\textbf{6}$+1$\times\textbf{7}$+1$\times\textbf{8}$&228&6,-8,-10&30&1$\times\textbf{2}$+2$\times\textbf{3}$+2$\times\textbf{4}$+2$\times\textbf{5}$+3$\times\textbf{6}$+1$\times\textbf{7}$+1$\times\textbf{9}$\\ 
154&2,-9,-11&23&2$\times\textbf{2}$+5$\times\textbf{3}$+1$\times\textbf{4}$+2$\times\textbf{5}$+1$\times\textbf{6}$+1$\times\textbf{7}$&229&6,9,-10&23&3$\times\textbf{2}$+4$\times\textbf{3}$+3$\times\textbf{5}$+1$\times\textbf{6}$+1$\times\textbf{7}$\\ 
155&2,-10,6&25&3$\times\textbf{2}$+2$\times\textbf{3}$+1$\times\textbf{4}$+2$\times\textbf{5}$+4$\times\textbf{6}$&230&6,9,12&24&2$\times\textbf{2}$+3$\times\textbf{3}$+1$\times\textbf{4}$+5$\times\textbf{5}$+1$\times\textbf{6}$\\ 
156&2,-10,-7&25&3$\times\textbf{2}$+2$\times\textbf{3}$+1$\times\textbf{4}$+3$\times\textbf{5}$+2$\times\textbf{6}$+1$\times\textbf{7}$&231&6,-9,1&22&2$\times\textbf{2}$+4$\times\textbf{3}$+2$\times\textbf{4}$+4$\times\textbf{5}$\\ 
157&2,-10,11&22&2$\times\textbf{2}$+6$\times\textbf{3}$+2$\times\textbf{5}$+2$\times\textbf{6}$&232&6,-9,-8&23&1$\times\textbf{2}$+6$\times\textbf{3}$+2$\times\textbf{4}$+1$\times\textbf{5}$+1$\times\textbf{6}$+1$\times\textbf{7}$\\
\hline
\end{tabular}}
\end{center}
%=================================================================
%=================================================================
\begin{center}
\tiny
\resizebox{1 \textwidth}{!}{
\begin{tabular}{ |c|c|c|c|c|c|c|c| } 
\hline
 Phase & Path & F & Fermi Multiplicities & Phase & Path & F & Fermi Multiplicities \\
\hline
301&1,2,-11,4&23&1$\times\textbf{2}$+4$\times\textbf{3}$+3$\times\textbf{4}$+4$\times\textbf{5}$&376&2,-4,-9,1&26&1$\times\textbf{2}$+4$\times\textbf{3}$+3$\times\textbf{4}$+1$\times\textbf{5}$+3$\times\textbf{7}$\\ 
302&1,3,2,-4&20&1$\times\textbf{2}$+8$\times\textbf{3}$+1$\times\textbf{4}$+2$\times\textbf{5}$&377&2,-4,-9,-8&22&3$\times\textbf{2}$+3$\times\textbf{3}$+4$\times\textbf{4}$+1$\times\textbf{5}$+1$\times\textbf{8}$\\ 
303&1,3,2,7&26&1$\times\textbf{2}$+4$\times\textbf{3}$+3$\times\textbf{4}$+2$\times\textbf{5}$+2$\times\textbf{8}$&378&2,-4,-9,-11&27&1$\times\textbf{2}$+5$\times\textbf{3}$+1$\times\textbf{4}$+1$\times\textbf{5}$+1$\times\textbf{6}$+2$\times\textbf{7}$+1$\times\textbf{8}$\\ 
304&1,3,2,8&28&1$\times\textbf{2}$+1$\times\textbf{3}$+4$\times\textbf{4}$+3$\times\textbf{5}$+1$\times\textbf{6}$+2$\times\textbf{7}$&379&2,-7,5,8&32&2$\times\textbf{2}$+3$\times\textbf{3}$+1$\times\textbf{4}$+1$\times\textbf{6}$+3$\times\textbf{7}$+1$\times\textbf{8}$+1$\times\textbf{12}$\\ 
305&1,3,2,-9&28&4$\times\textbf{3}$+4$\times\textbf{4}$+1$\times\textbf{5}$+1$\times\textbf{6}$+1$\times\textbf{8}$+1$\times\textbf{9}$&380&2,-7,5,-8&28&3$\times\textbf{2}$+3$\times\textbf{3}$+1$\times\textbf{4}$+1$\times\textbf{5}$+1$\times\textbf{6}$+1$\times\textbf{7}$+1$\times\textbf{8}$+1$\times\textbf{11}$\\ 
306&1,3,2,11&28&3$\times\textbf{3}$+5$\times\textbf{4}$+2$\times\textbf{5}$+1$\times\textbf{8}$+1$\times\textbf{9}$&381&2,-7,5,-10&30&1$\times\textbf{2}$+2$\times\textbf{3}$+3$\times\textbf{4}$+2$\times\textbf{5}$+2$\times\textbf{6}$+1$\times\textbf{7}$+1$\times\textbf{11}$\\ 
307&1,3,-4,7&24&7$\times\textbf{3}$+2$\times\textbf{4}$+1$\times\textbf{5}$+2$\times\textbf{7}$&382&2,-7,8,-10&26&1$\times\textbf{2}$+4$\times\textbf{3}$+3$\times\textbf{4}$+1$\times\textbf{5}$+1$\times\textbf{6}$+1$\times\textbf{7}$+1$\times\textbf{8}$\\
\hline
\end{tabular}}
\end{center}
%=================================================================
%=================================================================
\begin{center}
\tiny
\resizebox{1 \textwidth}{!}{
\begin{tabular}{ |c|c|c|c|c|c|c|c| } 
\hline
 Phase & Path & F & Fermi Multiplicities & Phase & Path & F & Fermi Multiplicities \\
\hline
451&4,5,10,2&28&3$\times\textbf{3}$+4$\times\textbf{4}$+2$\times\textbf{5}$+1$\times\textbf{6}$+1$\times\textbf{7}$+1$\times\textbf{8}$&526&6,-10,2,-11&30&2$\times\textbf{2}$+2$\times\textbf{3}$+1$\times\textbf{4}$+1$\times\textbf{5}$+3$\times\textbf{6}$+2$\times\textbf{7}$+1$\times\textbf{9}$\\ 
452&4,5,10,-11&22&2$\times\textbf{2}$+5$\times\textbf{3}$+2$\times\textbf{4}$+2$\times\textbf{5}$+1$\times\textbf{7}$&527&6,-10,2,12&30&2$\times\textbf{2}$+2$\times\textbf{3}$+1$\times\textbf{4}$+2$\times\textbf{5}$+2$\times\textbf{6}$+2$\times\textbf{7}$+1$\times\textbf{10}$\\ 
453&4,5,-11,1&23&3$\times\textbf{2}$+3$\times\textbf{3}$+3$\times\textbf{4}$+1$\times\textbf{5}$+1$\times\textbf{6}$+1$\times\textbf{8}$&528&6,-10,2,-12&26&3$\times\textbf{2}$+3$\times\textbf{3}$+3$\times\textbf{5}$+1$\times\textbf{6}$+1$\times\textbf{7}$+1$\times\textbf{9}$\\ 
454&4,5,-11,-8&24&2$\times\textbf{2}$+4$\times\textbf{3}$+2$\times\textbf{4}$+1$\times\textbf{5}$+2$\times\textbf{6}$+1$\times\textbf{7}$&529&6,-10,-11,4&24&6$\times\textbf{3}$+2$\times\textbf{4}$+2$\times\textbf{5}$+2$\times\textbf{6}$\\ 
455&4,5,-11,-10&24&2$\times\textbf{2}$+4$\times\textbf{3}$+2$\times\textbf{4}$+1$\times\textbf{5}$+2$\times\textbf{6}$+1$\times\textbf{7}$&530&6,-11,4,-8&26&5$\times\textbf{3}$+2$\times\textbf{4}$+2$\times\textbf{5}$+2$\times\textbf{6}$+1$\times\textbf{7}$\\ 
456&4,6,5,-10&26&3$\times\textbf{3}$+5$\times\textbf{4}$+1$\times\textbf{5}$+3$\times\textbf{6}$&531&6,-11,4,-12&18&2$\times\textbf{2}$+8$\times\textbf{3}$+2$\times\textbf{4}$\\ 
457&4,6,5,-11&24&1$\times\textbf{2}$+3$\times\textbf{3}$+4$\times\textbf{4}$+3$\times\textbf{5}$+1$\times\textbf{6}$&532&6,-11,-12,-5&20&4$\times\textbf{2}$+4$\times\textbf{3}$+1$\times\textbf{4}$+2$\times\textbf{5}$+1$\times\textbf{6}$\\
\hline
\end{tabular}}
\end{center}
%=================================================================
%=================================================================
\begin{center}
\tiny
\resizebox{1 \textwidth}{!}{
\begin{tabular}{ |c|c|c|c|c|c|c|c| } 
\hline
 Phase & Path & F & Fermi Multiplicities & Phase & Path & F & Fermi Multiplicities \\
\hline
601&1,3,2,-4,10&26&1$\times\textbf{2}$+5$\times\textbf{3}$+2$\times\textbf{5}$+3$\times\textbf{6}$+1$\times\textbf{7}$&676&2,-11,4,-1,-8&24&1$\times\textbf{2}$+5$\times\textbf{3}$+1$\times\textbf{4}$+3$\times\textbf{5}$+2$\times\textbf{6}$\\ 
602&1,3,2,7,11&30&1$\times\textbf{2}$+2$\times\textbf{3}$+4$\times\textbf{4}$+2$\times\textbf{5}$+1$\times\textbf{8}$+2$\times\textbf{9}$&677&3,2,4,6,10&28&4$\times\textbf{3}$+4$\times\textbf{4}$+1$\times\textbf{5}$+1$\times\textbf{6}$+1$\times\textbf{7}$+1$\times\textbf{10}$\\ 
603&1,3,2,7,-11&29&1$\times\textbf{2}$+3$\times\textbf{3}$+2$\times\textbf{4}$+3$\times\textbf{5}$+1$\times\textbf{7}$+1$\times\textbf{8}$+1$\times\textbf{9}$&678&3,2,4,6,12&25&5$\times\textbf{3}$+4$\times\textbf{4}$+1$\times\textbf{5}$+1$\times\textbf{6}$+1$\times\textbf{8}$\\ 
604&1,3,2,8,4&28&1$\times\textbf{2}$+1$\times\textbf{3}$+2$\times\textbf{4}$+6$\times\textbf{5}$+1$\times\textbf{6}$+1$\times\textbf{7}$&679&3,2,4,10,11&36&1$\times\textbf{3}$+4$\times\textbf{4}$+3$\times\textbf{5}$+2$\times\textbf{7}$+1$\times\textbf{10}$+1$\times\textbf{14}$\\ 
605&1,3,5,6,7&29&5$\times\textbf{3}$+2$\times\textbf{4}$+3$\times\textbf{5}$+1$\times\textbf{8}$+1$\times\textbf{12}$&680&3,2,-4,6,8&25&2$\times\textbf{2}$+4$\times\textbf{3}$+2$\times\textbf{4}$+1$\times\textbf{5}$+1$\times\textbf{6}$+1$\times\textbf{7}$+1$\times\textbf{8}$\\ 
606&1,3,7,-9,6&27&2$\times\textbf{2}$+4$\times\textbf{3}$+1$\times\textbf{4}$+3$\times\textbf{5}$+1$\times\textbf{6}$+1$\times\textbf{13}$&681&3,2,-4,6,10&27&6$\times\textbf{3}$+2$\times\textbf{4}$+1$\times\textbf{5}$+1$\times\textbf{6}$+1$\times\textbf{7}$+1$\times\textbf{10}$\\
\hline
\end{tabular}}
\end{center}
%=================================================================
%=================================================================
\begin{table}[H]
\begin{center}
\tiny
\resizebox{1 \textwidth}{!}{
\begin{tabular}{ |c|c|c|c|c|c|c|c| } 
\hline
 Phase & Path & F & Fermi Multiplicities & Phase & Path & F & Fermi Multiplicities \\
\hline
751&6,12,-11,-5,-7&24&5$\times\textbf{2}$+2$\times\textbf{3}$+1$\times\textbf{4}$+1$\times\textbf{5}$+1$\times\textbf{6}$+1$\times\textbf{8}$+1$\times\textbf{9}$&792&1,-7,3,-9,6,12&28&3$\times\textbf{2}$+1$\times\textbf{3}$+1$\times\textbf{4}$+4$\times\textbf{5}$+2$\times\textbf{6}$+1$\times\textbf{11}$\\ 
752&-7,2,5,-8,1&29&2$\times\textbf{2}$+3$\times\textbf{3}$+2$\times\textbf{4}$+1$\times\textbf{5}$+2$\times\textbf{6}$+1$\times\textbf{9}$+1$\times\textbf{11}$&793&1,-7,3,-9,6,-12&26&3$\times\textbf{2}$+2$\times\textbf{3}$+2$\times\textbf{4}$+3$\times\textbf{5}$+1$\times\textbf{7}$+1$\times\textbf{10}$\\ 
753&-7,2,5,-8,-10&34&1$\times\textbf{2}$+2$\times\textbf{3}$+3$\times\textbf{4}$+1$\times\textbf{5}$+1$\times\textbf{6}$+2$\times\textbf{8}$+1$\times\textbf{9}$+1$\times\textbf{12}$&794&1,-9,6,-11,-12,3&38&3$\times\textbf{4}$+4$\times\textbf{5}$+1$\times\textbf{6}$+1$\times\textbf{8}$+2$\times\textbf{9}$+1$\times\textbf{12}$\\ 
754&-7,2,-8,1,-10&30&2$\times\textbf{2}$+3$\times\textbf{3}$+3$\times\textbf{4}$+3$\times\textbf{8}$+1$\times\textbf{11}$&795&1,-9,6,-12,-11,-5&33&4$\times\textbf{4}$+4$\times\textbf{5}$+2$\times\textbf{7}$+2$\times\textbf{8}$\\ 
755&-7,2,-8,-1,-10&26&3$\times\textbf{2}$+4$\times\textbf{3}$+2$\times\textbf{4}$+2$\times\textbf{8}$+1$\times\textbf{10}$&796&1,-9,6,-12,-11,-7&34&2$\times\textbf{4}$+6$\times\textbf{5}$+2$\times\textbf{7}$+2$\times\textbf{8}$\\ 
756&-7,2,-8,-9,-10&28&2$\times\textbf{2}$+3$\times\textbf{3}$+2$\times\textbf{4}$+3$\times\textbf{6}$+1$\times\textbf{8}$+1$\times\textbf{9}$&797&1,-9,-12,-11,3,-5&40&2$\times\textbf{4}$+4$\times\textbf{5}$+2$\times\textbf{6}$+2$\times\textbf{9}$+2$\times\textbf{11}$\\
\hline
\end{tabular}}
\end{center}
\caption{Basic information regarding the 831 toric phases of Model 18.}
\label{Table 1 - Model 18}
\end{table}
%=================================================================

Table \ref{Table 2 - Model 18} summarizes the connection between the toric phases under triality.

%=================================================================
\begin{center}
\tiny
\resizebox{1 \textwidth}{!}{
\begin{tabular}{ |c|c|c|c|c|c|c|c|c|c|c|c|c|} 
\hline
N&1&2&3&4&5&6&7&8&9&10&11&12\\
\hline
1&2&3&4&5 , \underline{6}&7&8&\underline{9}&\underline{10}&11 , \underline{12}&\underline{13}&14 , \underline{15}&\underline{16}\\
2&\underline{1}&17&18&\underline{19}&20&&\underline{21}&&22 , \underline{23}&&24 , \underline{25}&\underline{26}\\
3&17&\underline{1}&27&28 , \underline{29}&&&\underline{30}&31 , \underline{32}&33 , \underline{34}&\underline{35}&36 , \underline{37}&\\
4&18&38&\underline{1}&39 , \underline{40}&41&42&&&\underline{43}&&44&\underline{45}\\
5&\underline{46}&28&39&6 , \underline{1}&47&48&\underline{49}&\underline{35}&50&\underline{51}&52 , \underline{53}&\\
6&19 , \underline{54}&29&40&1 , \underline{5}&55&56&\underline{57}&\underline{58}&\underline{59}&&\underline{60}&\\
7&20&&41&47 , \underline{55}&\underline{1}&61&\underline{62}&&&&\underline{63}&\\
\hline
\end{tabular}}
\end{center}                           
%=================================================================

%=================================================================
\begin{center}
\tiny
\resizebox{1 \textwidth}{!}{
\begin{tabular}{ |c|c|c|c|c|c|c|c|c|c|c|c|c|} 
\hline
N&1&2&3&4&5&6&7&8&9&10&11&12\\
\hline
31&144&&123&29 , \underline{133}&&&\underline{137}&32 , \underline{3}&\underline{145}&\underline{58}&&\\
32&146&\underline{10}&124&3 , \underline{31}&&&\underline{138}&3 , \underline{31}&147 , \underline{148}&\underline{10}&149&\\
33&93&\underline{11}&&129&&&\underline{139}&\underline{147}&34 , \underline{3}&\underline{150}&13 , \underline{80}&\\
34&151&\underline{12}&125&\underline{134}&&&\underline{140}&\underline{152}&3 , \underline{33}&\underline{153}&84 , \underline{154}&\\
35&&\underline{13}&&5&&155&\underline{156}&58 , \underline{10}&150 , \underline{153}&3&157 , \underline{158}&\\
36&95&\underline{14}&126&130&&&\underline{142}&\underline{149}&13 , \underline{84}&\underline{157}&37 , \underline{3}&\\
37&96&\underline{15}&&159 , \underline{135}&&&\underline{143}&&80 , \underline{154}&\underline{158}&3 , \underline{36}&\\
\hline
\end{tabular}}
\end{center}                           
%=================================================================
%=================================================================
\begin{center}
\tiny
\resizebox{1 \textwidth}{!}{
\begin{tabular}{ |c|c|c|c|c|c|c|c|c|c|c|c|c|} 
\hline
N&1&2&3&4&5&6&7&8&9&10&11&12\\
\hline
106&\underline{73}&\underline{21}&318&\underline{178}&280&&91&319 , \underline{320}&321 , \underline{322}&\underline{323}&221 , \underline{87}&\\
107&&318&\underline{21}&\underline{164}&165&&100&\underline{324}&\underline{325}&&220 , \underline{270}&\\
108&\underline{248}&320&324&&&&&21&326 , \underline{327}&&216 , \underline{224}&\\
109&230 , \underline{66}&321&&&&&22&\underline{326}&110 , \underline{21}&&72 , \underline{60}&\\
110&328 , \underline{250}&322&325&\underline{179}&&329&23&\underline{327}&21 , \underline{109}&&219 , \underline{86}&\\
111&\underline{22}&&&&&&&&114 , \underline{26}&&26 , \underline{114}&22\\
112&\underline{256}&&314&\underline{317}&&\underline{23}&\underline{329}&&&&281 , \underline{330}&\underline{331}\\
\hline
\end{tabular}}
\end{center}                           
%=================================================================
%=================================================================
\begin{center}
\tiny
\resizebox{1 \textwidth}{!}{
\begin{tabular}{ |c|c|c|c|c|c|c|c|c|c|c|c|c|} 
\hline
N&1&2&3&4&5&6&7&8&9&10&11&12\\
\hline
181&47&&427&177&\underline{46}&376 , \underline{134}&&&&443&\underline{444}&\\
182&51&46&&&445&374 , \underline{133}&&&&46&\underline{446}&\\
183&53 , \underline{200}&339&&\underline{274}&444&282 , \underline{120}&&\underline{447}&&\underline{446}&46&\\
184&\underline{376}&&428&202 , \underline{61}&\underline{187}&\underline{47}&&&&448&\underline{449}&450\\
185&\underline{443}&451&&203&\underline{197}&448&&&&\underline{47}&\underline{452}&\\
186&453 , \underline{444}&&&\underline{219}&\underline{53}&449&\underline{279}&\underline{454}&&452 , \underline{455}&47&\\
187&\underline{289}&&430&204 , \underline{64}&\underline{48}&184&&&&\underline{456}&\underline{457}&458\\
\hline
\end{tabular}}
\end{center}                           
%=================================================================
%=================================================================
\begin{center}
\tiny
\resizebox{1 \textwidth}{!}{
\begin{tabular}{ |c|c|c|c|c|c|c|c|c|c|c|c|c|} 
\hline
N&1&2&3&4&5&6&7&8&9&10&11&12\\
\hline
257&\underline{79}&401&425&\underline{492}&&554&\underline{549}&78&&&276&\underline{556}\\
258&\underline{81}&402&&\underline{493}&\underline{474}&555&\underline{273}&495&113&&78 , \underline{87}&\underline{557}\\
259&\underline{82}&&207&&&373&&\underline{558}&114&&117 , \underline{332}&78\\
260&\underline{559}&403&441&212 , \underline{79}&&522&\underline{550}&&\underline{252}&\underline{560}&275&\underline{561}\\
261&&252&&560&&562&&80&\underline{253}&79&159&\underline{563}\\
262&556&404&442&561 , \underline{494}&&493&&\underline{265}&\underline{254}&\underline{563}&273&79\\
263&&564&&&&\underline{80}&&\underline{562}&83 , \underline{139}&233&143 , \underline{392}&\underline{497}\\
\hline
\end{tabular}}
\end{center}                           
%=================================================================
%=================================================================
\begin{center}
\tiny
\resizebox{1 \textwidth}{!}{
\begin{tabular}{ |c|c|c|c|c|c|c|c|c|c|c|c|c|} 
\hline
N&1&2&3&4&5&6&7&8&9&10&11&12\\
\hline
334&\underline{417}&429 , \underline{175}&&172&206&116&\underline{213}&&&&&\\
335&\underline{573}&&628&233 , \underline{117}&236 , \underline{246}&&\underline{392}&&558&&&116\\
336&\underline{277}&&\underline{117}&628 , \underline{524}&525 , \underline{247}&&\underline{394}&&\underline{626}&&\underline{337}&\\
337&\underline{279}&&\underline{118}&&\underline{537}&&\underline{629}&&\underline{627}&&316&336\\
338&\underline{281}&&629&&\underline{136}&&118&&331 , \underline{625}&&&143\\
339&119&412&\underline{127}&183&&369 , \underline{131}&&183&&412&&\\
340&\underline{369}&414&&343 , \underline{121}&&\underline{119}&&343 , \underline{121}&&414&630&631\\
\hline
\end{tabular}}
\end{center}                           
%=================================================================
%=================================================================
\begin{center}
\tiny
\resizebox{1 \textwidth}{!}{
\begin{tabular}{ |c|c|c|c|c|c|c|c|c|c|c|c|c|} 
\hline
N&1&2&3&4&5&6&7&8&9&10&11&12\\
\hline
413&675 , \underline{676}&\underline{275}&&&&&\underline{670}&159&560&\underline{275}&159&\\
414&\underline{411}&\underline{429}&\underline{340}&417 , \underline{162}&&\underline{160}&&636&&677&605&678\\
415&\underline{676}&&\underline{160}&420&&677&&420&&\underline{160}&679&\\
416&&\underline{169}&\underline{342}&\underline{165}&&605&&&&679&\underline{160}&\\
417&\underline{273}&\underline{334}&\underline{343}&162 , \underline{414}&&\underline{161}&&680&\underline{487}&681&&495\\
418&540 , \underline{569}&&\underline{345}&163 , \underline{347}&&680&&\underline{161}&682 , \underline{538}&683&&\\
419&322 , \underline{242}&\underline{173}&\underline{346}&164&&487&&612 , \underline{319}&161&684&&\\
\hline
\end{tabular}}
\end{center}                           
%=================================================================
%=================================================================
\begin{center}
\tiny
\resizebox{1 \textwidth}{!}{
\begin{tabular}{ |c|c|c|c|c|c|c|c|c|c|c|c|c|} 
\hline
N&1&2&3&4&5&6&7&8&9&10&11&12\\
\hline
492&440 , \underline{212}&283&318&257&&689 , \underline{482}&\underline{303}&211&&&&\underline{694}\\
493&432 , \underline{213}&657&&258&\underline{516}&688 , \underline{434}&\underline{262}&&232&&211&\underline{686}\\
494&694 , \underline{438}&658&483&262 , \underline{561}&&434&&&\underline{491}&&&212\\
495&686 , \underline{422}&&637&258&&417&\underline{730}&&214&&&213\\
496&524 , \underline{646}&&\underline{214}&277 , \underline{576}&528 , \underline{616}&635&\underline{731}&&\underline{637}&&&\\
497&263&&731&278 , \underline{577}&268 , \underline{293}&729&214&&\underline{730}&&&\underline{264}\\
498&\underline{217}&&693&\underline{690}&\underline{509}&&&&&&\underline{732}&\underline{215}\\
\hline
\end{tabular}}
\end{center}                           
%=================================================================
%=================================================================
\begin{center}
\tiny
\resizebox{1 \textwidth}{!}{
\begin{tabular}{ |c|c|c|c|c|c|c|c|c|c|c|c|c|} 
\hline
N&1&2&3&4&5&6&7&8&9&10&11&12\\
\hline
571&764 , \underline{765}&275&&&766 , \underline{714}&713&&269&403&275&269&\\
572&574 , \underline{275}&675&&&533 , \underline{548}&\underline{681}&\underline{555}&273&561&\underline{764}&481&\\
573&575 , \underline{276}&&767&335&512 , \underline{542}&&\underline{665}&&556&\underline{717}&&273\\
574&275 , \underline{572}&676&&&535 , \underline{623}&601 , \underline{420}&&274&&\underline{765}&447&\\
575&276 , \underline{573}&&768&433&327 , \underline{617}&549&&&&\underline{545}&&274\\
576&767 , \underline{768}&&\underline{276}&496 , \underline{277}&733 , \underline{618}&703&\underline{769}&&&&&\\
577&665&&769&497 , \underline{278}&449 , \underline{294}&703&276&&554&&&\underline{300}\\
\hline
\end{tabular}}
\end{center}                           
%=================================================================
%=================================================================
\begin{center}
\tiny
\resizebox{1 \textwidth}{!}{
\begin{tabular}{ |c|c|c|c|c|c|c|c|c|c|c|c|c|} 
\hline
N&1&2&3&4&5&6&7&8&9&10&11&12\\
\hline
647&\underline{722}&686&&637 , \underline{352}&&&&805&\underline{747}&806&&\underline{631}\\
648&396&&\underline{379}&638 , \underline{671}&\underline{354}&&&649 , \underline{353}&&&&\\
649&656 , \underline{698}&&\underline{380}&353 , \underline{648}&\underline{355}&&&353 , \underline{648}&&&&\\
650&379&608&\underline{396}&586 , \underline{659}&&793&&651 , \underline{606}&&&&359\\
651&654&&\underline{399}&606 , \underline{650}&&792&&606 , \underline{650}&&&726&361\\
652&794&\underline{627}&&\underline{625}&\underline{368}&&&\underline{625}&&\underline{627}&365&\\
653&\underline{368}&\underline{722}&&\underline{669}&794&\underline{631}&369&\underline{669}&759&\underline{722}&&\\
\hline
\end{tabular}}
\end{center}                           
%=================================================================
%=================================================================
\begin{center}
\tiny
\resizebox{1 \textwidth}{!}{
\begin{tabular}{ |c|c|c|c|c|c|c|c|c|c|c|c|c|} 
\hline
N&1&2&3&4&5&6&7&8&9&10&11&12\\
\hline
723&\underline{475}&811&&\underline{555}&578&\underline{802}&481&474&742&&&\\
724&788&701&&\underline{732}&725&&477&\underline{732}&&701&&\\
725&\underline{642}&607&\underline{479}&\underline{708}&643&&&\underline{708}&\underline{786}&607&724&\\
726&\underline{651}&716&786&\underline{705}&&&&\underline{705}&479 , \underline{662}&716&663 , \underline{480}&800\\
727&821 , \underline{760}&&&\underline{484}&&&&\underline{484}&&&&\\
728&815&&777&532 , \underline{487}&487 , \underline{532}&\underline{612}&\underline{777}&&\underline{815}&&612&\\
729&&&801&533 , \underline{703}&705 , \underline{777}&\underline{497}&487&&\underline{681}&&684&\\
\hline
\end{tabular}}
\end{center}                           
%=================================================================
%=================================================================
\begin{table}[h]
\begin{center}
\tiny
\resizebox{1 \textwidth}{!}{
\begin{tabular}{ |c|c|c|c|c|c|c|c|c|c|c|c|c|} 
\hline
N&1&2&3&4&5&6&7&8&9&10&11&12\\
\hline
799&\underline{794}&746&&747&&&&747&631&746&820&\\
800&\underline{726}&815&&\underline{777}&&\underline{639}&&\underline{777}&661 , \underline{634}&815&&831\\
801&823 , \underline{704}&729&&&&\underline{640}&&830 , \underline{778}&636&\underline{635}&&731\\
802&\underline{723}&730&&\underline{806}&&&&&\underline{731}&\underline{637}&&\underline{636}\\
803&&791&&751 , \underline{809}&&&&804 , \underline{790}&&&&793 , \underline{641}\\
804&&&&790 , \underline{803}&&&&790 , \underline{803}&&&&792 , \underline{642}\\
805&\underline{827}&744&&&&&&\underline{647}&\underline{743}&744&&\underline{647}\\
\hline
\end{tabular}}
\end{center}
\caption{Triality connections between the 831 toric phases of Model 18.}
\label{Table 2 - Model 18}
\end{table}                           
%=================================================================

%=================================================================
\section{A Detailed Exploration of the Structure of the Triality Webs}
%=================================================================

\label{section detailed structure webs}

This section serves a dual purpose. First, it provides a concise overview of the general features that emerge from the detailed results in Section \sref{section triality webs}. Second, it introduces novel perspectives on triality webs, offering deeper insights into their structures and new tools that may prove valuable in future studies. It would be interesting to investigate whether some of the properties we observe can be inferred from the underlying geometry.

%=================================================================
\subsection{Toric Phases}
%=================================================================

Table \ref{number toric phases} summarizes the number of distinct toric phases for each of the models. Table \ref{min_phase} lists the minimal phase(s) for each of the models. The numerical label for phases is the one introduced in Section \sref{section triality webs}. We define a minimal phase as one with the lowest number of fields $N_{fields}$ defined as in \eqref{N_fields}. We observe that Models 1, 2, 3, 4, 5, 7, 9, 11 and 14 exhibit a single minimal phase. Models 8, 10, 12, 13, 15, 16, 17 and 18 display two minimal phases, while Model 6 uniquely possesses three minimal phases. Additionally, Models 1, 4 and 5 have a single phase. 

%=================================================================
\begin{table}[ht]
\begin{center}
\begin{tabular}{ |c|c||c|c||c|c|} 
\hline
Model & \# of Phases &Model & \# of Phases & Model & \# of Phases \\ \hline 
1 & 1 & 7 & 6 & 13 & 90 \\
2 & 2 & 8 & 4 & 14 & 120 \\
3 & 2 & 9 & 8 & 15 & 75 \\
4 & 1 & 10 & 8 & 16 & 143 \\
5 & 1 & 11 & 17 & 17 & 537 \\
6 & 6 & 12 & 14 & 18 & 831 \\ 
\hline
\end{tabular}
\end{center}
\caption{Number of toric phases for each of the models.}
\label{number toric phases}
\end{table}
%=================================================================

%=================================================================
\begin{table}[h]
\begin{center}
\begin{tabular}{ |c|c|c|c|c|c|} 
\hline
Model & Minimal Phase(s) & $N_G$ & $N_F$ & $N_\chi$ & $N_{fields}$ \\ \hline 
 1 & 1 & 4 & 12 & 16 & 32 \\
 2 & 2 & 6 & 12 & 18 & 36 \\
 3 & 1 & 6 & 12 & 18 & 36 \\
 4 & 1 & 6 & 14 & 20 & 40 \\
 5 & 1 & 6 & 15 & 21 & 42 \\
 6 & 1, 2, 4 & 8 & 16 & 24 & 48 \\
 7 & 2 & 8 & 15 & 23 & 46 \\
 8 & 1, 2 & 8 & 17 & 25 & 50 \\
 9 & 2 & 8 & 12 & 20 & 40 \\
 10 & 1, 2 & 8 & 14 & 22 & 44 \\
 11 & 8 & 8 & 12 & 20 & 40 \\
 12 & 3, 11 & 8 & 12 & 20 & 40 \\
 13 & 1, 8 & 10 & 16 & 26 & 52 \\
 14 & 3 & 10 & 14 & 24 & 48 \\
 15 & 2, 14 & 10 & 14 & 24 & 48 \\
 16 & 48, 58 & 10 & 14 & 24 & 48 \\
 17 & 18, 172 & 12 & 16 & 28 & 56 \\
 18 & 15, 86 & 12 & 17 & 29 & 58 \\
\hline
\end{tabular}
\end{center}
\caption{Minimal phases for each of the models and their field content.}
\label{min_phase}
\end{table}
%=================================================================

Given how we implemented our search of toric phases and our convention for labeling them,  Phase 1 of each model indicates the toric phase constructed in \cite{Franco:2022gvl}. From Table \ref{min_phase}, we note that only in 8 out of the 18 models, phase 1 is one of the minimal ones. This should not be surprising, since different approaches were used to determine the theories in \cite{Franco:2022gvl} and, quite often, the easiest to find phase is not necessarily the minimal one.

%=================================================================
\subsection{Field Content}
%=================================================================

In this section, we summarize the total numbers of fields for every toric phase of each of the models. We present this information using histograms to visualize key aspects such as the smallest and largest theories, the distribution of field content, and other relevant characteristics. It is important to note that the horizontal and vertical scales vary across the different plots. We do not include histograms for Models 1, 4 and 5, since each of them has a single toric phase.

%=================================================================
\begin{figure}[h]
    \centering
   \begin{minipage}{0.32\textwidth}
        \centering
        \includegraphics[width=\textwidth]{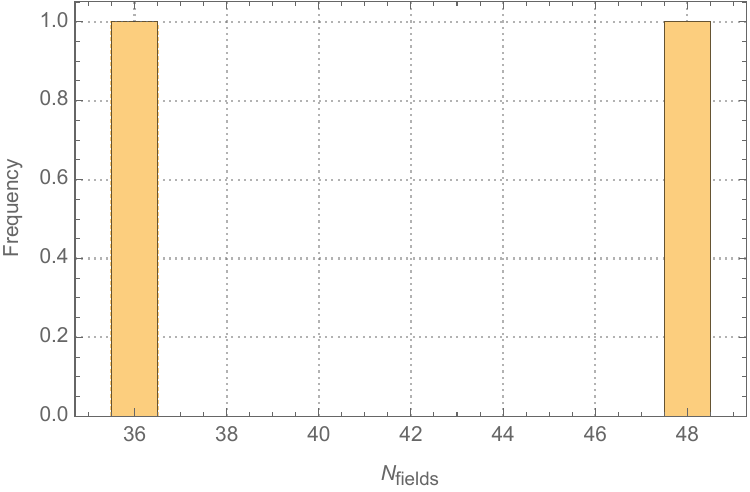}
        \vspace{0pt} % Add space between figure and label
        \makebox[0.95\textwidth]{\footnotesize{Model 2}}
    \end{minipage}
    \hfill
    \begin{minipage}{0.32\textwidth}
        \centering
        \includegraphics[width=\textwidth]{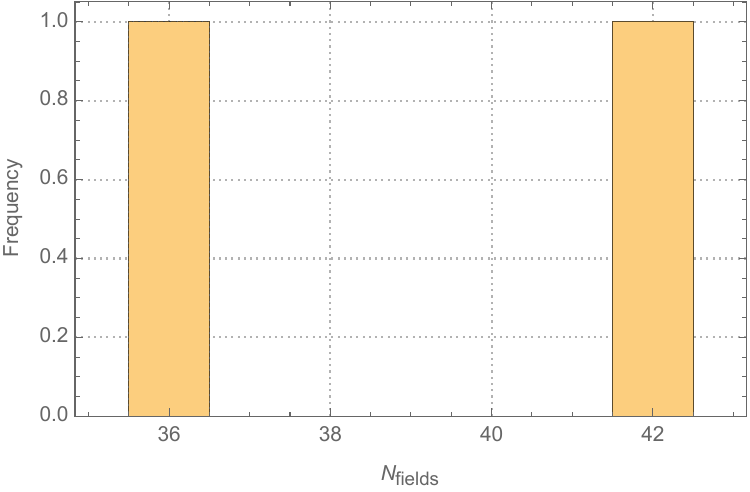}
        \vspace{0pt}\makebox[0.95\textwidth]{\footnotesize{Model 3}}
    \end{minipage}
    \hfill
    \begin{minipage}{0.32\textwidth}
        \centering
        \includegraphics[width=\textwidth]{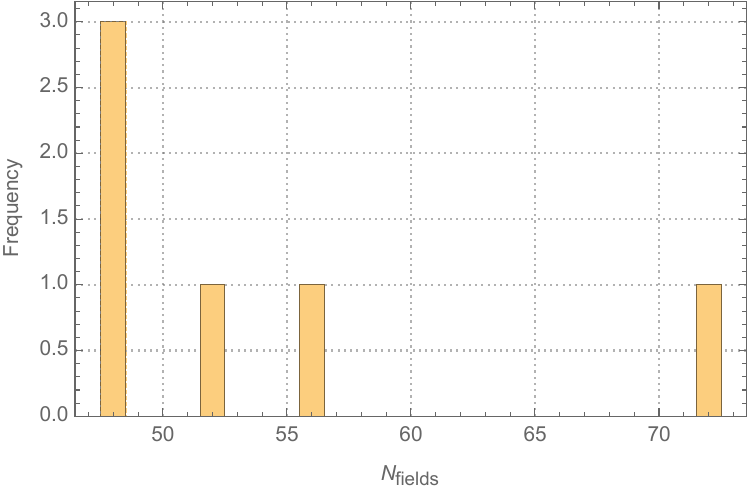}
        \vspace{0pt}
        \makebox[0.95\textwidth]{\footnotesize{Model 6}}
    \end{minipage}
    \caption{Distributions of the number of fields for the toric phases of Models 2, 3 and 6.}
    \label{Histograms 2, 3, 6}
\end{figure}
%=================================================================

%=================================================================
\begin{figure}[h]
    \centering
   \begin{minipage}{0.32\textwidth}
        \centering
        \includegraphics[width=\textwidth]{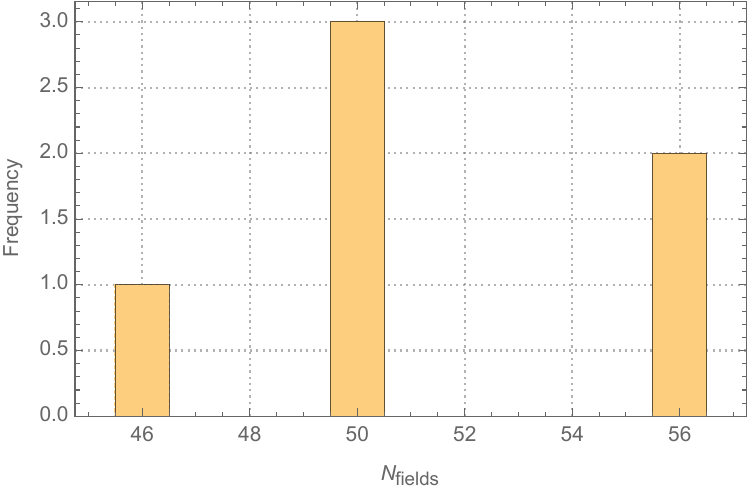}
        \vspace{0pt} % Add space between figure and label
        \makebox[0.95\textwidth]{\footnotesize{Model 7}}
    \end{minipage}
    \hfill
    \begin{minipage}{0.32\textwidth}
        \centering
        \includegraphics[width=\textwidth]{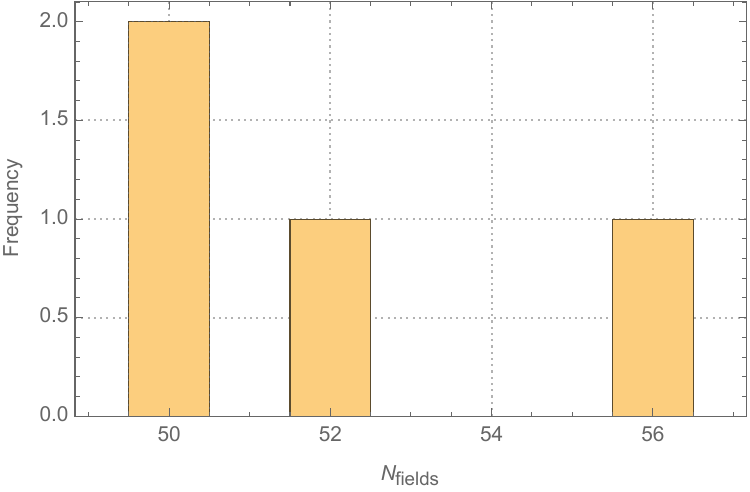}
        \vspace{0pt}\makebox[0.95\textwidth]{\footnotesize{Model 8}}
    \end{minipage}
    \hfill
    \begin{minipage}{0.32\textwidth}
        \centering
        \includegraphics[width=\textwidth]{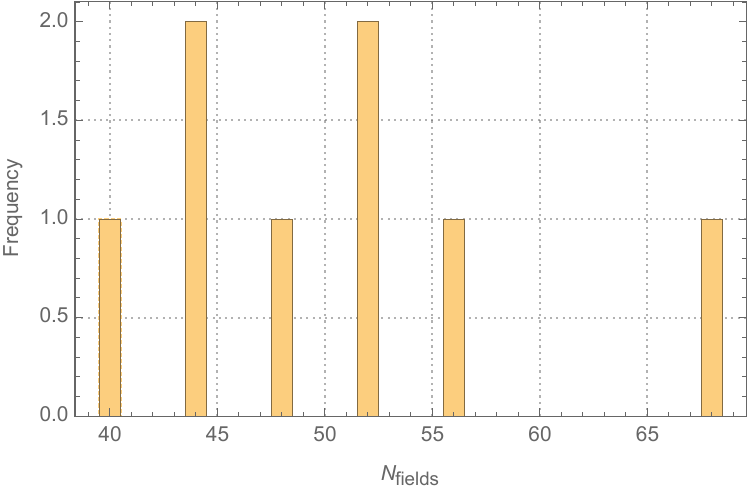}
        \vspace{0pt}
        \makebox[0.95\textwidth]{\footnotesize{Model 9}}
    \end{minipage}
    \caption{Distributions of the number of fields for the toric phases of Models 7, 8 and 9.}
    \label{Histograms 7 to 9}
\end{figure}
%=================================================================

%=================================================================
\begin{figure}[h]
    \centering
   \begin{minipage}{0.32\textwidth}
        \centering
        \includegraphics[width=\textwidth]{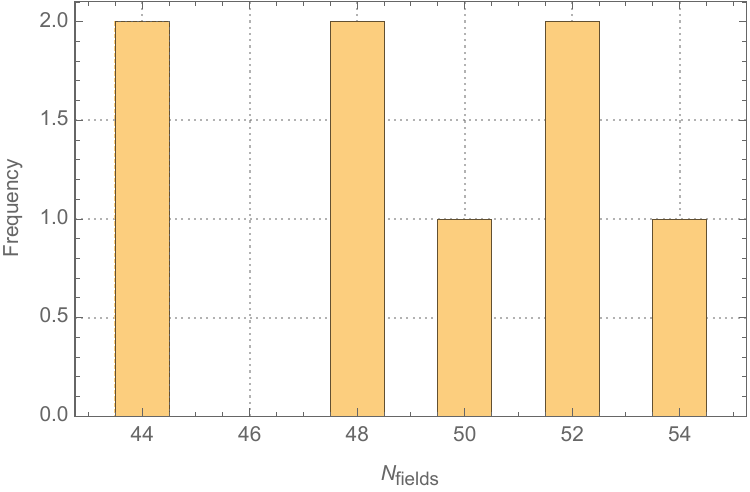}
        \vspace{0pt} % Add space between figure and label
        \makebox[0.95\textwidth]{\footnotesize{Model 10}}
    \end{minipage}
    \hfill
    \begin{minipage}{0.32\textwidth}
        \centering
        \includegraphics[width=\textwidth]{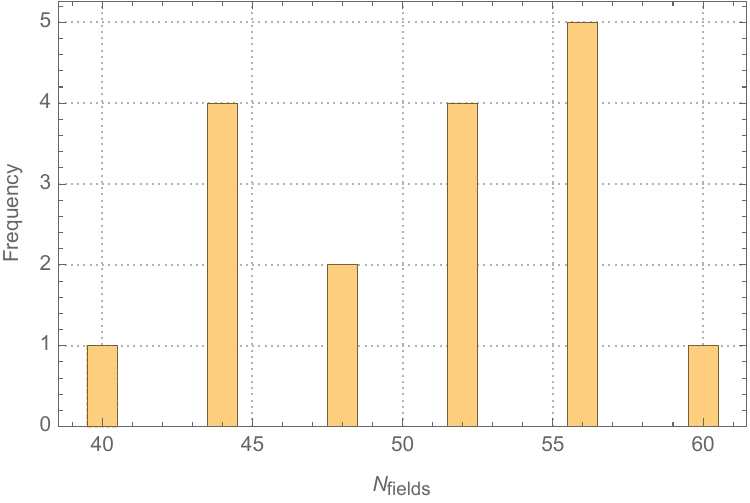}
        \vspace{0pt}\makebox[0.95\textwidth]{\footnotesize{Model 11}}
    \end{minipage}
    \hfill
    \begin{minipage}{0.32\textwidth}
        \centering
        \includegraphics[width=\textwidth]{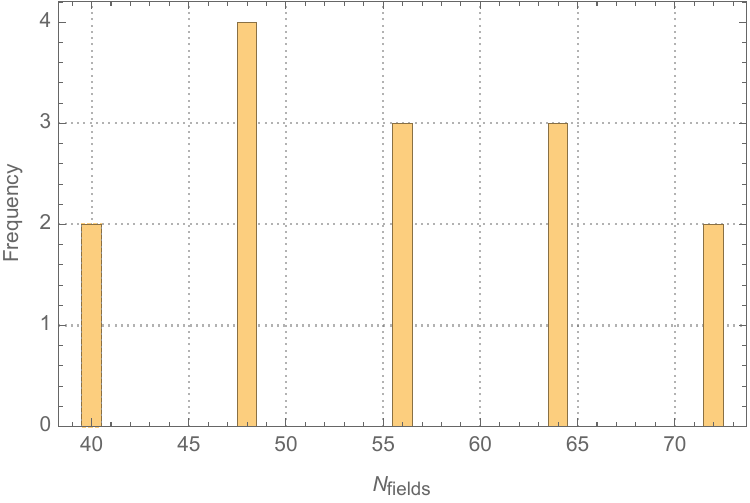}
        \vspace{0pt}
        \makebox[0.95\textwidth]{\footnotesize{Model 12}}
    \end{minipage}
    \caption{Distributions of the number of fields for the toric phases of Models 10, 11 and 12.}
    \label{Histograms 10 to 12}
\end{figure}
%=================================================================

\begin{figure}[h]
    \centering
    \begin{minipage}{0.32\textwidth}
        \centering
        \includegraphics[width=\textwidth]{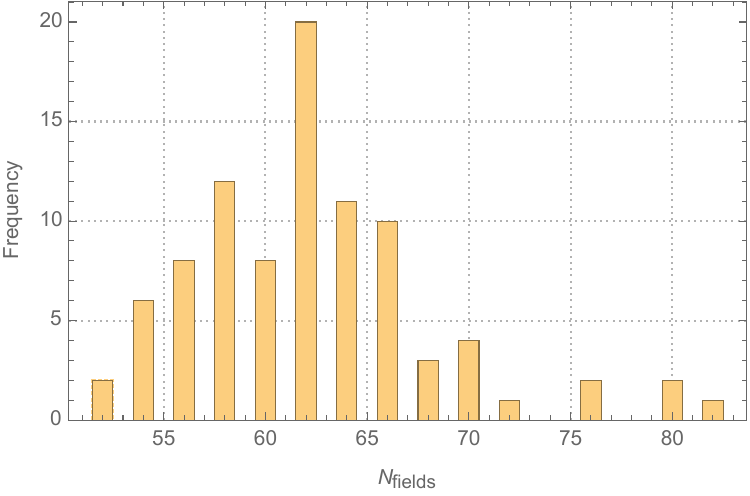}
        \vspace{0pt} % Add space between figure and label
        \makebox[0.95\textwidth]{\footnotesize{Model 13}}
    \end{minipage}
    \hfill
    \begin{minipage}{0.32\textwidth}
        \centering
        \includegraphics[width=\textwidth]{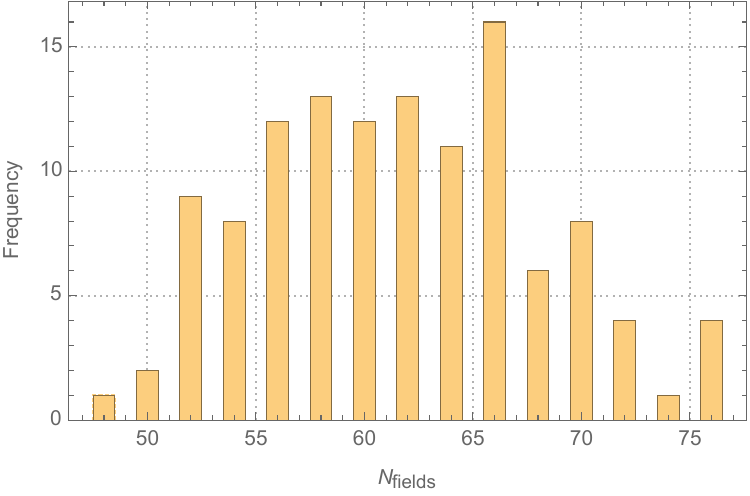}
        \vspace{0pt}\makebox[0.95\textwidth]{\footnotesize{Model 14}}
    \end{minipage}
    \hfill
    \begin{minipage}{0.32\textwidth}
        \centering
        \includegraphics[width=\textwidth]{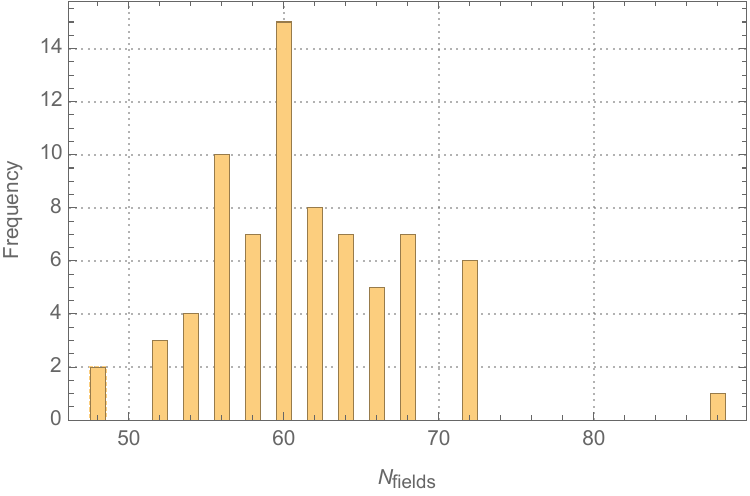}
        \vspace{0pt}
        \makebox[0.95\textwidth]{\footnotesize{Model 15}}
    \end{minipage}
    \caption{Distributions of the number of fields for the toric phases of Models 13, 14 and 15.}
    \label{Histograms 13 to 15}
\end{figure}

\begin{figure}[h]
    \centering
    \begin{minipage}{0.32\textwidth}
        \centering
        \includegraphics[width=\textwidth]{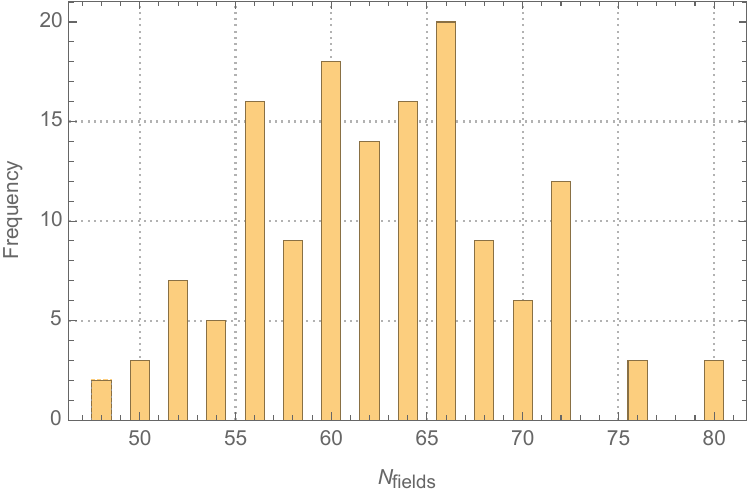}
        \vspace{0pt} % Add space between figure and label
        \makebox[0.95\textwidth]{\footnotesize{Model 16}}
    \end{minipage}
    \hfill
    \begin{minipage}{0.32\textwidth}
        \centering
        \includegraphics[width=\textwidth]{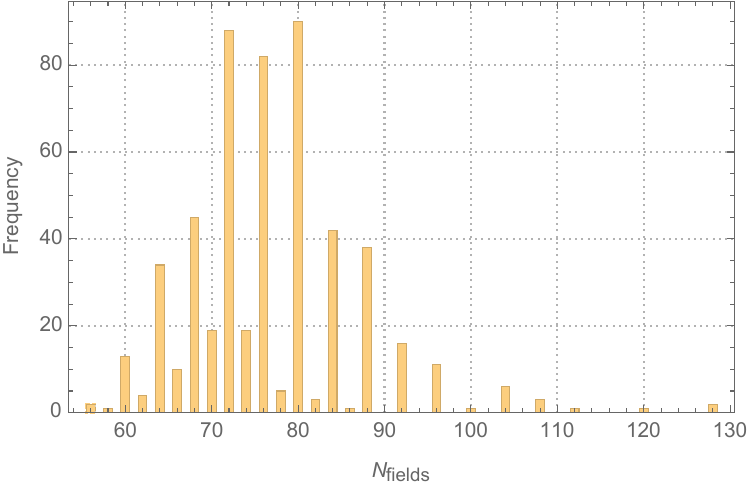}
        \vspace{0pt}\makebox[0.95\textwidth]{\footnotesize{Model 17}}
    \end{minipage}
    \hfill
    \begin{minipage}{0.32\textwidth}
        \centering
        \includegraphics[width=\textwidth]{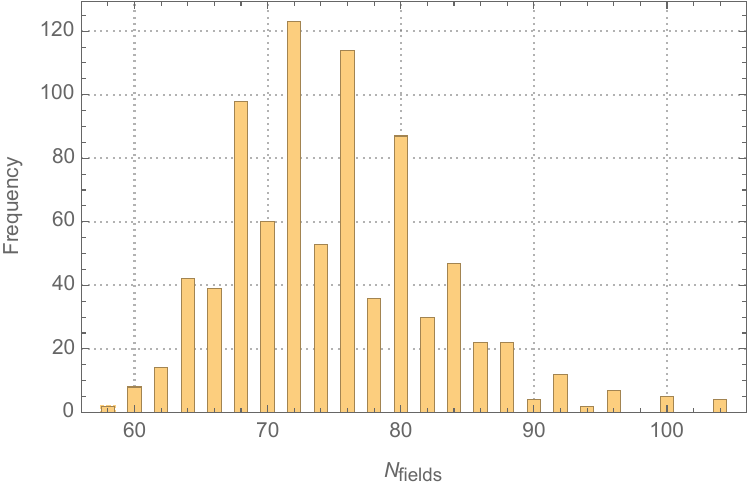}
        \vspace{0pt}
        \makebox[0.95\textwidth]{\footnotesize{Model 18}}
    \end{minipage}
    \caption{Distributions of the number of fields for the toric phases of Models 16, 17 and 18.}
    \label{Histograms 16 to 18}
\end{figure}

The distributions become more interesting for Models 13 to 18, as they exhibit significantly larger numbers of phases. These distributions are shown in Figures \ref{Histograms 13 to 15} and \ref{Histograms 16 to 18}. It is interesting to observe that Model 15 exhibits an outlier in Phase 75, which with 88 fields is significantly larger than the next phase in size.

%=================================================================
%=================================================================

%=================================================================
%=================================================================

%=================================================================
\subsection{Phase Multiplicity in Toric Islands}
%=================================================================

As mentioned earlier, we can explicitly generate the complete toric islands for all 18 models. This not only identifies the distinct toric phases present but also determines all instances in which a given toric phase appears within an island. The same toric phase may correspond to different nodes in the web, depending on node permutations or chiral conjugation. 

Model 1 has a single toric phase which, in addition, does not have toric nodes. Consequently, the toric island for this model consists of only one theory. In the tables below, we present the frequency of each toric phase appearing in a toric island for Models 2 to 12. Interestingly, these phases do not all appear with the same frequency. This fact has already been observed in Seiberg duality webs for toric CY 3-folds—for instance, in the toric island of $F_0$ (see, e.g., \cite{Franco:2003ja}). The island size indicates the total number of nodes within an island. 

Although Models 4 and 5 each have a single toric phase, these phases contain toric nodes, leading to non-trivial toric islands, each comprising six theories.

%=================================================================
\begin{table}[H]
\begin{center}
\renewcommand{\arraystretch}{1} % Optional: Adjust row height for better appearance
\setlength{\tabcolsep}{0pt} % Optional: Adjust column spacing
\begin{tabular}{ |>{\centering\arraybackslash}p{2.5cm} |>{\centering\arraybackslash}p{.7cm} |>{\centering\arraybackslash}p{.7cm} |} 
\hline
\multicolumn{3}{|c|}{\textbf{Island size: 12}} \\ % Spanning all columns
\hline
Phase & 1 & 2  \\
\hline
Multiplicity & 6 & 6 \\ 
\hline
\end{tabular}
\end{center}
\caption{Island size and toric phase multiplicity for Model 2.}
\label{Phase multiplicity - Model 2}
\end{table}
%=================================================================

%=================================================================
\begin{table}[H]
\begin{center}
\renewcommand{\arraystretch}{1} % Optional: Adjust row height for better appearance
\setlength{\tabcolsep}{0pt} % Optional: Adjust column spacing
\begin{tabular}{ |>{\centering\arraybackslash}p{2.5cm} |>{\centering\arraybackslash}p{.7cm} |>{\centering\arraybackslash}p{.7cm} |} 
\hline
\multicolumn{3}{|c|}{\textbf{Island size: 9}} \\ % Spanning all columns
\hline
Phase & 1 & 2  \\
\hline
Multiplicity & 3 & 6 \\ 
\hline
\end{tabular}
\end{center}
\caption{Island size and toric phase multiplicity for Model 3.}
\label{Phase multiplicity - Model 3}
\end{table}
%=================================================================

%=================================================================
\begin{table}[H]
\begin{center}
\renewcommand{\arraystretch}{1} % Optional: Adjust row height for better appearance
\setlength{\tabcolsep}{0pt} % Optional: Adjust column spacing
\begin{tabular}{ |>{\centering\arraybackslash}p{2.5cm} |>{\centering\arraybackslash}p{.7cm} |>{\centering\arraybackslash}p{.7cm} |} 
\hline
\multicolumn{2}{|c|}{\textbf{Island size: 6}} \\ % Spanning all columns
\hline
Phase & 1  \\
\hline
Multiplicity & 6  \\ 
\hline
\end{tabular}
\end{center}
\caption{Island size and toric phase multiplicity for Model 4.}
\label{Phase multiplicity - Model 4}
\end{table}
%=================================================================

%=================================================================
\begin{table}[H]
\begin{center}
\renewcommand{\arraystretch}{1} % Optional: Adjust row height for better appearance
\setlength{\tabcolsep}{0pt} % Optional: Adjust column spacing
\begin{tabular}{ |>{\centering\arraybackslash}p{2.5cm} |>{\centering\arraybackslash}p{.7cm} |>{\centering\arraybackslash}p{.7cm} |} 
\hline
\multicolumn{2}{|c|}{\textbf{Island size: 6}} \\ % Spanning all columns
\hline
Phase & 1  \\
\hline
Multiplicity & 6  \\ 
\hline
\end{tabular}
\end{center}
\caption{Island size and toric phase multiplicity for Model 5.}
\label{Phase multiplicity - Model 5}
\end{table}
%=================================================================

%=================================================================
\begin{table}[H]
\begin{center}
\renewcommand{\arraystretch}{1} % Optional: Adjust row height for better appearance
\setlength{\tabcolsep}{0pt} % Optional: Adjust column spacing
\begin{tabular}{ |>{\centering\arraybackslash}p{2.5cm} |>{\centering\arraybackslash}p{.7cm} |>{\centering\arraybackslash}p{.7cm} |>{\centering\arraybackslash}p{.7cm} |>{\centering\arraybackslash}p{.7cm} |>{\centering\arraybackslash}p{.7cm} |>{\centering\arraybackslash}p{.7cm} |} 
\hline
\multicolumn{7}{|c|}{\textbf{Island size: 80}} \\ % Spanning all columns
\hline
Phase & 1 & 2 & 3 & 4 & 5 & 6 \\
\hline
Multiplicity & 8 & 32 & 16 & 8 & 8 & 8 \\ 
\hline
\end{tabular}
\end{center}
\caption{Island size and toric phase multiplicity for Model 6.}
\label{Phase multiplicity - Model 6}
\end{table}
%=================================================================

%=================================================================
\begin{table}[H]
\begin{center}
\renewcommand{\arraystretch}{1} % Optional: Adjust row height for better appearance
\setlength{\tabcolsep}{0pt} % Optional: Adjust column spacing
\begin{tabular}{ |>{\centering\arraybackslash}p{2.5cm} |>{\centering\arraybackslash}p{.7cm} |>{\centering\arraybackslash}p{.7cm} |>{\centering\arraybackslash}p{.7cm} |>{\centering\arraybackslash}p{.7cm} |>{\centering\arraybackslash}p{.7cm} |>{\centering\arraybackslash}p{.7cm} |>{\centering\arraybackslash}p{.7cm} |>{\centering\arraybackslash}p{.7cm} |} 
\hline
\multicolumn{9}{|c|}{\textbf{Island size: 104}} \\ % Spanning all columns
\hline
Phase & 1 & 2 & 3 & 4 & 5 & 6 & 7 & 8 \\
\hline
Multiplicity & 32 & 8 & 16 & 8 & 8 & 32 & 16 & 16 \\ 
\hline
\end{tabular}
\end{center}
\caption{Island size and toric phase multiplicity for Model 9.}
\label{Phase multiplicity - Model 9}
\end{table}
%=================================================================

%=================================================================
\begin{table}[H]
\begin{center}
\renewcommand{\arraystretch}{1} % Optional: Adjust row height for better appearance
\setlength{\tabcolsep}{0pt} % Optional: Adjust column spacing
\begin{tabular}{ |>{\centering\arraybackslash}p{2.5cm} |>{\centering\arraybackslash}p{.7cm} |>{\centering\arraybackslash}p{.7cm} |>{\centering\arraybackslash}p{.7cm} |>{\centering\arraybackslash}p{.7cm} |>{\centering\arraybackslash}p{.7cm} |>{\centering\arraybackslash}p{.7cm} |>{\centering\arraybackslash}p{.7cm} |>{\centering\arraybackslash}p{.7cm} |} 
\hline
\multicolumn{9}{|c|}{\textbf{Island size: 104}} \\ % Spanning all columns
\hline
Phase & 1 & 2 & 3 & 4 & 5 & 6 & 7 & 8 \\
\hline
Multiplicity & 8 & 16 & 16 & 16 & 16 & 8 & 8 & 16 \\ 
\hline
\end{tabular}
\end{center}
\caption{Island size and toric phase multiplicity for Model 10.}
\label{Phase multiplicity - Model 10}
\end{table}
%=================================================================

%=================================================================
\begin{table}[H]
\begin{center}
\renewcommand{\arraystretch}{1} % Optional: Adjust row height for better appearance
\setlength{\tabcolsep}{0pt} % Optional: Adjust column spacing
\begin{tabular}{|>{\centering\arraybackslash}p{2.5cm} |>{\centering\arraybackslash}p{.7cm} |>{\centering\arraybackslash}p{.7cm} |>{\centering\arraybackslash}p{.7cm} |>{\centering\arraybackslash}p{.7cm} |>{\centering\arraybackslash}p{.7cm} |>{\centering\arraybackslash}p{.7cm} |>{\centering\arraybackslash}p{.7cm} |>{\centering\arraybackslash}p{.7cm} |>{\centering\arraybackslash}p{.7cm} |>{\centering\arraybackslash}p{.7cm} |>{\centering\arraybackslash}p{.7cm} |>{\centering\arraybackslash}p{.7cm} |>{\centering\arraybackslash}p{.7cm} |>{\centering\arraybackslash}p{.7cm} |>{\centering\arraybackslash}p{.7cm} |>{\centering\arraybackslash}p{.7cm} |>{\centering\arraybackslash}p{.7cm} |} 
\hline
\multicolumn{18}{|c|}{\textbf{Island size: 208}} \\ % Spanning all columns
\hline
Phase & 1 & 2 & 3 & 4 & 5 & 6 & 7 & 8 & 9 & 10 & 11 & 12 & 13 & 14 & 15 & 16 & 17 \\
\hline
Multiplicity & 16 & 16 & 16 & 8 & 16 & 8 & 16 & 8 & 8 & 16 & 8 & 8 & 16 & 8 & 16 & 16 & 8 \\ 
\hline
\end{tabular}
\end{center}
\caption{Island size and toric phase multiplicity for Model 11.}
\label{Phase multiplicity - Model 11}
\end{table}
%=================================================================

%=================================================================
\begin{table}[H]
\begin{center}
\renewcommand{\arraystretch}{1} % Optional: Adjust row height for better appearance
\setlength{\tabcolsep}{0pt} % Optional: Adjust column spacing
\begin{tabular}{ |>{\centering\arraybackslash}p{2.5cm} |>{\centering\arraybackslash}p{.7cm} |>{\centering\arraybackslash}p{.7cm} |>{\centering\arraybackslash}p{.7cm} |>{\centering\arraybackslash}p{.7cm} |>{\centering\arraybackslash}p{.7cm} |>{\centering\arraybackslash}p{.7cm} |>{\centering\arraybackslash}p{.7cm} |>{\centering\arraybackslash}p{.7cm} |>{\centering\arraybackslash}p{.7cm} |>{\centering\arraybackslash}p{.7cm} |>{\centering\arraybackslash}p{.7cm} |>{\centering\arraybackslash}p{.7cm} |>{\centering\arraybackslash}p{.7cm} |>{\centering\arraybackslash}p{.7cm} |} 
\hline
\multicolumn{15}{|c|}{\textbf{Island size: 290}} \\ % Spanning all columns
\hline
Phase & 1 & 2 & 3 & 4 & 5 & 6 & 7 & 8 & 9 & 10 & 11 & 12 & 13 & 14 \\
\hline
Multiplicity & 24 & 24 & 24 & 48 & 48 & 12 & 24 & 4 & 24 & 24 & 6 & 12 & 12 & 4 \\ 
\hline
\end{tabular}
\end{center}
\caption{Island size and toric phase multiplicity for Model 12.}
\label{Phase multiplicity - Model 12}
\end{table}
%=================================================================

While we were unable to fully construct the toric island for Models 7 and 8, our code suggests that each consists of over 100,000 theories. This is consistent with toric islands that comprise all possible permutations of every toric phase, a behavior that is not logically ruled out. It would be interesting to investigate what specific features of these theories give rise to this phenomenon.

\begin{figure}[H]
\begin{center}
\includegraphics[width=11cm]{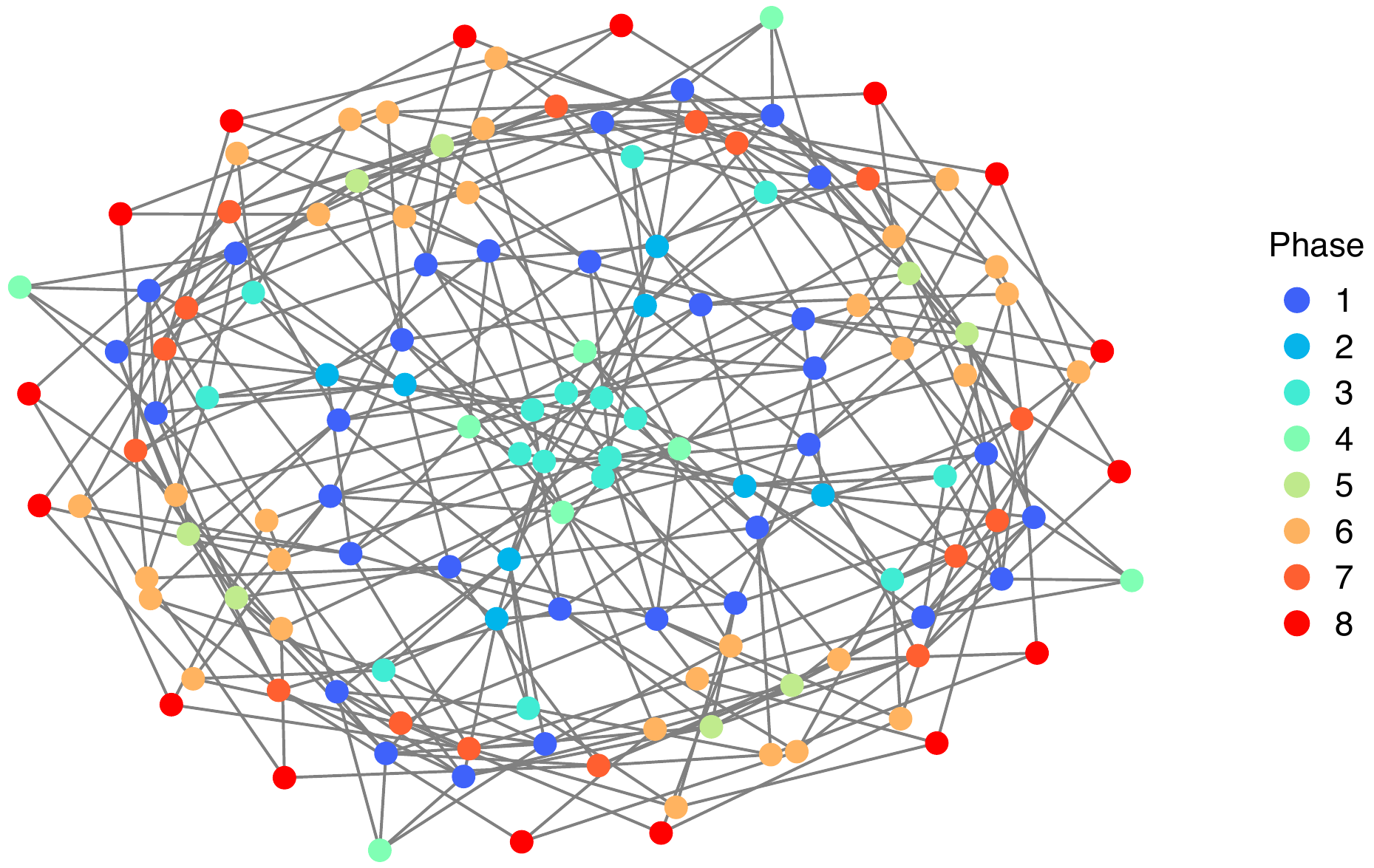}
\caption{Toric island for Model 9.}
\label{Web_Model_9}
 \end{center}
 \end{figure}

To illustrate the beautiful structure of some of these toric islands, we show the one for Model 9 in \fref{Web_Model_9}. To prevent clutter, we do not include arrows in the lines connecting theories.

%=================================================================
%=================================================================

%=================================================================
\subsection{Island radius}
%=================================================================

Here, we introduce an alternative quantity to characterize the size of an island, which we refer to as the {\it island radius}. This is defined as the distance, measured as the minimum number of trialities, between the first appearance of a toric phase and a minimal phase. Table \ref{island_radius} summarizes the radii for the different models.\footnote{In cases where multiple minimal phases exist, we take the largest of these distances. However, if a specific minimal phase is preferred, it is straightforward to recalculate the radius accordingly.} Note that the actual extent of an island may be larger, since each toric phase can appear multiple times.

%=================================================================
\begin{table}[ht]
\begin{center}
\begin{tabular}{ |c|c||c|c||c|c|} 
\hline
Model & \ $r$ \ & Model & \ $r$ \ & Model & \ $r$ \ \\ \hline 
1 & 0 & 7 & 2 & 13 & 4 \\
2 & 1 & 8 & 2 & 14 & 4 \\
3 & 1 & 9 & 3 & 15 & 5 \\
4 & 0 & 10 & 2 & 16 & 5 \\
5 & 0 & 11 & 3 & 17 & 9 \\
6 & 3 & 12 & 4 & 18 & 7 \\ 
\hline
\end{tabular}
\end{center}
\caption{Island radius for each of the models.}
\label{island_radius}
\end{table}
%=================================================================

Referring to Tables \ref{number toric phases} and \ref{island_radius}, we highlight how a remarkably large number of distinct toric phases can emerge within a relatively small distance—specifically, just a few triality steps—from a minimal phase. Consider, for instance, Model 18, which exhibits 831 distinct toric phases, all generated within at most 7 trialities from a minimal phase. This illustrates the inherently multidimensional nature of the triality web, where each node allows for multiple possible trialities.

%=================================================================
\section{Additional Consistency Checks Via the Forward Algorithm}
%=================================================================

\label{section checks forward algorithm}

Every toric phase we generated passes standard basic consistency checks, such as anomaly cancellation and the trace condition \cite{Franco:2015tna}. Furthermore, our algorithm produced the same toric phases through multiple sequences of trialities, confirming the self-consistency of our methods. 

In \cite{Franco:2022gvl}, the {\it forward algorithm} \cite{Franco:2015tna} was applied to Phase 1 of every model to confirm that their classical moduli spaces correspond to the appropriate toric CY 4-folds. As a further check of our construction, for every model we applied the forward algorithm to an additional toric phase, confirming that it corresponds to the correct geometry. The new phases to be tested were picked such that they maximize the number of trialities that separate them from Phase 1, a measure we refer to as {\it depth}.

Table \ref{td_test} lists the tested theories, many of which exhibit a significantly larger field content compared to their corresponding Phase 1. All cases successfully produced the correct toric diagrams. Due to space constraints, we do not include all the matrices involved in the Forward Algorithm. The quivers and $J$- and $E$-terms for all these theories are given in Appendix \sref{appendix additional toric phases}.

%=================================================================
\begin{table}[ht]
\begin{center}
\begin{tabular}{ |c|c|c|c|c|c|c|c|} 
\hline
 Model & Phase & Depth & Fermis & Model & Phase & Depth & Fermis  \\
\hline
2 & 2 & 1 & 12 & 12 & 11 & 2 & 12\\
3 & 2 & 1 & 15 & 13 & 54 & 3 & 18\\ 
6 & 4 & 2 & 16 & 14 & 76 & 3 & 18\\
7 & 5 & 2 & 17 & 15 & 56 & 3 & 22\\
8 & 3 & 1 & 20 & 16 & 140 & 4 & 28\\
9 & 7 & 1 & 15 & 17 & 490 & 6 & 24\\
10 & 7 & 2 & 16 & 18 & 502 & 4 & 26\\
11 & 17 & 3 & 16 &  &  &  & \\
\hline
\end{tabular}
\end{center}
\caption{Toric phases for every model whose moduli spaces were checked using the forward algorithm, in addition to the corresponding Phase 1. Models 1, 4 and 5 are excluded, as they have a single toric phase.
\label{td_test}}
\end{table}
%=================================================================

%=================================================================
\section{Conclusions}
%=================================================================

\label{section conclusions}

We determined all toric phases for the $2d$ $(0,2)$ theories on D1-branes probing the complex cones over the 18 smooth Fano 3-folds, whose toric diagrams correspond to the regular reflexive polytopes in 3 dimensions. Before our work, the entire set of toric phases for a CY 4-fold had been only determined for $Q^{1,1,1}/\mathbb{Z}_2$ \cite{Franco:2018qsc}, which is Model 12 in our classification.\footnote{This is, of course, beyond rather trivial examples of geometries with 1 or 2 toric phases.} These results significantly expand the list of explicitly known gauge theories on D1-branes over toric CY 4-folds.

We went beyond the classification of toric phases and mapped the corresponding toric islands of the triality webs, establishing how the toric phases are connected by triality. The size and complexity of the webs constructed in this paper far exceed anything previously known, both in the contexts of CY 3-folds and CY 4-folds. We proposed various new approaches for characterizing triality webs. Our work lays the foundation for a comprehensive exploration of the structure of triality webs. It would be interesting to investigate whether some of their features, e.g. number of toric phases, can be determined directly from the underlying geometry.

%======================================================================  
\acknowledgments
%======================================================================  

This work has been supported by the U.S. National Science Foundation grants PHY-2112729 and PHY-2412479.

\appendix

\newpage

%=================================================================
\section{Further Details on the Construction of the Webs}
%=================================================================

\label{appendix further details web construction}

Triality was performed at the level of the superpotential for each model as described in \cite{Franco:2017lpa}. A Python script was developed to manipulate the $J$- and $E$-terms of each model, systematically dualizing each toric node to identify all toric phases for a given model. The search algorithm proceeded as follows: starting from the initial phase, all toric nodes were identified and dualized at each stage. If the resulting phase was new—up to node relabeling and chiral conjugation—when compared to the stored phases, it was added to the dataset of toric phases and considered for further dualization. If dualization did not yield a new phase, the phase was still recorded but placed in a separate dataset to fully map the toric island.

Comparison between the phases was done at the level at the quiver. The quiver was read off of the superpotential at each step in order to determine whether the phase was new or not. This was the more computationally intensive part of the code. If brute force was used, it would require to check all possible node configurations in order to verify that each phase was unique or not up to a relabeling. In order to trim the relabeling requires to check, only relabels between nodes with similar field content were considered. This made sure that only similar nodes were compared and other iterations were not considered.

Explicitly, if a node had for instance 2 incoming chirals, 2 outgoing chirals, and 2 fermis for field content (and it was the only node with exactly that amount of each type of field) the code would only consider iterations that matches a node with the same field content on the other phase it was being compared to. This increased the runtime substantially.

%=================================================================
\section{Additional Toric Phases}
%=================================================================

\label{appendix additional toric phases}

In this appendix, we present the quivers and $J$- and $E$-terms for the additional toric phases that were tested using the Forward Algorithm in Section \sref{section checks forward algorithm}. The list of phases was summarized in Table \ref{td_test}. These effectively double the number of toric phases explicitly written down for $3d$ regular reflexive polytopes, substantially extending the findings of \cite{Franco:2022gvl}.\footnote{We do not present new phases for Models 1, 4 and 5, since each of them has a single toric phase.} It is interesting to note that four of these theories—Phase 2 of Model 2, Phase 3 of Model 8, Phase 7 of Model 9, and Phase 56 of Model 15—exhibit massless pairs of chiral and Fermi fields extending between the same pair of nodes. This phenomenon, previously observed in Phase 1 of Model 5, highlights the importance of having complete knowledge of a theory’s $J$- and $E$-terms.

%=================================================================
\subsection*{Model 2: Phase 2}
%=================================================================

\begin{figure}[H]
\begin{center}
\resizebox{0.25\hsize}{!}{
\includegraphics[height=6cm]{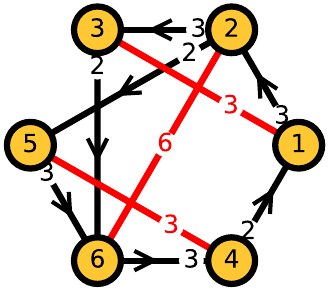} 
}
\caption{Quiver for Phase 2 of Model 2.
\label{model_2phase_2}}
 \end{center}
 \end{figure}

\fref{model_2phase_2} shows the quiver for this theory.

%=================================================================
%=================================================================

The $J$- and $E$-terms are
\beq
  {\footnotesize
\begin{array}{rrclrcl}
 &  & J &  & & E &  \\
\Lambda_{31}^{1}: & A_{12}X_{23} &-& C_{12}Y_{23} & P_{36}X_{64}P_{41} &-& Q_{36}X_{64}Q_{41}   \\ 
\Lambda_{31}^{2}: & A_{12}Z_{23} &-& B_{12}Y_{23} & Q_{36}Y_{64}Q_{41} &-& P_{36}Y_{64}P_{41} \\ 
\Lambda_{31}^{3}: & C_{12}Z_{23} &-& B_{12}X_{23} & P_{36}Z_{64}P_{41}&-& Q_{36}Z_{64}Q_{41}  \\ 
\Lambda_{26}^{1}: & X_{64}P_{41}A_{12} &-& Z_{64}P_{41}B_{12} & P_{25}X_{56} &-& X_{23}P_{36} \\ 
\Lambda_{26}^{2}: & Y_{64}P_{41}B_{12} &-& X_{64}P_{41}C_{12} & P_{25}Y_{56} &-& Y_{23}P_{36} \\ 
\Lambda_{26}^{3}: & Z_{64}P_{41}C_{12} &-& Y_{64}P_{41}A_{12} & P_{25}Z_{56} &-& Z_{23}P_{36} \\ 
\Lambda_{26}^{4}: & X_{64}Q_{41}A_{12} &-& Z_{64}Q_{41}B_{12} & X_{23}Q_{36} &-& Q_{25}X_{56} \\ 
\Lambda_{26}^{5}: & Y_{64}Q_{41}B_{12} &-& X_{64}Q_{41}C_{12} & Y_{23}Q_{36} &-&  Q_{25}Y_{56} \\ 
\Lambda_{26}^{6}: & Z_{64}Q_{41}C_{12} &-& Y_{64}Q_{41}A_{12} & Z_{23}Q_{36}&-& Q_{25}Z_{56} \\ 
\Lambda_{45}^{1}: & X_{56}X_{64} &-& Z_{56}Y_{64} & Q_{41}A_{12}Q_{25} &-& P_{41}A_{12}P_{25} \\ 
\Lambda_{45}^{2}: & Y_{56}Y_{64} &-& X_{56}Z_{64} & Q_{41}B_{12}Q_{25} &-& P_{41}B_{12}P_{25} \\ 
\Lambda_{45}^{3}: & Z_{56}Z_{64} &-& Y_{56}X_{64} & Q_{41}C_{12}Q_{25} &-& P_{41}C_{12}P_{25} \\ 
 
\end{array} 
 }~.~
\label{Model 2_Phase 2}
 \eeq

%=================================================================
\subsection*{Model 3: Phase 2}
%=================================================================

\begin{figure}[H]
\begin{center}
\resizebox{0.25\hsize}{!}{
\includegraphics[height=6cm]{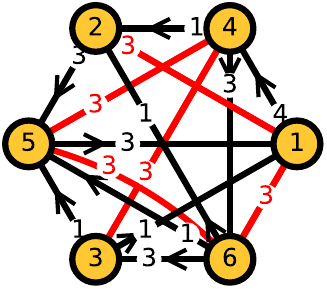} 
}
\caption{Quiver for Phase 2 of Model 3.
\label{model_3phase_2}}
 \end{center}
 \end{figure}

\fref{model_3phase_2} shows the quiver for this theory.

%=================================================================
%=================================================================

The $J$- and $E$-terms are
\beq
  {\footnotesize
\begin{array}{rrclrcl}
 &  & J &  & & E &  \\
\Lambda_{43}^{1}: & P_{35}Z_{51}Q_{14} &-& Q_{31}C_{14} & Z_{46}P_{63} &-& X_{46}X_{63} \\ 
\Lambda_{43}^{2}: & P_{35}X_{51}Q_{14} &-& Q_{31}B_{14} & X_{46}Z_{63} &-& Y_{46}P_{63} \\ 
\Lambda_{43}^{3}: & P_{35}Y_{51}Q_{14} &-& Q_{31}A_{14} & Y_{46}X_{63} &-& Z_{46}Z_{63} \\ 
\Lambda_{54}^{1}: & X_{46}Q_{65} &-& P_{42}G_{25} & Z_{51}B_{14} &-& X_{51}C_{14}  \\ 
\Lambda_{54}^{2}: & Y_{46}Q_{65} &-& P_{42}H_{25} & X_{51}A_{14} &-& Y_{51}B_{14} \\ 
\Lambda_{54}^{3}: & Z_{46}Q_{65} &-& P_{42}I_{25} & Y_{51}C_{14} &-& Z_{51}A_{14}  \\ 
\Lambda_{56}^{1}: & P_{63}P_{35} &-& X_{62}G_{25} & X_{51}Q_{14}Y_{46} &-& Z_{51}Q_{14}Z_{46} \\ 
\Lambda_{56}^{2}: & Z_{63}P_{35} &-& X_{62}H_{25} & Y_{51}Q_{14}Z_{46} &-& X_{51}Q_{14}X_{46} \\ 
\Lambda_{56}^{3}: & X_{62}I_{25} &-& X_{63}P_{35} & Y_{51}Q_{14}Y_{46} &-& Z_{51}Q_{14}X_{46}  \\ 
\Lambda_{12}^{1}: & I_{25}Y_{51} &-& G_{25}X_{51} & C_{14}P_{42} &-& Q_{14}Y_{46}X_{62} \\ 
\Lambda_{12}^{2}: & G_{25}Z_{51} &-& H_{25}Y_{51} & B_{14}P_{42} &-& Q_{14}Z_{46}X_{62} \\ 
\Lambda_{12}^{3}: & H_{25}X_{51} &-& I_{25}Z_{51} & A_{14}P_{42} &-& Q_{14}X_{46}X_{62} \\ 
\Lambda_{61}^{1}: & C_{14}X_{46} &-& A_{14}Y_{46} & Q_{65}X_{51} &-& X_{63}Q_{31} \\ 
\Lambda_{61}^{2}: & B_{14}Y_{46} &-& C_{14}Z_{46} & Q_{65}Y_{51} &-& P_{63}Q_{31} \\ 
\Lambda_{61}^{3}: & A_{14}Z_{46} &-& B_{14}X_{46} & Q_{65}Z_{51} &-& Z_{63}Q_{31} \\ 
 
\end{array}  }~.~
\label{E_J_Model 3_Phase 2}
 \eeq

%=================================================================
\subsection*{Model 6: Phase 4}
%=================================================================

\begin{figure}[H]
\begin{center}
\resizebox{0.33\hsize}{!}{
\includegraphics[height=6cm]{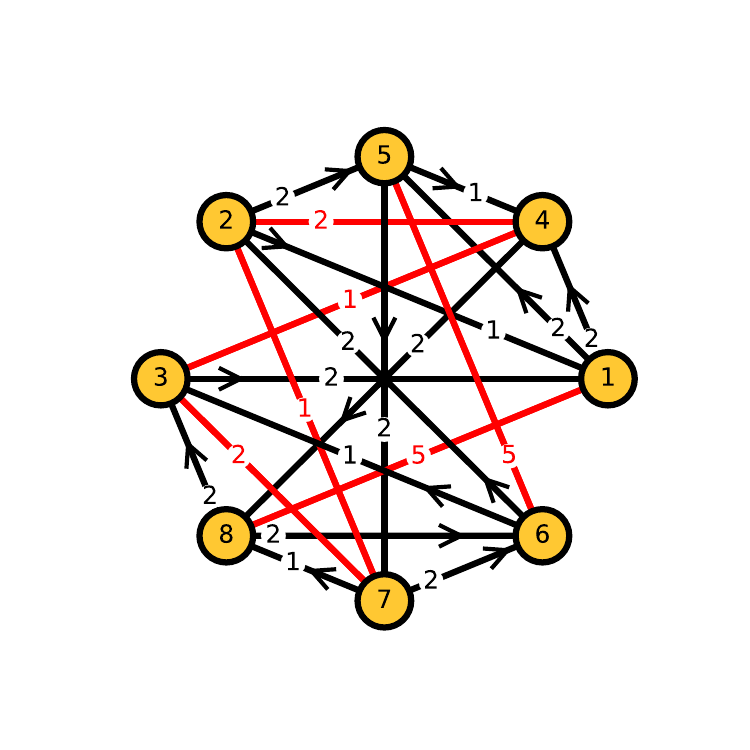} 
}
\caption{Quiver for Phase 4 of Model 6.
\label{model_6phase_4}}
 \end{center}
 \end{figure}

\fref{model_6phase_4} shows the quiver for this theory.

%=================================================================
%=================================================================

The $J$- and $E$-terms are
\beq
  {\footnotesize
\begin{array}{rrclrcl}
 &  & J &  & & E &  \\
\Lambda_{18}^{1}: & X_{86}I_{63}X_{31} &-& Y_{86}I_{63}Y_{31} & P_{15}X_{54}Q_{48} &-& Q_{15}X_{54}P_{48} \\ 
\Lambda_{18}^{2}: & Y_{86}P_{62}F_{21} &-& J_{83}X_{31} & P_{15}X_{57}X_{78} &-& B_{14}P_{48} \\ 
\Lambda_{18}^{3}: & X_{86}P_{62}F_{21} &-& J_{83}Y_{31} & A_{14}P_{48} &-& P_{15}Y_{57}X_{78} \\ 
\Lambda_{18}^{4}: & Y_{86}Q_{62}F_{21} &-& K_{83}X_{31} & B_{14}Q_{48} &-& Q_{15}X_{57}X_{78}  \\ 
\Lambda_{18}^{5}: & X_{86}Q_{62}F_{21} &-& K_{83}Y_{31} & Q_{15}Y_{57}X_{78} &-& A_{14}Q_{48} \\ 
\Lambda_{42}^{1}: & F_{21}B_{14} &-& H_{25}X_{54} & P_{48}Y_{86}P_{62} &-& Q_{48}Y_{86}Q_{62} \\ 
\Lambda_{42}^{2}: & F_{21}A_{14} &-& G_{25}X_{54} & Q_{48}X_{86}Q_{62} &-& P_{48}X_{86}P_{62} 
\end{array} 
 } \nonumber
\label{E_J_Model 6_Phase 4_1}
 \eeq

\beq
  {\footnotesize
\begin{array}{rrclrcl}
 &  & J &  & & E &  \\
\Lambda_{56}^{1}: & I_{63}X_{31}Q_{15} &-& P_{62}G_{25} & X_{54}P_{48}X_{86} &-& X_{57}A_{76} \\ 
\Lambda_{56}^{2}: & I_{63}X_{31}P_{15} &-& Q_{62}G_{25} & X_{57}B_{76} &-& X_{54}Q_{48}X_{86} \\ 
\Lambda_{56}^{3}: & I_{63}Y_{31}Q_{15} &-& P_{62}H_{25} & Y_{57}A_{76} &-& X_{54}P_{48}Y_{86} \\ 
\Lambda_{56}^{4}: & I_{63}Y_{31}P_{15} &-& Q_{62}H_{25} & X_{54}Q_{48}Y_{86} &-& Y_{57}B_{76}  \\ 
\Lambda_{72}^{1}: & H_{25}Y_{57} &-& G_{25}X_{57} & A_{76}P_{62} &-& B_{76}Q_{62} \\ 
\Lambda_{37}^{1}: & B_{76}I_{63} &-& X_{78}J_{83} & Y_{31}P_{15}Y_{57} &-& X_{31}P_{15}X_{57} \\ 
\Lambda_{37}^{2}: & A_{76}I_{63} &-& X_{78}K_{83} & X_{31}Q_{15}X_{57} &-& Y_{31}Q_{15}Y_{57} \\ 
\Lambda_{34}^{1}: & P_{48}J_{83} &-& Q_{48}K_{83} & Y_{31}A_{14} &-& X_{31}B_{14} \\ 
\Lambda_{65}^{1}: & X_{57}X_{78}Y_{86} &-& Y_{57}X_{78}X_{86} & Q_{62}F_{21}Q_{15} &-& P_{62}F_{21}P_{15} \\ 
 \end{array} 
 }~.~
\label{E_J_Model 6_Phase 4}
 \eeq

%=================================================================
\subsection*{Model 7: Phase 5}
%=================================================================

\begin{figure}[H]
\begin{center}
\resizebox{0.33\hsize}{!}{
\includegraphics[height=6cm]{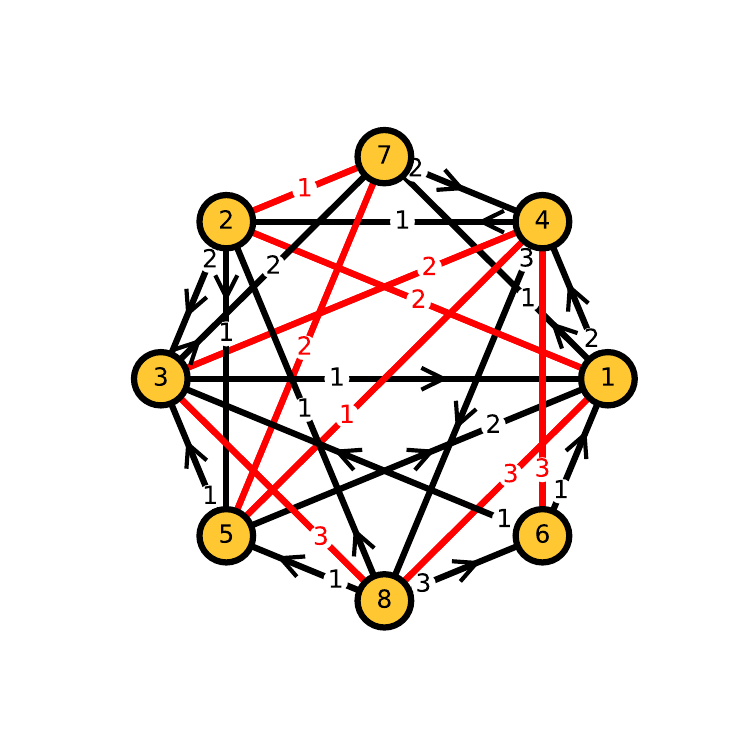} 
}
\caption{Quiver for Phase 5 of Model 7.
\label{model_7phase_5}}
 \end{center}
 \end{figure}

\fref{model_7phase_5} shows the quiver for this theory.

%=================================================================
%=================================================================

The $J$- and $E$-terms are
\beq
  {\footnotesize
\begin{array}{rrclrcl}
 &  & J &  & & E &  \\
\Lambda_{12}^{1}: & A_{25}X_{51} &-& Z_{23}D_{31} & Z_{17}Q_{74}Y_{48}X_{82} &-& B_{14}P_{42}  \\ 
\Lambda_{12}^{2}: & A_{25}Z_{51} &-& Y_{23}D_{31} & A_{14}P_{42} &-& Z_{17}Q_{74}Z_{48}X_{82} \\ 
\Lambda_{54}^{1}: & P_{42}A_{25} &-& X_{48}Q_{85} & X_{51}B_{14} &-& Z_{51}A_{14}  \\ 
\Lambda_{38}^{1}: & X_{82}Y_{23} &-& Z_{86}Y_{63} & J_{37}Q_{74}X_{48} &-& D_{31}Z_{17}Q_{74}Z_{48} \\ 
\Lambda_{38}^{2}: & X_{86}Y_{63} &-& X_{82}Z_{23} & K_{37}Q_{74}X_{48} &-& D_{31}Z_{17}Q_{74}Y_{48}  \\ 
\Lambda_{34}^{1}: & Y_{48}Q_{85}P_{53} &-& P_{42}Y_{23} & J_{37}A_{74} &-& D_{31}A_{14} \\ 
\Lambda_{34}^{2}: & Z_{48}Q_{85}P_{53} &-& P_{42}Z_{23} & D_{31}B_{14} &-& K_{37}A_{74} \\ 
\Lambda_{72}^{1}: & Y_{23}J_{37} &-& Z_{23}K_{37} & A_{74}P_{42} &-& Q_{74}X_{48}X_{82} \\ 
\Lambda_{64}^{1}: & X_{48}X_{86} &-& Z_{48}A_{86} & Q_{61}B_{14} &-& Y_{63}K_{37}Q_{74} \\ 
\Lambda_{64}^{2}: & X_{48}Z_{86} &-& Y_{48}A_{86} & Y_{63}J_{37}Q_{74} &-& Q_{61}A_{14} \\ 
\Lambda_{64}^{3}: & Z_{48}Z_{86} &-& Y_{48}X_{86} & Q_{61}Z_{17}A_{74} &-& Y_{63}D_{31}Z_{17}Q_{74} \\ 
\Lambda_{75}^{1}: & X_{51}Z_{17} &-& P_{53}J_{37} & A_{74}Y_{48}Q_{85} &-& Q_{74}Y_{48}X_{82}A_{25} \\ 
\Lambda_{75}^{2}: & Z_{51}Z_{17} &-& P_{53}K_{37} & Q_{74}Z_{48}X_{82}A_{25} &-& A_{74}Z_{48}Q_{85} \\ 
\Lambda_{83}^{1}: & J_{37}Q_{74}Y_{48} &-& K_{37}Q_{74}Z_{48} & A_{86}Y_{63} &-& X_{82}A_{25}P_{53} 
 \end{array} 
 } \nonumber
\label{E_J_Model 7_Phase 5_1}
 \eeq

\beq
  {\footnotesize
\begin{array}{rrclrcl}
 &  & J &  & & E &  \\
\Lambda_{81}^{1}: & Z_{17}A_{74}Y_{48} &-& B_{14}X_{48} & X_{86}Q_{61} &-& Q_{85}X_{51} \\ 
\Lambda_{81}^{2}: & A_{14}Y_{48} &-& B_{14}Z_{48} & Q_{85}P_{53}D_{31} &-& A_{86}Q_{61} \\ 
\Lambda_{81}^{3}: & Z_{17}A_{74}Z_{48} &-& A_{14}X_{48} & Q_{85}Z_{51} &-& Z_{86}Q_{61} 
 \end{array} 
 }~.~
\label{E_J_Model 7_Phase 5}
 \eeq

%=================================================================
\subsection*{Model 8: Phase 3}
%=================================================================

\begin{figure}[H]
\begin{center}
\resizebox{0.33\hsize}{!}{
\includegraphics[height=6cm]{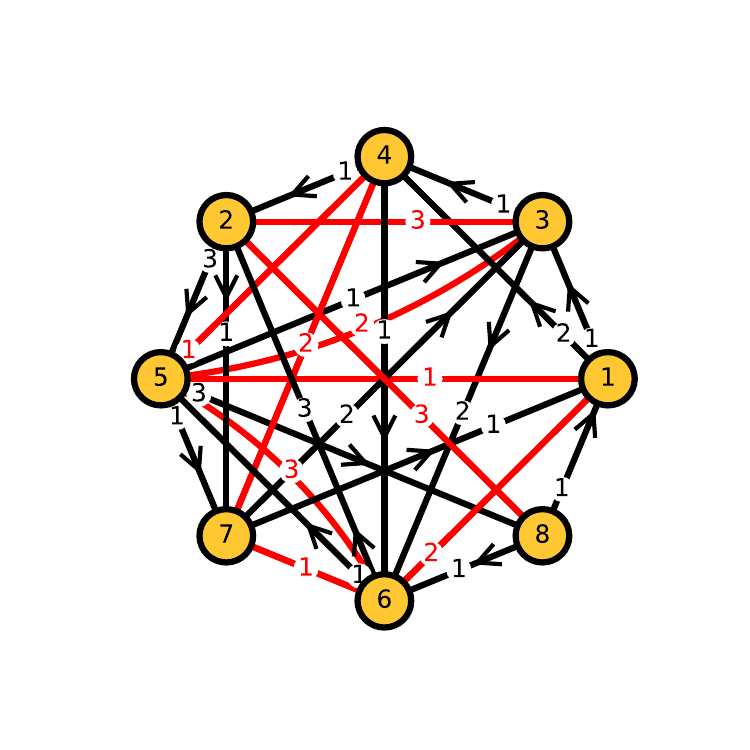} 
}
\caption{Quiver for Phase 3 of Model 8.
\label{model_8phase_3}}
 \end{center}
 \end{figure}

\fref{model_8phase_3} shows the quiver for this theory.

%=================================================================
%=================================================================

The $J$- and $E$-terms are
\beq
  {\footnotesize
\begin{array}{rrclrcl}
 &  & J &  & & E &  \\
\Lambda_{16}^{1}: & A_{65}Q_{58}X_{81} &-& X_{62}X_{27}C_{71} & P_{14}X_{46} &-& X_{13}P_{36}  \\ 
\Lambda_{16}^{2}: & A_{65}P_{58}X_{81} &-& R_{62}X_{27}C_{71} & X_{13}Q_{36} &-& Q_{14}X_{46} \\ 
\Lambda_{15}^{1}: & Y_{57}C_{71} &-& Y_{58}X_{81} & P_{14}Y_{42}Q_{25} &-& Q_{14}Y_{42}P_{25} \\ 
\Lambda_{47}^{1}: & C_{71}P_{14} &-& K_{73}X_{34} & X_{46}X_{62}X_{27} &-& Y_{42}Q_{25}Y_{57} \\ 
\Lambda_{47}^{2}: & C_{71}Q_{14} &-& J_{73}X_{34} & Y_{42}P_{25}Y_{57} &-& X_{46}R_{62}X_{27} \\ 
\Lambda_{28}^{1}: & X_{81}Q_{14}Y_{42} &-& X_{86}X_{62} & Y_{25}P_{58} &-& P_{25}Y_{58} \\ 
\Lambda_{28}^{2}: & X_{81}X_{13}X_{34}Y_{42} &-& X_{86}A_{62} & P_{25}Q_{58} &-& Q_{25}P_{58} \\ 
\Lambda_{28}^{3}: & X_{86}R_{62} &-& X_{81}P_{14}Y_{42} & Y_{25}Q_{58} &-& Q_{25}Y_{58} \\ 
\Lambda_{32}^{1}: & Y_{25}X_{53} &-& X_{27}C_{71}X_{13} & P_{36}X_{62} &-& Q_{36}R_{62} \\ 
\Lambda_{32}^{2}: & X_{27}J_{73} &-& Q_{25}X_{53} & P_{36}A_{62} &-& X_{34}X_{46}R_{62} \\ 
\Lambda_{32}^{3}: & P_{25}X_{53} &-& X_{27}K_{73} & Q_{36}A_{62} &-& X_{34}X_{46}X_{62} \\ 
\Lambda_{35}^{1}: & Q_{58}X_{81}X_{13} &-& Y_{57}J_{73} & P_{36}A_{65} &-& X_{34}Y_{42}P_{25} \\ 
\Lambda_{35}^{2}: & P_{58}X_{81}X_{13} &-& Y_{57}K_{73} & X_{34}Y_{42}Q_{25} &-& Q_{36}A_{65} \\ 
\Lambda_{67}^{1}: & J_{73}P_{36} &-& K_{73}Q_{36} & A_{65}Y_{57} &-& A_{62}X_{27} \\ 
\Lambda_{54}^{1}: & Y_{42}Y_{25} &-& X_{46}A_{65} & Q_{58}X_{81}P_{14} &-& P_{58}X_{81}Q_{14}  \\ 
\Lambda_{56}^{1}: & X_{62}Y_{25} &-& A_{62}Q_{25} & P_{58}X_{86} &-& X_{53}P_{36} \\ 
\Lambda_{56}^{2}: & R_{62}Y_{25} &-& A_{62}P_{25} & X_{53}Q_{36} &-& Q_{58}X_{86} \\ 
\Lambda_{65}^{1}: & Y_{58}X_{86} &-& X_{53}X_{34}X_{46} & R_{62}Q_{25} &-& X_{62}P_{25} \\ 
 
\end{array} 
 }~.~
\label{E_J_Model 8_Phase 3}
 \eeq

%=================================================================
\subsection*{Model 9: Phase 7}
%=================================================================

\begin{figure}[H]
\begin{center}
\resizebox{0.33\hsize}{!}{
\includegraphics[height=6cm]{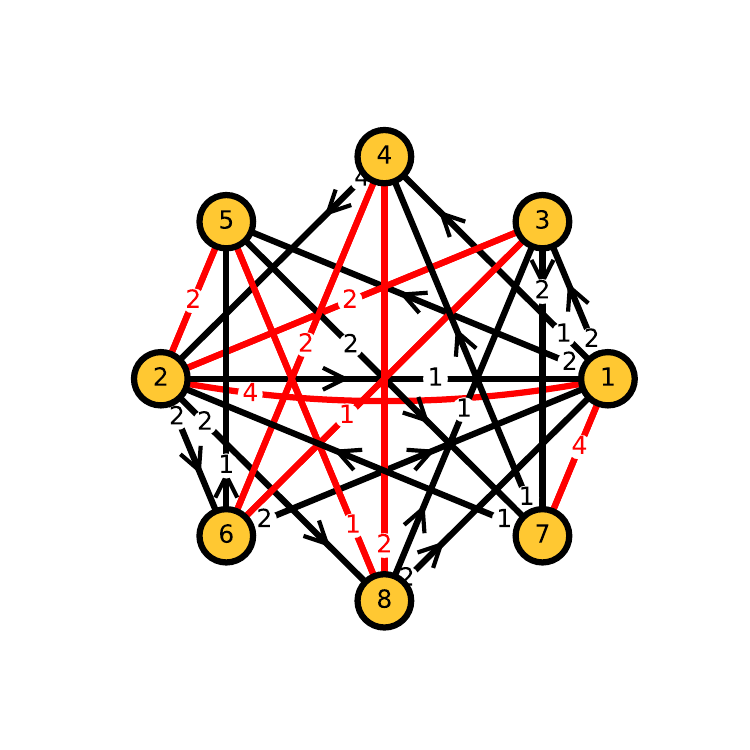} 
}
\caption{Quiver for Phase 7 of Model 9.
\label{model_9phase_7}}
 \end{center}
 \end{figure}

\fref{model_9phase_7} shows the quiver for this theory.

%=================================================================
%=================================================================

The $J$- and $E$-terms are
\beq
  {\footnotesize
\begin{array}{rrclrcl}
 &  & J &  & & E &  \\
\Lambda_{17}^{1}: & X_{74}A_{42}Y_{21} &-& X_{72}Q_{26}R_{61} & P_{15}X_{57} &-& X_{13}P_{37} \\ 
\Lambda_{17}^{2}: & X_{72}Q_{26}T_{61} &-& X_{74}C_{42}Y_{21} & P_{15}Y_{57} &-& Y_{13}P_{37} \\ 
\Lambda_{17}^{3}: & X_{74}B_{42}Y_{21} &-& X_{72}P_{26}R_{61} & X_{13}Q_{37} &-& Q_{15}X_{57} \\ 
\Lambda_{17}^{4}: & X_{72}P_{26}T_{61} &-& X_{74}D_{42}Y_{21} & Y_{13}Q_{37} &-& Q_{15}Y_{57} \\ 
\Lambda_{12}^{1}: & X_{28}E_{81} &-& Q_{26}R_{61} & X_{14}D_{42} &-& P_{15}X_{57}X_{72} \\ 
\Lambda_{12}^{2}: & Q_{26}T_{61} &-& Y_{28}E_{81} & X_{14}B_{42} &-& P_{15}Y_{57}X_{72} \\ 
\Lambda_{12}^{3}: & P_{26}R_{61} &-& X_{28}F_{81} & X_{14}C_{42} &-& Q_{15}X_{57}X_{72} \\ 
\Lambda_{12}^{4}: & Y_{28}F_{81} &-& P_{26}T_{61} & X_{14}A_{42} &-& Q_{15}Y_{57}X_{72} \\ 
\Lambda_{58}^{1}: & F_{81}Q_{15} &-& E_{81}P_{15} & Y_{57}X_{72}Y_{28} &-& X_{57}X_{72}X_{28} \\ 
\Lambda_{48}^{1}: & E_{81}X_{14} &-& L_{83}Q_{37}X_{74} & B_{42}Y_{28} &-& D_{42}X_{28} \\ 
\Lambda_{48}^{2}: & L_{83}P_{37}X_{74} &-& F_{81}X_{14} & A_{42}Y_{28} &-& C_{42}X_{28} \\ 
\Lambda_{25}^{1}: & X_{57}X_{74}A_{42} &-& Y_{57}X_{74}C_{42} & P_{26}Y_{65} &-& Y_{21}P_{15} \\ 
\Lambda_{25}^{2}: & X_{57}X_{74}B_{42} &-& Y_{57}X_{74}D_{42} & Y_{21}Q_{15} &-& Q_{26}Y_{65} \\ 
\Lambda_{36}^{1}: & T_{61}Y_{13} &-& R_{61}X_{13} & P_{37}X_{72}Q_{26} &-& Q_{37}X_{72}P_{26} \\ 
\Lambda_{32}^{1}: & Y_{28}L_{83} &-& Y_{21}X_{13} & Q_{37}X_{74}B_{42} &-& P_{37}X_{74}A_{42}  \\ 
\Lambda_{32}^{2}: & Y_{21}Y_{13} &-& X_{28}L_{83} & Q_{37}X_{74}D_{42} &-& P_{37}X_{74}C_{42}  \\ 
\Lambda_{46}^{1}: & R_{61}X_{14} &-& Y_{65}Y_{57}X_{74} & D_{42}Q_{26} &-& C_{42}P_{26} \\ 
\Lambda_{46}^{2}: & Y_{65}X_{57}X_{74} &-& T_{61}X_{14} & B_{42}Q_{26} &-& A_{42}P_{26} \\ 
 
\end{array} 
 }~.~
\label{E_J_Model 9_Phase 7}
 \eeq

\newpage

%=================================================================
\subsection*{Model 10: Phase 7}
%=================================================================

\begin{figure}[H]
\begin{center}
\resizebox{0.33\hsize}{!}{
\includegraphics[height=6cm]{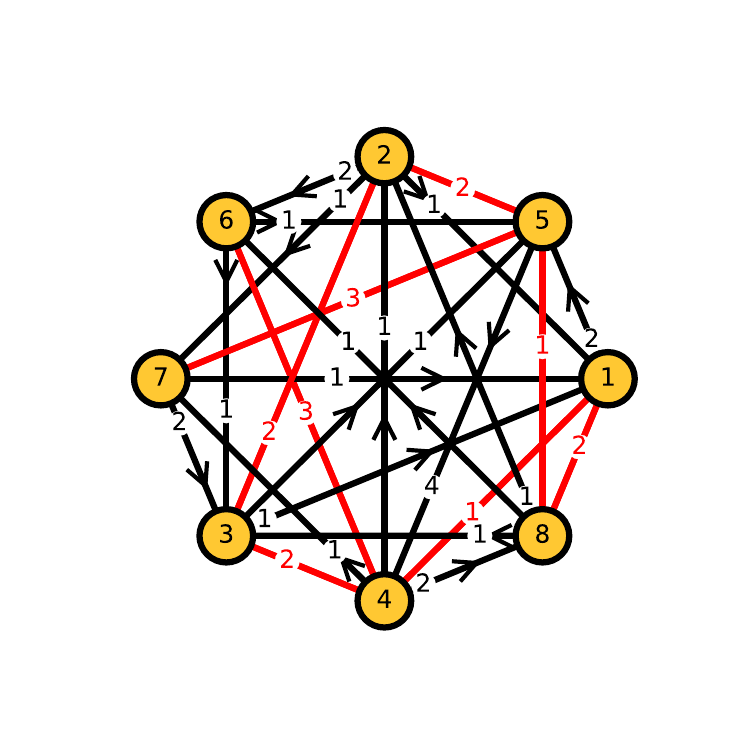} 
}
\caption{Quiver for Phase 7 of Model 10.
\label{model_10phase_7}}
 \end{center}
 \end{figure}

\fref{model_10phase_7} shows the quiver for this theory.

%=================================================================
%=================================================================

The $J$- and $E$-terms are
\beq
  {\footnotesize
\begin{array}{rrclrcl}
 &  & J &  & & E &  \\
\Lambda_{14}^{1}: & X_{42}Y_{21} &-& X_{47}B_{71} & Q_{15}C_{54} &-& P_{15}A_{54} \\ 
\Lambda_{57}^{1}: & Q_{73}Y_{35} &-& B_{71}P_{15} & A_{54}X_{47} &-& X_{54}Q_{48}Y_{82}X_{27}  \\ 
\Lambda_{57}^{2}: & B_{71}Q_{15} &-& P_{73}Y_{35} & C_{54}X_{47} &-& X_{54}P_{48}Y_{82}X_{27} \\ 
\Lambda_{57}^{3}: & P_{73}X_{31}P_{15} &-& Q_{73}X_{31}Q_{15} & B_{54}X_{47} &-& X_{54}X_{42}X_{27} \\ 
\Lambda_{18}^{1}: & X_{86}J_{63}X_{31} &-& Y_{82}X_{27}B_{71} & P_{15}X_{54}Q_{48} &-& Q_{15}X_{54}P_{48} \\ 
\Lambda_{18}^{2}: & K_{83}X_{31} &-& Y_{82}Y_{21} & Q_{15}B_{54}P_{48} &-& P_{15}B_{54}Q_{48} \\ 
\Lambda_{25}^{1}: & B_{54}Q_{48}Y_{82} &-& A_{54}X_{42} & P_{26}Y_{65} &-& Y_{21}P_{15} \\ 
\Lambda_{25}^{2}: & C_{54}X_{42} &-& B_{54}P_{48}Y_{82} & Q_{26}Y_{65}&-& Y_{21}Q_{15} \\ 
\Lambda_{32}^{1}: & Q_{26}J_{63} &-& X_{27}P_{73} & Y_{35}X_{54}P_{48}Y_{82} &-& X_{31}P_{15}X_{54}X_{42} \\ 
\Lambda_{32}^{2}: & P_{26}J_{63} &-& X_{27}Q_{73} & X_{31}Q_{15}X_{54}X_{42} &-& Y_{35}X_{54}Q_{48}Y_{82} \\ 
\Lambda_{34}^{1}: & Q_{48}K_{83} &-& X_{47}P_{73} & X_{31}P_{15}B_{54} &-& Y_{35}C_{54} \\ 
\Lambda_{34}^{2}: & X_{47}Q_{73} &-& P_{48}K_{83} & X_{31}Q_{15}B_{54} &-& Y_{35}A_{54} \\ 
\Lambda_{58}^{1}: & X_{86}Y_{65} &-& K_{83}Y_{35} & A_{54}P_{48} &-& C_{54}Q_{48} \\ 
\Lambda_{46}^{1}: & J_{63}X_{31}Q_{15}X_{54} &-& Y_{65}A_{54} & P_{48}X_{86} &-& X_{42}P_{26} \\ 
\Lambda_{46}^{2}: & Y_{65}B_{54} &-& J_{63}Y_{35}X_{54} & P_{48}Y_{82}Q_{26} &-& Q_{48}Y_{82}P_{26} \\ 
\Lambda_{46}^{3}: & J_{63}X_{31}P_{15}X_{54} &-& Y_{65}C_{54} & X_{42}Q_{26} &-& Q_{48}X_{86}  \\ 
 
\end{array} 
 }~.~
\label{E_J_Model 10_Phase 7}
 \eeq
 
 \newpage

%=================================================================
\subsection*{Model 11: Phase 17}
%=================================================================

\begin{figure}[H]
\begin{center}
\resizebox{0.33\hsize}{!}{
\includegraphics[height=6cm]{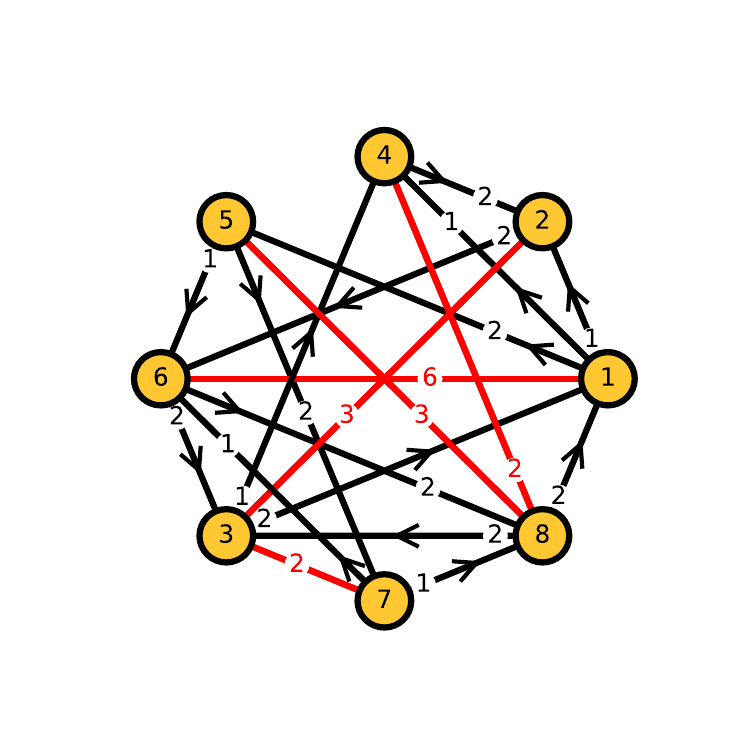} 
}
\caption{Quiver for Phase 17 of Model 11.
\label{model_11phase_17}}
 \end{center}
 \end{figure}

\fref{model_11phase_17} shows the quiver for this theory.

%=================================================================
%=================================================================

The $J$- and $E$-terms are
\beq
  {\footnotesize
\begin{array}{rrclrcl}
 &  & J &  & & E &  \\
\Lambda_{16}^{1}: & Y_{68}X_{81} &-& B_{63}M_{31} & P_{15}R_{57}X_{76} &-& O_{14}Y_{42}P_{26} \\ 
\Lambda_{16}^{2}: & B_{63}N_{31} &-& X_{68}X_{81} & P_{15}S_{57}X_{76} &-& O_{14}X_{42}P_{26} \\ 
\Lambda_{16}^{3}: & X_{68}B_{83}M_{31} &-& Y_{68}B_{83}N_{31} & P_{15}Z_{56} &-& Z_{12}P_{26} \\ 
\Lambda_{16}^{4}: & Y_{68}S_{81} &-& A_{63}M_{31} & O_{14}Y_{42}Q_{26} &-& Q_{15}R_{57}X_{76} \\ 
\Lambda_{16}^{5}: & X_{68}S_{81} &-& A_{63}N_{31} & Q_{15}S_{57}X_{76} &-& O_{14}X_{42}Q_{26} \\ 
\Lambda_{16}^{6}: & X_{68}A_{83}M_{31} &-& Y_{68}A_{83}N_{31} & Z_{12}Q_{26} &-& Q_{15}Z_{56} \\ 
\Lambda_{23}^{1}: & X_{34}X_{42} &-& M_{31}Z_{12} & Q_{26}X_{68}A_{83} &-& P_{26}X_{68}B_{83} \\ 
\Lambda_{23}^{2}: & X_{34}Y_{42} &-& N_{31}Z_{12} & P_{26}Y_{68}B_{83} &-& Q_{26}Y_{68}A_{83} \\ 
\Lambda_{23}^{3}: & M_{31}O_{14}Y_{42} &-& N_{31}O_{14}X_{42} & Q_{26}A_{63} &-& P_{26}B_{63} \\ 
\Lambda_{48}^{1}: & A_{83}X_{34} &-& S_{81}O_{14} & Y_{42}Q_{26}Y_{68} &-& X_{42}Q_{26}X_{68} \\ 
\Lambda_{48}^{2}: & B_{83}X_{34} &-& X_{81}O_{14} & X_{42}P_{26}X_{68} &-& Y_{42}P_{26}Y_{68} \\ 
\Lambda_{85}^{1}: & R_{57}X_{78} &-& Z_{56}X_{68} & B_{83}M_{31}P_{15} &-& A_{83}M_{31}Q_{15} \\ 
\Lambda_{85}^{2}: & S_{57}X_{78} &-& Z_{56}Y_{68} & A_{83}N_{31}Q_{15} &-& B_{83}N_{31}P_{15} \\ 
\Lambda_{85}^{3}: & R_{57}X_{76}Y_{68} &-& S_{57}X_{76}X_{68} & S_{81}Q_{15} &-& X_{81}P_{15} \\ 
\Lambda_{73}^{1}: & M_{31}Q_{15}R_{57} &-& N_{31}Q_{15}S_{57} & X_{78}A_{83} &-& X_{76}A_{63} \\ 
\Lambda_{73}^{2}: & M_{31}P_{15}R_{57} &-& N_{31}P_{15}S_{57} & X_{76}B_{63} &-& X_{78}B_{83} \\ 
 
\end{array} 
 }~.~
\label{E_J_Model 11_Phase 17}
 \eeq
 
 \newpage

%=================================================================
\subsection*{Model 12: Phase 11}
%=================================================================

\begin{figure}[H]
\begin{center}
\resizebox{0.33\hsize}{!}{
\includegraphics[height=6cm]{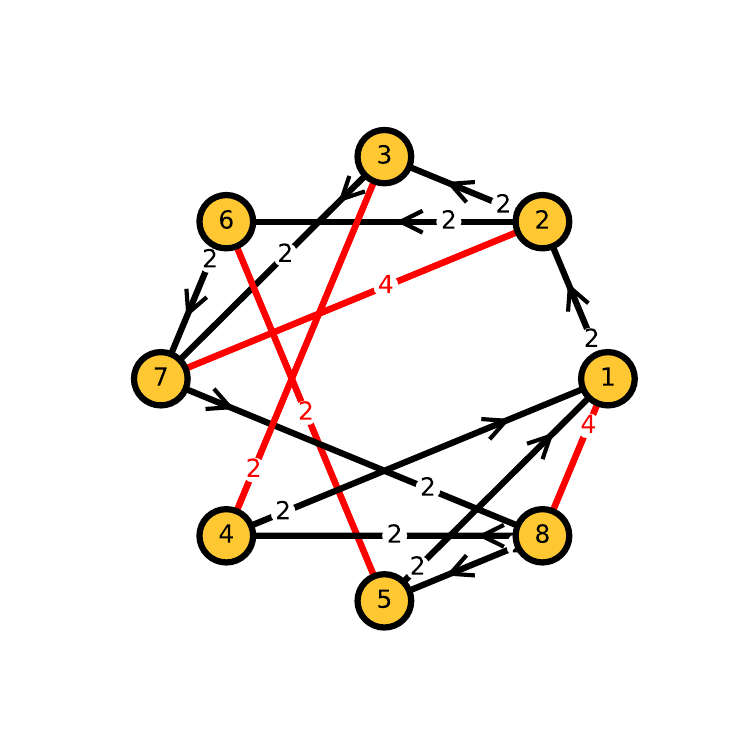} 
}
\caption{Quiver for Phase 11 of Model 12.
\label{model_12phase_11}}
 \end{center}
 \end{figure}

\fref{model_12phase_11} shows the quiver for this theory.

%=================================================================
%=================================================================

The $J$- and $E$-terms are
\beq
  {\footnotesize
\begin{array}{rrclrcl}
 &  & J &  & & E &  \\
\Lambda_{34}^{1}: & X_{41}A_{12}Y_{23} &-& Y_{41}A_{12}X_{23} & Q_{37}Y_{78}Q_{84} &-& P_{37}Y_{78}P_{84} \\ 
\Lambda_{34}^{2}: & X_{41}B_{12}Y_{23} &-& Y_{41}B_{12}X_{23} & P_{37}X_{78}P_{84} &-& Q_{37}X_{78}Q_{84} \\ 
\Lambda_{18}^{1}: & X_{85}Q_{51} &-& Q_{84}X_{41} & A_{12}Y_{23}Q_{37}Y_{78} &-& B_{12}Y_{23}Q_{37}X_{78}  \\ 
\Lambda_{18}^{2}: & P_{84}X_{41} &-& X_{85}P_{51} & A_{12}Y_{23}P_{37}Y_{78} &-& B_{12}Y_{23}P_{37}X_{78}  \\ 
\Lambda_{18}^{3}: & Y_{85}Q_{51} &-& Q_{84}Y_{41} & B_{12}X_{23}Q_{37}X_{78} &-& A_{12}X_{23}Q_{37}Y_{78} \\ 
\Lambda_{18}^{4}: & Y_{85}P_{51} &-& P_{84}Y_{41} & A_{12}X_{23}P_{37}Y_{78} &-& B_{12}X_{23}P_{37}X_{78} \\ 
\Lambda_{27}^{1}: & Y_{78}Y_{85}P_{51}A_{12} &-& X_{78}Y_{85}P_{51}B_{12} & P_{26}X_{67} &-& X_{23}P_{37} \\ 
\Lambda_{27}^{2}: & Y_{78}X_{85}P_{51}A_{12} &-& X_{78}X_{85}P_{51}B_{12} & Y_{23}P_{37}&-& P_{26}Y_{67} \\ 
\Lambda_{27}^{3}: & X_{78}Y_{85}Q_{51}B_{12} &-& Y_{78}Y_{85}Q_{51}A_{12} & Q_{26}X_{67} &-& X_{23}Q_{37} \\ 
\Lambda_{27}^{4}: & Y_{78}X_{85}Q_{51}A_{12} &-& X_{78}X_{85}Q_{51}B_{12} & Q_{26}Y_{67} &-& Y_{23}Q_{37}  \\ 
\Lambda_{56}^{1}: & X_{67}Y_{78}Y_{85} &-& Y_{67}Y_{78}X_{85} & Q_{51}A_{12}Q_{26} &-& P_{51}A_{12}P_{26}  \\ 
\Lambda_{56}^{2}: & Y_{67}X_{78}X_{85} &-& X_{67}X_{78}Y_{85} & Q_{51}B_{12}Q_{26} &-& P_{51}B_{12}P_{26} \\ 
 
\end{array} 
 }~.~
\label{E_J_Model 12_Phase 11}
 \eeq
 
 \newpage

%=================================================================
\subsection*{Model 13: Phase 54}
%=================================================================

\begin{figure}[H]
\begin{center}
\resizebox{0.4\hsize}{!}{
\includegraphics[height=6cm]{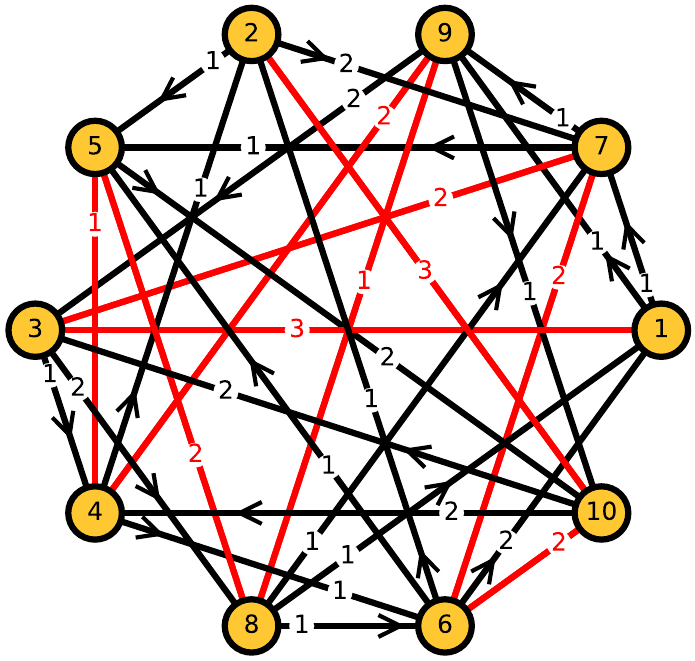} 
}
\caption{Quiver for Phase 54 of Model 13.
\label{model_13phase_54}}
 \end{center}
 \end{figure}

\fref{model_13phase_54} shows the quiver for this theory.

%=================================================================
%=================================================================

The $J$- and $E$-terms are
\beq
  {\footnotesize
\begin{array}{rrclrcl}
 &  & J &  & & E &  \\
\Lambda_{7.6}^{1}: & P_{61}X_{17} &-& D_{62}Q_{27} & X_{75}P_{5.10}X_{10.3}X_{34}X_{46} &-& Y_{79}A_{9.10}X_{10.3}P_{38}X_{86} \\ 
\Lambda_{7.6}^{2}: & Q_{61}X_{17} &-& D_{62}P_{27} & Y_{79}A_{9.10}X_{10.3}Q_{38}X_{86} &-& X_{75}Q_{5.10}X_{10.3}X_{34}X_{46} \\ 
\Lambda_{10.6}^{1}: & P_{61}C_{19}A_{9.10} &-& O_{65}Q_{5.10} & A_{10.3}P_{38}X_{86} &-& B_{10.4}X_{46} \\ 
\Lambda_{10.6}^{2}: & Q_{61}C_{19}A_{9.10} &-& O_{65}P_{5.10} & A_{10.4}X_{46} &-& A_{10.3}Q_{38}X_{86} \\ 
\Lambda_{1.3}^{1}: & P_{38}X_{86}P_{61} &-& Q_{38}X_{86}Q_{61} & X_{17}Y_{79}A_{9.10}X_{10.3} &-& C_{19}A_{9.10}A_{10.3} \\ 
\Lambda_{4.9}^{1}: & A_{9.10}B_{10.4} &-& J_{93}X_{34} & X_{46}P_{61}C_{19} &-& Y_{42}Q_{27}Y_{79} \\ 
\Lambda_{4.9}^{2}: & A_{9.10}A_{10.4} &-& I_{93}X_{34} & Y_{42}P_{27}Y_{79} &-& X_{46}Q_{61}C_{19} \\ 
\Lambda_{2.10}^{1}: & A_{10.3}X_{34}Y_{42} &-& X_{10.3}X_{34}X_{46}D_{62} & P_{27}X_{75}Q_{5.10} &-& Q_{27}X_{75}P_{5.10} \\ 
\Lambda_{2.10}^{2}: & X_{10.3}Q_{38}X_{86}D_{62} &-& A_{10.4}Y_{42} & P_{27}Y_{79}A_{9.10} &-& Y_{25}P_{5.10} \\ 
\Lambda_{2.10}^{3}: & X_{10.3}P_{38}X_{86}D_{62} &-& B_{10.4}Y_{42} & Y_{25}Q_{5.10} &-& Q_{27}Y_{79}A_{9.10} \\ 
\Lambda_{3.7}^{1}: & X_{75}Q_{5.10}A_{10.3} &-& Y_{79}I_{93} & P_{38}A_{87} &-& X_{34}Y_{42}P_{27} \\ 
\Lambda_{3.7}^{2}: & X_{75}P_{5.10}A_{10.3} &-& Y_{79}J_{93} & X_{34}Y_{42}Q_{27} &-& Q_{38}A_{87} \\ 
\Lambda_{3.1}^{1}: & C_{19}I_{93} &-& X_{17}X_{75}Q_{5.10}X_{10.3} & P_{38}A_{81} &-& X_{34}X_{46}Q_{61} \\ 
\Lambda_{3.1}^{2}: & X_{17}X_{75}P_{5.10}X_{10.3} &-& C_{19}J_{93} & Q_{38}A_{81} &-& X_{34}X_{46}P_{61} \\ 
\Lambda_{8.9}^{1}: & I_{93}P_{38} &-& J_{93}Q_{38} & A_{87}Y_{79} &-& A_{81}C_{19} \\ 
\Lambda_{5.8}^{1}: & A_{87}X_{75} &-& X_{86}O_{65} & P_{5.10}A_{10.3}Q_{38} &-& Q_{5.10}A_{10.3}P_{38} \\ 
\Lambda_{5.8}^{2}: & X_{86}D_{62}Y_{25} &-& A_{81}X_{17}X_{75} & P_{5.10}X_{10.3}Q_{38} &-& Q_{5.10}X_{10.3}P_{38} \\ 
\Lambda_{5.4}^{1}: & X_{46}O_{65} &-& Y_{42}Y_{25} & P_{5.10}A_{10.4} &-& Q_{5.10}B_{10.4} \\ 
 
\end{array} 
 }~.~
\label{E_J_Model 13_Phase 54}
 \eeq

 \newpage

%=================================================================
\subsection*{Model 14: Phase 76}
%=================================================================

\begin{figure}[H]
\begin{center}
\resizebox{0.4\hsize}{!}{
\includegraphics[height=6cm]{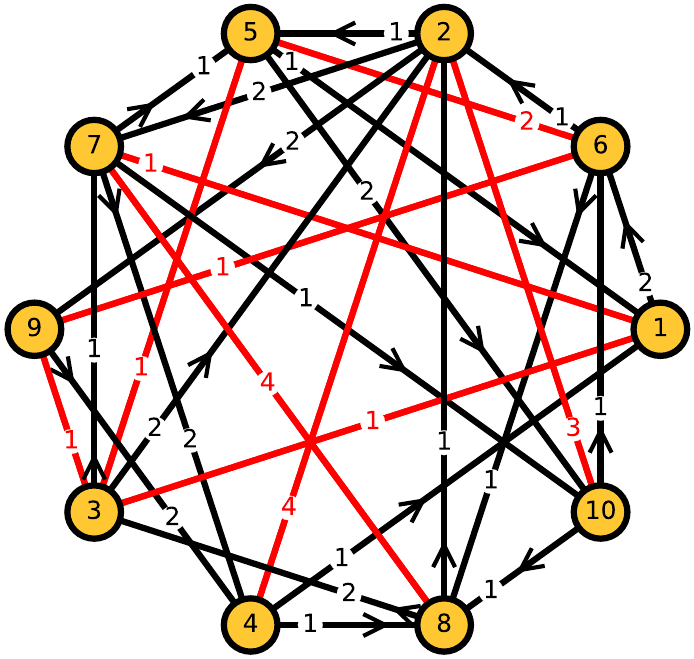} 
}
\caption{Quiver for Phase 76 of Model 14.
\label{model_14phase_76}}
 \end{center}
 \end{figure}

\fref{model_14phase_76} shows the quiver for this theory.

%=================================================================
%=================================================================

The $J$- and $E$-terms are
\beq
  {\footnotesize
\begin{array}{rrclrcl}
 &  & J &  & & E &  \\
\Lambda_{2.10}^{1}: & X_{10.6}X_{68}P_{83}B_{32} &-& X_{10.8}P_{83}A_{32} & P_{27}X_{7.10} &-& X_{25}P_{5.10} \\ 
\Lambda_{2.10}^{2}: & X_{10.8}X_{82} &-& X_{10.6}B_{62} & P_{27}Y_{75}Q_{5.10} &-& Q_{27}Y_{75}P_{5.10} \\ 
\Lambda_{2.10}^{3}: & X_{10.6}X_{68}Q_{83}B_{32} &-& X_{10.8}Q_{83}A_{32} & X_{25}Q_{5.10} &-& Q_{27}X_{7.10} \\ 
\Lambda_{5.3}^{1}: & I_{37}Y_{75} &-& A_{32}X_{25} & P_{5.10}X_{10.8}P_{83} &-& Q_{5.10}X_{10.8}Q_{83} \\ 
\Lambda_{7.8}^{1}: & X_{82}Q_{27} &-& P_{83}I_{37} & Y_{75}P_{5.10}X_{10.8} &-& K_{74}X_{41}P_{16}X_{68} \\ 
\Lambda_{7.8}^{2}: & X_{82}P_{27} &-& Q_{83}I_{37} & K_{74}X_{41}Q_{16}X_{68} &-& Y_{75}Q_{5.10}X_{10.8} \\ 
\Lambda_{1.3}^{1}: & I_{37}K_{74}X_{41} &-& B_{32}X_{25}X_{51} & Q_{16}X_{68}Q_{83} &-& P_{16}X_{68}P_{83} \\ 
\Lambda_{8.7}^{1}: & X_{7.10}X_{10.8} &-& K_{74}X_{48} & P_{83}A_{32}P_{27} &-& Q_{83}A_{32}Q_{27} \\ 
\Lambda_{8.7}^{2}: & X_{7.10}X_{10.6}X_{68} &-& L_{74}X_{48} & Q_{83}B_{32}Q_{27} &-& P_{83}B_{32}P_{27} \\ 
\Lambda_{4.2}^{1}: & Q_{27}K_{74} &-& X_{29}P_{94} & X_{41}P_{16}X_{68}X_{82} &-& X_{48}Q_{83}A_{32} \\ 
\Lambda_{4.2}^{2}: & Y_{29}P_{94} &-& Q_{27}L_{74} & X_{41}P_{16}B_{62} &-& X_{48}Q_{83}B_{32} \\ 
\Lambda_{4.2}^{3}: & P_{27}K_{74} &-& X_{29}Q_{94} & X_{48}P_{83}A_{32} &-& X_{41}Q_{16}X_{68}X_{82} \\ 
\Lambda_{4.2}^{4}: & P_{27}L_{74} &-& Y_{29}Q_{94} & X_{41}Q_{16}B_{62} &-& X_{48}P_{83}B_{32} \\ 
\Lambda_{3.9}^{1}: & P_{94}X_{48}Q_{83} &-& Q_{94}X_{48}P_{83} & B_{32}Y_{29} &-& A_{32}X_{29} \\ 
\Lambda_{6.9}^{1}: & P_{94}X_{41}P_{16} &-& Q_{94}X_{41}Q_{16} & X_{68}X_{82}X_{29} &-& B_{62}Y_{29} \\ 
\Lambda_{1.7}^{1}: & Y_{75}X_{51} &-& L_{74}X_{41} & Q_{16}B_{62}P_{27} &-& P_{16}B_{62}Q_{27} \\ 
\Lambda_{5.6}^{1}: & X_{68}P_{83}B_{32}X_{25} &-& B_{62}Q_{27}Y_{75} & P_{5.10}X_{10.6} &-& X_{51}P_{16} \\ 
\Lambda_{5.6}^{2}: & X_{68}Q_{83}B_{32}X_{25} &-& B_{62}P_{27}Y_{75} & X_{51}Q_{16} &-& Q_{5.10}X_{10.6} \\ 
 
\end{array} 
 }~.~
\label{E_J_Model 14_Phase 76}
 \eeq
 
  \newpage

%=================================================================
\subsection*{Model 15: Phase 56}
%=================================================================

\begin{figure}[H]
\begin{center}
\resizebox{0.4\hsize}{!}{
\includegraphics[height=6cm]{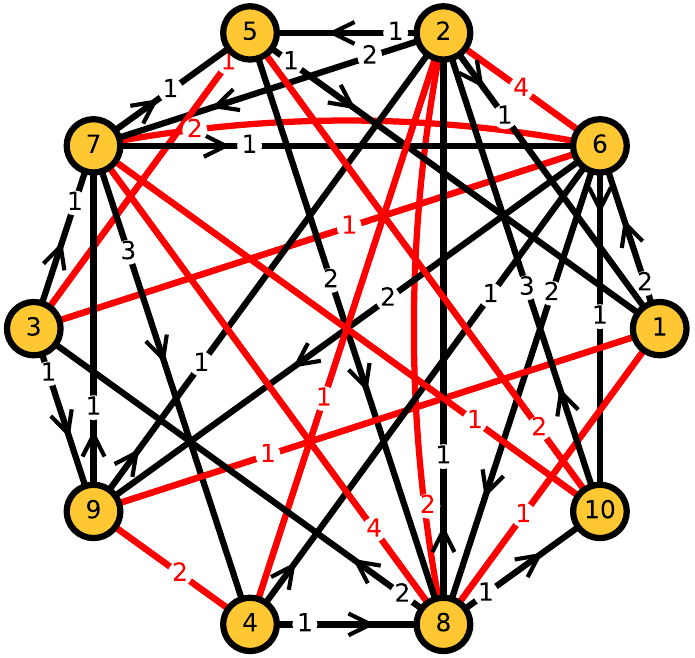} 
}
\caption{Quiver for Phase 56 of Model 15.
\label{model_15phase_56}}
 \end{center}
 \end{figure}

\fref{model_15phase_56} shows the quiver for this theory.

%=================================================================
%=================================================================

The $J$- and $E$-terms are
\beq
  {\footnotesize
\begin{array}{rrclrcl}
 &  & J &  & & E &  \\
\Lambda_{1.8}^{1}: & A_{82}X_{25}X_{51} &-& X_{8.10}F_{10.2}X_{21} & Q_{16}A_{68} &-& P_{16}B_{68} \\ 
\Lambda_{1.9}^{1}: & X_{92}X_{21} &-& Y_{97}Y_{75}X_{51} & Q_{16}B_{69} &-& P_{16}A_{69} \\ 
\Lambda_{2.6}^{1}: & A_{69}X_{92} &-& B_{68}X_{8.10}F_{10.2} & X_{21}P_{16} &-& P_{27}J_{74}X_{46} \\ 
\Lambda_{2.6}^{2}: & B_{69}X_{92} &-& A_{68}X_{8.10}F_{10.2} & Q_{27}J_{74}X_{46} &-& X_{21}Q_{16} \\ 
\Lambda_{2.6}^{3}: & B_{68}A_{82} &-& X_{6.10}E_{10.2} & X_{25}X_{51}P_{16} &-& P_{27}C_{76} \\ 
\Lambda_{2.6}^{4}: & A_{68}A_{82} &-& X_{6.10}G_{10.2} & Q_{27}C_{76} &-& X_{25}X_{51}Q_{16} \\ 
\Lambda_{2.8}^{1}: & P_{83}I_{39}X_{92} &-& X_{8.10}E_{10.2} & P_{27}J_{74}X_{48} &-& X_{25}B_{58} \\ 
\Lambda_{2.8}^{2}: & X_{8.10}G_{10.2} &-& Q_{83}I_{39}X_{92} & Q_{27}J_{74}X_{48} &-& X_{25}A_{58} \\ 
\Lambda_{7.10}^{1}: & E_{10.2}P_{27} &-& G_{10.2}Q_{27} & J_{74}X_{48}X_{8.10} &-& C_{76}X_{6.10} \\ 
\Lambda_{5.10}^{1}: & F_{10.2}Q_{27}Y_{75} &-& E_{10.2}X_{25} & B_{58}X_{8.10} &-& X_{51}P_{16}X_{6.10} \\ 
\Lambda_{5.10}^{2}: & F_{10.2}P_{27}Y_{75} &-& G_{10.2}X_{25} & X_{51}Q_{16}X_{6.10} &-& A_{58}X_{8.10} \\ 
\Lambda_{5.3}^{1}: & H_{37}Y_{75} &-& I_{39}X_{92}X_{25} & A_{58}Q_{83} &-& B_{58}P_{83} \\ 
\Lambda_{7.8}^{1}: & X_{8.10}F_{10.2}Q_{27} &-& P_{83}H_{37} & J_{74}X_{46}A_{68} &-& Y_{75}B_{58} \\ 
\Lambda_{7.8}^{2}: & Q_{83}H_{37} &-& X_{8.10}F_{10.2}P_{27} & J_{74}X_{46}B_{68} &-& Y_{75}A_{58} \\ 
\Lambda_{6.3}^{1}: & H_{37}J_{74}X_{46} &-& I_{39}Y_{97}C_{76} & A_{68}P_{83} &-& B_{68}Q_{83} \\ 
\Lambda_{4.9}^{1}: & X_{92}Q_{27}J_{74} &-& Y_{97}K_{74} & X_{48}Q_{83}I_{39} &-& X_{46}B_{69} \\ 
\Lambda_{4.9}^{2}: & X_{92}P_{27}J_{74} &-& Y_{97}L_{74} & X_{46}A_{69} &-& X_{48}P_{83}I_{39} \\ 
\Lambda_{8.7}^{1}: & C_{76}B_{68} &-& K_{74}X_{48} & A_{82}P_{27} &-& Q_{83}I_{39}Y_{97} \\ 
\Lambda_{8.7}^{2}: & C_{76}A_{68} &-& L_{74}X_{48} & P_{83}I_{39}Y_{97} &-& A_{82}Q_{27} \\ 
\Lambda_{6.7}^{1}: & K_{74}X_{46} &-& Y_{75}X_{51}Q_{16} & X_{6.10}F_{10.2}P_{27} &-& B_{69}Y_{97} \\ 
\Lambda_{6.7}^{2}: & Y_{75}X_{51}P_{16} &-& L_{74}X_{46} & X_{6.10}F_{10.2}Q_{27} &-& A_{69}Y_{97} \\ 
\Lambda_{4.2}^{1}: & P_{27}K_{74} &-& Q_{27}L_{74} & X_{48}A_{82} &-& X_{46}X_{6.10}F_{10.2} \\ 
 
\end{array} 
 }~.~
\label{E_J_Model 15_Phase 56}
 \eeq

%=================================================================
\subsection*{Model 16: Phase 140}
%=================================================================

\begin{figure}[H]
\begin{center}
\resizebox{0.4\hsize}{!}{
\includegraphics[height=6cm]{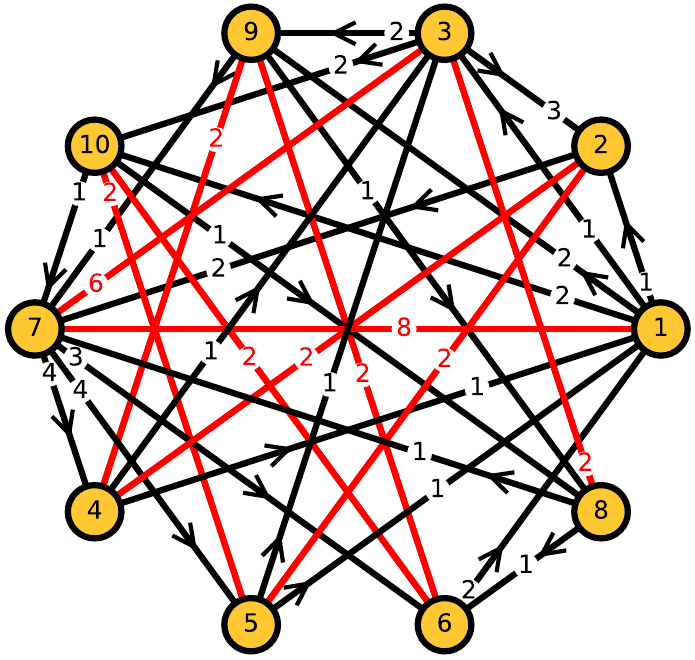} 
}
\caption{Quiver for Phase 140 of Model 16.
\label{model_16phase_140}}
 \end{center}
 \end{figure}

\fref{model_16phase_140} shows the quiver for this theory.

%=================================================================
%=================================================================

The $J$- and $E$-terms are
\beq
  {\footnotesize
\begin{array}{rrclrcl}
 &  & J &  & & E &  \\
\Lambda_{2.5}^{1}: & X_{53}B_{32} &-& X_{51}X_{13}A_{32} & P_{27}X_{75} &-& Q_{27}R_{75} \\ 
\Lambda_{2.5}^{2}: & X_{51}A_{12} &-& X_{53}X_{32} & P_{27}Y_{75} &-& Q_{27}S_{75} \\ 
\Lambda_{3.7}^{1}: & L_{74}X_{41}X_{13} &-& Y_{75}X_{53} & A_{3.10}V_{10.7} &-& X_{32}P_{27} \\ 
\Lambda_{3.7}^{2}: & N_{74}X_{41}X_{13} &-& S_{75}X_{53} & X_{32}Q_{27} &-& B_{3.10}V_{10.7} \\ 
\Lambda_{3.7}^{3}: & X_{75}X_{53} &-& L_{74}X_{43} & A_{39}X_{98}B_{87} &-& B_{32}P_{27} \\ 
\Lambda_{3.7}^{4}: & M_{74}X_{43} &-& X_{75}X_{51}X_{13} & A_{39}Y_{97} &-& A_{32}P_{27} \\ 
\Lambda_{3.7}^{5}: & R_{75}X_{53} &-& N_{74}X_{43} & B_{32}Q_{27} &-& B_{39}X_{98}B_{87} \\ 
\Lambda_{3.7}^{6}: & O_{74}X_{43} &-& R_{75}X_{51}X_{13} & A_{32}Q_{27} &-& B_{39}Y_{97} \\ 
\Lambda_{4.9}^{1}: & X_{98}B_{87}L_{74} &-& Y_{97}M_{74} & X_{43}A_{39} &-& X_{41}A_{19} \\ 
\Lambda_{4.9}^{2}: & Y_{97}O_{74} &-& X_{98}B_{87}N_{74} & X_{43}B_{39} &-& X_{41}B_{19} \\ 
\Lambda_{4.2}^{1}: & Q_{27}N_{74} &-& P_{27}L_{74} & X_{43}B_{32} &-& X_{41}X_{13}X_{32} \\ 
\Lambda_{4.2}^{2}: & P_{27}M_{74} &-& Q_{27}O_{74} & X_{43}A_{32} &-& X_{41}A_{12} \\ 
\Lambda_{1.7}^{1}: & Y_{75}X_{51} &-& M_{74}X_{41} & A_{19}Y_{97} &-& A_{12}P_{27} \\ 
\Lambda_{1.7}^{2}: & S_{75}X_{51} &-& O_{74}X_{41} & A_{12}Q_{27} &-& B_{19}Y_{97} \\ 
\Lambda_{1.7}^{3}: & T_{76}P_{61} &-& L_{74}X_{41} & X_{13}A_{3.10}V_{10.7} &-& A_{19}X_{98}B_{87} \\ 
\Lambda_{1.7}^{4}: & N_{74}X_{41} &-& T_{76}Q_{61} & X_{13}B_{3.10}V_{10.7} &-& B_{19}X_{98}B_{87} \\ 
\Lambda_{1.7}^{5}: & B_{76}P_{61} &-& X_{75}X_{51} & A_{1.10}X_{10.8}B_{87} &-& X_{13}A_{39}Y_{97} \\ 
\Lambda_{1.7}^{6}: & R_{75}X_{51} &-& B_{76}Q_{61} & B_{1.10}X_{10.8}B_{87} &-& X_{13}B_{39}Y_{97} \\ 
\Lambda_{1.7}^{7}: & A_{76}P_{61} &-& Y_{75}X_{51} & A_{19}Y_{97} &-& A_{1.10}V_{10.7} \\ 
\Lambda_{1.7}^{8}: & S_{75}X_{51} &-& A_{76}Q_{61} & B_{19}Y_{97} &-& B_{1.10}V_{10.7} \\ 
\Lambda_{6.10}^{1}: & V_{10.7}T_{76} &-& X_{10.8}X_{86} & Q_{61}X_{13}B_{3.10} &-& P_{61}X_{13}A_{3.10} \\ 
\Lambda_{6.10}^{2}: & V_{10.7}A_{76} &-& X_{10.8}B_{87}B_{76} & P_{61}A_{1.10} &-& Q_{61}B_{1.10} \\ 
\Lambda_{6.9}^{1}: & X_{98}X_{86} &-& Y_{97}B_{76} & Q_{61}X_{13}B_{39} &-& P_{61}X_{13}A_{39} \\ 
\Lambda_{6.9}^{2}: & X_{98}B_{87}T_{76} &-& Y_{97}A_{76} & P_{61}A_{19} &-& Q_{61}B_{19} 
 
\end{array} 
 } \nonumber
\label{E_J_Model 16_Phase 140_1}
 \eeq

\beq
  {\footnotesize
\begin{array}{rrclrcl}
 &  & J &  & & E &  \\
\Lambda_{3.8}^{1}: & B_{87}X_{75}X_{53} &-& X_{86}P_{61}X_{13} & A_{3.10}X_{10.8} &-& A_{39}X_{98} \\ 
\Lambda_{3.8}^{2}: & X_{86}Q_{61}X_{13} &-& B_{87}R_{75}X_{53} & B_{3.10}X_{10.8} &-& B_{39}X_{98} \\ 
\Lambda_{5.10}^{1}: & X_{10.8}B_{87}R_{75} &-& V_{10.7}S_{75} & X_{53}B_{3.10} &-& X_{51}B_{1.10} \\ 
\Lambda_{5.10}^{2}: & V_{10.7}Y_{75} &-& X_{10.8}B_{87}X_{75} & X_{53}A_{3.10} &-& X_{51}A_{1.10} \\ 
 
\end{array} 
 }~.~
\label{E_J_Model 16_Phase 140}
 \eeq

%=================================================================
\subsection*{Model 17: Phase 490}
%=================================================================

\begin{figure}[H]
\begin{center}
\resizebox{0.45\hsize}{!}{
\includegraphics[height=6cm]{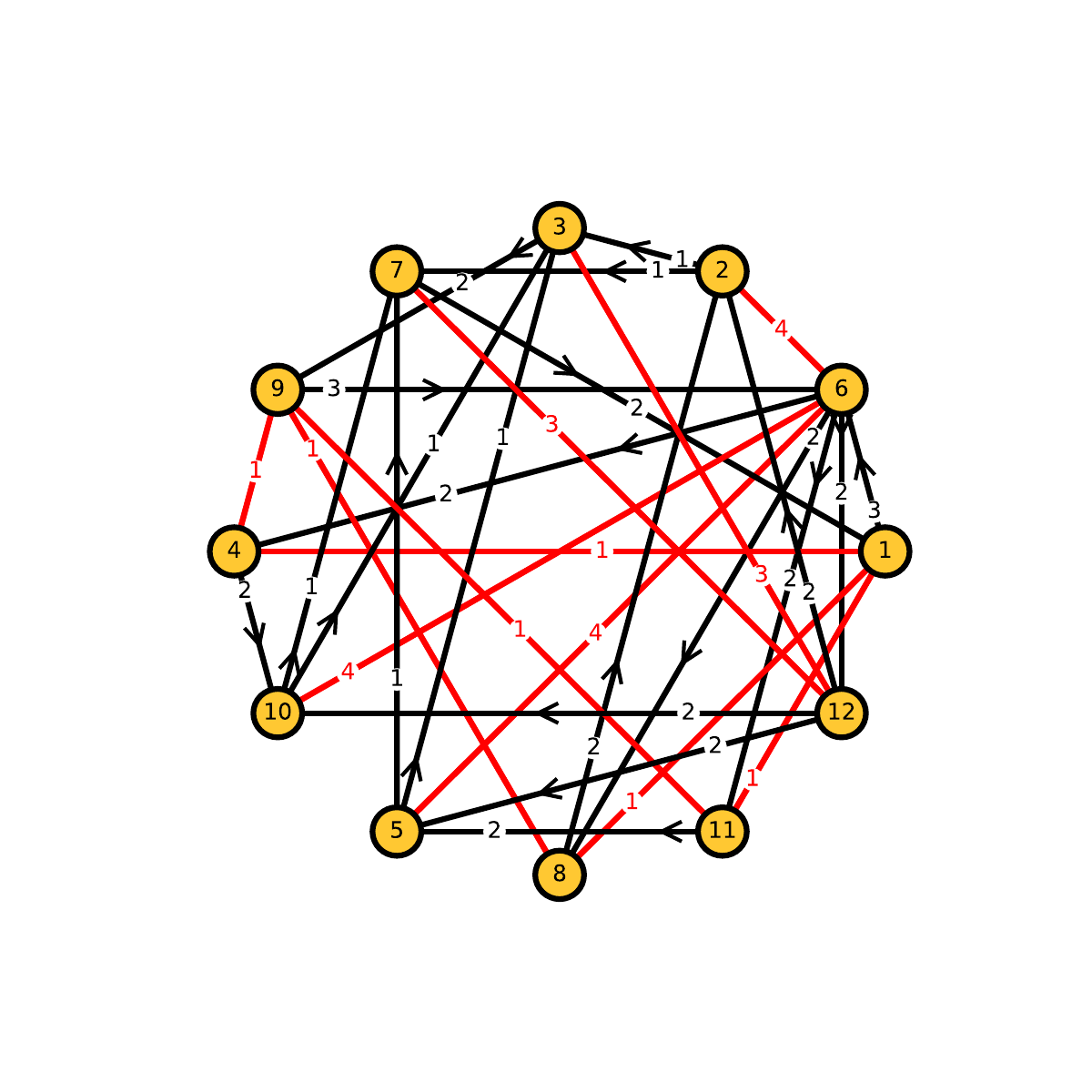} 
}
\caption{Quiver for Phase 490 of Model 17.
\label{model_17phase_490}}
 \end{center}
 \end{figure}

\fref{model_17phase_490} shows the quiver for this theory.

%=================================================================
%=================================================================

The $J$- and $E$-terms are
\beq
  {\footnotesize
\begin{array}{rrclrcl}
 &  & J &  & & E &  \\
\Lambda_{6.2}^{1}: & X_{23}Q_{39}C_{96} &-& N_{27}P_{71}X_{16} & A_{68}Q_{82} &-& P_{6.12}A_{12.2}  \\ 
\Lambda_{6.2}^{2}: & N_{27}Q_{71}X_{16} &-& X_{23}P_{39}C_{96} & A_{68}P_{82} &-& Q_{6.12}A_{12.2} \\ 
\Lambda_{6.10}^{1}: & X_{10.7}P_{71}X_{16} &-& X_{10.3}Q_{39}B_{96} & A_{64}P_{4.10} &-& P_{6.12}A_{12.10} \\ 
\Lambda_{6.10}^{2}: & X_{10.3}P_{39}B_{96} &-& X_{10.7}Q_{71}X_{16} & A_{64}Q_{4.10} &-& Q_{6.12}A_{12.10} \\ 
\Lambda_{6.10}^{3}: & X_{10.7}P_{71}B_{16} &-& X_{10.3}Q_{39}A_{96} & P_{6.12}X_{12.10} &-& X_{64}P_{4.10} \\ 
\Lambda_{6.10}^{4}: & X_{10.3}P_{39}A_{96} &-& X_{10.7}Q_{71}B_{16} & Q_{6.12}X_{12.10} &-& X_{64}Q_{4.10} \\ 
\Lambda_{1.8}^{1}: & P_{82}N_{27}Q_{71} &-& Q_{82}N_{27}P_{71} & A_{16}X_{68} &-& X_{16}A_{68} \\ 
\Lambda_{1.4}^{1}: & P_{4.10}X_{10.7}P_{71} &-& Q_{4.10}X_{10.7}Q_{71} & B_{16}X_{64} &-& X_{16}A_{64} \\ 
\Lambda_{12.7}^{1}: & Q_{71}X_{16}Q_{6.12} &-& P_{71}X_{16}P_{6.12} & A_{12.2}N_{27} &-& A_{12.10}X_{10.7} \\ 
\Lambda_{2.6}^{1}: & Q_{6.12}O_{12.2} &-& X_{68}P_{82} & N_{27}Q_{71}A_{16} &-& X_{23}P_{39}A_{96} \\ 
\Lambda_{2.6}^{2}: & P_{6.12}O_{12.2} &-& X_{68}Q_{82} & X_{23}Q_{39}A_{96} &-& N_{27}P_{71}A_{16} \\ 
\Lambda_{9.8}^{1}: & P_{82}X_{23}P_{39} &-& Q_{82}X_{23}Q_{39} & C_{96}A_{68} &-& A_{96}X_{68} \\ 
\Lambda_{3.12}^{1}: & O_{12.2}X_{23} &-& X_{12.10}X_{10.3} & P_{39}A_{96}Q_{6.12} &-& Q_{39}A_{96}P_{6.12} \\ 
\Lambda_{3.12}^{2}: & A_{12.5}X_{53} &-& A_{12.10}X_{10.3} & Q_{39}B_{96}P_{6.12} &-& P_{39}B_{96}Q_{6.12} \\ 
\Lambda_{3.12}^{3}: & A_{12.2}X_{23} &-& R_{12.5}X_{53} & Q_{39}C_{96}P_{6.12} &-& P_{39}C_{96}Q_{6.12}
\end{array} 
 } \nonumber
\label{E_J_Model 17_Phase 490_1}
 \eeq

\beq
  {\footnotesize
\begin{array}{rrclrcl}
 &  & J &  & & E &  \\
\Lambda_{7.12}^{1}: & O_{12.2}N_{27} &-& A_{12.5}A_{57} & P_{71}A_{16}P_{6.12} &-& Q_{71}A_{16}Q_{6.12} \\ 
\Lambda_{7.12}^{2}: & R_{12.5}A_{57} &-& X_{12.10}X_{10.7} & P_{71}B_{16}P_{6.12} &-& Q_{71}B_{16}Q_{6.12} \\ 
\Lambda_{5.6}^{1}: & P_{6.12}A_{12.5} &-& A_{6.11}Q_{11.5} & A_{57}P_{71}A_{16} &-& X_{53}Q_{39}B_{96} \\ 
\Lambda_{5.6}^{2}: & P_{6.12}R_{12.5} &-& X_{6.11}Q_{11.5} & X_{53}Q_{39}C_{96} &-& A_{57}P_{71}B_{16} \\ 
\Lambda_{5.6}^{3}: & Q_{6.12}A_{12.5} &-& A_{6.11}P_{11.5} & X_{53}P_{39}B_{96} &-& A_{57}Q_{71}A_{16} \\ 
\Lambda_{5.6}^{4}: & Q_{6.12}R_{12.5} &-& X_{6.11}P_{11.5} & A_{57}Q_{71}B_{16} &-& X_{53}P_{39}C_{96} \\ 
\Lambda_{9.11}^{1}: & Q_{11.5}X_{53}Q_{39} &-& P_{11.5}X_{53}P_{39} & C_{96}X_{6.11} &-& B_{96}A_{6.11} \\ 
\Lambda_{1.11}^{1}: & Q_{11.5}A_{57}P_{71} &-& P_{11.5}A_{57}Q_{71} & A_{16}A_{6.11} &-& B_{16}X_{6.11} \\ 
\Lambda_{4.9}^{1}: & A_{96}X_{64} &-& B_{96}A_{64} & Q_{4.10}X_{10.3}P_{39} &-& P_{4.10}X_{10.3}Q_{39}  
\end{array} 
 }~.~
\label{E_J_Model 17_Phase 490}
 \eeq

%=================================================================
\subsection*{Model 18: Phase 502}
%=================================================================

\begin{figure}[H]
\begin{center}
\resizebox{0.45\hsize}{!}{
\includegraphics[height=6cm]{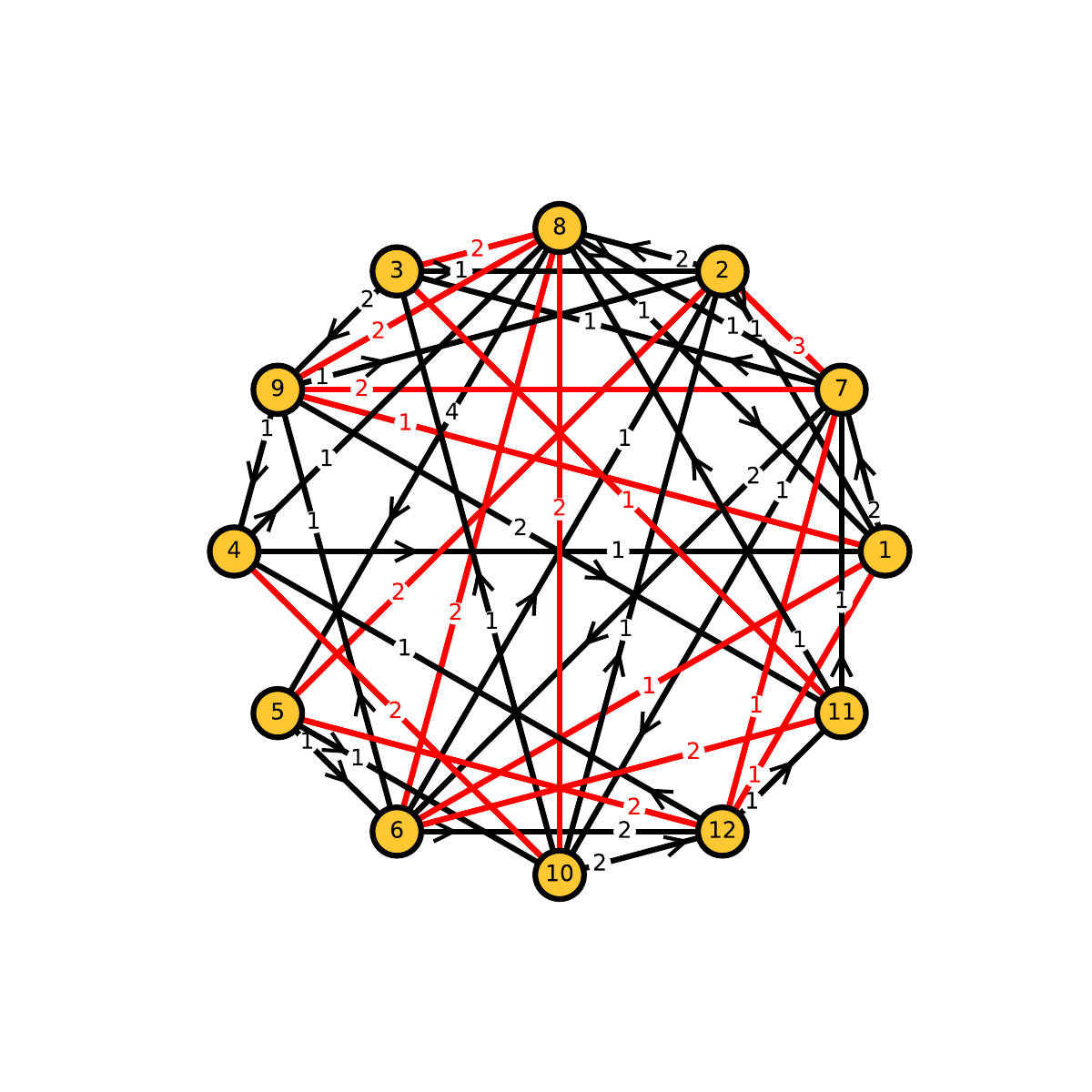} 
}
\caption{Quiver for Phase 502 of Model 18.
\label{model_18phase_502}}
 \end{center}
 \end{figure}

\fref{model_18phase_502} shows the quiver for this theory.

%=================================================================
%=================================================================

The $J$- and $E$-terms are
\beq
  {\footnotesize
\begin{array}{rrclrcl}
 &  & J &  & & E &  \\
\Lambda_{2.7}^{1}: & X_{7.10}A_{10.2} &-& X_{73}X_{32} & P_{28}X_{81}Q_{17}  &-& Q_{28}X_{81}P_{17} \\ 
\Lambda_{12.7}^{1}: & P_{76}P_{6.12} &-& Q_{76}Q_{6.12} & X_{12.11}A_{11.7} &-& X_{12.4}A_{48}A_{87} \\ 
\Lambda_{2.7}^{2}: & X_{7.10}X_{10.3}Q_{39}A_{92} &-& P_{76}X_{62} & A_{21}P_{17} &-& P_{28}A_{87} \\ 
\Lambda_{2.7}^{3}: & Q_{76}X_{62} &-& X_{7.10}X_{10.3}P_{39}A_{92} & A_{21}Q_{17} &-& Q_{28}A_{87} \\ 
\Lambda_{3.8}^{1}: & B_{85}X_{5.10}X_{10.3} &-& X_{81}Q_{17}X_{73} & P_{39}J_{94}A_{48} &-& X_{32}P_{28} \\ 
\Lambda_{3.8}^{2}: & A_{85}X_{5.10}X_{10.3} &-& X_{81}P_{17}X_{73} & X_{32}Q_{28} &-& Q_{39}J_{94}A_{48} \\ 
\Lambda_{3.11}^{1}: & A_{11.7}X_{73} &-& X_{11.8}A_{87}X_{7.10}X_{10.3} & P_{39}A_{9.11} &-& Q_{39}B_{9.11} \\ 
\Lambda_{1.9}^{1}: & J_{94}X_{41} &-& A_{92}A_{21} & P_{17}X_{7.10}X_{10.3}Q_{39} &-& Q_{17}X_{7.10}X_{10.3}P_{39} \\ 
\Lambda_{4.10}^{1}: & X_{10.3}Q_{39}J_{94} &-& B_{10.12}X_{12.4} & A_{48}A_{85}X_{5.10} &-& X_{41}P_{17}X_{7.10} \\ 
\Lambda_{4.10}^{2}: & X_{10.3}P_{39}J_{94} &-& A_{10.12}X_{12.4} & X_{41}Q_{17}X_{7.10}  &-& A_{48}B_{85}X_{5.10} 
\end{array} 
 } \nonumber
\label{E_J_Model 18_Phase 502_1}
 \eeq

\beq
  {\footnotesize
\begin{array}{rrclrcl}
 &  & J &  & & E &  \\
\Lambda_{10.8}^{1}: & X_{81}Q_{17}X_{7.10} &-& L_{85}X_{5.10} & A_{10.12}X_{12.11}X_{11.8} &-& A_{10.2}P_{28} \\ 
\Lambda_{10.8}^{2}: & X_{81}P_{17}X_{7.10} &-& M_{85}X_{5.10} & A_{10.2}Q_{28} &-& B_{10.12}X_{12.11}X_{11.8} \\ 
\Lambda_{9.8}^{1}: & A_{87}X_{7.10}X_{10.3}Q_{39} &-& L_{85}X_{56}A_{69} & A_{92}P_{28} &-& B_{9.11}X_{11.8} \\ 
\Lambda_{9.8}^{2}: & M_{85}X_{56}A_{69} &-& A_{87}X_{7.10}X_{10.3}P_{39} & A_{92}Q_{28} &-& A_{9.11}X_{11.8} \\ 
\Lambda_{5.2}^{1}: & Q_{28}M_{85} &-& P_{28}L_{85} & X_{5.10}A_{10.2} &-& X_{56}A_{69}A_{92} \\ 
\Lambda_{1.12}^{1}: & X_{12.11}X_{11.8}X_{81} &-& X_{12.4}X_{41} & P_{17}X_{7.10}B_{10.12} &-& Q_{17}X_{7.10}A_{10.12} \\ 
\Lambda_{12.5}^{1}: & X_{5.10}A_{10.12} &-& X_{56}P_{6.12} & X_{12.11}X_{11.8}L_{85} &-& X_{12.4}A_{48}B_{85} \\ 
\Lambda_{12.5}^{2}: & X_{56}Q_{6.12} &-& X_{5.10}B_{10.12} & X_{12.11}X_{11.8}M_{85} &-& X_{12.4}A_{48}A_{85} \\ 
\Lambda_{11.6}^{1}: & A_{69}B_{9.11} &-& P_{6.12}X_{12.11} & A_{11.7}P_{76} &-& X_{11.8}L_{85}X_{56} \\ 
\Lambda_{11.6}^{2}: & A_{69}A_{9.11} &-& Q_{6.12}X_{12.11} & X_{11.8}M_{85}X_{56} &-& A_{11.7}Q_{76} \\ 
\Lambda_{8.6}^{1}: & X_{62}P_{28} &-& P_{6.12}X_{12.4}A_{48} & B_{85}X_{56} &-& A_{87}P_{76} \\ 
\Lambda_{8.6}^{2}: & X_{62}Q_{28} &-& Q_{6.12}X_{12.4}A_{48} & A_{87}Q_{76} &-& A_{85}X_{56} \\ 
\Lambda_{2.5}^{1}: & X_{5.10}X_{10.3}X_{32} &-& X_{56}X_{62} & P_{28}B_{85} &-& Q_{28}A_{85} \\ 
\Lambda_{9.7}^{1}: & X_{73}Q_{39} &-& P_{76}A_{69} & B_{9.11}A_{11.7} &-& J_{94}A_{48}X_{81}P_{17} \\ 
\Lambda_{9.7}^{2}: & X_{73}P_{39} &-& Q_{76}A_{69} & J_{94}A_{48}X_{81}Q_{17} &-& A_{9.11}A_{11.7} \\ 
\Lambda_{6.1}^{1}: & P_{17}P_{76} &-& Q_{17}Q_{76} & X_{62}A_{21} &-& A_{69}J_{94}A_{48}X_{81} \\ 
 
\end{array} 
 }~.~
\label{E_J_Model 18_Phase 502}
 \eeq

%======================================================================
\bibliographystyle{JHEP}
\bibliography{mybib}
%======================================================================

\end{document}